\documentclass[a4paper,11pt]{article}
\pdfoutput=1 % if your are submitting a pdflatex (i.e. if you have
\usepackage{jheppub} % for details on the use of the package, please
\usepackage{graphicx}
\usepackage{hyperref}
\usepackage{pdfpages}
\usepackage[section]{placeins}
\usepackage{aligned-overset}
\usepackage{epstopdf}
\usepackage{float} % improved interface for floating objects
\usepackage[colorinlistoftodos]{todonotes}
\usepackage{epsfig}
\usepackage{graphicx,color}
\usepackage{cancel}
\usepackage[utf8]{inputenc}
\usepackage{dsfont}

\newcommand{\be}{\begin{equation}}
\newcommand{\ee}{\end{equation}}

\newcommand{\bea}{\begin{aligned}}
\newcommand{\eea}{\end{aligned}}

\newcommand{\bgamma}{{ \boldsymbol{\gamma}}}

\newcommand{\sT}{{\scriptscriptstyle T}}

\newcommand{\beq}{\begin{eqnarray}}
\newcommand{\eeq}{\end{eqnarray}}

\definecolor{red}{rgb}{1,0,0}
\definecolor{gray}{rgb}{0.5,0.5,0.5}

\usepackage{bbold}
\usepackage{tensor}
\usepackage[normalem]{ulem} % \sout{old text} for strikeout
\renewcommand\sout{\bgroup  \ULdepth=-.5ex \ULset}

\usepackage{pifont}% http://ctan.org/pkg/pifont

\title{Forward production of a Drell-Yan pair and a jet at small $x$ at next-to-leading order}
\author[]{Pieter Taels}

\affiliation[]{Universiteit Antwerpen, Departement fysica, Groenenborgerlaan
171, 2020 Antwerpen, Belgium}

\emailAdd{pieter.taels@uantwerpen.be}

\date{\today}

\preprint{}

\abstract{We perform the analytical next-to-leading order calculation of the process $p+A\to \gamma^*+\mathrm{jet}+X$, at forward rapidities and low $x$. These kinematics justify a hybrid approach, where a quark from the ‘projectile' proton scatters off the gluon distribution of the ‘target', which can be a nucleus or a highly boosted proton. By using the Color Glass Condensate effective theory approach, this gluon distribution is allowed to be so dense that the quark undergoes multiple scattering. Moreover, large high-energy logarithms in the ratio of the hard scale and the center-of-mass energy are resummed by the Balitsky, Kovchegov, Jalilian-Marian, Iancu, McLerran, Weigert, Leonidov, Kovner or BK-JIMWLK evolution equations. We demonstrate that all ultraviolet divergences encountered in the calculation cancel, while the high-energy divergences are absorbed into BK-JIMWLK. The remaining singularities are collinear in nature and can be either absorbed into the Dokshitzer-Gribov-Lipatov-Altarelli-Parisi evolution of the incoming quark, when they stem from initial-state radiation, or else can be treated by introduction of a jet function in case they are caused by final-state emissions. The resulting cross section is completely finite and expressed in function of only a small set of color operators.}

\begin{document}

\maketitle

\section{Introduction}
In order to apply Quantum Chromodynamics (QCD) to the analysis of collider experiments with hadrons, the standard approach is to rely on collinear factorization. A hard scale $\mu^{2}$ justifies a perturbative treatment of the underlying partonic hard scattering process, while the hadron structure is parameterized by parton distribution functions (PDFs) or fragmentation functions (FFs). Using the Dokshitzer-Gribov-Lipatov-Altarelli-Parisi (DGLAP) equations~\cite{Gribov:1972ri,Dokshitzer:1977sg,Altarelli:1977zs}, the scale dependence of the PDFs and FFs can be perturbatively calculated, resumming large collinear logarithms in the ratio $\mu^{2}/\mu^2_{0}$ of the hard scale and the hadronic one.

However, collinear factorization is known to break down whenever there is an additional large ratio of scales in the process under consideration. For example, it is implicitly assumed that the center-of-mass energy $\sqrt{s}$ is of the same order as the hard scale: $s\sim\mu^{2}$. At high energies or low $x\sim \mu^2/s$, this condition can be violated, and it becomes necessary to perform an additional resummation of large high-energy- or rapidity logarithms $\alpha_{s}\ln s/\mu^{2}\sim \alpha_{s}\ln 1/x\gg1$. A commonly used framework to do so is High-Energy Factorization (HEF)~\cite{Catani:1990xk,Catani:1990eg,Catani:1994sq}, in which the Balitsky-Fadin-Kuraev-Lipatov (BFKL) equations~\cite{Kuraev:1977fs,Balitsky:1978ic} are used to resum the high-energy logarithms on top of the collinear ones already resummed by DGLAP.

In this paper, we work within another framework applicable at small $x$, known as the Color Glass Condensate (CGC)~\cite{McLerran:1993ni,McLerran:1993ka,McLerran:1994vd,Jalilian-Marian:1997qno,Jalilian-Marian:1997jhx,Jalilian-Marian:1997ubg,Kovner:2000pt,Weigert:2000gi,Iancu:2000hn,Iancu:2001ad,Ferreiro:2001qy}. The CGC is an effective theory based upon the separation of fast and slow gluon fields in a highly boosted proton or nucleus (a shockwave). Only the former are regarded as quantum fields, while the latter are integrated out and treated as a semiclassical background field with a characteristic scale $Q_{s}$. This scale is known as the saturation scale, because it marks the onset of a regime in which nonlinear gluon recombinations counteract the exponential growth of the gluon density predicted by BFKL~\cite{Gribov:1983ivg}.  The renormalization group evolution associated with the CGC effective theory leads to the nonlinear Balitsky, Kovchegov, Jalilian-Marian, Iancu, McLerran, Weigert, Leonidov, Kovner or BK-JIMWLK evolution equations~\cite{Balitsky:1995ub, Balitsky:1998kc, Balitsky:1998ya, Kovchegov:1999yj, Jalilian-Marian:1997qno, Jalilian-Marian:1997jhx, Jalilian-Marian:1997ubg, Kovner:2000pt, Weigert:2000gi, Iancu:2000hn, Iancu:2001ad, Ferreiro:2001qy}. In a sense, JIMWLK can be regarded as the generalization to the nonlinear saturation regime of the linear BFKL evolution, to which it reduces in the hard scattering limit $k_\sT^2\gg Q_{s}^2$.

Although there have been many hints, conclusive experimental proof of saturation is still lacking, see for instance~\cite{Morreale:2021pnn} and references therein. After all, it is an effect hidden in a corner of phase space. Precise theoretical predictions are, therefore, essential, and since the last decade a lot of progress has been made to compute several relevant processes in the saturation regime at next-to-leading order (NLO) accuracy. Nowadays, the impact factors are known for single inclusive hadron production in proton-nucleus collisions~\cite{Chirilli:2011km,Chirilli:2012jd}, inclusive deep-inelastic scattering (DIS)~\cite{Balitsky:2010ze, Balitsky:2012bs, Beuf:2011xd, Beuf:2016wdz, Beuf:2017bpd, Hanninen:2017ddy}, DIS with massive quarks~\cite{Beuf:2021qqa, Beuf:2021srj,Beuf:2022ndu}, exclusive vector meson production in DIS~\cite{Boussarie:2016bkq,Mantysaari:2022bsp,Mantysaari:2021ryb,Mantysaari:2022kdm}, photon plus dijet production in DIS~\cite{Roy:2019hwr}, diffractive dijet or dihadron production in DIS~\cite{Boussarie:2016ogo,Fucilla:2022wcg}, inclusive dijet~\cite{Caucal:2021ent} or dihadron~\cite{Bergabo:2022tcu,Bergabo:2023wed} production in DIS, semi-inclusive DIS~\cite{Bergabo:2022zhe}, and inclusive dijet photoproduction~\cite{Taels:2022tza}. The next-to-leading logarithmic extension of the JIMWLK equations has been studied in refs.~\cite{Balitsky:2013fea,Kovner:2013ona,Kovner:2014lca,Lublinsky:2016meo}. Finally, results for multiparticle production at tree level were obtained in~\cite{Ayala:2016lhd,Iancu:2018hwa,Altinoluk:2018byz,Altinoluk:2020qet,Iancu:2020mos,Iancu:2022gpw}.

In this work, we contribute to the above-mentioned theory effort by presenting the next-to-leading order calculation of the process $p+A\to\gamma^{*}+\mathrm{jet}+X$ at forward rapidities and at high energies. The resulting cross section is easily promoted to the inclusive production of a Drell-Yan pair~\cite{Drell:1970wh} plus a jet by multiplying with the $\gamma^*\to\ell^-+\ell^+$ decay rate, provided the azimuthal angle between the two leptons is integrated out~\cite{Berger:2001wr}. At leading order in the CGC, the calculation was first performed in ref.~\cite{Gelis:2002fw} and the process further studied in, e.g., ref.~\cite{Stasto:2012ru,Basso:2015pba}. Within the HEF framework, low-$x$ inclusive Drell-Yan plus jet production has been studied in refs.~\cite{vanHameren:2015uia,Deak:2018obv}.

In the kinematics under consideration, a quark or gluon carrying a longitudinal momentum fraction $x_{p}$ from the ‘projectile' proton, probes the low-$x_{\scriptscriptstyle{A}}$ gluon distribution of the ‘target' proton or nucleus. This motivates us to work in an approximation known as the hybrid or dilute-dense factorization scheme \cite{Dumitru:2005gt,Elastic_vs_Inelastic}, in which the partonic structure of the projectile proton can be parameterized with collinear PDFs, while the CGC is applied to the gluon structure of the target.\footnote{In the recently appeared work~\cite{Altinoluk:2023hfz}, the hybrid approach has been generalized to include transverse-momentum dependence also on the projectile side.} The rationale behind this is that the typical transverse momenta of the target constituents are of the order of the saturation scale $Q_{s}^{2}$, while those of the projectile have approximately the size of the hadronic one $\mu_{0}^{2}$. Therefore, one expects the error corresponding to this approximation to be parametrically $\mu_{0}^{2}/Q_{s}^{2}$, which at very low $x$ can indeed become negligible (around $1\%$ for $x\sim10^{-5}$). In spirit, this hybrid scheme is in fact very similar to the dipole picture~\cite{Mueller:1989st,Nikolaev:1990ja} used in deep-inelastic scattering. In particular, it allows one to formulate the $p+A\to\gamma^{*}+\mathrm{jet}+X$ cross section as a convolution of the projectile quark PDF with the perturbative $q\to\gamma^{*}+q(+g)$ splitting inside the semiclassical color background field of the target, described by the CGC. In this work, we limit ourselves to the quark channel, and perform the calculation using light-cone perturbation theory (LCPT)~\cite{Kogut:1969xa,Bjorken:1970ah,Brodsky:1997de}. We refer to~\cite{Beuf:2016wdz} for a very clear presentation of LCPT in the context of the dipole picture and the CGC.

As in most higher-order calculations, we will encounter different classes of divergences. We will use the standard approach of dimensional regularization \cite{tHooft:1972tcz} to treat ultraviolet (UV) and infrared (IR) singularities. At the present perturbative order, the former only appear in the loop diagrams, and will be shown to cancel in the total virtual NLO contribution. Infrared or, in this case, collinear divergences, appear both in virtual diagrams (more specifically in the asymptotic quark self-energy corrections), and on the cross section level after integrating over the momentum of the radiated gluon. The poles stemming from collinear gluon radiation in the initial state will be shown to cancel with the DGLAP evolution of the incoming quark, while those in the final state will be regulated by a jet definition. Finally, we will regularize high-energy or rapidity divergences with a cutoff method, as is customary in CGC calculations, and demonstrate how they are absorbed in the JIMWLK evolution of the target.

The paper is organized as follows: we start with the derivation of the cross section at leading order in section~\ref{sec:LO}, which allows us to introduce the conventions of our light-cone perturbation theory approach to the CGC. In section~\ref{sec:virtual}, we list all the one-loop virtual diagrams and their corresponding amplitudes. The cancellation of the ultraviolet divergences which they contain is then treated in section~\ref{sec:UV}. In the next section,~\ref{sec:real}, we present the results of all the real radiative corrections to the cross section. The treatment of the corresponding collinear singularities in the initial and final state is discussed in sections~\ref{sec:DGLAP} and~\ref{sec:jet}, respectively. In section~\ref{sec:JIMWLK}, we show how the remaining rapidity singularities can be absorbed into the JIMWLK evolution of the leading-order cross section, after which we present the completely finite total cross section in section~\ref{sec:NLO}. We conclude with a discussion of the applications of our calculation and further research directions. Readers interested in the technicalities of deriving the one-loop amplitudes in our framework are referred to the appendix, section~\ref{sec:Example-calculation}, where the calculation of one virtual diagram is presented in full detail.

\section{\label{sec:LO}Leading-order cross section }

\begin{figure}[t]
\begin{centering}
\includegraphics[scale=0.3]{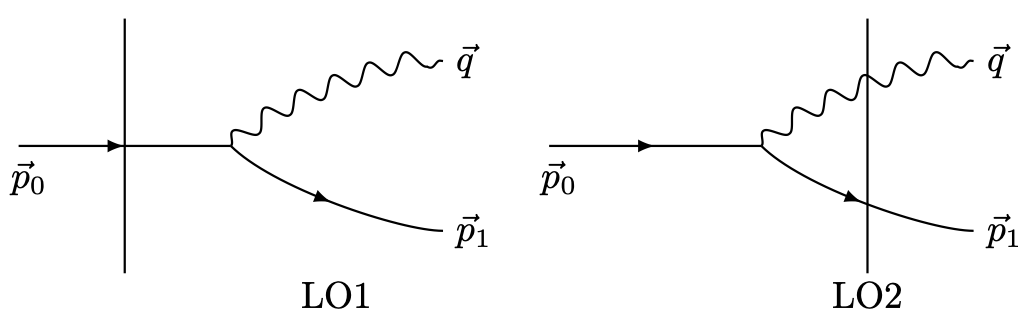}
\par\end{centering}
\centering{}\caption{\label{fig:LO}The two Feynman graphs for the LO partonic $q+A\to\gamma^{*}+q$
process. The highly boosted target $A$: a shockwave, is represented
by the full vertical line.}
\end{figure}

In this first section, we briefly review the leading-order (LO) calculation
of inclusive forward virtual photon plus jet production ($p+A\to\gamma^{*}+\mathrm{jet}+X$).
Along the way, we will specify the notations and conventions that
we use throughout this work. First and foremost, let us introduce
our convention for light-cone (LC) momenta, namely:
\begin{equation}
k^{+}=\frac{k_{0}+k_{3}}{\sqrt{2}}\quad\mathrm{and}\quad k^{-}=\frac{k_{0}-k_{3}}{\sqrt{2}}\;,
\end{equation}
where $3$ is the direction along the beam line. Transverse momenta
or coordinates are always written in boldface, and directly evaluated
in Euclidean space. The metric is, therefore, given by the product:
\begin{equation}
k\cdot p=k^{+}p^{-}+k^{-}p^{+}-\mathbf{k}\cdot\mathbf{p}\;.
\end{equation}
In light-cone perturbation theory, all particle momenta obey
the mass-shell condition: $k^{-}=(\mathbf{k}^{2}+m^{2})/2k^{+}$.
For this reason, we can use the compact notation $\vec{k}=(k^{+},\mathbf{k})$
for a generic momentum vector. 

Figure \ref{fig:LO} depicts the two Feynman diagrams corresponding
to the partonic $q+A\to\gamma^{*}+q$ process. All fermions are taken
to be massless, and we choose a frame in which the momentum of the
incoming quark is oriented along the positive light-cone direction:
\begin{equation}
\vec{p}_{0}=(p_{0}^{+},\mathbf{p}_{0}=0)\;.
\end{equation}
As is conventional, the gluon fields are described using the light-cone
gauge $A^{+}=0.$ Such a gauge condition cannot, however, be set for
the virtual photon field, which has a virtual mass $M$ and is described
by the Proca equation. Taking $\vec{q}=(q^{+},\mathbf{q})$ to be
the momentum of the virtual photon, we make the following choice for
its transverse ($\lambda=1,2$) and longitudinal polarization vectors
$\epsilon_{\lambda}^{\mu}$ resp. $\epsilon_{0}^{\mu}$:
\begin{equation}
\begin{aligned}\epsilon_{\lambda}^{\mu} & =\big(0,\frac{\mathbf{q}\cdot\boldsymbol{\epsilon}_{\lambda}}{q^{+}},\boldsymbol{\epsilon}_{\lambda}\big)\qquad\mathrm{and}\qquad\epsilon_{0}^{\mu}=\big(\frac{q^{+}}{M},\frac{\mathbf{q}^{2}-M^{2}}{2q^{+}M},\frac{\mathbf{q}}{M}\big)\;.\end{aligned}
\label{eq:vbosonpol}     
\end{equation}
Furthermore, in what follows we work with linear transverse polarization
vectors $\boldsymbol{\epsilon}_{\lambda}^{i}=\delta^{i\lambda}$.

In $D$ dimensions, the leading-order partonic cross section is given
by:
\begin{equation}
\begin{aligned}\mathrm{d}\hat{\sigma}_{\mathrm{LO}} & \!=\!\frac{1}{2p_{0}^{+}}\frac{\mathrm{d}p_{1}^{+}\mathrm{\mathrm{d}}^{D\!-\!2}\mathbf{p}_{1}\,\theta(p_{1}^{+})}{(2\pi)^{D\!-\!1}2p_{1}^{+}}\frac{\mathrm{d}q^{+}\mathrm{\mathrm{d}}^{D\!-\!2}\mathbf{q}\,\theta(q^{+})}{(2\pi)^{D\!-\!1}2q^{+}}2\pi\delta(p_{0}^{+}\!-\!p_{1}^{+}\!-\!q^{+})\frac{1}{D\!-\!2}\big|\mathcal{M}_{\mathrm{LO1\!+\!2}}\big|^{2}\;,\end{aligned}
\label{eq:crosssectiondef}
\end{equation}
where $\vec{p}_{1}=(p_{1}^{+},\mathbf{p}_{1})$ is the momentum of
the outgoing quark, and where $\mathcal{M}_{\mathrm{LO1+2}}$ is the sum of the 
two amplitudes associated with the diagrams in figure \ref{fig:LO}.
From eq. (\ref{eq:crosssectiondef}), the $pp$- or $pA$ cross section
is obtained after convolving with the quark PDF and averaging over
the semiclassical gluon fields in the target (an operation we denote
with $\langle...\rangle$):
\begin{equation}
\begin{aligned}\mathrm{d}\sigma_{\mathrm{LO}} & =\int\frac{\mathrm{d}p_{0}^{+}}{p_{0}^{+}}\frac{p_{0}^{+}}{p_{p}^{+}}f_{q}\big(\frac{p_{0}^{+}}{p_{p}^{+}},\mu^{2}\big)\langle\mathrm{d}\hat{\sigma}_{\mathrm{LO}}\rangle\;,\\
 & =x_{p}f_{q}(x_{p},\mu^{2})\frac{2\pi}{2(p_{0}^{+})^{2}}\frac{\mathrm{d}p_{1}^{+}\mathrm{\mathrm{d}}^{D-2}\mathbf{p}_{1}\,\theta(p_{1}^{+})}{(2\pi)^{D-1}2p_{1}^{+}}\frac{\mathrm{d}q^{+}\mathrm{\mathrm{d}}^{D-2}\mathbf{q}\,\theta(q^{+})}{(2\pi)^{D-1}2q^{+}}\\
 &\times\frac{1}{D-2}\big\langle\big|\mathcal{M}_{\mathrm{LO1}}+\mathcal{M}_{\mathrm{LO2}}\big|^{2}\big\rangle\Big|_{p^+_0=p^+_1+q^+}\;.
\end{aligned}
\label{eq:parton2hadron}
\end{equation}
In the above equation, $x_{p}=p_{0}^{+}/p_{p}^{+}$ is the plus-momentum
fraction of the quark with respect to its parent (projectile) proton. 

As already mentioned, we work in the dipole picture, formulated in
LCPT. In this approach, the projectile dynamics take place on a much
shorter timescale than those of the target. Therefore, on the partonic
level the projectile can be described by a Fock state with perturbatively
calculable dynamics, interacting with a static external potential
provided by the target. The amplitudes $\mathcal{M}$ are defined
as follows:
\begin{equation}
\begin{aligned}\,_{f}\big\langle\mathbf{q}(\vec{p}_{1})\boldsymbol{\gamma}^{*}(\vec{q})\big|\hat{F}-1\big|\mathbf{q}(\vec{p}_{0})\big\rangle_{i} & =2\pi\delta(p_{0}^{+}-p_{1}^{+}-q^{+})\mathcal{M}\;,\end{aligned}
\label{eq:Mdef-1}
\end{equation}
where the external potential $\hat{F}$ ($-1$ for reasons of unitarity)
is evaluated between the Fock states of the incoming ($i$) quark,
and outgoing ($f$) quark together with the virtual photon. The first step is
to calculate the perturbative evolution of the asymptotic Fock states
to and from the position of the shockwave at $x^{+}=0$. Indeed, the
two interacting or \textquoteleft dressed' Fock states $|...\rangle_{i,f}$
are related to eigenstates $|...\rangle$ of the free Hamiltonian
through the time evolution:
\begin{equation}
\begin{aligned}\big|\mathbf{q}(\vec{p}_{0})\big\rangle_{i} & =\hat{\mathcal{U}}(-\infty,0)\big|\mathbf{q}(\vec{p}_{0})\big\rangle\;,\\
\,_{f}\big\langle\mathbf{q}(\vec{p}_{1})\boldsymbol{\gamma}^{*}(\vec{q})\big| & =\big\langle\mathbf{q}(\vec{p}_{1})\boldsymbol{\gamma}^{*}(\vec{q})\big|\hat{\mathcal{U}}(0,+\infty)\;,
\end{aligned}
\end{equation}
where $\hat{\mathcal{U}}$ is the LC-time evolution operator defined
as:
\begin{equation}
\begin{aligned}\hat{\mathcal{U}}(a,b) & \equiv\hat{T}\exp\Bigl(-i\int_{a}^{b}\mathrm{d}x^{+}\hat{\mathcal{H}}(x^{+})\Bigr)\;,\end{aligned}
\label{eq:timeevo}
\end{equation}
with $\hat{T}$ the (LC)-time ordering operator and $\hat{\mathcal{H}}(x^{+})$
the LC Hamiltonian. In the interaction picture, the time-evolution
of the latter is given by:
\begin{equation}
\begin{aligned}\hat{\mathcal{H}}(x^{+}) & =e^{i\hat{H}_{0}x^{+}}\hat{V}e^{-i\hat{H}_{0}x^{+}}\;,\end{aligned}
\end{equation}
with $\hat{H}_{0}$ the free Hamiltonian, and $\hat{V}$ the collection
of interaction terms. Acting with (\ref{eq:timeevo}) on the Fock
state $\big|\mathbf{q}(\vec{p}_{0})\big\rangle$, which is per definition
an eigenstate of the free Hamiltonian with $\hat{H}_{0}\big|\mathbf{q}(\vec{p}_{0})\big\rangle=p_{0}^{-}\big|\mathbf{q}(\vec{p}_{0})\big\rangle$,
we obtain:
\begin{equation}
\begin{aligned}\big|\mathbf{q}(\vec{p}_{0})\big\rangle_{i} & =\hat{\mathcal{U}}(-\infty,0)\big|\mathbf{q}(\vec{p}_{0})\big\rangle\\
 & =\big|\mathbf{q}(\vec{p}_{0})\big\rangle-i\int\mathrm{PS}(\vec{\ell},\vec{k})\int_{-\infty}^{0}\mathrm{d}x^{+}\big|\mathbf{q}(\vec{\ell})\boldsymbol{\gamma}^{*}(\vec{k})\big\rangle\\
 & \times\big\langle\mathbf{q}(\vec{\ell})\boldsymbol{\gamma}^{*}(\vec{k})\big|e^{i(\ell^{-}+k^{-})x^{+}}\hat{V}e^{-ip_{0}^{-}x^{+}}\big|\mathbf{q}(\vec{p}_{0})\big\rangle\\
 & =\big|\mathbf{q}(\vec{p}_{0})\big\rangle+\int\mathrm{PS}(\vec{\ell},\vec{k})\frac{\big\langle\mathbf{q}(\vec{\ell})\boldsymbol{\gamma}^{*}(\vec{k})\big|\hat{V}\big|\mathbf{q}(\vec{p}_{0})\big\rangle}{p_{0}^{-}-\ell^{-}-k^{-}+i0^{+}}\big|\mathbf{q}(\vec{\ell})\boldsymbol{\gamma}^{*}(\vec{k})\big\rangle\;,
\end{aligned}
\label{eq:forwardsevo-1}
\end{equation}
up to the perturbative order we are interested in, and where we have
introduced the notation:
\begin{equation}
\int\mathrm{PS}(\vec{\ell})=\int\frac{\mathrm{d}^{D-1}\vec{\ell}\,\theta(\ell^{+})}{(2\pi)^{D-1}2\ell^{+}}\;.\label{eq:phasespace-1}
\end{equation}
Likewise, at leading order:
\begin{equation}
\begin{aligned} & \big\langle\mathbf{q}(\vec{p}_{1})\boldsymbol{\gamma}_{\lambda}^{*}(\vec{q})\big|\hat{\mathcal{U}}(0,+\infty)=\big\langle\mathbf{q}(\vec{p}_{1})\boldsymbol{\gamma}_{\lambda}^{*}(\vec{q})\big|\\
 & +\int\mathrm{PS}(\vec{\ell})\big\langle\mathbf{q}(\vec{p}_{1})\boldsymbol{\gamma}_{\lambda}^{*}(\vec{q})\big|\exp\Bigl(-i\int_{0}^{+\infty}\mathrm{d}x^{+}\hat{\mathcal{H}}(x^{+})\Bigr)\big|\mathbf{q}(\vec{\ell})\big\rangle\big\langle\mathbf{q}(\vec{\ell})\big|\;,\\
 & =\big\langle\mathbf{q}(\vec{p}_{1})\boldsymbol{\gamma}_{\lambda}^{*}(\vec{q})\big|+\int\mathrm{PS}(\vec{\ell})\frac{\big\langle\mathbf{q}(\vec{p}_{1})\boldsymbol{\gamma}_{\lambda}^{*}(\vec{q})\big|\hat{V}\big|\mathbf{q}(\vec{\ell})\big\rangle}{p_{1}^{+}+q^{-}-\ell_{0}^{-}+i0^{+}}\big\langle\mathbf{q}(\vec{\ell})\big|\;.
\end{aligned}
\label{eq:backwardsevo-1}
\end{equation}
 Combining (\ref{eq:Mdef-1}), (\ref{eq:forwardsevo-1}), and (\ref{eq:backwardsevo-1}),
we obtain:
\begin{equation}
\begin{aligned} & \,_{f}\big\langle\mathbf{q}(\vec{p}_{1})\boldsymbol{\gamma}_{\lambda}^{*}(\vec{q})\big|\hat{F}-1\big|\mathbf{q}(\vec{p}_{0})\big\rangle_{i}=\big\langle\mathbf{q}(\vec{p}_{1})\boldsymbol{\gamma}_{\lambda}^{*}(\vec{q})\big|\hat{F}-1\big|\mathbf{q}(\vec{p}_{0})\big\rangle\\
 & +\int\mathrm{PS}(\vec{\ell})\frac{\big\langle\mathbf{q}(\vec{p}_{1})\boldsymbol{\gamma}_{\lambda}^{*}(\vec{q})\big|\hat{V}\big|\mathbf{q}(\vec{\ell})\big\rangle}{p_{1}^{-}+q^{-}-\ell^{-}+i0^{+}}\big\langle\mathbf{q}(\vec{\ell})\big|\hat{F}-1\big|\mathbf{q}(\vec{p}_{0})\big\rangle\\
 & +\int\mathrm{PS}(\vec{\ell},\vec{k})\frac{\big\langle\mathbf{q}(\vec{\ell})\boldsymbol{\gamma}_{\lambda}^{*}(\vec{k})\big|\hat{V}\big|\mathbf{q}(\vec{p}_{0})\big\rangle}{p_{0}^{-}-\ell^{-}-k^{-}+i0^{+}}\big\langle\mathbf{q}(\vec{p}_{1})\boldsymbol{\gamma}_{\lambda}^{*}(\vec{q})\big|\hat{F}-1\big|\mathbf{q}(\vec{\ell})\boldsymbol{\gamma}^{*}(\vec{k})\big\rangle\;.
\end{aligned}
\label{eq:Fockint-1}
\end{equation}
In the eikonal approximation we are using, the potential $\hat{F}$
can never change the particle content of the Fock states, which are
orthogonal. The first line in the above equation, therefore, disappears.
The second and third lines correspond to the first and second diagram
in figure \ref{fig:LO}, respectively. In the former, the incoming
quark scatters off the external potential after which it emits a virtual
photon. In the latter, the photon is emitted before the interaction
with the potential. We will now evaluate eq. (\ref{eq:Fockint-1})
further, starting with the perturbative interaction terms. The relevant
piece of the interaction part of the QED LC Hamiltonian is, with the
short-hand notations $\int\mathrm{d}^{D-3}\vec{x}=\mu^{4-D}\int\mathrm{d}x^{-}\mathrm{d}^{D-2}\mathbf{x}$
and $\vec{x}=(x^{-},\mathbf{x})$:
\begin{equation}
\begin{aligned}\hat{V} & =\int\mathrm{d}^{D-3}\vec{x}:g_{\mathrm{em}}\bar{\psi}(\vec{x})\cancel{A}(\vec{x})\psi(\vec{x}):\;,\end{aligned}
\end{equation}
where $:\;:$ denotes normal ordering. Using the definitions of the
fermion- and boson fields in terms of the creation- and annihilation
operators:
\begin{equation}
\begin{aligned}\psi(\vec{x}) & =\int\mathrm{PS}(\vec{k})\big(e^{-i\vec{k}\cdot\vec{x}}u(\vec{k})b_{\vec{k}}+e^{i\vec{k}\cdot\vec{x}}v(\vec{k})d_{\vec{k}}^{\dagger}\big)\;,\\
\cancel{A}(\vec{x}) & =\int\mathrm{PS}(\vec{k})\big(e^{i\vec{k}\cdot\vec{x}}\cancel{\epsilon}^{*}(\vec{k})a_{\vec{k}}^{\dagger}+e^{-i\vec{k}\cdot\vec{x}}\cancel{\epsilon}(\vec{k})a_{\vec{k}}\big)\;,
\end{aligned}
\end{equation}
it is then a straightforward exercise to show that:
\begin{equation}
\begin{aligned}\big\langle\mathbf{q}(\vec{\ell})\boldsymbol{\gamma}_{\lambda}^{*}(\vec{k})\big|\hat{V}\big|\mathbf{q}(\vec{p})\big\rangle & =\big\langle0\big|b_{\vec{\ell}}\,a_{\vec{k}}^{\lambda}\,\hat{V}\,b_{\vec{p}}^{\dagger}\big|0\big\rangle\;,\\
 & =(2\pi)^{D-1}\delta^{(D-1)}\big(\vec{p}-\vec{\ell}-\vec{k}\big)g_{\mathrm{em}}\bar{u}(\vec{\ell})\cancel{\epsilon}_{\lambda}^{*}(\vec{k})u(\vec{p})\;.
\end{aligned}
\label{eq:Vaction1-1}
\end{equation}
We note that the Fock states are normalized the usual way with the
help of the (anti)-commutation relations:
\begin{equation}
\begin{aligned}\{b_{\vec{k}},b_{\vec{p}}^{\dagger}\} & =\{d_{\vec{k}},d_{\vec{p}}^{\dagger}\}=(2k^{+})(2\pi)^{D-1}\delta^{(D-1)}(\vec{k}-\vec{p})\;,\\{}
[a_{\vec{k}},a_{\vec{p}}^{\dagger}] & =(2k^{+})(2\pi)^{D-1}\delta^{(D-1)}(\vec{k}-\vec{p})\;,
\end{aligned}
\label{eq:anticommutation}
\end{equation}
such that, demanding that for empty states $\langle0|0\rangle=1$,
it follows that, e.g.:
\begin{equation}
\begin{aligned}\big\langle\mathbf{q}(\vec{\ell}_{2})\boldsymbol{\gamma}^{*}(\vec{k}_{2})\big|\mathbf{q}(\vec{\ell}_{1})\boldsymbol{\gamma}^{*}(\vec{k}_{1})\big\rangle & \!=\!2k_{1}^{+}(2\pi)^{D\!-\!1}\delta^{(D\!-\!1)}(\vec{k}_{1}\!-\!\vec{k}_{2})2\ell_{1}^{+}(2\pi)^{D\!-\!1}\delta^{(D\!-\!1)}(\vec{\ell}_{1}\!-\!\vec{\ell}_{2})\,.\end{aligned}
\label{eq:trivialinteraction-1}
\end{equation}
In order not to clutter the formulas, we will always suppress spinor
indices, fundamental color indices, and fermion helicity indices. 

We now proceed by simplifying the Dirac algebra in (\ref{eq:Fockint-1}).
Keeping the anticommutation relation $\{\gamma^{\mu},\gamma^{\nu}\}=2g^{\mu\nu}$
in mind, we introduce the projectors $\mathcal{P}_{G}=\gamma^{-}\gamma^{+}/2$
and $\mathcal{P}_{B}=\gamma^{+}\gamma^{-}/2$ which allow us to define
\textquoteleft good' and \textquoteleft bad' spinor components:
\begin{equation}
\begin{aligned}u(\vec{k}) & =\big(\mathcal{P}_{G}+\mathcal{P}_{B}\big)u(\vec{k})\equiv u_{G}(\vec{k})+u_{B}(\vec{k})\;,\\
\bar{u}(\vec{k}) & =\vec{u}(\vec{k})\big(\mathcal{P}_{G}+\mathcal{P}_{B}\big)\equiv\bar{u}_{B}(\vec{k})+\bar{u}_{G}(\vec{k})\;.
\end{aligned}
\end{equation}
The same definitions hold for the antifermion spinors $v(\vec{k})$.
The Dirac equation $\cancel{k}u(\vec{k})=0$ then introduces a dependence
between both components:
\begin{equation}
\begin{aligned}u_{B}(\vec{k}) & =\frac{\gamma^{+}}{2k^{+}}\mathbf{k}\cdot\boldsymbol{\gamma}\,u_{G}(k^{+})\;,\end{aligned}
\label{eq:bad2good}
\end{equation}
with exactly the same equation holding for the antifermion spinors,
at least in the massless case. Note that the good spinors only depend
on the plus-momentum component.

Relation (\ref{eq:bad2good}) allows us to rewrite a generic spinor
product $\bar{u}(\vec{k}_{1})\cancel{\epsilon}_{\lambda}(\vec{k}_{3})u(\vec{k}_{2})$
as:
\begin{equation}
\begin{aligned} & \bar{u}(\vec{k}_{1})\cancel{\epsilon}_{\lambda}(\vec{k}_{3})u(\vec{k}_{2}) \\ & =\bar{u}_{G}(k_{1}^{+})\Big(1\!+\!\frac{\mathbf{k}_{1}\!\cdot\!\boldsymbol{\gamma}\,\gamma^{+}}{2k_{1}^{+}}\Big)\Big(\gamma^{+}\epsilon_{\lambda}^{-}(\vec{k}_{3})\!+\!\gamma^{-}\epsilon_{\lambda}^{+}(\vec{k}_{3})\!-\!\boldsymbol{\gamma}\!\cdot\!\boldsymbol{\epsilon}_{\lambda}(\vec{k}_{3})\Big)\Big(\frac{\gamma^{+}\,\mathbf{k}_{2}\!\cdot\!\boldsymbol{\gamma}}{2k_{2}^{+}}\!+\!1\Big)\,u_{G}(k_{2}^{+})\,.\end{aligned}
\label{eq:spinor_0}
\end{equation}
Applying the above identity to the spinor product in (\ref{eq:Vaction1-1}),
using the polarization vectors (\ref{eq:vbosonpol}), we find in the
transverse case:
\begin{equation}
\begin{aligned}\big\langle\mathbf{q}(\vec{\ell})\boldsymbol{\gamma}_{\lambda}^{*}(\vec{k})\big|\hat{V}\big|\mathbf{q}(\vec{p})\big\rangle & =g_{\mathrm{em}}(2\pi)^{D-1}\delta^{(D-1)}\big(\vec{p}-\vec{\ell}-\vec{k}\big)\\
 & \times\bar{u}_{G}(\ell^{+})\gamma^{+}\Big(\frac{\mathbf{k}^{\lambda}}{k^{+}}+\frac{\boldsymbol{\ell}^{\bar{\lambda}}}{2\ell^{+}}\boldsymbol{\gamma}^{\bar{\lambda}}\boldsymbol{\gamma}^{\lambda}+\frac{\mathbf{p}^{\bar{\lambda}}}{2p^{+}}\boldsymbol{\gamma}^{\lambda}\boldsymbol{\gamma}^{\bar{\lambda}}\Big)\,u_{G}(p^{+})\;,
\end{aligned}
\end{equation}
since from $\{\gamma^{\mu},\gamma^{\nu}\}=2g^{\mu\nu}$ it follows
that $\gamma^{+}\gamma^{+}=0$ and $\{\gamma^{\pm},\boldsymbol{\gamma}^{i}\}=0$.
For this same reason, it is easy to see that $\bar{u}_{G}\boldsymbol{\gamma}^{i}u_{G}=0$.
Finally, for the momentum configurations in (\ref{eq:Fockint-1}),
we end up with:
\begin{equation}
\begin{aligned}\big\langle\mathbf{q}(\vec{p}_{1})\boldsymbol{\gamma}_{\lambda}^{*}(\vec{q})\big|\hat{V}\big|\mathbf{q}(\vec{\ell})\big\rangle & =g_{\mathrm{em}}(2\pi)^{D-1}\delta^{(D-1)}\big(\vec{p}_{1}+\vec{q}-\vec{\ell}\big)\\
 & \times\frac{p_{1}^{+}\mathbf{q}^{\bar{\lambda}}-q^{+}\mathbf{p}_{1}^{\bar{\lambda}}}{2p_{1}^{+}(p_{1}^{+}+q^{+})}\bar{u}_{G}(p_{1}^{+})\gamma^{+}\Big[\Big(1+\frac{2p_{1}^{+}}{q^{+}}\Big)\delta^{\lambda\bar{\lambda}}-i\sigma^{\lambda\bar{\lambda}}\Big]u_{G}(\ell^{+})\;,
\end{aligned}
\label{eq:numLO1}
\end{equation}
and:
\begin{equation}
\begin{aligned}\big\langle\mathbf{q}(\vec{\ell})\boldsymbol{\gamma}_{\lambda}^{*}(\vec{k})\big|\hat{V}\big|\mathbf{q}(\vec{p})\big\rangle & =g_{\mathrm{em}}(2\pi)^{D-1}\delta^{(D-1)}\big(\vec{p}-\vec{\ell}-\vec{k}\big)\\
 & \times\frac{-\boldsymbol{\ell}^{\bar{\lambda}}}{2\ell^{+}}\bar{u}_{G}(\ell^{+})\gamma^{+}\Big[\Big(1+\frac{2\ell^{+}}{k^{+}}\Big)\delta^{\lambda\bar{\lambda}}-i\sigma^{\lambda\bar{\lambda}}\Big]u_{G}(p^{+})\;,
\end{aligned}
\label{eq:numLO2}
\end{equation}
In the above formulas, we replaced the products of two transverse
gamma matrices with:
\begin{equation}
\boldsymbol{\gamma}^{i}\boldsymbol{\gamma}^{j}=-\delta^{ij}-i\sigma^{ij}\;,
\end{equation}
where the Dirac sigma defined as $\sigma^{ij}=(i/2)[\gamma^{i},\gamma^{j}]$.

Let us now turn to the interaction of the projectile with the target.
This interaction is, as explained above, described by an external
potential. Moreover, the Color Glass Condensate approach asserts that,
in the eikonal approximation, this potential is built from Wilson
lines along the projectile direction $x^{+}$: 
\begin{equation}
\begin{aligned}U_{\mathbf{x}} & \equiv\mathcal{P}\exp\Big(-ig_{s}\int_{-\infty}^{+\infty}\mathrm{d}x^{+}A_{a}^{-}(x^{+},\mathbf{x})t^{a}\Big)\;,\\
W_{\mathbf{x}} & \equiv\mathcal{P}\exp\Big(-ig_{s}\int_{-\infty}^{+\infty}\mathrm{d}x^{+}A_{a}^{-}(x^{+},\mathbf{x})T^{a}\Big)\;,
\end{aligned}
\label{eq:WLdef}
\end{equation}
where $\mathcal{P}$ is the path-ordering operator, $g_{s}$ is the
strong coupling, $A_{a}^{-}$ the minus-component of the gluon field,
and where $t^{a}$ and $T^{a}$ are the generators of $\mathrm{SU}(N_{c})$
in the fundamental and adjoint representation, respectively. The precise
form of the potential depends on the Fock state. In transverse coordinate
space, each (anti)quark and gluon in the projectile adds a Wilson
line to the potential that inherits the quark or gluon transverse
coordinate and color representation. In the present case, the projectile-target
interactions in (\ref{eq:Fockint-1}) are given by:
\begin{equation}
\begin{aligned}\big\langle\mathbf{q}(\vec{\ell})\big|\hat{F}-1\big|\mathbf{q}(\vec{p}_{0})\big\rangle & =2p_{0}^{+}2\pi\delta(p_{0}^{+}-\ell^{+})\int_{\mathbf{x}}e^{-i\mathbf{x}\cdot\boldsymbol{\ell}}\big(U_{\mathbf{x}}-1\big)\;,\\
\big\langle\mathbf{q}(\vec{p}_{1})\boldsymbol{\gamma}^{*}(\vec{q})\big|\hat{F}\!-\!1\big|\mathbf{q}(\vec{\ell})\boldsymbol{\gamma}^{*}(\vec{k})\big\rangle & =2p_{1}^{+}2\pi\delta(p_{1}^{+}-\ell^{+})2q^{+}2\pi\delta(q^{+}-k^{+}) \\ & \times \int_{\mathbf{x}}e^{-i\mathbf{x}\cdot(\mathbf{p}_{1}-\mathbf{k})}\big(U_{\mathbf{x}}-1\big)\;,
\end{aligned}
\label{eq:LOpotential}
\end{equation}
where we introduced the short-hand notation $\int_{\mathbf{\mathbf{x}}}=\int\mathrm{d}^{D-2}\mathbf{x}$
for the integration over $D-2$ transverse coordinate space. Moreover,
in the present eikonal approximation, the interaction preserves plus-momentum
and the helicity. Combining (\ref{eq:Mdef-1}), (\ref{eq:Fockint-1}),
(\ref{eq:numLO1}), (\ref{eq:numLO2}), and (\ref{eq:LOpotential}),
we can finally write down the expressions for the leading-order amplitudes:
\begin{equation}
\begin{aligned}\mathcal{M}_{\mathrm{LO1}}^{\lambda} & =\frac{-g_{\mathrm{em}}q^{+}\mathbf{P}_{\perp}^{\bar{\lambda}}}{p_{0}^{+}\mathbf{P}_{\perp}^{2}+p_{1}^{+}M^{2}}\bar{u}_{G}(p_{1}^{+})\gamma^{+}\mathrm{Dirac}^{\lambda\bar{\lambda}}\big(1+\frac{2p_{1}^{+}}{q^{+}}\big)u_{G}(p_{0}^{+})\int_{\mathbf{x}}e^{-i\mathbf{k}_{\perp}\cdot\mathbf{x}}\big(U_{\mathbf{x}}-1\big)\;,\\
\mathcal{M}_{\mathrm{LO2}}^{\lambda} & =\frac{-g_{\mathrm{em}}q^{+}\mathbf{q}^{\bar{\lambda}}}{p_{0}^{+}\mathbf{q}^{2}+p_{1}^{+}M^{2}}\bar{u}_{G}(p_{1}^{+})\gamma^{+}\mathrm{Dirac}^{\lambda\bar{\lambda}}\big(1+\frac{2p_{1}^{+}}{q^{+}}\big)u_{G}(p_{0}^{+})\int_{\mathbf{x}}e^{-i\mathbf{k}_{\perp}\cdot\mathbf{x}}\big(U_{\mathbf{x}}-1\big)\;,
\end{aligned}
\label{eq:MLOT}
\end{equation}
where we have introduced the following momentum combinations, which will be
used throughout this work:
\begin{equation}
\begin{aligned}\mathbf{P}_{\perp} & \equiv\frac{q^{+}\mathbf{p}_{1}-p_{1}^{+}\mathbf{q}}{p_{0}^{+}}\qquad\mathrm{and}\qquad\mathbf{k}_{\perp}\equiv\mathbf{p}_{1}+\mathbf{q}\;,\end{aligned}
\label{eq:defPkperp}
\end{equation}
as well as the definition:
\begin{equation}
\begin{aligned}\mathrm{Dirac}^{\lambda\bar{\lambda}}(\xi) & \equiv\xi\delta^{\lambda\bar{\lambda}}\mathds{1}_{4}-i\sigma^{\lambda\bar{\lambda}}\;.\end{aligned}
\label{eq:Diracdef-1}
\end{equation}
A similar calculation for the longitudinally polarized virtual photon
gives:
\begin{equation}
\begin{aligned}\mathcal{M}_{\mathrm{LO1}}^{0} & =\frac{g_{\mathrm{em}}}{M}\frac{p_{0}^{+}\mathbf{P}_{\perp}^{2}-p_{1}^{+}M^{2}}{p_{0}^{+}\mathbf{P}_{\perp}^{2}+p_{1}^{+}M^{2}}\bar{u}_{G}(p_{1}^{+})\gamma^{+}u_{G}(p_{0}^{+})\int_{\mathbf{x}}e^{-i\mathbf{k}_{\perp}\cdot\mathbf{x}}\big(U_{\mathbf{x}}-1\big)\;,\\
\mathcal{M}_{\mathrm{LO2}}^{0} & =-\frac{g_{\mathrm{em}}}{M}\frac{p_{0}^{+}\mathbf{q}^{2}-p_{1}^{+}M^{2}}{p_{0}^{+}\mathbf{q}^{2}+p_{1}^{+}M^{2}}\bar{u}_{G}(p_{1}^{+})\gamma^{+}u_{G}(p_{0}^{+})\int_{\mathbf{x}}e^{-i\mathbf{k}_{\perp}\cdot\mathbf{x}}\big(U_{\mathbf{x}}-1\big)\;.
\end{aligned}
\label{eq:MLOL}
\end{equation}
Multiplying the amplitudes (\ref{eq:MLOT}) and (\ref{eq:MLOL}) with
their complex conjugate, and summing over the polarization indices
$\lambda$ of the transversely polarized virtual photon or vector
boson, we arrive at:
\begin{equation}
\begin{aligned}\sum_{\lambda}\big|\mathcal{M}_{\mathrm{LO1}}^{\lambda}+\mathcal{M}_{\mathrm{LO2}}^{\lambda}\big|^{2} & =g_{\mathrm{em}}^{2}N_{c}\Big(\frac{q^{+}\mathbf{q}}{p_{0}^{+}\mathbf{q}^{2}+p_{1}^{+}M^{2}}+\frac{q^{+}\mathbf{P}_{\perp}}{p_{0}^{+}\mathbf{P}_{\perp}^{2}+p_{1}^{+}M^{2}}\Big)^{2}|\mathrm{Dirac}_{\mathrm{LO}}^{\mathrm{T}}|^{2}\\
 & \times\int_{\mathbf{x},\mathbf{x}^{\prime}}e^{-i\mathbf{k}_{\perp}\cdot(\mathbf{x}-\mathbf{x}^{\prime})}\big(s_{\mathbf{x}\mathbf{x}^{\prime}}+1\big)\;,
\end{aligned}
\label{eq:LOTsquared}
\end{equation}
and:
\begin{equation}
\begin{aligned}|\mathcal{M}_{\mathrm{LO1}}^{0}+\mathcal{M}_{\mathrm{LO2}}^{0}|^{2} & =g_{\mathrm{em}}^{2}\frac{N_{c}}{M^{2}}\Big(\frac{p_{0}^{+}\mathbf{P}_{\perp}^{2}-p_{1}^{+}M^{2}}{p_{0}^{+}\mathbf{P}_{\perp}^{2}+p_{1}^{+}M^{2}}-\frac{p_{0}^{+}\mathbf{q}^{2}-p_{1}^{+}M^{2}}{p_{0}^{+}\mathbf{q}^{2}+p_{1}^{+}M^{2}}\Big)^{2}|\mathrm{Dirac}_{\mathrm{LO}}^{0}|^{2}\\
 & \times\int_{\mathbf{x},\mathbf{x}^{\prime}}e^{-i\mathbf{k}_{\perp}\cdot(\mathbf{x}-\mathbf{x}^{\prime})}\big(s_{\mathbf{x}\mathbf{x}^{\prime}}+1\big)\;.
\end{aligned}
\label{eq:LOLsquared}
\end{equation}
In the above results, we used $s_{\mathbf{x}\mathbf{x}^{\prime}}$
to denote the dipole color operator:
\begin{equation}
s_{\mathbf{x}\mathbf{x}^{\prime}}=\frac{1}{N_{c}}\mathrm{Tr}\big(U_{\mathbf{x}}U_{\mathbf{x}^{\prime}}^{\dagger}\big)\;,\label{eq:dipoledef}
\end{equation}
where we take the trace over the fundamental $\mathrm{SU}(N_{c})$
color indices (which we do not write explicitly). The traces over
spinor indices are given by:
\begin{equation}
\begin{aligned} & |\mathrm{Dirac}_{\mathrm{LO}}^{\mathrm{T}}|^{2} \\ & \equiv\mathrm{Tr}\Bigg[\bar{u}_{G}(p_{0}^{+})\gamma^{+}\mathrm{Dirac}^{\lambda\lambda^{\prime}}\big(1+\frac{2p_{1}^{+}}{q^{+}}\big)^{\dagger}u_{G}(p_{1}^{+})\bar{u}_{G}(p_{1}^{+})\gamma^{+}\mathrm{Dirac}^{\lambda\bar{\lambda}}\big(1+\frac{2p_{1}^{+}}{q^{+}}\big)u_{G}(p_{0}^{+})\Bigg]\;,\\
& |\mathrm{Dirac}_{\mathrm{LO}}^{0}|^{2}  \equiv\mathrm{Tr}\Bigg[\bar{u}_{G}(p_{0}^{+})\gamma^{+}u_{G}(p_{1}^{+})\bar{u}_{G}(p_{1}^{+})\gamma^{+}u_{G}(p_{0}^{+})\Bigg]\;.
\end{aligned}
\label{eq:DiracLO-2}
\end{equation}
Applying the cyclic permutation property of the trace as well as the
completeness relation (a summation over fermion spin is understood):
\begin{equation}
\begin{aligned}u_{G}(q^{+})\bar{u}_{G}(q^{+})\gamma^{+} & =2q^{+}\mathcal{P}_{G}\;,\end{aligned}
\end{equation}
we obtain:
\begin{equation}
\begin{aligned}|\mathrm{Dirac}_{\mathrm{LO}}^{\mathrm{T}}|^{2} & =\mathrm{Tr}\Bigg[2p_{0}^{+}\mathcal{P}_{G}\mathrm{Dirac}^{\lambda\lambda^{\prime}}\big(1+\frac{2p_{1}^{+}}{q^{+}}\big)^{\dagger}2p_{1}^{+}\mathcal{P}_{G}\mathrm{Dirac}^{\lambda\bar{\lambda}}\big(1+\frac{2p_{1}^{+}}{q^{+}}\big)\Bigg]\;.\end{aligned}
\end{equation}
Since the projector $\mathcal{P}_{G}$ commutes with transverse gamma
matrices, from which the structures $\mathrm{Dirac}^{\lambda\lambda^{\prime}}$
are built, the above expression further simplifies to:
\begin{equation}
\begin{aligned}|\mathrm{Dirac}_{\mathrm{LO}}^{\mathrm{T}}|^{2} & =4p_{0}^{+}p_{1}^{+}\mathrm{Tr}\Bigg[\mathcal{P}_{G}\Big(\big(1\!+\!\frac{2p_{1}^{+}}{q^{+}}\big)\delta^{\lambda\lambda^{\prime}}\mathds{1}_{4}\!+\!i\sigma^{\lambda\lambda^{\prime}}\Big)\Big(\big(1\!+\!\frac{2p_{1}^{+}}{q^{+}}\big)\delta^{\lambda\bar{\lambda}}\mathds{1}_{4}\!-\!i\sigma^{\lambda\bar{\lambda}}\Big)\Bigg]\;,\\
 & =4p_{0}^{+}p_{1}^{+}\mathrm{Tr}\Bigg[\mathcal{P}_{G}\Big(\big(1+\frac{2p_{1}^{+}}{q^{+}}\big)^{2}\delta^{\lambda^{\prime}\bar{\lambda}}\mathds{1}_{4}+\sigma^{\lambda\lambda^{\prime}}\sigma^{\lambda\bar{\lambda}}+2i\sigma^{\bar{\lambda}\lambda^{\prime}}\big(1+\frac{2p_{1}^{+}}{q^{+}}\big)\Big)\Bigg]\;.
\end{aligned}
\end{equation}
Finally, with the help of the identities:
\begin{equation}
\begin{aligned}\sigma^{\lambda\lambda^{\prime}}\sigma^{\lambda\bar{\lambda}} & =(D-3)\delta^{\lambda^{\prime}\bar{\lambda}}\mathds{1}_{4}+i(D-4)\sigma^{\lambda^{\prime}\bar{\lambda}}\;,\end{aligned}
\end{equation}
as well as:
\begin{equation}
\begin{aligned}\mathrm{Tr}\big(\mathcal{P}_{G}\big) & =2\;,\qquad\mathrm{Tr}\big(\mathcal{P}_{G}\sigma^{ij}\big)=0\;,\end{aligned}
\end{equation}
we end up with:
\begin{equation}
\begin{aligned}|\mathrm{Dirac}_{\mathrm{LO}}^{\mathrm{T}}|^{2} & =8p_{0}^{+}p_{1}^{+}\delta^{\lambda^{\prime}\bar{\lambda}}\Big(\big(1+\frac{2p_{1}^{+}}{q^{+}}\big)^{2}+D-3\Big)\;,\end{aligned}
\label{eq:DiracLOT}
\end{equation}
and similarly 
\begin{equation}
\begin{aligned}|\mathrm{Dirac}_{\mathrm{LO}}^{0}|^{2} & =8p_{1}^{+}p_{0}^{+}\;.\end{aligned}
\label{eq:DiracLOL}
\end{equation}
In appendix~\ref{sec:Dirac}, the most important gamma-matrix
identities used in this work are collected.

Combining (\ref{eq:crosssectiondef}), (\ref{eq:LOTsquared}), (\ref{eq:LOLsquared}),
and (\ref{eq:DiracLOL}), we eventually arrive at the following result
for the partonic cross sections:
\begin{equation}
\begin{aligned}\frac{\mathrm{d}\hat{\sigma}_{\mathrm{LO}}^{\mathrm{T}}}{\mathrm{d}z\mathrm{d}\bar{z}\mathrm{\mathrm{d}}^{2}\mathbf{P}_{\perp}\mathrm{\mathrm{d}}^{2}\mathbf{k}_{\perp}} & =\frac{g_{\mathrm{em}}^{2}N_{c}}{(2\pi)^{5}}\delta(1-z-\bar{z})\frac{1+(1-z)^{2}}{z}\Big(\frac{\mathbf{P}_{\perp}}{\mathbf{P}_{\perp}^{2}+\bar{z}M^{2}}+\frac{\mathbf{q}}{\mathbf{q}^{2}+\bar{z}M^{2}}\Big)^{2}\\
 & \times\int_{\mathbf{x},\mathbf{x}^{\prime}}e^{-i\mathbf{k}_{\perp}\cdot(\mathbf{x}-\mathbf{x}^{\prime})}\big(s_{\mathbf{x}\mathbf{x}^{\prime}}+1\big)\;,
\end{aligned}
\label{eq:crossLOT}
\end{equation}
\begin{equation}
\begin{aligned}\frac{\mathrm{d}\hat{\sigma}_{\mathrm{LO}}^{\mathrm{L}}}{\mathrm{d}z\mathrm{d}\bar{z}\mathrm{\mathrm{d}}^{2}\mathbf{P}_{\perp}\mathrm{\mathrm{d}}^{2}\mathbf{k}_{\perp}} & =\frac{g_{\mathrm{em}}^{2}N_{c}}{(2\pi)^{5}}\delta(1-z-\bar{z})\frac{1}{2zM^{2}}\Big(\frac{\mathbf{P}_{\perp}^{2}-\bar{z}M^{2}}{\mathbf{P}_{\perp}^{2}+\bar{z}M^{2}}-\frac{\mathbf{q}^{2}-\bar{z}M^{2}}{\mathbf{q}^{2}+\bar{z}M^{2}}\Big)^{2}\\
 & \times\int_{\mathbf{x},\mathbf{x}^{\prime}}e^{-i\mathbf{k}_{\perp}\cdot(\mathbf{x}-\mathbf{x}^{\prime})}\big(s_{\mathbf{x}\mathbf{x}^{\prime}}+1\big)\;,
\end{aligned}
\label{eq:crossLOL}
\end{equation}
defining the momentum fractions $z\equiv q^{+}/p_{0}^{+}$ and $\bar{z}=1-z$. 

In the results (\ref{eq:crossLOT}) and (\ref{eq:crossLOL}), the
term containing the dipole $s_{\mathbf{x}\mathbf{x}^{\prime}}$ quantifies
the potential provided by the target and experienced by the projectile.
However, it is built from Wilson lines that depend on the semi-classical
gauge fields $A^{-}$ of the highly boosted target \textquoteleft shockwave'.
These fields carry the information on the target gluon structure,
which is at least partially nonperturbative. To indicate that, on
the hadronic level, the fields still need to be related to the target
properties, for instance using a model or by linking them to gluon
transverse momentum dependent PDFs, we write the above mentioned \textquoteleft target
average' $\langle...\rangle$. Likewise, the incoming quark can be
related to its parent proton by convolving the cross section with the
quark PDF, and a factor $\frac{\alpha_{\mathrm{em}}}{3\pi}\mathrm{d}\ln M^{2}$
takes the $\gamma^{*}\to\ell^{+}\ell^{-}$ splitting into account
in the simple scenario where only the total three-momentum and the invariant mass of the lepton
pair is measured \cite{Berger:2001wr}. We finally obtain: 
\begin{equation}
\begin{aligned} & \frac{\mathrm{d}\sigma_{\mathrm{LO}}}{\mathrm{d}z\mathrm{d}\bar{z}\mathrm{\mathrm{d}}^{2}\mathbf{P}_{\perp}\mathrm{\mathrm{d}}^{2}\mathbf{k}_{\perp}\mathrm{d}\ln M^{2}} \\ & =\frac{2\alpha_{\mathrm{em}}^{2}N_{c}}{3\pi}\frac{1}{(2\pi)^{4}}x_{p}f_{q}\big(x_{p},\mu^{2}\big)\Bigg[\frac{1+(1-z)^{2}}{z}\Big(\frac{\mathbf{P}_{\perp}}{\mathbf{P}_{\perp}^{2}+\bar{z}M^{2}}+\frac{\mathbf{q}}{\mathbf{q}^{2}+\bar{z}M^{2}}\Big)^{2}\\
 & +\frac{1}{2zM^{2}}\Big(\frac{\mathbf{P}_{\perp}^{2}-\bar{z}M^{2}}{\mathbf{P}_{\perp}^{2}+\bar{z}M^{2}}-\frac{\mathbf{q}^{2}-\bar{z}M^{2}}{\mathbf{q}^{2}+\bar{z}M^{2}}\Big)^{2}\Bigg]\int_{\mathbf{x},\mathbf{x}^{\prime}}e^{-i\mathbf{k}_{\perp}\cdot(\mathbf{x}-\mathbf{x}^{\prime})}\big\langle s_{\mathbf{x}\mathbf{x}^{\prime}}+1\big\rangle\;,
\end{aligned}
\label{eq:crossLO}
\end{equation}
in agreement with earlier results in the literature \cite{Gelis:2002fw,Stasto:2012ru}.

Before turning to the next-to-leading order calculation, we remark
that, apart from the universal coupling constant $g_{\mathrm{em}}$, the amplitudes
(\ref{eq:MLOT}) and (\ref{eq:MLOL}) all contain a spinor structure
$\bar{u}_{G}(p_{1}^{+})\gamma^{+}(...)u_{G}(p_{0}^{+})$. In fact,
this will also be true for all the NLO amplitudes. For ease of notation,
we therefore define the \textquoteleft reduced amplitudes' $\tilde{\mathcal{M}}$:
\begin{equation}
\mathcal{M}=g_{\mathrm{em}}\bar{u}_{G}(p_{1}^{+})\gamma^{+}\tilde{\mathcal{M}}u_{G}(p_{0}^{+})\;,\label{eq:reduceddef}
\end{equation}
with which we will work in what follows. We also define, for later
convenience:
\begin{equation}
\begin{aligned}\mathrm{Dirac}_{\mathrm{LO}}^{\lambda\bar{\lambda}} & \equiv\mathrm{Dirac}^{\lambda\bar{\lambda}}\Big(1+\frac{2p_{1}^{+}}{q^{+}}\Big)\;.\end{aligned}
\label{eq:DiracLO}
\end{equation}

\section{\label{sec:virtual}Virtual next-to-leading order corrections}

In this section, we review all the one-loop Feynman graphs for the
channel $q+A\to\gamma^{*}+q$. The corresponding amplitudes are presented
in general $D$ dimensions, as many of them contain ultraviolet divergences
which will be regulated using dimensional regularization. The resulting
UV poles are subtracted from the amplitudes with the help of counterterms
that will, in turn, shown to cancel in the next section. Readers who
are interested how the amplitudes themselves are obtained within our
LCPT approach to the CGC are referred to appendix \ref{sec:Example-calculation},
where a representative amplitude is calculated from start to finish.

\subsection{\label{subsec:SE}Self-energy corrections}

\begin{figure}[t]
\begin{centering}
\includegraphics[scale=0.25]{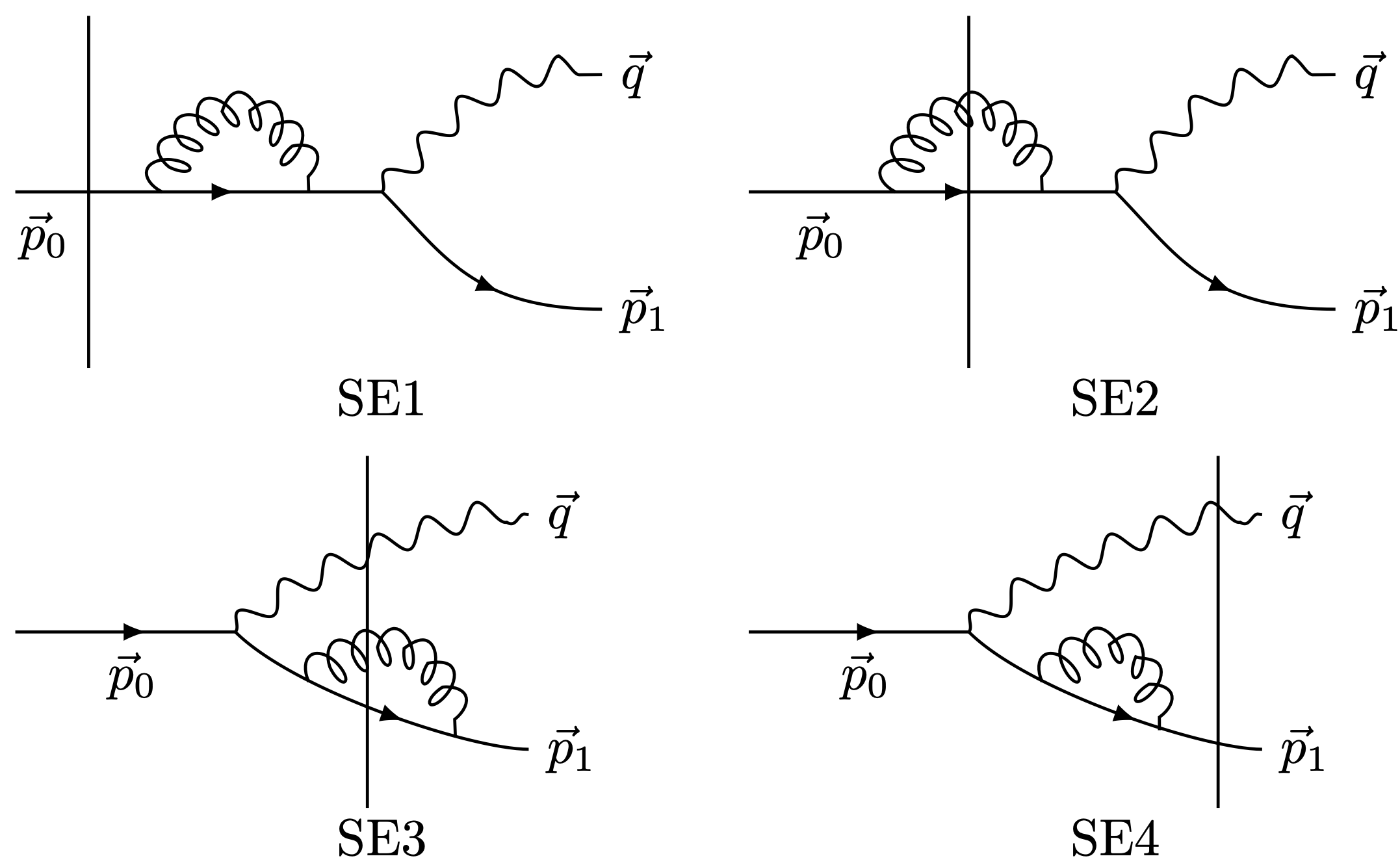}
\par\end{centering}
\caption{\label{fig:SE}The four virtual contributions with a gluon loop on
the quark, which in two cases ($\mathrm{SE2}$ and $\mathrm{SE3}$)
scatters off the shockwave. Similar one-loop corrections on the asymptotic
incoming or outgoing quark are treated separately in subsection \ref{subsec:Z}.}
\end{figure}

\paragraph{Diagram SE1}

The reduced amplitude corresponding to a quark self-energy correction
after having hit the shockwave, but before the emission of the photon
(see fig. \ref{fig:SE}), reads:
\begin{equation}
\begin{aligned}\tilde{\mathcal{M}}_{\mathrm{SE1}}^{0} & =\frac{\alpha_{s}C_{F}}{D-2}\frac{1}{M}\frac{p_{0}^{+}\mathbf{P}_{\perp}^{2}-p_{1}^{+}M^{2}}{p_{0}^{+}\mathbf{P}_{\perp}^{2}+p_{1}^{+}M^{2}}\int_{k_{\mathrm{min}}^{+}}^{p_{0}^{+}}\frac{\mathrm{d}k^{+}}{k^{+}}\Big(\frac{k^{+}}{p_{0}^{+}}\Big)^{2}\mathrm{Dirac}_{\mathrm{SE}}^{jj}(p_{0}^{+})\\
 & \times\mathcal{A}_{0}(\Delta_{\mathrm{P}})\int_{\mathbf{x}}e^{-i\mathbf{k}_{\perp}\cdot\mathbf{x}}\big(U_{\mathbf{x}}-1\big)\;,
\end{aligned}
\label{eq:MSE1divL}
\end{equation}
for a longitudinally polarized virtual photon, and
\begin{equation}
\begin{aligned}\tilde{\mathcal{M}}_{\mathrm{SE1}}^{\lambda} & =-\frac{\alpha_{s}C_{F}}{D-2}\frac{q^{+}\mathbf{P}_{\perp}^{\bar{\lambda}}}{p_{0}^{+}\mathbf{P}_{\perp}^{2}+p_{1}^{+}M^{2}}\int_{k_{\mathrm{min}}^{+}}^{p_{0}^{+}}\frac{\mathrm{d}k^{+}}{k^{+}}\Big(\frac{k^{+}}{p_{0}^{+}}\Big)^{2}\mathrm{Dirac}_{\mathrm{LO}}^{\lambda\bar{\lambda}}\mathrm{Dirac}_{\mathrm{SE}}^{jj}(p_{0}^{+})\\
 & \times\mathcal{A}_{0}(\Delta_{\mathrm{P}})\int_{\mathbf{x}}e^{-i\mathbf{k}_{\perp}\cdot\mathbf{x}}\big(U_{\mathbf{x}}-1\big)\;,
\end{aligned}
\label{eq:MSE1div}
\end{equation}
when the photon is transversely polarized. In the above formulas,
$C_{F}=(N_{c}^{2}-1)/2N_{c}$, and we defined the Dirac structures:
\begin{equation}
\begin{aligned}\mathrm{Dirac}_{\mathrm{SE}}^{\bar{\eta}\eta^{\prime}}(p_{0}^{+}) & \equiv\Big[\big(1-\frac{2p_{0}^{+}}{k^{+}}\big)\delta^{\eta\bar{\eta}}\mathds{1}_{4}-i\sigma^{\eta\bar{\eta}}\Big]\Big[\big(\frac{2p_{0}^{+}}{k^{+}}-1\big)\delta^{\eta\eta^{\prime}}\mathds{1}_{4}-i\sigma^{\eta\eta^{\prime}}\Big]\;,\end{aligned}
\label{eq:DiracSE}
\end{equation}
as well as:
\begin{equation}
\Delta_{\mathrm{P}}\equiv-\frac{k^{+}(p_{0}^{+}-k^{+})}{p_{0}^{+}p_{1}^{+}q^{+}}\big(p_{0}^{+}\mathbf{P}_{\perp}^{2}+p_{1}^{+}M^{2}\big)-i0^{+}\;.\label{eq:DeltaP}
\end{equation}
At a later stage we will evaluate the $k^{+}$-integral and encounter
logarithmic branch cuts. Therefore, we keep track of the infinitesimal
imaginary part $i0^{+}$ in the above expression, stemming from the
energy denominator from which this term stems. 

The quantity $\mathcal{A}_{0}(\Delta_{\mathrm{P}})$ in eqs. (\ref{eq:MSE1divL})
and (\ref{eq:MSE1div}) results from the integration over the gluon
transverse loop momentum, which we have evaluated using integral identity
eq. (\ref{eq:A0}). As could be expected, the result is divergent
in the ultraviolet, and we use dimensional regularization with in
$D=4-2\epsilon_{\mathrm{UV}}$ to regulate it. Finally, $\tilde{\mathcal{M}}_{\mathrm{SE1}}^{\lambda}$
and $\tilde{\mathcal{M}}_{\mathrm{SE1}}^{0}$diverge in the limit
of vanishing gluon plus-momentum $k^{+}\to0$. This rapidity divergence
is regulated using a cutoff $k_{\mathrm{min}}^{+}$.

We now introduce a counterterm to cancel the UV-pole in expression
(\ref{eq:MSE1divL}):
\begin{equation}
\begin{aligned}\tilde{\mathcal{M}}_{\mathrm{SE1,UV}}^{0} & =\frac{\alpha_{s}C_{F}}{D-2}\frac{1}{M}\frac{p_{0}^{+}\mathbf{P}_{\perp}^{2}-p_{1}^{+}M^{2}}{p_{0}^{+}\mathbf{P}_{\perp}^{2}+p_{1}^{+}M^{2}}\int_{k_{\mathrm{min}}^{+}}^{p_{0}^{+}}\frac{\mathrm{d}k^{+}}{k^{+}}\Big(\frac{k^{+}}{p_{0}^{+}}\Big)^{2}\mathrm{Dirac}_{\mathrm{SE}}^{jj}(p_{0}^{+})\\
 & \times\mathcal{A}_{0}(\Delta_{\mathrm{UV}})\int_{\mathbf{x}}e^{-i\mathbf{k}_{\perp}\cdot\mathbf{x}}\big(U_{\mathbf{x}}-1\big)\;.
\end{aligned}
\label{eq:MSE1L_UV}
\end{equation}
The UV-subtracted amplitude can then be evaluated in $D=4$ dimensions:
\begin{equation}
\begin{aligned}\tilde{\mathcal{M}}_{\mathrm{SE1,sub}}^{0} & =\tilde{\mathcal{M}}_{\mathrm{LO1}}^{0}\frac{\alpha_{s}C_{F}}{\pi}\int_{k_{\mathrm{min}}^{+}}^{p_{0}^{+}}\frac{\mathrm{d}k^{+}}{8k^{+}}\Big(\frac{k^{+}}{p_{0}^{+}}\Big)^{2}\mathrm{Dirac}_{\mathrm{SE}}^{jj}(p_{0}^{+})\ln\frac{\Delta_{\mathrm{UV}}}{\Delta_{\mathrm{P}}}\\
 & =\tilde{\mathcal{M}}_{\mathrm{LO1}}^{0}\frac{\alpha_{s}C_{F}}{\pi}\int_{k_{\mathrm{min}}^{+}}^{p_{0}^{+}}\frac{\mathrm{d}k^{+}}{4k^{+}}\Big(\frac{k^{+}}{p_{0}^{+}}\Big)^{2}\Big(\big(1-\frac{2p_{0}^{+}}{k^{+}}\big)^{2}+1\Big)\ln\frac{\Delta_{\mathrm{P}}}{\Delta_{\mathrm{UV}}}\;.
\end{aligned}
\label{eq:MSE1Lsub_def}
\end{equation}
The exact same steps we just took can be followed in the transversely
polarized case, defining a counterterm:
\begin{equation}
\begin{aligned}\tilde{\mathcal{M}}_{\mathrm{SE1,UV}}^{\lambda} & =-\frac{\alpha_{s}C_{F}}{D-2}\frac{q^{+}\mathbf{P}_{\perp}^{\bar{\lambda}}}{p_{0}^{+}\mathbf{P}_{\perp}^{2}+p_{1}^{+}M^{2}}\int_{k_{\mathrm{min}}^{+}}^{p_{0}^{+}}\frac{\mathrm{d}k^{+}}{k^{+}}\Big(\frac{k^{+}}{p_{0}^{+}}\Big)^{2}\mathrm{Dirac}_{\mathrm{LO}}^{\lambda\bar{\lambda}}\mathrm{Dirac}_{\mathrm{SE}}^{jj}(p_{0}^{+})\\
 & \times\mathcal{A}_{0}(\Delta_{\mathrm{UV}})\int_{\mathbf{x}}e^{-i\mathbf{k}_{\perp}\cdot\mathbf{x}}\big(U_{\mathbf{x}}-1\big)\;,
\end{aligned}
\label{eq:MSE1div_UV}
\end{equation}
such that the UV-subtracted amplitude can be evaluated in $D=4$ dimensions,
yielding:
\begin{equation}
\begin{aligned}\tilde{\mathcal{M}}_{\mathrm{SE1,sub}}^{\lambda} & =\tilde{\mathcal{M}}_{\mathrm{LO1}}^{\lambda}\frac{\alpha_{s}C_{F}}{\pi}\int_{k_{\mathrm{min}}^{+}}^{p_{0}^{+}}\frac{\mathrm{d}k^{+}}{4k^{+}}\Big(\frac{k^{+}}{p_{0}^{+}}\Big)^{2}\Big(\big(1-\frac{2p_{0}^{+}}{k^{+}}\big)^{2}+1\Big)\ln\frac{\Delta_{\mathrm{P}}}{\Delta_{\mathrm{UV}}}\;.\end{aligned}
\label{eq:MSE1Tsub}
\end{equation}
Comparing the above result with (\ref{eq:MSE1Lsub_def}), it is clear
that the loop correction calculated here is independent of the polarization
of the virtual photon.

\paragraph{Diagram SE2}

Using the expressions (\ref{eq:WWfield}) and (\ref{eq:WWfieldmass})
for the so-called Weizsäcker-Williams fields $A^{i}(\mathbf{x})$
and $A^{i}(\mathbf{x},\Delta)$, as well as the property (\ref{eq:WWlinear}),
one obtains for the diagrams where the quark self-energy loop scatters
off the shockwave before emitting a virtual photon (see fig. \ref{fig:SE}):
\begin{equation}
\begin{aligned}\tilde{\mathcal{M}}_{\mathrm{SE2}}^{0} & =-\frac{\alpha_{s}}{D-2}\frac{1}{M}\frac{p_{0}^{+}\mathbf{P}_{\perp}^{2}-p_{1}^{+}M^{2}}{p_{0}^{+}\mathbf{P}_{\perp}^{2}+p_{1}^{+}M^{2}}\int_{k_{\mathrm{min}}^{+}}^{p_{0}^{+}}\frac{\mathrm{d}k^{+}}{k^{+}}\Big(\frac{k^{+}}{p_{0}^{+}}\Big)^{2}\mathrm{Dirac}_{\mathrm{SE}}^{jj}(p_{0}^{+})\\
 & \times\int_{\mathbf{x},\mathbf{z}}A^{i}(\mathbf{x}-\mathbf{z})A^{i}\big(\mathbf{x}-\mathbf{z},\Delta_{\mathrm{P}}\big)\\
 & \times e^{-i\mathbf{k}_{\perp}\cdot\Big(\frac{p_{0}^{+}-k^{+}}{p_{0}^{+}}\mathbf{x}+\frac{k^{+}}{p_{0}^{+}}\mathbf{z}\Big)}\big(t^{c}U_{\mathbf{x}}U_{\mathbf{z}}^{\dagger}t^{c}U_{\mathbf{z}}-C_{F}\big)\;,
\end{aligned}
\label{eq:MSE2L}
\end{equation}
and:
\begin{equation}
\begin{aligned}\tilde{\mathcal{M}}_{\mathrm{SE2}}^{\lambda} & =\frac{\alpha_{s}}{D-2}\frac{q^{+}\mathbf{P}_{\perp}^{\bar{\lambda}}}{p_{0}^{+}\mathbf{P}_{\perp}^{2}+p_{1}^{+}M^{2}}\int_{k_{\mathrm{min}}^{+}}^{p_{0}^{+}}\frac{\mathrm{d}k^{+}}{k^{+}}\Big(\frac{k^{+}}{p_{0}^{+}}\Big)^{2}\mathrm{Dirac}_{\mathrm{LO}}^{\lambda\bar{\lambda}}\mathrm{Dirac}_{\mathrm{SE}}^{jj}(p_{0}^{+})\\
 & \times\int_{\mathbf{x},\mathbf{z}}A^{i}(\mathbf{x}-\mathbf{z})A^{i}\big(\mathbf{x}-\mathbf{z},\Delta_{\mathrm{P}}\big)\\
 & \times e^{-i\mathbf{k}_{\perp}\cdot\Big(\frac{p_{0}^{+}-k^{+}}{p_{0}^{+}}\mathbf{x}+\frac{k^{+}}{p_{0}^{+}}\mathbf{z}\Big)}\big(t^{c}U_{\mathbf{x}}U_{\mathbf{z}}^{\dagger}t^{c}U_{\mathbf{z}}-C_{F}\big)\;.
\end{aligned}
\label{eq:MSE2}
\end{equation}
Note that we have used the Fierz identity $t^{a}W^{ab}=U^{\dagger}t^{b}U$
to transform Wilson lines $W$ in the adjoint representation, which
parameterize the virtual gluon interacting with the shockwave, into
the usual Wilson lines $U$ in the fundamental representation (\ref{eq:WLdef}).

In the limit $\mathbf{z}\to\mathbf{x}$, or equivalently when the
virtual gluon obtains an infinitely large transverse momentum, the
amplitudes (\ref{eq:MSE2}) and (\ref{eq:MSE2L}) exhibit an UV divergence.
Indeed, we have from definitions (\ref{eq:WWfield}) and (\ref{eq:WWfieldmass}):
\begin{equation}
\begin{aligned} & A^{i}(\mathbf{x}-\mathbf{z})A^{i}(\mathbf{x}-\mathbf{z},\Delta) \\ & =-\int_{\mathbf{k},\boldsymbol{\ell}}e^{-i(\mathbf{k}+\boldsymbol{\ell})\cdot(\mathbf{x}-\mathbf{z})}\frac{\mathbf{k}^{i}}{\mathbf{k}^{2}}\frac{\mathbf{\boldsymbol{\ell}}^{i}}{\mathbf{\boldsymbol{\ell}}^{2}+\Delta}\;,\\
 & =-\frac{\mu^{2(4-D)}}{2(2\pi^{2})^{\frac{D-2}{2}}}\frac{1}{|\mathbf{x}-\mathbf{z}|^{\frac{3D-10}{4}}}\Gamma(\frac{D-2}{2})(\sqrt{\Delta})^{\frac{D-2}{2}}\mathrm{K}_{\frac{D-2}{2}}\big(|\mathbf{x}-\mathbf{z}|\sqrt{\Delta}\big)\;,
\end{aligned}
\label{eq:SE2UVdiv}
\end{equation}
which is singular for $\mathbf{x}-\mathbf{z}\to0$ but tends to zero
for $\mathbf{x}-\mathbf{z}\to\infty$. Moreover, since $\lim_{\Delta\to0}\sqrt{\Delta}\mathrm{K}_{1}\big(\sqrt{\Delta}\big)=1$,
there are no extra divergences generated when integrating over $k^{+}$
(since $\Delta_{\mathrm{P}}(k^{+}\to p_{0}^{+})\to0$). 

A counterterm for this UV divergence can be constructed from (\ref{eq:MSE2})
and (\ref{eq:MSE2L}) by setting $\mathbf{z}=\mathbf{x}$ everywhere
except in the singular part, after which the integral over $\mathbf{z}$
can be evaluated by reverting to momentum space and using the standard
result (\ref{eq:A0}):
\begin{equation}
\begin{aligned}\tilde{\mathcal{M}}_{\mathrm{SE2,UV}}^{0} & =-\frac{\alpha_{s}C_{F}}{D-2}\frac{1}{M}\frac{p_{0}^{+}\mathbf{P}_{\perp}^{2}-p_{1}^{+}M^{2}}{p_{0}^{+}\mathbf{P}_{\perp}^{2}+p_{1}^{+}M^{2}}\int_{k_{\mathrm{min}}^{+}}^{p_{0}^{+}}\frac{\mathrm{d}k^{+}}{k^{+}}\Big(\frac{k^{+}}{p_{0}^{+}}\Big)^{2}\mathrm{Dirac}_{\mathrm{SE}}^{jj}(p_{0}^{+})\\
 & \times\mathcal{A}_{0}(\Delta_{\mathrm{UV}})\int_{\mathbf{x}}e^{-i\mathbf{k}_{\perp}\cdot\mathbf{x}}\big(U_{\mathbf{x}}-1\big)=-\tilde{\mathcal{M}}_{\mathrm{SE1,UV}}^{0}\;,
\end{aligned}
\label{eq:MSE2L_UV}
\end{equation}
and similarly for the transverse case:
\begin{equation}
\begin{aligned}\tilde{\mathcal{M}}_{\mathrm{SE2,UV}}^{\lambda} & =-\tilde{\mathcal{M}}_{\mathrm{SE1,\mathrm{UV}}}^{\lambda}\;.\end{aligned}
\label{eq:MSE2TUV}
\end{equation}
Therefore, the UV-counterterms of the amplitudes $\tilde{\mathcal{M}}_{\mathrm{SE2}}^{\lambda,0}$
happen to be the exact opposite of $\tilde{\mathcal{M}}_{\mathrm{SE1}}^{\lambda,0}$
(\ref{eq:MSE1div}). 

The UV-subtracted amplitudes read:
\begin{equation}
\begin{aligned}\tilde{\mathcal{M}}_{\mathrm{SE2,sub}}^{0} & =\alpha_{s}\frac{1}{M}\frac{p_{0}^{+}\mathbf{P}_{\perp}^{2}-p_{1}^{+}M^{2}}{p_{0}^{+}\mathbf{P}_{\perp}^{2}+p_{1}^{+}M^{2}}\int_{k_{\mathrm{min}}^{+}}^{p_{0}^{+}}\frac{\mathrm{d}k^{+}}{k^{+}}\Big(\frac{k^{+}}{p_{0}^{+}}\Big)^{2}\Big(\big(1-\frac{2p_{0}^{+}}{k^{+}}\big)^{2}+1\Big)\\
 & \times\Bigg[\int_{\mathbf{x},\mathbf{z}}A^{i}(\mathbf{x}-\mathbf{z})A^{i}\big(\mathbf{x}-\mathbf{z},\Delta_{\mathrm{P}}\big)e^{-i\mathbf{k}_{\perp}\cdot\Big(\frac{p_{0}^{+}-k^{+}}{p_{0}^{+}}\mathbf{x}+\frac{k^{+}}{p_{0}^{+}}\mathbf{z}\Big)}\big(t^{c}U_{\mathbf{x}}U_{\mathbf{z}}^{\dagger}t^{c}U_{\mathbf{z}}-C_{F}\big)\\
 & -\mathcal{A}_{0}(\Delta_{\mathrm{UV}})\int_{\mathbf{x}}e^{-i\mathbf{k}_{\perp}\cdot\mathbf{x}}C_{F}\big(U_{\mathbf{x}}-1\big)\Bigg]\;,
\end{aligned}
\label{eq:MSE2L_sub}
\end{equation}
and:
\begin{equation}
\begin{aligned} & \tilde{\mathcal{M}}_{\mathrm{SE2,sub}}^{\lambda} \\ & =-\alpha_{s}\frac{q^{+}\mathbf{P}_{\perp}^{\bar{\lambda}}}{p_{0}^{+}\mathbf{P}_{\perp}^{2}+p_{1}^{+}M^{2}}\mathrm{Dirac}^{\lambda\bar{\lambda}}\big(1+2\frac{p_{1}^{+}}{q^{+}}\big)\int_{k_{\mathrm{min}}^{+}}^{p_{0}^{+}}\frac{\mathrm{d}k^{+}}{k^{+}}\Big(\frac{k^{+}}{p_{0}^{+}}\Big)^{2}\Big(\big(1-\frac{2p_{0}^{+}}{k^{+}}\big)^{2}+1\Big)\\
 & \times\Bigg[\int_{\mathbf{x},\mathbf{z}}A^{i}(\mathbf{x}-\mathbf{z})A^{i}\big(\mathbf{x}-\mathbf{z},\Delta_{\mathrm{P}}\big)e^{-i\mathbf{k}_{\perp}\cdot\Big(\frac{p_{0}^{+}-k^{+}}{p_{0}^{+}}\mathbf{x}+\frac{k^{+}}{p_{0}^{+}}\mathbf{z}\Big)}\big(t^{c}U_{\mathbf{x}}U_{\mathbf{z}}^{\dagger}t^{c}U_{\mathbf{z}}-C_{F}\big)\\
 & -\mathcal{A}_{0}(\Delta_{\mathrm{UV}})\int_{\mathbf{x}}e^{-i\mathbf{k}_{\perp}\cdot\mathbf{x}}C_{F}\big(U_{\mathbf{x}}-1\big)\Bigg]\;.
\end{aligned}
\label{eq:MSE2T_sub}
\end{equation}

\paragraph{Diagram SE3}

This diagram is the counterpart of graph SE2, when the emission of
the virtual photon takes place before the quark self-energy loop (see
fig. \ref{fig:SE}). The corresponding amplitudes read:
\begin{equation}
\begin{aligned}\tilde{\mathcal{M}}_{\mathrm{SE3}}^{0} & =\frac{\alpha_{s}(-1)^{4-D}}{D-2}\frac{p_{0}^{+}\mathbf{q}^{2}-p_{1}^{+}M^{2}}{p_{0}^{+}\mathbf{q}^{2}+p_{1}^{+}M^{2}}\frac{1}{M}\int_{k_{\mathrm{min}}^{+}}^{p_{1}^{+}}\frac{\mathrm{d}k^{+}}{k^{+}}\Big(\frac{k^{+}}{p_{1}^{+}}\Big)^{2}\mathrm{Dirac}_{\mathrm{SE}}^{jj}(p_{1}^{+})\\
 & \times\int_{\mathbf{x},\mathbf{z}}A^{i}(\mathbf{z}-\mathbf{x})A^{i}(\mathbf{z}-\mathbf{x},\Delta_{\mathrm{q}})\\
 & \times e^{-i\mathbf{k}_{\perp}\cdot\Big(\frac{p_{1}^{+}-k^{+}}{p_{1}^{+}}\mathbf{x}+\frac{k^{+}}{p_{1}^{+}}\mathbf{z}\Big)}\big(t^{c}U_{\mathbf{x}}U_{\mathbf{z}}^{\dagger}t^{c}U_{\mathbf{z}}-C_{F}\big)\;,
\end{aligned}
\label{eq:MSE3L}
\end{equation}
and:
\begin{equation}
\begin{aligned}\tilde{\mathcal{M}}_{\mathrm{SE3}}^{\lambda} & =\frac{\alpha_{s}(-1)^{4-D}}{D-2}\frac{q^{+}\mathbf{q}^{\bar{\lambda}}}{p_{0}^{+}\mathbf{q}^{2}+p_{1}^{+}M^{2}}\int_{k_{\mathrm{min}}^{+}}^{p_{1}^{+}}\frac{\mathrm{d}k^{+}}{k^{+}}\Big(\frac{k^{+}}{p_{1}^{+}}\Big)^{2}\mathrm{Dirac}_{\mathrm{SE}}^{jj}(p_{1}^{+})\mathrm{Dirac}_{\mathrm{LO}}^{\lambda\bar{\lambda}}\\
 & \times\int_{\mathbf{x},\mathbf{z}}A^{i}(\mathbf{z}-\mathbf{x})A^{i}(\mathbf{z}-\mathbf{x},\Delta_{\mathrm{q}})\\
 & \times e^{-i\mathbf{k}_{\perp}\cdot\Big(\frac{p_{1}^{+}-k^{+}}{p_{1}^{+}}\mathbf{x}+\frac{k^{+}}{p_{1}^{+}}\mathbf{z}\Big)}\big(t^{c}U_{\mathbf{x}}U_{\mathbf{z}}^{\dagger}t^{c}U_{\mathbf{z}}-C_{F}\big)\;.
\end{aligned}
\label{eq:MSE3}
\end{equation}
In the above, we defined:
\begin{equation}
\begin{aligned}\Delta_{\mathrm{q}} & \equiv\frac{k^{+}(p_{1}^{+}-k^{+})}{(p_{1}^{+})^{2}q^{+}}(p_{0}^{+}\mathbf{q}^{2}+p_{1}^{+}M^{2})-i0^{+}\;.\end{aligned}
\label{eq:Deltaq}
\end{equation}
Just like the amplitudes $\tilde{\mathcal{M}}_{\mathrm{SE2}}^{\lambda,0}$,
also (\ref{eq:MSE3}) and (\ref{eq:MSE3L}) are UV-divergent in the
limit $\mathbf{z}\to\mathbf{x}$. Again, this divergence can be extracted
by setting $\mathbf{z}=\mathbf{x}$ everywhere except in the singular
part, defining the counterterms:
\begin{equation}
\begin{aligned}\tilde{\mathcal{M}}_{\mathrm{SE3,UV}}^{0} & =-\alpha_{s}C_{F}\frac{p_{0}^{+}\mathbf{q}^{2}-p_{1}^{+}M^{2}}{p_{0}^{+}\mathbf{q}^{2}+p_{1}^{+}M^{2}}\frac{1}{M}\int_{k_{\mathrm{min}}^{+}}^{p_{1}^{+}}\frac{\mathrm{d}k^{+}}{k^{+}}\Big(\frac{k^{+}}{p_{1}^{+}}\Big)^{2}\Big(\big(1-\frac{2p_{1}^{+}}{k^{+}}\big)^{2}+D-3\Big)\\
 & \times\mathcal{A}_{0}(\Delta_{\mathrm{UV}})\int_{\mathbf{x}}e^{-i\mathbf{k}_{\perp}\cdot\mathbf{x}}\big(U_{\mathbf{x}}-1\big)\;,\\
\tilde{\mathcal{M}}_{\mathrm{SE3,UV}}^{\lambda} & =-\alpha_{s}C_{F}\frac{q^{+}\mathbf{q}^{\bar{\lambda}}}{p_{0}^{+}\mathbf{q}^{2}+p_{1}^{+}M^{2}}\mathrm{Dirac}_{\mathrm{LO}}^{\lambda\bar{\lambda}}\int_{k_{\mathrm{min}}^{+}}^{p_{1}^{+}}\frac{\mathrm{d}k^{+}}{k^{+}}\Big(\frac{k^{+}}{p_{1}^{+}}\Big)^{2}\Big(\big(1\!-\!\frac{2p_{1}^{+}}{k^{+}}\big)^{2}\!+\!D\!-\!3\Big)\\
 & \times\mathcal{A}_{0}(\Delta_{\mathrm{UV}})\int_{\mathbf{x}}e^{-i\mathbf{k}_{\perp}\cdot\mathbf{x}}\big(U_{\mathbf{x}}-1\big)\;.
\end{aligned}
\label{eq:MSE3UV}
\end{equation}
The subtracted amplitudes then become:
\begin{equation}
\begin{aligned}\tilde{\mathcal{M}}_{\mathrm{SE3,sub}}^{0} & =-\alpha_{s}\frac{p_{0}^{+}\mathbf{q}^{2}-p_{1}^{+}M^{2}}{p_{0}^{+}\mathbf{q}^{2}+p_{1}^{+}M^{2}}\frac{1}{M}\int_{k_{\mathrm{min}}^{+}}^{p_{1}^{+}}\frac{\mathrm{d}k^{+}}{k^{+}}\Big(\frac{k^{+}}{p_{1}^{+}}\Big)^{2}\Big(\big(1-\frac{2p_{1}^{+}}{k^{+}}\big)^{2}+1\Big)\\
 & \times\Bigg[\int_{\mathbf{x},\mathbf{z}}A^{i}(\mathbf{z}-\mathbf{x})A^{i}(\mathbf{z}-\mathbf{x},\Delta_{\mathrm{q}})e^{-i\mathbf{k}_{\perp}\cdot\Big(\frac{p_{1}^{+}-k^{+}}{p_{1}^{+}}\mathbf{x}+\frac{k^{+}}{p_{1}^{+}}\mathbf{z}\Big)}\big(t^{c}U_{\mathbf{x}}U_{\mathbf{z}}^{\dagger}t^{c}U_{\mathbf{z}}-C_{F}\big)\\
 & -\mathcal{A}_{0}(\Delta_{\mathrm{UV}})\int_{\mathbf{x}}e^{-i\mathbf{k}_{\perp}\cdot\mathbf{x}}C_{F}\big(U_{\mathbf{x}}-1\big)\Bigg]\;,
\end{aligned}
\label{eq:MSE3L_sub}
\end{equation}
and:
\begin{equation}
\begin{aligned}\tilde{\mathcal{M}}_{\mathrm{SE3,sub}}^{\lambda} & =-\alpha_{s}\frac{q^{+}\mathbf{q}^{\bar{\lambda}}}{p_{0}^{+}\mathbf{q}^{2}+p_{1}^{+}M^{2}}\mathrm{Dirac}_{\mathrm{LO}}^{\lambda\bar{\lambda}}\int_{k_{\mathrm{min}}^{+}}^{p_{1}^{+}}\frac{\mathrm{d}k^{+}}{k^{+}}\Big(\frac{k^{+}}{p_{1}^{+}}\Big)^{2}\Big(\big(1-\frac{2p_{1}^{+}}{k^{+}}\big)^{2}+1\Big)\\
 & \times\Bigg[\int_{\mathbf{x},\mathbf{z}}A^{i}(\mathbf{z}-\mathbf{x})A^{i}(\mathbf{z}-\mathbf{x},\Delta_{\mathrm{q}})e^{-i\mathbf{k}_{\perp}\cdot\Big(\frac{p_{1}^{+}-k^{+}}{p_{1}^{+}}\mathbf{x}+\frac{k^{+}}{p_{1}^{+}}\mathbf{z}\Big)}\big(t^{c}U_{\mathbf{x}}U_{\mathbf{z}}^{\dagger}t^{c}U_{\mathbf{z}}-C_{F}\big)\\
 & -\mathcal{A}_{0}(\Delta_{\mathrm{UV}})\int_{\mathbf{x}}e^{-i\mathbf{k}_{\perp}\cdot\mathbf{x}}C_{F}\big(U_{\mathbf{x}}-1\big)\Bigg]\;.
\end{aligned}
\label{eq:MSE3T_sub}
\end{equation}

\paragraph{Diagram SE4}

Evaluating the Feynman diagrams corresponding to a quark self-energy
correction after the photon emission, but before the scattering off
the shockwave (see fig. \ref{fig:SE}), we find:
\begin{equation}
\begin{aligned}\tilde{\mathcal{M}}_{\mathrm{SE4}}^{0} & =\alpha_{s}C_{F}\frac{p_{0}^{+}\mathbf{q}^{2}-p_{1}^{+}M^{2}}{p_{0}^{+}\mathbf{q}^{2}+p_{1}^{+}M^{2}}\frac{1}{M}\int_{k_{\mathrm{min}}^{+}}^{p_{1}^{+}}\frac{\mathrm{d}k^{+}}{k^{+}}\Big(\frac{k^{+}}{p_{1}^{+}}\Big)^{2}\Big(\big(1-\frac{2p_{1}^{+}}{k^{+}}\big)^{2}+D-3\Big)\\
 & \times\mathcal{A}_{0}(\Delta_{\mathrm{q}})\int_{\mathbf{x}}e^{-i\mathbf{k}_{\perp}\cdot\mathbf{x}}\big(U_{\mathbf{x}}-1\big)\;,\\
\tilde{\mathcal{M}}_{\mathrm{SE4}}^{\lambda} & \!=\!\alpha_{s}C_{F}\frac{q^{+}\mathbf{q}^{\bar{\lambda}}}{p_{0}^{+}\mathbf{q}^{2}\!+\!p_{1}^{+}M^{2}}\mathrm{Dirac}^{\lambda\bar{\lambda}}\big(1\!+\!2\frac{p_{1}^{+}}{q^{+}}\big)\!\int_{k_{\mathrm{min}}^{+}}^{p_{1}^{+}}\frac{\mathrm{d}k^{+}}{k^{+}}\!\Big(\!\frac{k^{+}}{p_{1}^{+}}\! \Big)^{2}\!\Big(\!\big(\!1\!-\!\frac{2p_{1}^{+}}{k^{+}}\! \big)^{2}\!+\!D\!-\!3\Big)\\
 & \times\mathcal{A}_{0}(\Delta_{\mathrm{q}})\int_{\mathbf{x}}e^{-i\mathbf{k}_{\perp}\cdot\mathbf{x}}\big(U_{\mathbf{x}}-1\big)\;.
\end{aligned}
\label{eq:MSE4}
\end{equation}
The counterterms are simply defined by setting $\mathcal{A}_{0}(\Delta_{\mathrm{q}})\to\mathcal{A}_{0}(\Delta_{\mathrm{UV}})$,
and are once again opposite to those of $\tilde{\mathcal{M}}_{\mathrm{SE3}}^{0,\lambda}$:
\begin{equation}
\begin{aligned}\tilde{\mathcal{M}}_{\mathrm{SE4,UV}}^{0,\lambda} & =-\tilde{\mathcal{M}}_{\mathrm{SE3,UV}}^{0,\lambda}\;.\end{aligned}
\label{eq:MSE4_UV}
\end{equation}
 In $D=4$ dimensions, the subtracted amplitudes are then given by:
\begin{equation}
\begin{aligned}\tilde{\mathcal{M}}_{\mathrm{SE4,sub}}^{0,\lambda} & =\tilde{\mathcal{M}}_{\mathrm{LO2}}^{0,\lambda}\frac{\alpha_{s}C_{F}}{4\pi}\int_{k_{\mathrm{min}}^{+}}^{p_{1}^{+}}\frac{\mathrm{d}k^{+}}{k^{+}}\Big(\frac{k^{+}}{p_{1}^{+}}\Big)^{2}\Big(\big(1-\frac{2p_{1}^{+}}{k^{+}}\big)^{2}+1\Big)\ln\frac{\Delta_{\mathrm{q}}}{\Delta_{\mathrm{UV}}}\;.\end{aligned}
\label{eq:MSE4_sub}
\end{equation}

\subsection{\label{subsec:V}Vertex corrections}

\begin{figure}[t]
\begin{centering}
\includegraphics[scale=0.35]{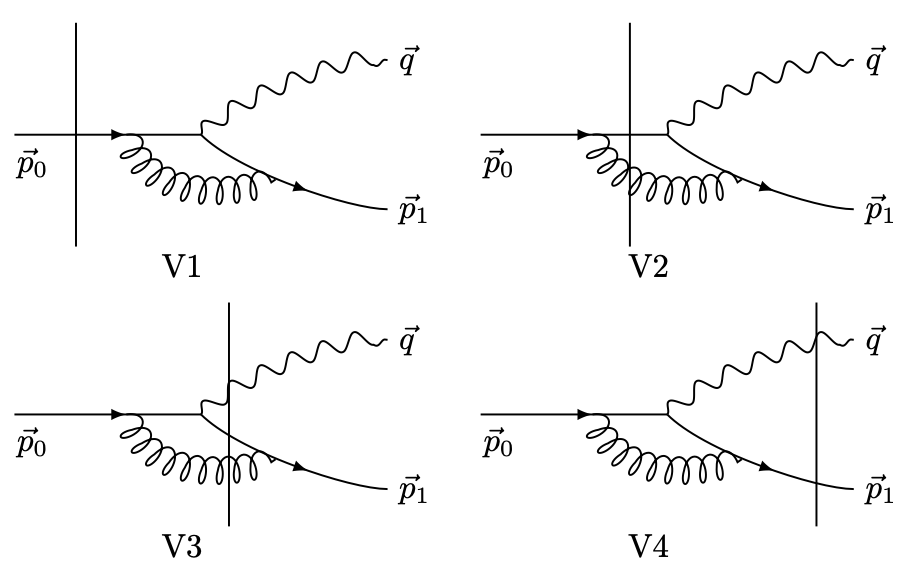}
\par\end{centering}
\caption{\label{fig:V}The four virtual contributions with a vertex correction,
which in two cases ($\mathrm{V2}$ and $\mathrm{V3}$) scatters off
the shockwave. }
\end{figure}

\paragraph{Diagram V1}

The amplitude for Feynman diagram V1 (see fig. \ref{fig:V}) reads,
when the produced virtual photon is longitudinally polarized:
\begin{equation}
\begin{aligned}\tilde{\mathcal{M}}_{\mathrm{V1}}^{0} & =-\frac{\alpha_{s}C_{F}}{p_{0}^{+}\mathbf{P}_{\perp}^{2}+p_{1}^{+}M^{2}}\frac{1}{M}\frac{1}{D-2}\int_{k_{\mathrm{min}}^{+}}^{p_{1}^{+}}\frac{\mathrm{d}k^{+}}{k^{+}}\frac{(k^{+}q^{+})^{2}}{p_{0}^{+}(p_{1}^{+}-k^{+})(p_{0}^{+}-k^{+})}\mathrm{Dirac}_{\mathrm{V}}^{j0j}\\
 & \times\Bigg\{\Bigg[-\Delta_{\mathrm{P}}+\frac{(p_{0}^{+}-k^{+})(p_{1}^{+}-k^{+})}{p_{1}^{+}(q^{+})^{2}}\big(p_{0}^{+}\mathbf{P}_{\perp}^{2}-p_{1}^{+}M^{2}\big)\Bigg]\mathcal{A}_{0}(\Delta_{\mathrm{P}})\\
 & -2\hat{M}^{2}\Big(\frac{k^{+}}{p_{1}^{+}}\Big)^{2}\mathbf{P}_{\perp}^{2}\mathcal{B}_{1}\big(0,\Delta_{\mathrm{P}},\frac{k^{+}}{p_{1}^{+}}\mathbf{P}_{\perp}\big)\Bigg\}\int_{\mathbf{x}}e^{-i\mathbf{k}_{\perp}\cdot\mathbf{x}}\big(U_{\mathbf{x}}-1\big)\;,
\end{aligned}
\label{eq:MV1L_final}
\end{equation}
with the Dirac structure (simplified using the spinor relations in
sec.~\ref{sec:Dirac}):
\begin{equation}
\begin{aligned}\mathrm{Dirac}_{\mathrm{V}}^{\bar{\eta}0\eta^{\prime}} & \equiv\Big[\big(1-2\frac{p_{1}^{+}}{k^{+}}\big)\delta^{\eta\bar{\eta}}\mathds{1}_{4}-i\sigma^{\eta\bar{\eta}}\Big]\Big[\big(2\frac{p_{0}^{+}}{k^{+}}-1\big)\delta^{\eta\eta^{\prime}}\mathds{1}_{4}-i\sigma^{\eta\eta^{\prime}}\Big]\;,\\
\mathrm{Dirac}_{\mathrm{V}}^{j0j} & =-(D-2)\Big(\big(2\frac{p_{1}^{+}}{k^{+}}-1\big)\big(2\frac{p_{0}^{+}}{k^{+}}-1\big)+D-3\Big)\;,
\end{aligned}
\label{eq:DiracVL}
\end{equation}
and the definition:
\begin{equation}
\hat{M}^{2}\equiv\frac{(p_{1}^{+}-k^{+})(p_{0}^{+}-k^{+})}{(q^{+})^{2}}M^{2}+i0^{+}\;.\label{eq:defM}
\end{equation}
Expression (\ref{eq:MV1L_final}) features the UV-divergent integral
$\mathcal{A}_{0}(\Delta_{\mathrm{P}})$, as well as the integral $\mathcal{B}_{1}$
whose expression is a finite albeit complicated sum of hypergeometric
functions, see eqs.~\eqref{eq:IntegralIdentitiesB} and \eqref{eq:B1}
in the appendix.

When the virtual photon is transversely polarized, the amplitude reads:
\begin{equation}
\begin{aligned}\tilde{\mathcal{M}}_{\mathrm{V1}}^{\lambda} & =-\frac{\alpha_{s}C_{F}}{p_{0}^{+}\mathbf{P}_{\perp}^{2}+p_{1}^{+}M^{2}}\mathbf{P}_{\perp}^{\eta^{\prime}}\frac{1}{D-2}\int_{k_{\mathrm{min}}^{+}}^{p_{1}^{+}}\frac{\mathrm{d}k^{+}}{k^{+}}\frac{(k^{+}q^{+})^{2}}{p_{0}^{+}(p_{1}^{+}-k^{+})(p_{0}^{+}-k^{+})}\frac{k^{+}}{p_{1}^{+}}\\
 & \times\Bigg\{\Bigg[\Big(\frac{1}{2}-p_{1}^{+}\frac{p_{0}^{+}-k^{+}}{q^{+}k^{+}}\Big)\delta^{\bar{\lambda}\eta^{\prime}}\mathrm{Dirac}_{\mathrm{V}}^{j\bar{\lambda}j}+\epsilon^{ij}\mathrm{Dirac}_{\mathrm{V}}^{i\bar{\lambda}j}\frac{\epsilon^{\bar{\lambda}\eta^{\prime}}}{D-3}\frac{1}{2}\Bigg]\mathcal{A}_{0}(\Delta_{\mathrm{P}})\\
 & +\Bigg[-\Bigg(\frac{p_{0}^{+}(p_{1}^{+}-k^{+})}{k^{+}q^{+}}\Big(\frac{k^{+}}{p_{1}^{+}}\Big)^{2}\mathbf{P}_{\perp}^{2}+\frac{1}{2}\Big(\Delta_{\mathrm{P}}+\Big(\frac{k^{+}}{p_{1}^{+}}\Big)^{2}\mathbf{P}_{\perp}^{2}\Big)\Bigg)\delta^{\bar{\lambda}\eta^{\prime}}\mathrm{Dirac}_{\mathrm{V}}^{j\bar{\lambda}j}\\
 & +\epsilon^{ij}\mathrm{Dirac}_{\mathrm{V}}^{i\bar{\lambda}j}\frac{\epsilon^{\bar{\lambda}\eta^{\prime}}}{D-3}\frac{1}{2}\Big(\Delta_{\mathrm{P}}+\Big(\frac{k^{+}}{p_{1}^{+}}\Big)^{2}\mathbf{P}_{\perp}^{2}\Big)\Bigg]\mathcal{B}_{1}\big(0,\Delta_{\mathrm{P}},\frac{k^{+}}{p_{1}^{+}}\mathbf{P}_{\perp}\big)\Bigg\}\\
 & \times\int_{\mathbf{x}}e^{-i\mathbf{k}_{\perp}\cdot\mathbf{x}}\big(U_{\mathbf{x}}-1\big)\;,
\end{aligned}
\label{eq:MV1_final}
\end{equation}
with:
\begin{equation}
\begin{aligned}\mathrm{Dirac}_{\mathrm{V}}^{\bar{\eta}\bar{\lambda}\eta^{\prime}} & =\Big[\big(1-2\frac{p_{1}^{+}}{k^{+}}\big)\delta^{\eta\bar{\eta}}\mathds{1}_{4}-i\sigma^{\eta\bar{\eta}}\Big]\Big[\big(1-2\frac{k^{+}-p_{1}^{+}}{q^{+}}\big)\delta^{\lambda\bar{\lambda}}\mathds{1}_{4}-i\sigma^{\lambda\bar{\lambda}}\Big]\\
 & \times\Big[\big(2\frac{p_{0}^{+}}{k^{+}}-1\big)\delta^{\eta\eta^{\prime}}\mathds{1}_{4}-i\sigma^{\eta\eta^{\prime}}\Big]\;,\\
\mathrm{Dirac}_{\mathrm{V}}^{j\bar{\lambda}j} & =(D-2)\big(1-2\frac{k^{+}-p_{1}^{+}}{q^{+}}\big)\Big(\big(1-2\frac{p_{1}^{+}}{k^{+}}\big)\big(2\frac{p_{0}^{+}}{k^{+}}-1\big)-(D-3)\Big)\delta^{\lambda\bar{\lambda}}\mathds{1}_{4}\\
 & -i\Bigg[(D-2)\Big(\big(1-2\frac{p_{1}^{+}}{k^{+}}\big)\big(2\frac{p_{0}^{+}}{k^{+}}-1\big)-(D-3)\Big)+8(D-4\big)\Bigg]\sigma^{\lambda\bar{\lambda}}\;.
\end{aligned}
\label{eq:DiracV}
\end{equation}
We introduce the following counterterms to absorb the UV divergences:
\begin{equation}
\begin{aligned}\tilde{\mathcal{M}}_{\mathrm{V1,\mathrm{UV}}}^{0} & =-\frac{\alpha_{s}C_{F}}{p_{0}^{+}\mathbf{P}_{\perp}^{2}+p_{1}^{+}M^{2}}\frac{1}{M}\frac{1}{D-2}\int_{k_{\mathrm{min}}^{+}}^{p_{1}^{+}}\frac{\mathrm{d}k^{+}}{k^{+}}\frac{(k^{+}q^{+})^{2}}{p_{0}^{+}(p_{1}^{+}-k^{+})(p_{0}^{+}-k^{+})}\mathrm{Dirac}_{\mathrm{V}}^{j0j}\\
 & \!\times\!\Bigg[\!-\!\Delta_{\mathrm{P}}\!+\!\frac{(p_{0}^{+}\!-\!k^{+})(p_{1}^{+}\!-\!k^{+})}{p_{1}^{+}(q^{+})^{2}}\big(p_{0}^{+}\mathbf{P}_{\perp}^{2}\!-\!p_{1}^{+}M^{2}\big)\!\Bigg]\!\mathcal{A}_{0}(\Delta_{\mathrm{UV}})\!\int_{\mathbf{x}}e^{-i\mathbf{k}_{\perp}\cdot\mathbf{x}}\!\big(\!U_{\mathbf{x}}\!-\!1\!\big)\!\,,
\end{aligned}
\label{eq:MV1L_UV}
\end{equation}
and
\begin{equation}
\begin{aligned}\tilde{\mathcal{M}}_{\mathrm{V1,UV}}^{\lambda} & =-\frac{\alpha_{s}C_{F}}{p_{0}^{+}\mathbf{P}_{\perp}^{2}+p_{1}^{+}M^{2}}\mathbf{P}_{\perp}^{\eta^{\prime}}\frac{1}{D-2}\int_{k_{\mathrm{min}}^{+}}^{p_{1}^{+}}\frac{\mathrm{d}k^{+}}{k^{+}}\frac{(k^{+}q^{+})^{2}}{p_{0}^{+}(p_{1}^{+}-k^{+})(p_{0}^{+}-k^{+})}\frac{k^{+}}{p_{1}^{+}}\\
 & \times\Bigg[\Big(\frac{1}{2}-p_{1}^{+}\frac{p_{0}^{+}-k^{+}}{q^{+}k^{+}}\Big)\delta^{\bar{\lambda}\eta^{\prime}}\mathrm{Dirac}_{\mathrm{V}}^{j\bar{\lambda}j}+\epsilon^{ij}\mathrm{Dirac}_{\mathrm{V}}^{i\bar{\lambda}j}\frac{\epsilon^{\bar{\lambda}\eta^{\prime}}}{D-3}\frac{1}{2}\Bigg]\mathcal{A}_{0}(\Delta_{\mathrm{UV}})\\
 & \times\int_{\mathbf{x}}e^{-i\mathbf{k}_{\perp}\cdot\mathbf{x}}\big(U_{\mathbf{x}}-1\big)\;.
\end{aligned}
\label{eq:MV1T_UV}
\end{equation}
The subtracted amplitudes, therefore, become:
\begin{equation}
\begin{aligned}\tilde{\mathcal{M}}_{\mathrm{V1,sub}}^{0} & \!=\!\frac{\alpha_{s}C_{F}}{p_{0}^{+}\mathbf{P}_{\perp}^{2}\!+\!p_{1}^{+}M^{2}}\!\frac{1}{M}\!\int_{k_{\mathrm{min}}^{+}}^{p_{1}^{+}}\!\frac{\mathrm{d}k^{+}}{k^{+}}\!\frac{(k^{+}q^{+})^{2}}{p_{0}^{+}(p_{1}^{+}\!-\!k^{+})(p_{0}^{+}\!-\!k^{+})}\!\Big(\!\big(2\frac{p_{1}^{+}}{k^{+}}-\!1\!\big)\!\big(2\frac{p_{0}^{+}}{k^{+}}\!-\!1\big)\!+\!1\!\Big)\!\\
 & \times\Bigg\{\Bigg[-\Delta_{\mathrm{P}}+\frac{(p_{0}^{+}-k^{+})(p_{1}^{+}-k^{+})}{p_{1}^{+}(q^{+})^{2}}\big(p_{0}^{+}\mathbf{P}_{\perp}^{2}-p_{1}^{+}M^{2}\big)\Bigg]\frac{1}{4\pi}\ln\frac{\Delta_{\mathrm{UV}}}{\Delta_{\mathrm{P}}}\\
 & -2\hat{M}^{2}\Big(\frac{k^{+}}{p_{1}^{+}}\Big)^{2}\mathbf{P}_{\perp}^{2}\mathcal{B}_{1}\big(0,\Delta_{\mathrm{P}},\frac{k^{+}}{p_{1}^{+}}\mathbf{P}_{\perp}\big)\Bigg\}\int_{\mathbf{x}}e^{-i\mathbf{k}_{\perp}\cdot\mathbf{x}}\big(U_{\mathbf{x}}-1\big)\;,
\end{aligned}
\label{eq:MV1L_sub}
\end{equation}
and:
\begin{equation}
\begin{aligned}\tilde{\mathcal{M}}_{\mathrm{V1,sub}}^{\lambda} & =-\frac{\alpha_{s}C_{F}}{p_{0}^{+}\mathbf{P}_{\perp}^{2}+p_{1}^{+}M^{2}}\mathbf{P}_{\perp}^{\eta^{\prime}}\frac{1}{D-2}\int_{k_{\mathrm{min}}^{+}}^{p_{1}^{+}}\frac{\mathrm{d}k^{+}}{k^{+}}\frac{(k^{+}q^{+})^{2}}{p_{0}^{+}(p_{1}^{+}-k^{+})(p_{0}^{+}-k^{+})}\frac{k^{+}}{p_{1}^{+}}\\
 & \times\Bigg\{\Bigg[\Big(\frac{1}{2}-p_{1}^{+}\frac{p_{0}^{+}-k^{+}}{q^{+}k^{+}}\Big)\delta^{\bar{\lambda}\eta^{\prime}}\mathrm{Dirac}_{\mathrm{V}}^{j\bar{\lambda}j}+\epsilon^{ij}\mathrm{Dirac}_{\mathrm{V}}^{i\bar{\lambda}j}\frac{\epsilon^{\bar{\lambda}\eta^{\prime}}}{D-3}\frac{1}{2}\Bigg]\frac{1}{4\pi}\ln\frac{\Delta_{\mathrm{UV}}}{\Delta_{\mathrm{P}}}\\
 & +\Bigg[-\Bigg(\frac{p_{0}^{+}(p_{1}^{+}-k^{+})}{k^{+}q^{+}}\Big(\frac{k^{+}}{p_{1}^{+}}\Big)^{2}\mathbf{P}_{\perp}^{2}+\frac{1}{2}\Big(\Delta_{\mathrm{P}}+\Big(\frac{k^{+}}{p_{1}^{+}}\Big)^{2}\mathbf{P}_{\perp}^{2}\Big)\Bigg)\delta^{\bar{\lambda}\eta^{\prime}}\mathrm{Dirac}_{\mathrm{V}}^{j\bar{\lambda}j}\\
 & +\epsilon^{ij}\mathrm{Dirac}_{\mathrm{V}}^{i\bar{\lambda}j}\frac{\epsilon^{\bar{\lambda}\eta^{\prime}}}{D-3}\frac{1}{2}\Big(\Delta_{\mathrm{P}}+\Big(\frac{k^{+}}{p_{1}^{+}}\Big)^{2}\mathbf{P}_{\perp}^{2}\Big)\Bigg]\mathcal{B}_{1}\big(0,\Delta_{\mathrm{P}},\frac{k^{+}}{p_{1}^{+}}\mathbf{P}_{\perp}\big)\Bigg\}\\
 & \times\int_{\mathbf{x}}e^{-i\mathbf{k}_{\perp}\cdot\mathbf{x}}\big(U_{\mathbf{x}}-1\big)\;.
\end{aligned}
\label{eq:MV1T_sub}
\end{equation}

\paragraph{Diagram V2}

The amplitude for Feynman diagram V2 (see fig. \ref{fig:V}) reads,
in case of a longitudinally polarized virtual photon:
\begin{equation}
\begin{aligned}\tilde{\mathcal{M}}_{\mathrm{V2}}^{0} & =\frac{\alpha_{s}}{M}\int_{k_{\mathrm{min}}^{+}}^{p_{1}^{+}}\frac{\mathrm{d}k^{+}}{k^{+}}\frac{(k^{+})^{3}q^{+}}{(p_{0}^{+})^{2}p_{1}^{+}(p_{1}^{+}-k^{+})}\mathrm{Dirac}_{\mathrm{V}}^{\bar{\eta}0\eta^{\prime}}\\
 & \times\int_{\mathbf{x},\mathbf{z}}\int_{\mathbf{k}_{1}}e^{-i\mathbf{k}_{1}\cdot(\mathbf{x}-\mathbf{z})}\frac{\mathbf{k}_{1}^{\eta^{\prime}}}{\mathbf{k}_{1}^{2}}\int_{\boldsymbol{\ell}}\frac{\boldsymbol{\ell}^{\bar{\eta}}}{\boldsymbol{\ell}^{2}}\frac{e^{i\big(\boldsymbol{\ell}+\frac{k^{+}}{p_{1}^{+}}\mathbf{P}_{\perp}\big)\cdot(\mathbf{x}-\mathbf{z})}}{\big(\boldsymbol{\ell}+\frac{k^{+}}{p_{1}^{+}}\mathbf{P}_{\perp}\big)^{2}+\Delta_{\mathrm{P}}}\\
 & \times\Bigg[\big(\boldsymbol{\ell}-\frac{p_{0}^{+}(p_{1}^{+}-k^{+})}{q^{+}p_{1}^{+}}\mathbf{P}_{\perp}\big)^{2}-\frac{(p_{1}^{+}-k^{+})(p_{0}^{+}-k^{+})}{(q^{+})^{2}}M^{2}\Bigg]\\
 & \times e^{-i\mathbf{k}_{\perp}\cdot\Big(\frac{p_{0}^{+}-k^{+}}{p_{0}^{+}}\mathbf{x}+\frac{k^{+}}{p_{0}^{+}}\mathbf{z}\Big)}[U_{\mathbf{z}}t^{c}U_{\mathbf{z}}^{\dagger}U_{\mathbf{x}}t^{c}-C_{F}]\;.
\end{aligned}
\label{eq:MV2L}
\end{equation}
To the best of our knowledge, the transverse integral over $\boldsymbol{\ell}$
does not allow a general analytic solution.\footnote{See \cite{Caucal:2021ent,Taels:2022tza} for very similar integrals
in NLO $\gamma^{(*)}+A\to\mathrm{dijet}+X$ calculations.} However, we expect amplitude $\tilde{\mathcal{M}}_{\mathrm{V2}}^{0}$
to contain a UV divergence in the limit $\mathbf{z}\to\mathbf{x}$.
With the same procedure as for the counterterms in sec. \ref{subsec:SE},
namely setting $\mathbf{z}=\mathbf{x}$ in the phase and the Wilson-line
structures, but not within the divergent integrations, the integral
over $\mathbf{z}$ can be carried out, after which one obtains:
\begin{equation}
\begin{aligned}\lim_{\mathbf{z}\to\mathbf{x}}\tilde{\mathcal{M}}_{\mathrm{V2}}^{0} & =\frac{\alpha_{s}C_{F}}{M}\frac{1}{D-2}\int_{k_{\mathrm{min}}^{+}}^{p_{1}^{+}}\frac{\mathrm{d}k^{+}}{k^{+}}\frac{(k^{+})^{3}q^{+}}{(p_{0}^{+})^{2}p_{1}^{+}(p_{1}^{+}-k^{+})}\mathrm{Dirac}_{\mathrm{V}}^{j0j}\\
 & \times\int_{\boldsymbol{\ell}}\frac{\boldsymbol{\ell}\cdot\big(\boldsymbol{\ell}+\frac{k^{+}}{p_{1}^{+}}\mathbf{P}_{\perp}\big)}{\boldsymbol{\ell}^{2}\big(\boldsymbol{\ell}+\frac{k^{+}}{p_{1}^{+}}\mathbf{P}_{\perp}\big)^{2}}\frac{1}{\big(\boldsymbol{\ell}+\frac{k^{+}}{p_{1}^{+}}\mathbf{P}_{\perp}\big)^{2}+\Delta_{\mathrm{P}}}\\
 & \times\Bigg[\big(\boldsymbol{\ell}-\frac{p_{0}^{+}(p_{1}^{+}-k^{+})}{q^{+}p_{1}^{+}}\mathbf{P}_{\perp}\big)^{2}-\frac{(p_{1}^{+}-k^{+})(p_{0}^{+}-k^{+})}{(q^{+})^{2}}M^{2}\Bigg]\\
 & \times\int_{\mathbf{x}}e^{-i\mathbf{k}_{\perp}\cdot\mathbf{x}}\big(U_{\mathbf{x}}-1\big)\;.
\end{aligned}
\label{eq:MV2L_limit}
\end{equation}
The loop integral can now be evaluated with the help of the identities
(\ref{eq:IntegralIdentitiesC}):
\begin{equation}
\begin{aligned}\mathcal{I}_{\mathrm{V2}}^{0} & =\int_{\boldsymbol{\ell}}\frac{\boldsymbol{\ell}\cdot\big(\boldsymbol{\ell}+\frac{k^{+}}{p_{1}^{+}}\mathbf{P}_{\perp}\big)}{\boldsymbol{\ell}^{2}\big(\boldsymbol{\ell}+\frac{k^{+}}{p_{1}^{+}}\mathbf{P}_{\perp}\big)^{2}}\frac{1}{\big(\boldsymbol{\ell}+\frac{k^{+}}{p_{1}^{+}}\mathbf{P}_{\perp}\big)^{2}+\Delta_{\mathrm{P}}}\\
 & \times\Bigg[\big(\boldsymbol{\ell}-\frac{p_{0}^{+}(p_{1}^{+}-k^{+})}{q^{+}p_{1}^{+}}\mathbf{P}_{\perp}\big)^{2}-\frac{(p_{1}^{+}-k^{+})(p_{0}^{+}-k^{+})}{(q^{+})^{2}}M^{2}\Bigg]\\
 & =\mathcal{A}_{0}(\Delta_{\mathrm{P}})-p_{0}^{+}\frac{p_{1}^{+}-k^{+}}{k^{+}q^{+}}\Big(\frac{k^{+}}{p_{1}^{+}}\Big)^{2}\mathbf{P}_{\perp}^{2}\mathcal{B}_{1}\big(0,\Delta_{\mathrm{P}},-\frac{k^{+}}{p_{1}^{+}}\mathbf{P}_{\perp}\big)\\
 & +\frac{1}{2}\frac{(p_{1}^{+}-k^{+})(p_{0}^{+}-k^{+})}{(q^{+})^{2}p_{1}^{+}}\Big(p_{0}^{+}\mathbf{P}_{\perp}^{2}-p_{1}^{+}M^{2}\Big)\\
 & \times\Bigg[\mathcal{B}_{0}(\Delta_{\mathrm{P}})+\mathcal{B}_{0}\big(\Delta_{\mathrm{P}},-\frac{k^{+}}{p_{1}^{+}}\mathbf{P}_{\perp}\big)-\Big(\frac{k^{+}}{p_{1}^{+}}\Big)^{2}\mathbf{P}_{\perp}^{2}\mathcal{C}_{0}\big(\Delta_{\mathrm{P}},-\frac{k^{+}}{p_{1}^{+}}\mathbf{P}_{\perp}\big)\Big)\Bigg]\;.
\end{aligned}
\label{eq:integralV2L}
\end{equation}
The infrared poles contained in the structures $\mathcal{B}_{0}(\Delta_{\mathrm{P}})$,
$\mathcal{B}_{0}\big(0,\Delta_{\mathrm{P}},-\frac{k^{+}}{p_{1}^{+}}\mathbf{P}_{\perp}\big)$,
and $\mathcal{C}_{0}\big(\Delta_{\mathrm{P}},-\frac{k^{+}}{p_{1}^{+}}\mathbf{P}_{\perp}\big)$
all cancel (see eqs. (\ref{eq:B0_explicit}) and (\ref{eq:C0})).
Since $\mathcal{B}_{1}\big(0,\Delta_{\mathrm{P}},-\frac{k^{+}}{p_{1}^{+}}\mathbf{P}_{\perp}\big)$
is finite, the only remaining singularity is the ultraviolet one contained
in $\mathcal{A}_{0}(\Delta_{\mathrm{P}})$. We can, therefore, define the counterterm
\begin{equation}
\begin{aligned}\tilde{\mathcal{M}}_{\mathrm{V2,UV}}^{0} & =\frac{\alpha_{s}C_{F}}{M}\frac{1}{D-2}\int_{k_{\mathrm{min}}^{+}}^{p_{1}^{+}}\frac{\mathrm{d}k^{+}}{k^{+}}\frac{(k^{+})^{3}q^{+}}{(p_{0}^{+})^{2}p_{1}^{+}(p_{1}^{+}-k^{+})}\mathrm{Dirac}_{\mathrm{V}}^{j0j}\\
 & \times\mathcal{A}_{0}(\Delta_{\mathrm{UV}})\int_{\mathbf{x}}e^{-i\mathbf{k}_{\perp}\cdot\mathbf{x}}\big(U_{\mathbf{x}}-1\big)\;.
\end{aligned}
\label{eq:MV2L_UV}
\end{equation}
Finally, the subtracted longitudinal amplitude reads
\begin{equation}
\begin{aligned}\tilde{\mathcal{M}}_{\mathrm{V2,sub}}^{0} & =\frac{\alpha_{s}}{M}\int_{k_{\mathrm{min}}^{+}}^{p_{1}^{+}}\frac{\mathrm{d}k^{+}}{k^{+}}\frac{(k^{+})^{3}q^{+}}{(p_{0}^{+})^{2}p_{1}^{+}(p_{1}^{+}-k^{+})}\mathrm{Dirac}_{\mathrm{V}}^{\bar{\eta}0\eta^{\prime}}\\
 & \times\Bigg[\int_{\mathbf{x},\mathbf{z}}iA^{\eta^{\prime}}(\mathbf{x}-\mathbf{z})\int_{\boldsymbol{\ell}}\frac{\boldsymbol{\ell}^{\bar{\eta}}}{\boldsymbol{\ell}^{2}}\frac{e^{i\big(\boldsymbol{\ell}+\frac{k^{+}}{p_{1}^{+}}\mathbf{P}_{\perp}\big)\cdot(\mathbf{x}-\mathbf{z})}}{\big(\boldsymbol{\ell}+\frac{k^{+}}{p_{1}^{+}}\mathbf{P}_{\perp}\big)^{2}+\Delta_{\mathrm{P}}}\\
 & \times\Big(\big(\boldsymbol{\ell}-\frac{p_{0}^{+}(p_{1}^{+}-k^{+})}{q^{+}p_{1}^{+}}\mathbf{P}_{\perp}\big)^{2}-\frac{(p_{1}^{+}-k^{+})(p_{0}^{+}-k^{+})}{(q^{+})^{2}}M^{2}\Big)\\
 & \times e^{-i\mathbf{k}_{\perp}\cdot\Big(\frac{p_{0}^{+}-k^{+}}{p_{0}^{+}}\mathbf{x}+\frac{k^{+}}{p_{0}^{+}}\mathbf{z}\Big)}[U_{\mathbf{z}}t^{c}U_{\mathbf{z}}^{\dagger}U_{\mathbf{x}}t^{c}-C_{F}]\\
 & -\frac{\delta^{\bar{\eta}\eta^{\prime}}}{2}\mathcal{A}_{0}(\Delta_{\mathrm{UV}})\int_{\mathbf{x}}e^{-i\mathbf{k}_{\perp}\cdot\mathbf{x}}C_{F}\big(U_{\mathbf{x}}-1\big)\Bigg]\;.
\end{aligned}
\label{eq:MV2L_sub}
\end{equation}
The amplitude for the emission of a transversely polarized virtual
photon is equal to:
\begin{equation}
\begin{aligned}\tilde{\mathcal{M}}_{\mathrm{V2}}^{\lambda} & =\alpha_{s}\int_{k_{\mathrm{min}}^{+}}^{p_{1}^{+}}\frac{\mathrm{d}k^{+}}{k^{+}}\frac{(k^{+})^{3}q^{+}}{p_{1}^{+}(p_{0}^{+})^{2}(p_{1}^{+}-k^{+})}\mathrm{Dirac}_{\mathrm{V}}^{\bar{\eta}\bar{\lambda}\eta^{\prime}}\\
 & \times\int_{\mathbf{x},\mathbf{z}}iA^{\eta^{\prime}}(\mathbf{x}-\mathbf{z})\int_{\boldsymbol{\ell}}e^{i\boldsymbol{\ell}\cdot(\mathbf{x}-\mathbf{z})}\frac{\boldsymbol{\ell}^{\bar{\eta}}-\frac{k^{+}}{p_{1}^{+}}\mathbf{P}_{\perp}^{\bar{\eta}}}{\big(\boldsymbol{\ell}-\frac{k^{+}}{p_{1}^{+}}\mathbf{P}_{\perp}\big)^{2}}\frac{\boldsymbol{\ell}^{\bar{\lambda}}-\frac{p_{0}^{+}-k^{+}}{q^{+}}\mathbf{P}_{\perp}^{\bar{\lambda}}}{\boldsymbol{\ell}^{2}+\Delta_{\mathrm{P}}}\\
 & \times e^{-i\mathbf{k}_{\perp}\cdot\Big(\frac{p_{0}^{+}-k^{+}}{p_{0}^{+}}\mathbf{x}+\frac{k^{+}}{p_{0}^{+}}\mathbf{z}\Big)}[U_{\mathbf{z}}t^{c}U_{\mathbf{z}}^{\dagger}U_{\mathbf{x}}t^{c}-C_{F}]\;.
\end{aligned}
\label{eq:MV2}
\end{equation}
Like in the longitudinal case (\ref{eq:MV2L}), we are not aware of
an analytic solution for the integral over $\boldsymbol{\ell}$. The
integral, however, can be shown to be free from UV divergences (see
appendix~\ref{sec:MV2_UV}.

\paragraph{Diagram V3}

We obtain for diagram V3 (see fig. \ref{fig:V}):
\begin{equation}
\begin{aligned}\tilde{\mathcal{M}}_{\mathrm{V3}}^{0} & =-\frac{\alpha_{s}}{M}\int_{k_{\mathrm{min}}^{+}}^{p_{1}^{+}}\frac{\mathrm{d}k^{+}}{k^{+}}\frac{(k^{+})^{3}q^{+}}{(p_{1}^{+})^{2}p_{0}^{+}(p_{0}^{+}-k^{+})}\mathrm{Dirac}_{\mathrm{V}}^{\bar{\eta}0\eta^{\prime}}\\
 & \times\int_{\mathbf{x},\mathbf{z}}iA^{\bar{\eta}}(\mathbf{x}-\mathbf{z})\int_{\boldsymbol{\mathbf{\ell}}}\frac{\boldsymbol{\ell}^{\eta^{\prime}}}{\boldsymbol{\ell}^{2}}\frac{e^{i\big(\boldsymbol{\ell}+\frac{k^{+}}{p_{1}^{+}}\mathbf{q}\big)\cdot(\mathbf{z}-\mathbf{x})}}{\big(\boldsymbol{\ell}+\frac{k^{+}}{p_{1}^{+}}\mathbf{q}\big)^{2}+\Delta_{\mathrm{q}}}\\
 & \times\Bigg[\big(\boldsymbol{\ell}+\frac{p_{0}^{+}-k^{+}}{q^{+}}\mathbf{q}\big)^{2}-\frac{(p_{1}^{+}-k^{+})(p_{0}^{+}-k^{+})}{(q^{+})^{2}}M^{2}\Bigg]\\
 & \times e^{-i\mathbf{k}_{\perp}\cdot\Big(\frac{p_{1}^{+}-k^{+}}{p_{1}^{+}}\mathbf{x}+\frac{k^{+}}{p_{1}^{+}}\mathbf{z}\Big)}[U_{\mathbf{z}}t^{c}U_{\mathbf{z}}^{\dagger}U_{\mathbf{x}}t^{c}-C_{F}]\;.
\end{aligned}
\label{eq:MV3L}
\end{equation}
Similarly to the amplitude $\tilde{\mathcal{M}}_{\mathrm{V2}}^{0}$,
eq. (\ref{eq:MV3L}) contains a UV divergence in the limit $\mathbf{z}\to\mathbf{x}$:
\begin{equation}
\begin{aligned}\lim_{\mathbf{z}\to\mathbf{x}}\tilde{\mathcal{M}}_{\mathrm{V3}}^{0} & =\frac{\alpha_{s}C_{F}}{M}\frac{1}{D-2}\int_{k_{\mathrm{min}}^{+}}^{p_{1}^{+}}\frac{\mathrm{d}k^{+}}{k^{+}}\frac{(k^{+})^{3}q^{+}}{(p_{1}^{+})^{2}p_{0}^{+}(p_{0}^{+}-k^{+})}\mathrm{Dirac}_{\mathrm{V}}^{j0j}\\
 & \times\int_{\boldsymbol{\mathbf{\ell}}}\frac{\boldsymbol{\mathbf{\ell}}\cdot\big(\boldsymbol{\mathbf{\ell}}+\frac{k^{+}}{p_{1}^{+}}\mathbf{q}\big)}{\boldsymbol{\mathbf{\ell}}^{2}\big(\boldsymbol{\mathbf{\ell}}+\frac{k^{+}}{p_{1}^{+}}\mathbf{q}\big)^{2}}\frac{1}{\big(\boldsymbol{\mathbf{\ell}}+\frac{k^{+}}{p_{1}^{+}}\mathbf{q}\big)^{2}+\Delta_{\mathrm{q}}}\\
 & \times\Bigg[\big(\boldsymbol{\mathbf{\ell}}+\frac{p_{0}^{+}-k^{+}}{q^{+}}\mathbf{q}\big)^{2}-\frac{(p_{1}^{+}-k^{+})(p_{0}^{+}-k^{+})}{(q^{+})^{2}}M^{2}\Bigg]\\
 & \times\int_{\mathbf{x}}e^{-i\mathbf{k}_{\perp}\cdot\mathbf{x}}\big(U_{\mathbf{x}}-1\big)\;.
\end{aligned}
\label{eq:MV3L_limit}
\end{equation}
The integral over $\boldsymbol{\ell}$ can be calculated with the
help of the identities \ref{eq:IntegralIdentitiesB} and (\ref{eq:IntegralIdentitiesC})
in the appendix:
\begin{equation}
\begin{aligned}\mathcal{I}_{\mathrm{V3}}^{0} & =\int_{\boldsymbol{\mathbf{\ell}}}\frac{\boldsymbol{\mathbf{\ell}}\cdot\big(\boldsymbol{\mathbf{\ell}}+\frac{k^{+}}{p_{1}^{+}}\mathbf{q}\big)}{\boldsymbol{\mathbf{\ell}}^{2}\big(\boldsymbol{\mathbf{\ell}}+\frac{k^{+}}{p_{1}^{+}}\mathbf{q}\big)^{2}}\frac{1}{\big(\boldsymbol{\mathbf{\ell}}+\frac{k^{+}}{p_{1}^{+}}\mathbf{q}\big)^{2}+\Delta_{\mathrm{q}}}\\
 & \times\Bigg[\big(\boldsymbol{\mathbf{\ell}}+\frac{p_{0}^{+}-k^{+}}{q^{+}}\mathbf{q}\big)^{2}-\frac{(p_{1}^{+}-k^{+})(p_{0}^{+}-k^{+})}{(q^{+})^{2}}M^{2}\Bigg]\\
 & =\mathcal{A}_{0}(\Delta_{\mathrm{q}})+p_{1}^{+}\frac{p_{0}^{+}-k^{+}}{q^{+}k^{+}}\Big(\frac{k^{+}}{p_{1}^{+}}\Big)^{2}\mathbf{q}^{2}\mathcal{B}_{1}\big(0,\Delta_{\mathrm{q}},\frac{k^{+}}{p_{1}^{+}}\mathbf{q}\big)\\
 & +\frac{1}{2}\frac{(p_{1}^{+}-k^{+})(p_{0}^{+}-k^{+})}{(q^{+})^{2}p_{1}^{+}}\Big(p_{0}^{+}\mathbf{q}^{2}-p_{1}^{+}M^{2}\Big)\\
 & \times\Bigg[\mathcal{B}_{0}(\Delta_{\mathrm{q}})+\mathcal{B}_{0}\big(\Delta_{\mathrm{q}},\frac{k^{+}}{p_{1}^{+}}\mathbf{q}\big)-\Big(\frac{k^{+}}{p_{1}^{+}}\Big)^{2}\mathbf{q}^{2}\mathcal{C}_{0}\big(\Delta_{\mathrm{q}},\frac{k^{+}}{p_{1}^{+}}\mathbf{q}\big)\Bigg]\;.
\end{aligned}
\label{eq:IV3L}
\end{equation}
Due to the cancellation of IR divergences in the last line of (\ref{eq:IV3L}),
the only remaining divergence is the UV-one in $\mathcal{A}_{0}(\Delta_{\mathrm{q}})$.
We can, therefore, define the counterterm:
\begin{equation}
\begin{aligned}\tilde{\mathcal{M}}_{\mathrm{V3,UV}}^{0} & =\frac{\alpha_{s}C_{F}}{M}\frac{1}{D-2}\int_{k_{\mathrm{min}}^{+}}^{p_{1}^{+}}\frac{\mathrm{d}k^{+}}{k^{+}}\frac{(k^{+})^{3}q^{+}}{(p_{1}^{+})^{2}p_{0}^{+}(p_{0}^{+}-k^{+})}\mathrm{Dirac}_{\mathrm{V}}^{j0j}\\
 & \times\mathcal{A}_{0}(\Delta_{\mathrm{UV}})\int_{\mathbf{x}}e^{-i\mathbf{k}_{\perp}\cdot\mathbf{x}}\big(U_{\mathbf{x}}-1\big)\;,
\end{aligned}
\label{eq:MV3L_UV}
\end{equation}
which results in the following UV-subtracted amplitude:
\begin{equation}
\begin{aligned}\tilde{\mathcal{M}}_{\mathrm{V3,sub}}^{0} & =-\frac{\alpha_{s}}{M}\int_{0}^{p_{1}^{+}}\frac{\mathrm{d}k^{+}}{k^{+}}\frac{(k^{+})^{3}q^{+}}{(p_{1}^{+})^{2}p_{0}^{+}(p_{0}^{+}-k^{+})}\mathrm{Dirac}_{\mathrm{V}}^{\bar{\eta}0\eta^{\prime}}\\
 & \times\Bigg[\int_{\mathbf{x},\mathbf{z}}iA^{\bar{\eta}}(\mathbf{x}-\mathbf{z})\int_{\boldsymbol{\mathbf{\ell}}}\frac{\boldsymbol{\ell}^{\eta^{\prime}}}{\boldsymbol{\ell}^{2}}\frac{e^{i\big(\boldsymbol{\ell}+\frac{k^{+}}{p_{1}^{+}}\mathbf{q}\big)\cdot(\mathbf{z}-\mathbf{x})}}{\big(\boldsymbol{\ell}+\frac{k^{+}}{p_{1}^{+}}\mathbf{q}\big)^{2}+\Delta_{\mathrm{q}}}\\
 & \times\Big(\big(\boldsymbol{\ell}+\frac{p_{0}^{+}-k^{+}}{q^{+}}\mathbf{q}\big)^{2}-\frac{(p_{1}^{+}-k^{+})(p_{0}^{+}-k^{+})}{(q^{+})^{2}}M^{2}\Big)\\
 & \times e^{-i\mathbf{k}_{\perp}\cdot\Big(\frac{p_{1}^{+}-k^{+}}{p_{1}^{+}}\mathbf{x}+\frac{k^{+}}{p_{1}^{+}}\mathbf{z}\Big)}[U_{\mathbf{z}}t^{c}U_{\mathbf{z}}^{\dagger}U_{\mathbf{x}}t^{c}-C_{F}]\\
 & +\frac{\delta^{\bar{\eta}\eta^{\prime}}}{2}\mathcal{A}_{0}(\Delta_{\mathrm{UV}})\int_{\mathbf{x}}e^{-i\mathbf{k}_{\perp}\cdot\mathbf{x}}C_{F}\big(U_{\mathbf{x}}-1\big)\Bigg]\;.
\end{aligned}
\label{eq:MV3L_sub}
\end{equation}
In the transversely polarized case, the amplitude is given by:
\begin{equation}
\begin{aligned}\tilde{\mathcal{M}}_{\mathrm{V3}}^{\lambda} & =\alpha_{s}\int_{k_{\mathrm{min}}^{+}}^{p_{1}^{+}}\frac{\mathrm{d}k^{+}}{k^{+}}\frac{q^{+}(k^{+})^{3}}{p_{0}^{+}(p_{1}^{+})^{2}(p_{0}^{+}-k^{+})}\mathrm{Dirac}_{\mathrm{V}}^{\bar{\eta}\bar{\lambda}\eta^{\prime}}\\
 & \times\int_{\mathbf{x},\mathbf{z}}iA^{\bar{\eta}}(\mathbf{z}-\mathbf{x})\int_{\boldsymbol{\ell}}e^{-i\big(\boldsymbol{\ell}+\frac{k^{+}}{p_{1}^{+}}\mathbf{q}\big)\cdot(\mathbf{x}-\mathbf{z})}\frac{\boldsymbol{\ell}^{\eta^{\prime}}}{\boldsymbol{\ell}^{2}}\frac{\boldsymbol{\ell}^{\bar{\lambda}}+\frac{p_{0}^{+}-k^{+}}{q^{+}}\mathbf{q}^{\bar{\lambda}}}{\big(\boldsymbol{\ell}+\frac{k^{+}}{p_{1}^{+}}\mathbf{q}\big)^{2}+\Delta_{\mathrm{q}}}\\
 & \times e^{-i\mathbf{k}_{\perp}\cdot\Big(\frac{p_{1}^{+}-k^{+}}{p_{1}^{+}}\mathbf{x}+\frac{k^{+}}{p_{1}^{+}}\mathbf{z}\Big)}[U_{\mathbf{z}}t^{c}U_{\mathbf{z}}^{\dagger}U_{\mathbf{x}}t^{c}-C_{F}]\;.
\end{aligned}
\label{eq:MV3}
\end{equation}
We can show that, although we cannot analytically
evaluate the above transverse integration, it is free from UV divergences. 

\paragraph{Diagram V4}

One obtains for the vertex correction where the outgoing quark scatters
off the shockwave \ref{fig:V}:
\begin{equation}
\begin{aligned}\tilde{\mathcal{M}}_{\mathrm{V4}}^{0} & =\frac{1}{p_{0}^{+}\mathbf{q}^{2}+p_{1}^{+}M^{2}}\frac{\alpha_{s}C_{F}}{M}\frac{1}{D-2}\int_{k_{\mathrm{min}}^{+}}^{p_{1}^{+}}\frac{\mathrm{d}k^{+}}{k^{+}}\frac{(q^{+})^{2}(k^{+})^{2}}{p_{0}^{+}(p_{1}^{+}-k^{+})(p_{0}^{+}-k^{+})}\mathrm{Dirac}_{\mathrm{V}}^{j0j}\\
 & \times\Bigg\{\Bigg[-\Delta_{\mathrm{q}}+\frac{(p_{1}^{+}-k^{+})(p_{0}^{+}-k^{+})}{p_{1}^{+}(q^{+})^{2}}\Big(p_{0}^{+}\mathbf{q}^{2}-p_{1}^{+}M^{2}\Big)\Bigg]\mathcal{A}_{0}(\Delta_{\mathrm{q}})\\
 & -2\Big(\frac{k^{+}}{p_{1}^{+}}\Big)^{2}\mathbf{q}^{2}\hat{M}^{2}\mathcal{B}_{1}\big(0,\Delta_{\mathrm{q}},\frac{k^{+}}{p_{1}^{+}}\mathbf{q}\big)\Bigg\}\\
 & \times\int_{\mathbf{x}}e^{-i\mathbf{k}_{\perp}\cdot\mathbf{x}}\big(U_{\mathbf{x}}-1\big)\;,
\end{aligned}
\label{eq:MV4L_final}
\end{equation}
and
\begin{equation}
\begin{aligned}\tilde{\mathcal{M}}_{\mathrm{V4}}^{\lambda} & =\frac{\mathbf{q}^{\rho}}{p_{0}^{+}\mathbf{q}^{2}+p_{1}^{+}M^{2}}\frac{\alpha_{s}C_{F}}{D-2}\int_{k_{\mathrm{min}}^{+}}^{p_{1}^{+}}\frac{\mathrm{d}k^{+}}{k^{+}}\frac{k^{+}}{p_{1}^{+}}\frac{(k^{+}q^{+})^{2}}{p_{0}^{+}(p_{1}^{+}-k^{+})(p_{0}^{+}-k^{+})}\\
 & \times\Bigg\{\Bigg[\Big(\frac{1}{2}+p_{0}^{+}\frac{p_{1}^{+}-k^{+}}{k^{+}q^{+}}\Big)\mathrm{Dirac}_{\mathrm{V}}^{j\rho j}+\epsilon^{\rho\bar{\lambda}}\frac{1}{2}\frac{1}{D-3}\epsilon^{ij}\mathrm{Dirac}_{\mathrm{V}}^{i\bar{\lambda}j}\Bigg]\mathcal{A}_{0}(\Delta_{\mathrm{q}})\\
 & +\Bigg[\Big(\frac{1}{2}\big(\Delta_{\mathrm{q}}+\Big(\frac{k^{+}}{p_{1}^{+}}\Big)^{2}\mathbf{q}^{2}\big)-\frac{k^{+}(p_{1}^{+}-k^{+})M^{2}}{p_{1}^{+}q^{+}}\Big)\mathrm{Dirac}_{\mathrm{V}}^{j\rho j}\\
 & +\epsilon^{\rho\bar{\lambda}}\frac{1}{D-3}\frac{1}{2}\big(\Delta_{\mathrm{q}}+\Big(\frac{k^{+}}{p_{1}^{+}}\Big)^{2}\mathbf{q}^{2}\big)\epsilon^{ij}\mathrm{Dirac}_{\mathrm{V}}^{i\bar{\lambda}j}\Bigg]\mathcal{B}_{1}\big(0,\Delta_{\mathrm{q}},\frac{k^{+}}{p_{1}^{+}}\mathbf{q}\big)\Bigg\}\\
 & \times\int_{\mathbf{x}}e^{-i\mathbf{k}_{\perp}\cdot\mathbf{x}}\big(U_{\mathbf{x}}-1\big)\;.
\end{aligned}
\label{eq:MV4_final}
\end{equation}
The UV-counterterms read:
\begin{equation}
\begin{aligned}\tilde{\mathcal{M}}_{\mathrm{V4,UV}}^{0} & =\frac{1}{p_{0}^{+}\mathbf{q}^{2}+p_{1}^{+}M^{2}}\frac{\alpha_{s}C_{F}}{M}\frac{1}{D-2}\int_{k_{\mathrm{min}}^{+}}^{p_{1}^{+}}\frac{\mathrm{d}k^{+}}{k^{+}}\frac{(q^{+})^{2}(k^{+})^{2}}{p_{0}^{+}(p_{1}^{+}-k^{+})(p_{0}^{+}-k^{+})}\mathrm{Dirac}_{\mathrm{V}}^{j0j}\\
 & \times\Bigg[-\Delta_{\mathrm{q}}+\frac{(p_{1}^{+}-k^{+})(p_{0}^{+}-k^{+})}{p_{1}^{+}(q^{+})^{2}}\Big(p_{0}^{+}\mathbf{q}^{2}-p_{1}^{+}M^{2}\Big)\Bigg]\mathcal{A}_{0}(\Delta_{\mathrm{UV}})\\
 & \times\int_{\mathbf{x}}e^{-i\mathbf{k}_{\perp}\cdot\mathbf{x}}\big(U_{\mathbf{x}}-1\big)\;,
\end{aligned}
\label{eq:MV4L_UV}
\end{equation}
and
\begin{equation}
\begin{aligned}\tilde{\mathcal{M}}_{\mathrm{V4,UV}}^{\lambda} & =\frac{\mathbf{q}^{\rho}}{p_{0}^{+}\mathbf{q}^{2}+p_{1}^{+}M^{2}}\frac{\alpha_{s}C_{F}}{D-2}\int_{k_{\mathrm{min}}^{+}}^{p_{1}^{+}}\frac{\mathrm{d}k^{+}}{k^{+}}\frac{k^{+}}{p_{1}^{+}}\frac{(k^{+}q^{+})^{2}}{p_{0}^{+}(p_{1}^{+}-k^{+})(p_{0}^{+}-k^{+})}\\
 & \times\Bigg[\Big(\frac{1}{2}+p_{0}^{+}\frac{p_{1}^{+}-k^{+}}{k^{+}q^{+}}\Big)\mathrm{Dirac}_{\mathrm{V}}^{j\rho j}+\epsilon^{\rho\bar{\lambda}}\frac{1}{2}\frac{1}{D-3}\epsilon^{ij}\mathrm{Dirac}_{\mathrm{V}}^{i\bar{\lambda}j}\Bigg]\mathcal{A}_{0}(\Delta_{\mathrm{UV}})\\
 & \times\int_{\mathbf{x}}e^{-i\mathbf{k}_{\perp}\cdot\mathbf{x}}\big(U_{\mathbf{x}}-1\big)\;.
\end{aligned}
\label{eq:MV4_UV}
\end{equation}
Therefore, the subtracted amplitudes become:
\begin{equation}
\begin{aligned}\tilde{\mathcal{M}}_{\mathrm{V4,sub}}^{0} & =\frac{1}{p_{0}^{+}\mathbf{q}^{2}+p_{1}^{+}M^{2}}\frac{\alpha_{s}C_{F}}{M}\frac{1}{2}\int_{k_{\mathrm{min}}^{+}}^{p_{1}^{+}}\frac{\mathrm{d}k^{+}}{k^{+}}\frac{(q^{+})^{2}(k^{+})^{2}}{p_{0}^{+}(p_{1}^{+}-k^{+})(p_{0}^{+}-k^{+})}\mathrm{Dirac}_{\mathrm{V}}^{j0j}\\
 & \times\Bigg\{\Bigg[-\Delta_{\mathrm{q}}+\frac{(p_{1}^{+}-k^{+})(p_{0}^{+}-k^{+})}{p_{1}^{+}(q^{+})^{2}}\Big(p_{0}^{+}\mathbf{q}^{2}-p_{1}^{+}M^{2}\Big)\Bigg]\frac{1}{4\pi}\ln\frac{\Delta_{\mathrm{UV}}}{\Delta_{\mathrm{q}}}\\
 & -2\Big(\frac{k^{+}}{p_{1}^{+}}\Big)^{2}\mathbf{q}^{2}\hat{M}^{2}\mathcal{B}_{1}\big(0,\Delta_{\mathrm{q}},\frac{k^{+}}{p_{1}^{+}}\mathbf{q}\big)\Bigg\}\\
 & \times\int_{\mathbf{x}}e^{-i\mathbf{k}_{\perp}\cdot\mathbf{x}}\big(U_{\mathbf{x}}-1\big)\;,
\end{aligned}
\label{eq:MV4L_sub}
\end{equation}
and:
\begin{equation}
\begin{aligned}\tilde{\mathcal{M}}_{\mathrm{V4,sub}}^{\lambda} & =\frac{\mathbf{q}^{\rho}}{p_{0}^{+}\mathbf{q}^{2}+p_{1}^{+}M^{2}}\frac{\alpha_{s}C_{F}}{2}\int_{k_{\mathrm{min}}^{+}}^{p_{1}^{+}}\frac{\mathrm{d}k^{+}}{k^{+}}\frac{k^{+}}{p_{1}^{+}}\frac{(k^{+}q^{+})^{2}}{p_{0}^{+}(p_{1}^{+}-k^{+})(p_{0}^{+}-k^{+})}\\
 & \times\Bigg\{\Bigg[\Big(\frac{1}{2}+p_{0}^{+}\frac{p_{1}^{+}-k^{+}}{k^{+}q^{+}}\Big)\mathrm{Dirac}_{\mathrm{V}}^{j\rho j}+\epsilon^{\rho\bar{\lambda}}\frac{1}{2}\epsilon^{ij}\mathrm{Dirac}_{\mathrm{V}}^{i\bar{\lambda}j}\Bigg]\frac{1}{4\pi}\ln\frac{\Delta_{\mathrm{UV}}}{\Delta_{\mathrm{q}}}\\
 & +\Bigg[\Big(-\frac{1}{2}\big(\Delta_{\mathrm{q}}+\Big(\frac{k^{+}}{p_{1}^{+}}\Big)^{2}\mathbf{q}^{2}\big)+\frac{k^{+}(p_{0}^{+}-k^{+})\mathbf{q}^{2}}{p_{1}^{+}q^{+}}\Big)\mathrm{Dirac}_{\mathrm{V}}^{j\rho j}\\
 & +\epsilon^{\rho\bar{\lambda}}\frac{1}{D-3}\frac{1}{2}\big(\Delta_{\mathrm{q}}+\Big(\frac{k^{+}}{p_{1}^{+}}\Big)^{2}\mathbf{q}^{2}\big)\epsilon^{ij}\mathrm{Dirac}_{\mathrm{V}}^{i\bar{\lambda}j}\Bigg]\mathcal{B}_{1}\big(0,\Delta_{\mathrm{q}},\frac{k^{+}}{p_{1}^{+}}\mathbf{q}\big)\Bigg\}\\
 & \times\int_{\mathbf{x}}e^{-i\mathbf{k}_{\perp}\cdot\mathbf{x}}\big(U_{\mathbf{x}}-1\big)\;.
\end{aligned}
\label{eq:MV4T_sub}
\end{equation}

\subsection{\label{subsec:antiq}Antiquark vertex corrections }

\begin{figure}[t]
\begin{centering}
\includegraphics[scale=0.3]{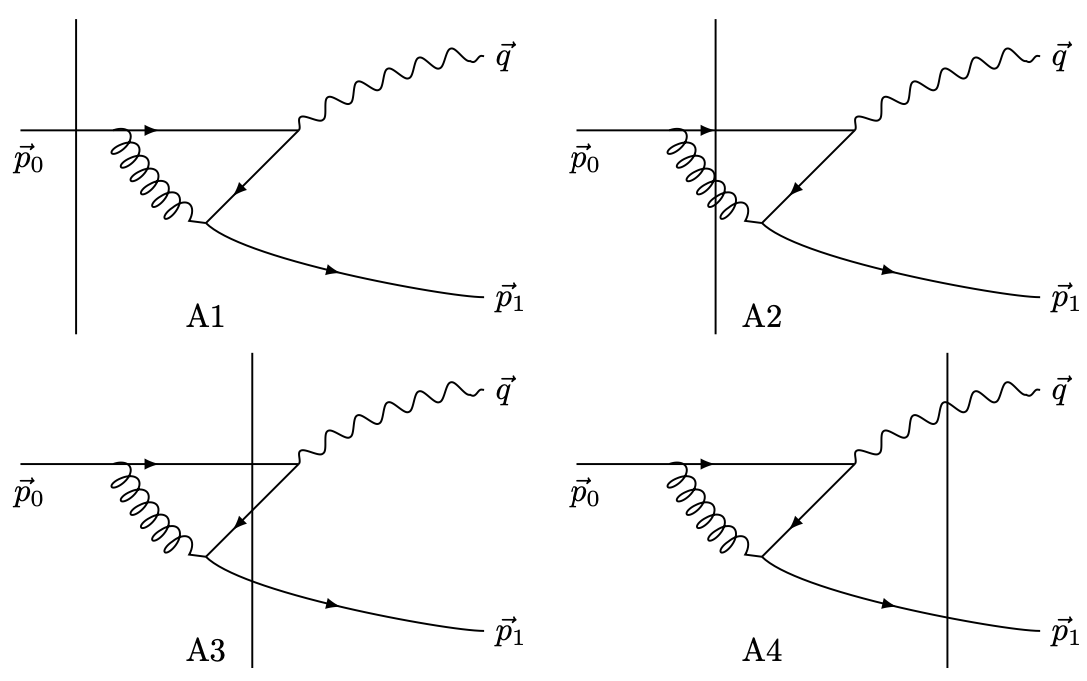}
\par\end{centering}
\caption{\label{fig:A}The four virtual contributions with a virtual gluon
and a virtual antiquark.}
\end{figure}

\paragraph{Diagram A1}

In the case of longitudinal polarization, one obtains for the amplitude
corresponding to diagram A1 in fig. \ref{fig:A}:
\begin{equation}
\begin{aligned}\tilde{\mathcal{M}}_{\mathrm{A1}}^{0} & =\frac{1}{p_{0}^{+}\mathbf{P}_{\perp}^{2}+p_{1}^{+}M^{2}}\frac{\alpha_{s}C_{F}}{M}\frac{1}{D-2}\int_{p_{1}^{+}}^{p_{0}^{+}}\frac{\mathrm{d}k^{+}}{k^{+}}\frac{k^{+}p_{1}^{+}(p_{0}^{+}-k^{+})}{p_{0}^{+}q^{+}}\mathrm{Dirac}_{\mathrm{V}}^{j0j}\\
 & \times\Bigg[2M^{2}\mathcal{A}_{0}(\hat{M}^{2})+\frac{(q^{+})^{2}\Delta_{\mathrm{P}}}{(p_{1}^{+}-k^{+})(p_{0}^{+}-k^{+})}\mathcal{A}_{0}(\Delta_{\mathrm{P}})\\
 & -2M^{2}\Delta_{\mathrm{P}}\mathcal{B}_{0}\big(\Delta_{\mathrm{P}},\hat{M}^{2},\frac{p_{0}^{+}-k^{+}}{q^{+}}\mathbf{P}_{\perp}\big)\\
 & +2\frac{k^{+}}{p_{1}^{+}}\frac{p_{0}^{+}-k^{+}}{q^{+}}M^{2}\mathbf{P}_{\perp}^{2}\mathcal{B}_{1}\big(\Delta_{\mathrm{P}},\hat{M}^{2},\frac{p_{0}^{+}-k^{+}}{q^{+}}\mathbf{P}_{\perp}\big)\Bigg]\\
 & \times\int_{\mathbf{x}}e^{-i\mathbf{k}_{\perp}\cdot\mathbf{x}}\big(U_{\mathbf{x}}-1\big)\;.
\end{aligned}
\label{eq:MA1L_final}
\end{equation}
Note that $\hat{M}^{2}$ is defined in (\ref{eq:Mdef-1}). In the transverse case, we find:
\begin{equation}
\begin{aligned}\tilde{\mathcal{M}}_{\mathrm{A1}}^{\lambda} & =\frac{\mathbf{P}_{\perp}^{\rho}}{p_{0}^{+}\mathbf{P}_{\perp}^{2}+p_{1}^{+}M^{2}}\frac{\alpha_{s}C_{F}}{D-2}\int_{p_{1}^{+}}^{p_{0}^{+}}\frac{\mathrm{d}k^{+}}{k^{+}}\frac{k^{+}p_{1}^{+}q^{+}}{p_{0}^{+}(p_{1}^{+}-k^{+})}\\
 & \times\Bigg\{\frac{k^{+}}{2p_{1}^{+}}\mathcal{A}_{0}(\hat{M}^{2})\Big(\mathrm{Dirac}_{\mathrm{V}}^{j\rho j}+\frac{1}{D-3}\epsilon^{\rho\bar{\lambda}}\epsilon^{ij}\mathrm{Dirac}_{\mathrm{V}}^{i\bar{\lambda}j}\Big)\\
 & -\Big(\frac{p_{0}^{+}-k^{+}}{q^{+}}\mathrm{Dirac}_{\mathrm{V}}^{j\rho j}+\frac{k^{+}}{p_{1}^{+}}\frac{1}{D-3}\epsilon^{\rho\bar{\lambda}}\epsilon^{ij}\mathrm{Dirac}_{\mathrm{V}}^{i\bar{\lambda}j}\Big)\Delta_{\mathrm{P}}\mathcal{B}_{0}\big(\Delta_{\mathrm{P}},\hat{M}^{2},\frac{p_{0}^{+}-k^{+}}{q^{+}}\mathbf{P}_{\perp}\big)\\
 & +\frac{k^{+}(p_{0}^{+}-k^{+})}{2p_{1}^{+}(q^{+})^{2}}\Bigg(\frac{2p_{1}^{+}(p_{0}^{+}-k^{+})-k^{+}q^{+}}{p_{0}^{+}p_{1}^{+}}\Big(p_{0}^{+}\mathbf{P}_{\perp}^{2}+p_{1}^{+}M^{2}\Big)\\
 & +\Big((p_{0}^{+}-k^{+}\big)\mathbf{P}_{\perp}^{2}-(p_{1}^{+}-k^{+})M^{2}\Big)\Bigg)\mathrm{Dirac}_{\mathrm{V}}^{j\rho j}\mathcal{B}_{1}\big(\Delta_{\mathrm{P}},\hat{M}^{2},\frac{p_{0}^{+}-k^{+}}{q^{+}}\mathbf{P}_{\perp}\big)\\
 & \!-\!\frac{k^{+}}{2p_{1}^{+}}\!\Big(\!\Delta_{\mathrm{P}}\!-\!\hat{M}^{2}\!-\!\Big(\!\frac{p_{0}^{+}\!-\!k^{+}}{q^{+}}\mathbf{P}_{\perp}\!\Big)^{\!2}\Big)\!\frac{1}{D\!-\!3}\epsilon^{\rho\bar{\lambda}}\epsilon^{ij}\mathrm{Dirac}_{\mathrm{V}}^{i\bar{\lambda}j}\mathcal{B}_{1}\big(\Delta_{\mathrm{P}},\!\hat{M}^{2},\!\frac{p_{0}^{+}\!-\!k^{+}}{q^{+}}\!\mathbf{P}_{\perp}\big)\!\Bigg\}\\
 & \times\int_{\mathbf{x}}e^{-i\mathbf{k}_{\perp}\cdot\mathbf{x}}\big(U_{\mathbf{x}}-1\big)\;.
\end{aligned}
\label{eq:MA1-1}
\end{equation}
Note that the spinor structures appearing here are the same as the
\textquoteleft vertex' ones (\ref{eq:DiracV}) and (\ref{eq:DiracVL}).
Moreover, the amplitudes are all divergent in the ultraviolet, and
the singularities are contained in the structures $\mathcal{A}_{0}$.
We can extract it by defining the counterterms:
\begin{equation}
\begin{aligned}\tilde{\mathcal{M}}_{\mathrm{A1,UV}}^{0} & =\frac{1}{p_{0}^{+}\mathbf{P}_{\perp}^{2}+p_{1}^{+}M^{2}}\frac{\alpha_{s}C_{F}}{M}\frac{1}{D-2}\int_{p_{1}^{+}}^{p_{0}^{+}}\frac{\mathrm{d}k^{+}}{k^{+}}\frac{k^{+}p_{1}^{+}(p_{0}^{+}-k^{+})}{p_{0}^{+}q^{+}}\mathrm{Dirac}_{\mathrm{V}}^{j0j}\\
 & \times\Big(2M^{2}+\frac{(q^{+})^{2}\Delta_{\mathrm{P}}}{(p_{1}^{+}-k^{+})(p_{0}^{+}-k^{+})}\Big)\mathcal{A}_{0}(\Delta_{\mathrm{UV}})\\
 & \times\int_{\mathbf{x}}e^{-i\mathbf{k}_{\perp}\cdot\mathbf{x}}\big(U_{\mathbf{x}}-1\big)\;,
\end{aligned}
\label{eq:MA1L_UV}
\end{equation}
and:
\begin{equation}
\begin{aligned}\tilde{\mathcal{M}}_{\mathrm{A1,UV}}^{\lambda} & =\frac{\mathbf{P}_{\perp}^{\rho}}{p_{0}^{+}\mathbf{P}_{\perp}^{2}+p_{1}^{+}M^{2}}\frac{\alpha_{s}C_{F}}{D-2}\int_{p_{1}^{+}}^{p_{0}^{+}}\frac{\mathrm{d}k^{+}}{k^{+}}\frac{(k^{+})^{2}q^{+}}{2p_{0}^{+}(p_{1}^{+}-k^{+})}\\
 & \times\Big(\mathrm{Dirac}_{\mathrm{V}}^{j\rho j}+\frac{1}{D-3}\epsilon^{\rho\bar{\lambda}}\epsilon^{ij}\mathrm{Dirac}_{\mathrm{V}}^{i\bar{\lambda}j}\Big)\mathcal{A}_{0}(\Delta_{\mathrm{UV}})\\
 & \times\int_{\mathbf{x}}e^{-i\mathbf{k}_{\perp}\cdot\mathbf{x}}\big(U_{\mathbf{x}}-1\big)\;.
\end{aligned}
\label{eq:MA1T_UV}
\end{equation}
The subtracted amplitudes read:
\begin{equation}
\begin{aligned}\tilde{\mathcal{M}}_{\mathrm{A1,sub}}^{0} & \!=\!-\!\frac{1}{p_{0}^{+}\mathbf{P}_{\perp}^{2}\!+\!p_{1}^{+}M^{2}}\!\frac{\alpha_{s}C_{F}}{M}\!\int_{p_{1}^{+}}^{p_{0}^{+}}\frac{\mathrm{d}k^{+}}{k^{+}}\frac{k^{+}p_{1}^{+}(p_{0}^{+}\!-\!k^{+})}{p_{0}^{+}q^{+}}\Big(\!\big(2\frac{p_{1}^{+}}{k^{+}}\!-1\!\big)\big(2\frac{p_{0}^{+}}{k^{+}}\!-\!1\big)\!+\!1\!\Big)\\
 & \times\Bigg[\Big(2M^{2}\frac{1}{4\pi}\ln\frac{\Delta_{\mathrm{UV}}}{\hat{M}^{2}}+\frac{(q^{+})^{2}\Delta_{\mathrm{P}}}{(p_{1}^{+}-k^{+})(p_{0}^{+}-k^{+})}\frac{1}{4\pi}\ln\frac{\Delta_{\mathrm{UV}}}{\Delta_{\mathrm{P}}}\\
 & -2M^{2}\Delta_{\mathrm{P}}\mathcal{B}_{0}\big(\Delta_{\mathrm{P}},\hat{M}^{2},\frac{p_{0}^{+}-k^{+}}{q^{+}}\mathbf{P}_{\perp}\big)\\
 & +2\frac{k^{+}}{p_{1}^{+}}\frac{p_{0}^{+}-k^{+}}{q^{+}}M^{2}\mathbf{P}_{\perp}^{2}\mathcal{B}_{1}\big(\Delta_{\mathrm{P}},\hat{M}^{2},\frac{p_{0}^{+}-k^{+}}{q^{+}}\mathbf{P}_{\perp}\big)\Bigg] \int_{\mathbf{x}}e^{-i\mathbf{k}_{\perp}\cdot\mathbf{x}}\big(U_{\mathbf{x}}-1\big)\;,
\end{aligned}
\label{eq:MA1L_sub}
\end{equation}
and:
\begin{equation}
\begin{aligned} & \tilde{\mathcal{M}}_{\mathrm{A1,sub}}^{\lambda}  =\frac{\mathbf{P}_{\perp}^{\rho}}{p_{0}^{+}\mathbf{P}_{\perp}^{2}+p_{1}^{+}M^{2}}\frac{\alpha_{s}C_{F}}{D-2}\int_{p_{1}^{+}}^{p_{0}^{+}}\frac{\mathrm{d}k^{+}}{k^{+}}\frac{k^{+}p_{1}^{+}q^{+}}{p_{0}^{+}(p_{1}^{+}-k^{+})}\\
 & \times\Bigg\{\frac{k^{+}}{2p_{1}^{+}}\Big(\mathrm{Dirac}_{\mathrm{V}}^{j\rho j}+\frac{1}{D-3}\epsilon^{\rho\bar{\lambda}}\epsilon^{ij}\mathrm{Dirac}_{\mathrm{V}}^{i\bar{\lambda}j}\Big)\frac{1}{4\pi}\ln\frac{\Delta_{\mathrm{UV}}}{\hat{M}^{2}}\\
 & -\Big(\frac{p_{0}^{+}-k^{+}}{q^{+}}\mathrm{Dirac}_{\mathrm{V}}^{j\rho j}+\frac{k^{+}}{p_{1}^{+}}\frac{1}{D-3}\epsilon^{\rho\bar{\lambda}}\epsilon^{ij}\mathrm{Dirac}_{\mathrm{V}}^{i\bar{\lambda}j}\Big)\Delta_{\mathrm{P}}\mathcal{B}_{0}\big(\Delta_{\mathrm{P}},\hat{M}^{2},\frac{p_{0}^{+}-k^{+}}{q^{+}}\mathbf{P}_{\perp}\big)\\
 & +\frac{k^{+}(p_{0}^{+}-k^{+})}{2p_{1}^{+}(q^{+})^{2}}\Bigg(\frac{2p_{1}^{+}(p_{0}^{+}-k^{+})-k^{+}q^{+}}{p_{0}^{+}p_{1}^{+}}\Big(p_{0}^{+}\mathbf{P}_{\perp}^{2}+p_{1}^{+}M^{2}\Big)\\
 & +\Big((p_{0}^{+}-k^{+}\big)\mathbf{P}_{\perp}^{2}-(p_{1}^{+}-k^{+})M^{2}\Big)\Bigg)\mathrm{Dirac}_{\mathrm{V}}^{j\rho j}\mathcal{B}_{1}\big(\Delta_{\mathrm{P}},\hat{M}^{2},\frac{p_{0}^{+}-k^{+}}{q^{+}}\mathbf{P}_{\perp}\big)\\
 & \!-\!\frac{k^{+}}{2p_{1}^{+}}\!\Big(\!\Delta_{\mathrm{P}}\!-\!\hat{M}^{2}\!-\!\Big(\!\frac{p_{0}^{+}\!-\!k^{+}}{q^{+}}\mathbf{P}_{\perp}\!\Big)^{\!2}\Big)\!\frac{1}{D\!-\!3}\epsilon^{\rho\bar{\lambda}}\epsilon^{ij}\mathrm{Dirac}_{\mathrm{V}}^{i\bar{\lambda}j}\mathcal{B}_{1}\big(\Delta_{\mathrm{P}},\!\hat{M}^{2},\!\frac{p_{0}^{+}\!-\!k^{+}}{q^{+}}\!\mathbf{P}_{\perp}\big)\!\Bigg\}\\
 & \times\int_{\mathbf{x}}e^{-i\mathbf{k}_{\perp}\cdot\mathbf{x}}\big(U_{\mathbf{x}}-1\big)\;.\end{aligned}
\label{eq:MA1_T_sub}
\end{equation}

\paragraph{Diagram A2}

For the production amplitudes of a longitudinally or transversely
polarized virtual photon, we obtain, respectively:
\begin{equation}
\begin{aligned}\tilde{\mathcal{M}}_{\mathrm{A2}}^{0} & =-\frac{\alpha_{s}}{M}\int_{p_{1}^{+}}^{p_{0}^{+}}\frac{\mathrm{d}k^{+}}{k^{+}}\frac{(k^{+})^{2}(p_{0}^{+}-k^{+})}{(p_{0}^{+})^{2}(p_{1}^{+}-k^{+})}\mathrm{Dirac}_{\mathrm{V}}^{\bar{\eta}0\eta^{\prime}}\\
 & \times\int_{\mathbf{x},\mathbf{z}}iA^{\eta^{\prime}}(\mathbf{x}-\mathbf{z})\int_{\boldsymbol{\ell}}e^{-i\boldsymbol{\ell}\cdot(\mathbf{x}-\mathbf{z})}\frac{\boldsymbol{\ell}^{\bar{\eta}}+\frac{k^{+}}{p_{1}^{+}}\mathbf{P}_{\perp}^{\bar{\eta}}}{\boldsymbol{\ell}^{2}+\Delta_{\mathrm{P}}}\frac{\Big(\boldsymbol{\ell}+\frac{p_{0}^{+}-k^{+}}{q^{+}}\mathbf{P}_{\perp}\Big)^{2}-\hat{M}^{2}}{\Big(\boldsymbol{\ell}+\frac{p_{0}^{+}-k^{+}}{q^{+}}\mathbf{P}_{\perp}\Big)^{2}+\hat{M}^{2}}\\
 & \times e^{-i\mathbf{k}_{\perp}\cdot\Big(\frac{p_{0}^{+}-k^{+}}{p_{0}^{+}}\mathbf{x}+\frac{k^{+}}{p_{0}^{+}}\mathbf{z}\Big)}\big(t^{c}U_{\mathbf{x}}U_{\mathbf{z}}^{\dagger}t^{c}U_{\mathbf{z}}-C_{F}\big)\;,
\end{aligned}
\label{eq:MA2L}
\end{equation}
and:
\begin{equation}
\begin{aligned}\tilde{\mathcal{M}}_{\mathrm{A2}}^{\lambda} & =\alpha_{s}\int_{p_{1}^{+}}^{p_{0}^{+}}\frac{\mathrm{d}k^{+}}{k^{+}}\frac{(k^{+})^{2}(p_{0}^{+}-k^{+})}{(p_{0}^{+})^{2}(p_{1}^{+}-k^{+})}\mathrm{Dirac}_{\mathrm{V}}^{\bar{\eta}\bar{\lambda}\eta^{\prime}}\\
 & \times\int_{\mathbf{x},\mathbf{z}}iA^{\eta^{\prime}}(\mathbf{x}-\mathbf{z})\int_{\boldsymbol{\ell}}e^{-i\boldsymbol{\ell}\cdot(\mathbf{x}-\mathbf{z})}\frac{\boldsymbol{\ell}^{\bar{\eta}}+\frac{k^{+}}{p_{1}^{+}}\mathbf{P}_{\perp}^{\bar{\eta}}}{\boldsymbol{\ell}^{2}+\Delta_{\mathrm{P}}}\frac{\Big(\boldsymbol{\ell}^{\bar{\lambda}}+\frac{p_{0}^{+}-k^{+}}{q^{+}}\mathbf{P}_{\perp}^{\bar{\lambda}}\Big)}{\Big(\boldsymbol{\ell}+\frac{p_{0}^{+}-k^{+}}{q^{+}}\mathbf{P}_{\perp}\Big)^{2}+\hat{M}^{2}}\\
 & \times e^{-i\mathbf{k}_{\perp}\cdot\Big(\frac{p_{0}^{+}-k^{+}}{p_{0}^{+}}\mathbf{x}+\frac{k^{+}}{p_{0}^{+}}\mathbf{z}\Big)}\big(t^{c}U_{\mathbf{x}}U_{\mathbf{z}}^{\dagger}t^{c}U_{\mathbf{z}}-C_{F}\big)\;.
\end{aligned}
\label{eq:MA2}
\end{equation}
Similarly to the amplitudes $\tilde{\mathcal{M}}_{\mathrm{V2,3}}$,
the transverse integrals over $\boldsymbol{\ell}$ do not admit an
analytic solution, at least to the best of our knowledge. It is, however,
possible to study their behavior in the limit $\lim_{\mathbf{z}\to\mathbf{x}}$
where they might exhibit a UV divergence, which turns out to be the
case for the \textquoteleft longitudinal' amplitude \ref{eq:MA2L}:
\begin{equation}
\begin{aligned}\lim_{\mathbf{z}\to\mathbf{x}}\tilde{\mathcal{M}}_{\mathrm{A2}}^{0} & =\frac{\alpha_{s}C_{F}}{M}\frac{1}{D-2}\int_{p_{1}^{+}}^{p_{0}^{+}}\frac{\mathrm{d}k^{+}}{k^{+}}\frac{(k^{+})^{2}(p_{0}^{+}-k^{+})}{(p_{0}^{+})^{2}(p_{1}^{+}-k^{+})}\mathrm{Dirac}_{\mathrm{V}}^{j0j}\\
 & \times\Bigg[\mathcal{A}_{0}(\Delta_{\mathrm{P}})-2\hat{M}^{2}\mathcal{B}_{0}\big(\Delta_{\mathrm{P}},\hat{M}^{2},\frac{p_{0}^{+}-k^{+}}{q^{+}}\mathbf{P}_{\perp}\big)\\
 & -2\hat{M}^{2}\frac{p_{0}^{+}-k^{+}}{q^{+}}\frac{k^{+}}{p_{1}^{+}}\mathbf{P}_{\perp}^{2}\mathcal{C}_{1}\big(\Delta_{\mathrm{P}},\hat{M}^{2},\frac{p_{0}^{+}-k^{+}}{q^{+}}\mathbf{P}_{\perp}\big)\Bigg]\\
 & \times\int_{\mathbf{x}}e^{-i\mathbf{k}_{\perp}\cdot\mathbf{x}}\big(U_{\mathbf{x}}-1\big)\;.
\end{aligned}
\label{eq:MA2L_limit_final}
\end{equation}
The counterterm, therefore, reads:
\begin{equation}
\begin{aligned}\tilde{\mathcal{M}}_{\mathrm{A2,UV}}^{0} & =\frac{\alpha_{s}C_{F}}{M}\frac{1}{D-2}\int_{p_{1}^{+}}^{p_{0}^{+}}\frac{\mathrm{d}k^{+}}{k^{+}}\frac{(k^{+})^{2}(p_{0}^{+}-k^{+})}{(p_{0}^{+})^{2}(p_{1}^{+}-k^{+})}\mathrm{Dirac}_{\mathrm{V}}^{j0j}\\
 & \times\mathcal{A}_{0}(\Delta_{\mathrm{UV}})\int_{\mathbf{x}}e^{-i\mathbf{k}_{\perp}\cdot\mathbf{x}}\big(U_{\mathbf{x}}-1\big)\;.
\end{aligned}
\label{eq:MA2L_UV}
\end{equation}
The subtracted amplitude becomes:
\begin{equation}
\begin{aligned}\tilde{\mathcal{M}}_{\mathrm{A2,sub}}^{0} & =-\frac{\alpha_{s}}{M}\int_{p_{1}^{+}}^{p_{0}^{+}}\frac{\mathrm{d}k^{+}}{k^{+}}\frac{(k^{+})^{2}(p_{0}^{+}-k^{+})}{(p_{0}^{+})^{2}(p_{1}^{+}-k^{+})}\mathrm{Dirac}_{\mathrm{V}}^{\bar{\eta}0\eta^{\prime}}\\
 & \times\Bigg[\int_{\mathbf{x},\mathbf{z}}iA^{\eta^{\prime}}(\mathbf{x}-\mathbf{z})\int_{\boldsymbol{\ell}}e^{-i\boldsymbol{\ell}\cdot(\mathbf{x}-\mathbf{z})}\frac{\boldsymbol{\ell}^{\bar{\eta}}+\frac{k^{+}}{p_{1}^{+}}\mathbf{P}_{\perp}^{\bar{\eta}}}{\boldsymbol{\ell}^{2}+\Delta_{\mathrm{P}}}\frac{\Big(\boldsymbol{\ell}+\frac{p_{0}^{+}-k^{+}}{q^{+}}\mathbf{P}_{\perp}\Big)^{2}-\hat{M}^{2}}{\Big(\boldsymbol{\ell}+\frac{p_{0}^{+}-k^{+}}{q^{+}}\mathbf{P}_{\perp}\Big)^{2}+\hat{M}^{2}}\\
 & \times e^{-i\mathbf{k}_{\perp}\cdot\Big(\frac{p_{0}^{+}-k^{+}}{p_{0}^{+}}\mathbf{x}+\frac{k^{+}}{p_{0}^{+}}\mathbf{z}\Big)}\big(t^{c}U_{\mathbf{x}}U_{\mathbf{z}}^{\dagger}t^{c}U_{\mathbf{z}}-C_{F}\big)\\
 & +\frac{\delta^{\bar{\eta}\eta^{\prime}}}{D-2}\mathcal{A}_{0}(\Delta_{\mathrm{UV}})\int_{\mathbf{x}}e^{-i\mathbf{k}_{\perp}\cdot\mathbf{x}}C_{F}\big(U_{\mathbf{x}}-1\big)\Bigg]\;.
\end{aligned}
\label{eq:MA2L_sub}
\end{equation}
Amplitude $\tilde{\mathcal{M}}_{\mathrm{A2}}^{\lambda}$ turns out
to be UV-finite.

\paragraph{Diagram A3}

We obtain for the amplitudes where both the virtual quark and antiquark,
as well as the outgoing quark scatter off the shockwave (fig. \ref{fig:A}):
\begin{equation}
\begin{aligned}\tilde{\mathcal{M}}_{\mathrm{A3}}^{0} & =\frac{\alpha_{s}}{M}\int_{p_{1}^{+}}^{p_{0}^{+}}\frac{\mathrm{d}k^{+}}{k^{+}}\frac{2k^{+}(p_{0}^{+}-k^{+})}{q^{+}p_{0}^{+}}\mathrm{Dirac}_{\mathrm{V}}^{\bar{\eta}0\eta^{\prime}}\int_{\mathbf{x}_{1},\mathbf{x}_{2},\mathbf{x}_{3}}\\
 & \times\hat{M}^{2}\mathcal{K}\big(\mathbf{x}_{1}-\mathbf{x}_{2},\hat{M}^{2}\big)\\
 & \times\int_{\boldsymbol{\ell},\boldsymbol{\ell}_{2}}e^{-i\boldsymbol{\ell}\cdot\mathbf{x}_{12}}e^{-i\boldsymbol{\ell}_{2}\cdot\mathbf{x}_{23}}\frac{\boldsymbol{\ell}^{\eta^{\prime}}}{\mathbf{\boldsymbol{\ell}}^{2}}\frac{\boldsymbol{\ell}^{\bar{\eta}}-\frac{k^{+}}{p_{1}^{+}}\boldsymbol{\ell}_{2}^{\bar{\eta}}}{\Big(\boldsymbol{\ell}-\frac{p_{0}^{+}-k^{+}}{q^{+}}\boldsymbol{\ell}_{2}\Big)^{2}-\frac{p_{0}^{+}(p_{0}^{+}-k^{+})(p_{1}^{+}-k^{+})}{p_{1}^{+}(q^{+})^{2}}\boldsymbol{\ell}_{2}^{2}}\\
 & \times e^{-i\mathbf{p}_{1}\cdot\mathbf{x}_{3}}e^{-i\mathbf{q}\cdot\Big(\frac{p_{0}^{+}-k^{+}}{q^{+}}\mathbf{x}_{1}-\frac{p_{1}^{+}-k^{+}}{q^{+}}\mathbf{x}_{2}\Big)}[U_{\mathbf{x}_{3}}t^{c}U_{\mathbf{x}_{2}}^{\dagger}U_{\mathbf{x}_{1}}t^{c}-C_{F}]\;,
\end{aligned}
\label{eq:MA3L}
\end{equation}
and
\begin{equation}
\begin{aligned}\tilde{\mathcal{M}}_{\mathrm{A3}}^{\lambda} & =\alpha_{s}\int_{p_{1}^{+}}^{p_{0}^{+}}\frac{\mathrm{d}k^{+}}{k^{+}}\frac{k^{+}(p_{0}^{+}-k^{+})}{q^{+}p_{0}^{+}}\mathrm{Dirac}_{\mathrm{V}}^{\bar{\eta}\bar{\lambda}\eta^{\prime}}\int_{\mathbf{x}_{1},\mathbf{x}_{2},\mathbf{x}_{3}}\\
 & \times iA^{\bar{\lambda}}\big(\mathbf{x}_{1}-\mathbf{x}_{2},\hat{M}^{2}\big)\\
 & \times\int_{\boldsymbol{\ell},\boldsymbol{\ell}_{2}}e^{-i\boldsymbol{\ell}\cdot\mathbf{x}_{12}}e^{-i\boldsymbol{\ell}_{2}\cdot\mathbf{x}_{23}}\frac{\boldsymbol{\ell}^{\eta^{\prime}}}{\mathbf{\boldsymbol{\ell}}^{2}}\frac{\boldsymbol{\ell}^{\bar{\eta}}-\frac{k^{+}}{p_{1}^{+}}\boldsymbol{\ell}_{2}^{\bar{\eta}}}{\Big(\boldsymbol{\ell}-\frac{p_{0}^{+}-k^{+}}{q^{+}}\boldsymbol{\ell}_{2}\Big)^{2}-\frac{p_{0}^{+}(p_{0}^{+}-k^{+})(p_{1}^{+}-k^{+})}{p_{1}^{+}(q^{+})^{2}}\boldsymbol{\ell}_{2}^{2}}\\
 & \times e^{-i\mathbf{p}_{1}\cdot\mathbf{x}_{3}}e^{-i\mathbf{q}\cdot\Big(\frac{p_{0}^{+}-k^{+}}{q^{+}}\mathbf{x}_{1}-\frac{p_{1}^{+}-k^{+}}{q^{+}}\mathbf{x}_{2}\Big)}[U_{\mathbf{x}_{3}}t^{c}U_{\mathbf{x}_{2}}^{\dagger}U_{\mathbf{x}_{1}}t^{c}-C_{F}]\;.
\end{aligned}
\label{eq:MA3}
\end{equation}
In the above expressions, we used the short-hand notation $\mathbf{x}_{ij}=\mathbf{x}_{i}-\mathbf{x}_{j}$.
Investigating for possible UV divergences in the limit $\mathbf{x}_{1}\to\mathbf{x}_{2}\to\mathbf{x}_{3}$,
it turns out that $\tilde{\mathcal{M}}_{\mathrm{A3}}^{\lambda}$ is
finite while $\tilde{\mathcal{M}}_{\mathrm{A3}}^{0}$ contains an
ultraviolet pole, which can be absorbed into the counterterm:
\begin{equation}
\begin{aligned}\tilde{\mathcal{M}}_{\mathrm{A3,UV}}^{0} & =-\frac{\alpha_{s}C_{F}}{M}\frac{1}{D-2}\int_{p_{1}^{+}}^{p_{0}^{+}}\frac{\mathrm{d}k^{+}}{k^{+}}\frac{k^{+}(p_{0}^{+}-k^{+})}{q^{+}p_{0}^{+}}\mathrm{Dirac}_{\mathrm{V}}^{j0j}\\
 & \times\mathcal{A}_{0}(\Delta_{\mathrm{UV}})\int_{\mathbf{x}}e^{-i\mathbf{k}_{\perp}\cdot\mathbf{x}}\big(U_{\mathbf{x}}-1\big)\;.
\end{aligned}
\label{eq:MA3L_UV}
\end{equation}
The subtracted amplitude becomes:
\begin{equation}
\begin{aligned}\tilde{\mathcal{M}}_{\mathrm{A3,sub}}^{0} & =\frac{\alpha_{s}}{M}\mathrm{Dirac}_{\mathrm{V}}^{\bar{\eta}0\eta^{\prime}}\int_{p_{1}^{+}}^{p_{0}^{+}}\frac{\mathrm{d}k^{+}}{k^{+}}\frac{k^{+}(p_{0}^{+}-k^{+})}{q^{+}p_{0}^{+}}\\
 & \times\Bigg[\int_{\mathbf{x}_{1},\mathbf{x}_{2},\mathbf{x}_{3}}2\hat{M}^{2}\mathcal{K}\big(\mathbf{x}_{1}-\mathbf{x}_{2},\hat{M}^{2}\big)\\
 & \times\int_{\boldsymbol{\ell},\boldsymbol{\ell}_{2}}e^{-i\boldsymbol{\ell}\cdot\mathbf{x}_{12}}e^{-i\boldsymbol{\ell}_{2}\cdot\mathbf{x}_{23}}\frac{\boldsymbol{\ell}^{\eta^{\prime}}}{\mathbf{\boldsymbol{\ell}}^{2}}\frac{\boldsymbol{\ell}^{\bar{\eta}}-\frac{k^{+}}{p_{1}^{+}}\boldsymbol{\ell}_{2}^{\bar{\eta}}}{\Big(\boldsymbol{\ell}-\frac{p_{0}^{+}-k^{+}}{q^{+}}\boldsymbol{\ell}_{2}\Big)^{2}-\frac{p_{0}^{+}(p_{0}^{+}-k^{+})(p_{1}^{+}-k^{+})}{p_{1}^{+}(q^{+})^{2}}\boldsymbol{\ell}_{2}^{2}}\\
 & \times e^{-i\mathbf{p}_{1}\cdot\mathbf{x}_{3}}e^{-i\mathbf{q}\cdot\Big(\frac{p_{0}^{+}-k^{+}}{q^{+}}\mathbf{x}_{1}-\frac{p_{1}^{+}-k^{+}}{q^{+}}\mathbf{x}_{2}\Big)}[U_{\mathbf{x}_{3}}t^{c}U_{\mathbf{x}_{2}}^{\dagger}U_{\mathbf{x}_{1}}t^{c}-C_{F}]\\
 & +\frac{\delta^{\bar{\eta}\eta^{\prime}}}{D-2}\mathcal{A}_{0}(\Delta_{\mathrm{UV}})\int_{\mathbf{x}}e^{-i\mathbf{k}_{\perp}\cdot\mathbf{x}}C_{F}\big(U_{\mathbf{x}}-1\big)\Bigg]\;.
\end{aligned}
\label{eq:MA3L_sub}
\end{equation}

\paragraph{Diagram A4}

In the longitudinal case, the amplitude is given by:

\begin{equation}
\begin{aligned}\tilde{\mathcal{M}}_{\mathrm{A4}}^{0} & =\frac{p_{0}^{+}\mathbf{q}^{2}-p_{1}^{+}M^{2}}{p_{0}^{+}\mathbf{q}^{2}+p_{1}^{+}M^{2}}\frac{\alpha_{s}C_{F}}{M}\frac{1}{D-2}\int_{p_{1}^{+}}^{p_{0}^{+}}\frac{\mathrm{d}k^{+}}{k^{+}}\frac{k^{+}(p_{0}^{+}-k^{+})}{p_{0}^{+}q^{+}}\mathrm{Dirac}_{\mathrm{V}}^{j0j}\\
 & \times\Big(\mathcal{A}_{0}(\hat{Q}^{2})+\frac{k^{+}}{p_{1}^{+}}\frac{p_{0}^{+}-k^{+}}{q^{+}}\mathbf{q}^{2}\mathcal{B}_{1}\big(0,\hat{Q}^{2},\frac{p_{0}^{+}-k^{+}}{q^{+}}\mathbf{q}\big)\Big)\\
 & \times\int_{\mathbf{x}}e^{-i\mathbf{k}_{\perp}\cdot\mathbf{x}}\big(U_{\mathbf{x}}-1\big)\;,
\end{aligned}
\label{eq:MA4L_final}
\end{equation}
where we have defined:
\begin{equation}
\hat{Q}^{2}\equiv-\frac{p_{0}^{+}(p_{0}^{+}-k^{+})(p_{1}^{+}-k^{+})}{p_{1}^{+}(q^{+})^{2}}\mathbf{q}^{2}+i0^{+}\;.\label{eq:Qhatdef}
\end{equation}
Likewise, in the transverse case, we find:
\begin{equation}
\begin{aligned}\tilde{\mathcal{M}}_{\mathrm{A4}}^{\lambda} & =\frac{\mathbf{q}^{\rho}}{p_{0}^{+}\mathbf{q}^{2}+p_{1}^{+}M^{2}}\frac{\alpha_{s}C_{F}}{D-2}\int_{p_{1}^{+}}^{p_{0}^{+}}\frac{\mathrm{d}k^{+}}{k^{+}}\frac{q^{+}(k^{+})^{2}}{2p_{0}^{+}(p_{1}^{+}-k^{+})}\\
 & \times\Bigg\{\Bigg[\mathrm{Dirac}_{\mathrm{V}}^{j\rho j}+\epsilon^{ij}\mathrm{Dirac}_{\mathrm{V}}^{i\bar{\lambda}j}\epsilon^{\rho\bar{\lambda}}\frac{1}{D-3}\Bigg]\mathcal{A}_{0}(\hat{Q}^{2})\\
 & +\Bigg[\!-\!\Big(\hat{Q}^{2}\!-\!\Big(\frac{p_{0}^{+}\!-\!k^{+}}{q^{+}}\mathbf{q}\Big)^{2}\Big)\mathrm{Dirac}_{\mathrm{V}}^{j\rho j}\!+\!\frac{k^{+}(p_{0}^{+}\!-\!k^{+})}{p_{1}^{+}q^{+}}\mathbf{q}^{2}\epsilon^{ij}\mathrm{Dirac}_{\mathrm{V}}^{i\bar{\lambda}j}\epsilon^{\rho\bar{\lambda}}\frac{1}{D\!-\!3}\Bigg]\\
 & \times\mathcal{B}_{1}\big(0,\hat{Q}^{2},\frac{p_{0}^{+}-k^{+}}{q^{+}}\mathbf{q}\big)\Bigg\}\int_{\mathbf{x}}e^{-i\mathbf{k}_{\perp}\cdot\mathbf{x}}\big(U_{\mathbf{x}}-1\big)\;.
\end{aligned}
\label{eq:MA4_final}
\end{equation}
Clearly, both $\tilde{\mathcal{M}}_{\mathrm{A4}}^{0}$ and $\tilde{\mathcal{M}}_{\mathrm{A4}}^{\lambda}$
are divergent in the ultraviolet. The poles are extracted by defining
the counterterms:
\begin{equation}
\begin{aligned}\tilde{\mathcal{M}}_{\mathrm{A4,UV}}^{0} & =\frac{p_{0}^{+}\mathbf{q}^{2}-p_{1}^{+}M^{2}}{p_{0}^{+}\mathbf{q}^{2}+p_{1}^{+}M^{2}}\frac{\alpha_{s}C_{F}}{M}\frac{1}{D-2}\int_{p_{1}^{+}}^{p_{0}^{+}}\frac{\mathrm{d}k^{+}}{k^{+}}\frac{k^{+}(p_{0}^{+}-k^{+})}{p_{0}^{+}q^{+}}\mathrm{Dirac}_{\mathrm{V}}^{j0j}\\
 & \times\mathcal{A}_{0}(\Delta_{\mathrm{UV}})\int_{\mathbf{x}}e^{-i\mathbf{k}_{\perp}\cdot\mathbf{x}}\big(U_{\mathbf{x}}-1\big)\;,
\end{aligned}
\label{eq:MA4L_UV}
\end{equation}
and:
\begin{equation}
\begin{aligned}\tilde{\mathcal{M}}_{\mathrm{A4,UV}}^{\lambda} & =\frac{\mathbf{q}^{\rho}}{p_{0}^{+}\mathbf{q}^{2}+p_{1}^{+}M^{2}}\frac{\alpha_{s}C_{F}}{D-2}\int_{p_{1}^{+}}^{p_{0}^{+}}\frac{\mathrm{d}k^{+}}{k^{+}}\frac{q^{+}(k^{+})^{2}}{2p_{0}^{+}(p_{1}^{+}-k^{+})}\\
 & \times\Bigg[\mathrm{Dirac}_{\mathrm{V}}^{j\rho j}+\epsilon^{ij}\mathrm{Dirac}_{\mathrm{V}}^{i\bar{\lambda}j}\epsilon^{\rho\bar{\lambda}}\frac{1}{D-3}\Bigg]\\
 & \times\mathcal{A}_{0}(\Delta_{\mathrm{UV}})\int_{\mathbf{x}}e^{-i\mathbf{k}_{\perp}\cdot\mathbf{x}}\big(U_{\mathbf{x}}-1\big)\;.
\end{aligned}
\label{eq:MA4T_UV}
\end{equation}
The subtracted amplitudes, therefore, become:
\begin{equation}
\begin{aligned}\tilde{\mathcal{M}}_{\mathrm{A4,sub}}^{0} & =\tilde{\mathcal{M}}_{\mathrm{LO2}}^{0}\times\frac{\alpha_{s}C_{F}}{\pi}\int_{p_{1}^{+}}^{p_{0}^{+}}\frac{\mathrm{d}k^{+}}{4k^{+}}\frac{k^{+}(p_{0}^{+}-k^{+})}{p_{0}^{+}q^{+}}\Big(\big(2\frac{p_{1}^{+}}{k^{+}}-1\big)\big(2\frac{p_{0}^{+}}{k^{+}}-1\big)+1\Big)\\
 & \times\Big(\ln\frac{\Delta_{\mathrm{UV}}}{\hat{Q}^{2}}+4\pi\frac{k^{+}}{p_{1}^{+}}\frac{p_{0}^{+}-k^{+}}{q^{+}}\mathbf{q}^{2}\mathcal{B}_{1}\big(0,\hat{Q}^{2},\frac{p_{0}^{+}-k^{+}}{q^{+}}\mathbf{q}\big)\Big)\;,
\end{aligned}
\label{eq:MA4L_final-2}
\end{equation}
in the longitudinally polarized case, and:
\begin{equation}
\begin{aligned}\tilde{\mathcal{M}}_{\mathrm{A4,sub}}^{\lambda} & =\frac{\mathbf{q}^{\rho}}{p_{0}^{+}\mathbf{q}^{2}+p_{1}^{+}M^{2}}\frac{\alpha_{s}C_{F}}{D-2}\int_{p_{1}^{+}}^{p_{0}^{+}}\frac{\mathrm{d}k^{+}}{k^{+}}\frac{q^{+}(k^{+})^{2}}{2p_{0}^{+}(p_{1}^{+}-k^{+})}\\
 & \times\Bigg\{\Bigg[\mathrm{Dirac}_{\mathrm{V}}^{j\rho j}+\epsilon^{ij}\mathrm{Dirac}_{\mathrm{V}}^{i\bar{\lambda}j}\epsilon^{\rho\bar{\lambda}}\frac{1}{D-3}\Bigg]\frac{1}{4\pi}\ln\frac{\Delta_{\mathrm{UV}}}{\hat{Q}^{2}}\\
 & +\!\Bigg[\!-\!\Big(\hat{Q}^{2}\!-\!\Big(\frac{p_{0}^{+}\!-\!k^{+}}{q^{+}}\mathbf{q}\Big)^{2}\Big)\mathrm{Dirac}_{\mathrm{V}}^{j\rho j}\!+\!\frac{k^{+}(p_{0}^{+}\!-\!k^{+})}{p_{1}^{+}q^{+}}\mathbf{q}^{2}\epsilon^{ij}\mathrm{Dirac}_{\mathrm{V}}^{i\bar{\lambda}j}\epsilon^{\rho\bar{\lambda}}\frac{1}{D\!-\!3}\Bigg]\\
 & \times\mathcal{B}_{1}\big(0,\hat{Q}^{2},\frac{p_{0}^{+}-k^{+}}{q^{+}}\mathbf{q}\big)\Bigg\}\int_{\mathbf{x}}e^{-i\mathbf{k}_{\perp}\cdot\mathbf{x}}\big(U_{\mathbf{x}}-1\big)\;,
\end{aligned}
\label{eq:MA4_final-1}
\end{equation}
when the photon is transversally polarized.

\subsection{\label{subsec:4q}Instantaneous four-fermion interaction}

\begin{figure}[t]
\begin{centering}
\includegraphics[scale=0.25]{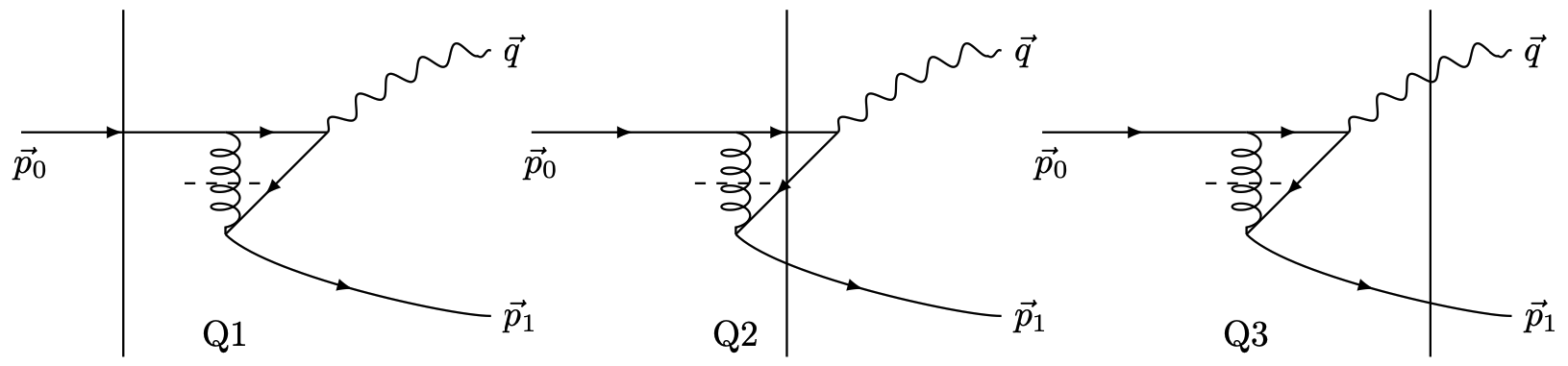}
\par\end{centering}
\caption{\label{fig:Q}The three virtual contributions with an instantaneous
$qq\bar{q}q$ vertex or, equivalently, a fictitious instantaneous
gluon in the $t$-channel.}
\end{figure}

\paragraph{Diagram Q1}

We obtain the following amplitudes for the first diagram in figure
\ref{fig:Q}:
\begin{equation}
\begin{aligned}\tilde{\mathcal{M}}_{\mathrm{Q1}}^{0} & \!=\!-\!\frac{\alpha_{s}C_{F}M}{p_{0}^{+}\mathbf{P}_{\perp}^{2}\!+\!p_{1}^{+}M^{2}}\!\int_{0}^{q^{+}}\mathrm{d}\ell_{1}^{+}\frac{8p_{1}^{+}\ell_{1}^{+}(q^{+}\!-\!\ell_{1}^{+})}{q^{+}(p_{0}^{+}\!-\!\ell_{1}^{+})^{2}}\mathcal{A}_{0}(\tilde{M}^{2})\int_{\mathbf{x}}e^{-i\mathbf{k}_{\perp}\cdot\mathbf{x}}\big(U_{\mathbf{x}}\!-\!1\big)\;,
\end{aligned}
\label{eq:MQ1L}
\end{equation}
with:
\begin{equation}
\tilde{M}^{2}\equiv-\frac{\ell_{1}^{+}(q^{+}-\ell_{1}^{+})}{(q^{+})^{2}}M^{2}-i0^{+}\;.\label{eq:Mtildedef}
\end{equation}
In the transverse case, the amplitude simply disappears (it is proportional
to an integral of the form $\int_{\boldsymbol{\ell}}\boldsymbol{\ell}^{i}/(\boldsymbol{\ell}^{2}+\Delta)$):
\begin{equation}
\begin{aligned}\tilde{\mathcal{M}}_{\mathrm{Q1}}^{\lambda} & =0\;.\end{aligned}
\label{eq:MQ1}
\end{equation}
The UV counterterm for $\tilde{\mathcal{M}}_{\mathrm{Q1}}^{0}$ is
given by:
\begin{equation}
\begin{aligned}\tilde{\mathcal{M}}_{\mathrm{Q1,UV}}^{0} & \!=\!-\frac{\alpha_{s}C_{F}M}{p_{0}^{+}\mathbf{P}_{\perp}^{2}\!+\!p_{1}^{+}M^{2}}\!\int_{0}^{q^{+}}\!\mathrm{d}\ell_{1}^{+}\frac{8p_{1}^{+}\ell_{1}^{+}(q^{+}\!-\!\ell_{1}^{+})}{q^{+}(p_{0}^{+}\!-\!\ell_{1}^{+})^{2}}\mathcal{A}_{0}\big(\Delta_{\mathrm{UV}}\big)\!\int_{\mathbf{x}}e^{-i\mathbf{k}_{\perp}\cdot\mathbf{x}}\big(U_{\mathbf{x}}-1\big)
\end{aligned}
\label{eq:MQ1L_UV}
\end{equation}
such that the subtracted amplitude is equal to:
\begin{equation}
\begin{aligned}\tilde{\mathcal{M}}_{\mathrm{Q1\!,sub}}^{0} & \!=\!-\frac{\alpha_{s}C_{F}M}{p_{0}^{+}\mathbf{P}_{\perp}^{2}\!+\!p_{1}^{+}M^{2}}\!\int_{0}^{q^{+}}\!\mathrm{d}\ell_{1}^{+}\frac{8p_{1}^{+}\ell_{1}^{+}(q^{+}\!-\!\ell_{1}^{+})}{q^{+}(p_{0}^{+}\!-\!\ell_{1}^{+})^{2}}\frac{1}{4\pi}\!\ln\frac{\Delta_{\mathrm{UV}}}{\tilde{M}^{2}}\!\int_{\mathbf{x}}e^{-i\mathbf{k}_{\perp}\cdot\mathbf{x}}\big(U_{\mathbf{x}}\!-\!1\big).\end{aligned}
\label{eq:MQ1L_sub}
\end{equation}

\paragraph{Diagram Q2}

We obtain for the amplitude in which the outgoing photon is longitudinally
polarized:
\begin{equation}
\begin{aligned}\tilde{\mathcal{M}}_{\mathrm{Q2}}^{0} & =-\frac{8\alpha_{s}}{M}\int_{0}^{q^{+}}\mathrm{d}\ell_{1}^{+}\frac{1}{(p_{0}^{+}-\ell_{1}^{+})^{2}}\frac{p_{1}^{+}\ell_{1}^{+}}{p_{1}^{+}+\ell_{1}^{+}}\\
 & \times\int_{\mathbf{x}_{1},\mathbf{x}_{2},\mathbf{x}_{3}}\tilde{M}^{2}\mathcal{K}\big(\mathbf{x}_{12},\tilde{M}^{2}\big)\\
 & \times\int_{\boldsymbol{\ell},\boldsymbol{\ell}_{2}}\frac{e^{i\boldsymbol{\ell}\cdot\mathbf{x}_{13}}e^{i\boldsymbol{\ell}_{2}\cdot\mathbf{x}_{23}}}{\Big(\boldsymbol{\ell}+\frac{\ell_{1}^{+}}{\ell_{1}^{+}+p_{1}^{+}}\boldsymbol{\ell}_{2}\Big)^{2}+\frac{p_{1}^{+}\ell_{1}^{+}p_{0}^{+}\boldsymbol{\ell}_{2}^{2}}{(q^{+}-\ell_{1}^{+})(p_{1}^{+}+\ell_{1}^{+})^{2}}}\\
 & \times e^{-i\mathbf{q}\cdot\Big(\frac{q^{+}-\ell_{1}^{+}}{q^{+}}\mathbf{x}_{2}+\frac{\ell_{1}^{+}}{q^{+}}\mathbf{x}_{1}\Big)}e^{-i\mathbf{p}_{1}\cdot\mathbf{x}_{3}}[U_{\mathbf{x}_{3}}t^{a}U_{\mathbf{x}_{2}}^{\dagger}U_{\mathbf{x}_{1}}t^{a}-C_{F}]\;,
\end{aligned}
\label{eq:MQ2L}
\end{equation}
and for the transverse case:
\begin{equation}
\begin{aligned}\tilde{\mathcal{M}}_{\mathrm{Q2}}^{\lambda} & =-4\alpha_{s}\int_{0}^{q^{+}}\mathrm{d}\ell_{1}^{+}\frac{1}{(p_{0}^{+}-\ell_{1}^{+})^{2}}\frac{p_{1}^{+}\ell_{1}^{+}}{p_{1}^{+}+\ell_{1}^{+}}\mathrm{Dirac}^{\lambda\bar{\lambda}}\Big(\frac{2\ell_{1}^{+}}{q^{+}}-1\Big)\\
 & \times\int_{\mathbf{x}_{1},\mathbf{x}_{2},\mathbf{x}_{3}}iA^{\bar{\lambda}}\big(\mathbf{x}_{12},\tilde{M}^{2}\big)\\
 & \times\int_{\boldsymbol{\ell},\boldsymbol{\ell}_{2}}\frac{e^{i\boldsymbol{\ell}\cdot\mathbf{x}_{13}}e^{i\boldsymbol{\ell}_{2}\cdot\mathbf{x}_{23}}}{\Big(\boldsymbol{\ell}+\frac{\ell_{1}^{+}}{\ell_{1}^{+}+p_{1}^{+}}\boldsymbol{\ell}_{2}\Big)^{2}+\frac{p_{1}^{+}\ell_{1}^{+}p_{0}^{+}\boldsymbol{\ell}_{2}^{2}}{(q^{+}-\ell_{1}^{+})(p_{1}^{+}+\ell_{1}^{+})^{2}}}\\
 & \times e^{-i\mathbf{q}\cdot\Big(\frac{q^{+}-\ell_{1}^{+}}{q^{+}}\mathbf{x}_{2}+\frac{\ell_{1}^{+}}{q^{+}}\mathbf{x}_{1}\Big)}e^{-i\mathbf{p}_{1}\cdot\mathbf{x}_{3}}[U_{\mathbf{x}_{3}}t^{a}U_{\mathbf{x}_{2}}^{\dagger}U_{\mathbf{x}_{1}}t^{a}-C_{F}]\;.
\end{aligned}
\label{eq:MQ2}
\end{equation}
To investigate whether these amplitudes contain any divergences, we
will take the $\mathbf{x}_{3}\to\mathbf{x}_{2}\to\mathbf{x}_{1}\to\mathbf{x}$
limit in the non-divergent piece, i.e. the last line, which allows
us to evaluate the integrals over $\mathbf{x}_{2}$ and $\mathbf{x}_{3}$
in the transverse momentum integrations. We easily obtain:
\begin{equation}
\begin{aligned}\lim_{\mathbf{x}_{i}\to\mathbf{x}}\tilde{\mathcal{M}}_{\mathrm{Q2}}^{0} & =\frac{4\alpha_{s}C_{F}}{M}\int_{0}^{q^{+}}\mathrm{d}\ell_{1}^{+}\frac{(q^{+}-\ell_{1}^{+})}{(p_{0}^{+}-\ell_{1}^{+})^{2}}\frac{\ell_{1}^{+}}{q^{+}}\mathcal{A}_{0}(\tilde{M}^{2})\int_{\mathbf{x}}e^{-i\mathbf{k}_{\perp}\cdot\mathbf{x}}\big(U_{\mathbf{x}}-1\big)\\
 & +\mathrm{finite}\;\mathrm{terms}\;,
\end{aligned}
\label{eq:MQ2L_limit}
\end{equation}
where we made us of identities and (\ref{eq:A0}) \ref{eq:SEloop}.
The counterterm for $\tilde{\mathcal{M}}_{\mathrm{Q2,UV}}^{0}$, therefore,
reads:
\begin{equation}
\begin{aligned}\tilde{\mathcal{M}}_{\mathrm{Q2,UV}}^{0} & =\frac{4\alpha_{s}C_{F}}{M}\int_{0}^{q^{+}}\mathrm{d}\ell_{1}^{+}\frac{(q^{+}-\ell_{1}^{+})}{(p_{0}^{+}-\ell_{1}^{+})^{2}}\frac{\ell_{1}^{+}}{q^{+}}\mathcal{A}_{0}(\Delta_{\mathrm{UV}})\int_{\mathbf{x}}e^{-i\mathbf{k}_{\perp}\cdot\mathbf{x}}\big(U_{\mathbf{x}}-1\big)\;,\\
 & =\frac{4\alpha_{s}C_{F}}{M}\Big(-2+\frac{(p_{0}^{+}+p_{1}^{+})}{q^{+}}\ln\frac{p_{0}^{+}}{p_{1}^{+}}\Big)\mathcal{A}_{0}(\Delta_{\mathrm{UV}})\int_{\mathbf{x}}e^{-i\mathbf{k}_{\perp}\cdot\mathbf{x}}\big(U_{\mathbf{x}}-1\big)\;,
\end{aligned}
\label{eq:MQ2L_UV}
\end{equation}
and the subtracted amplitude is given by:
\begin{equation}
\begin{aligned}\tilde{\mathcal{M}}_{\mathrm{Q2,sub}}^{0} & =-\frac{8\alpha_{s}}{M}\Bigg[\int_{0}^{q^{+}}\mathrm{d}\ell_{1}^{+}\frac{1}{(p_{0}^{+}-\ell_{1}^{+})^{2}}\frac{p_{1}^{+}\ell_{1}^{+}}{p_{1}^{+}+\ell_{1}^{+}}\int_{\mathbf{x}_{1},\mathbf{x}_{2},\mathbf{x}_{3}}\tilde{M}^{2}\mathcal{K}\big(\mathbf{x}_{12},\tilde{M}^{2}\big)\\
 & \times\int_{\boldsymbol{\ell},\boldsymbol{\ell}_{2}}\frac{e^{i\boldsymbol{\ell}\cdot\mathbf{x}_{13}}e^{i\boldsymbol{\ell}_{2}\cdot\mathbf{x}_{23}}}{\Big(\boldsymbol{\ell}+\frac{\ell_{1}^{+}}{\ell_{1}^{+}+p_{1}^{+}}\boldsymbol{\ell}_{2}\Big)^{2}+\frac{p_{1}^{+}\ell_{1}^{+}p_{0}^{+}\boldsymbol{\ell}_{2}^{2}}{(q^{+}-\ell_{1}^{+})(p_{1}^{+}+\ell_{1}^{+})^{2}}}\\
 & \times e^{-i\mathbf{q}\cdot\Big(\frac{q^{+}-\ell_{1}^{+}}{q^{+}}\mathbf{x}_{2}+\frac{\ell_{1}^{+}}{q^{+}}\mathbf{x}_{1}\Big)}e^{-i\mathbf{p}_{1}\cdot\mathbf{x}_{3}}[U_{\mathbf{x}_{3}}t^{a}U_{\mathbf{x}_{2}}^{\dagger}U_{\mathbf{x}_{1}}t^{a}-C_{F}]\\
 & -\Big(1-\frac{p_{0}^{+}+p_{1}^{+}}{2q^{+}}\ln\frac{p_{0}^{+}}{p_{1}^{+}}\Big)\mathcal{A}_{0}(\Delta_{\mathrm{UV}})\int_{\mathbf{x}}e^{-i\mathbf{k}_{\perp}\cdot\mathbf{x}}C_{F}\big(U_{\mathbf{x}}-1\big)\Bigg]\;.
\end{aligned}
\label{eq:MQ2L_sub}
\end{equation}
In the transversely polarized case, the amplitude turns out to be
free from UV divergences:
\begin{equation}
\begin{aligned}\lim_{\mathbf{x}_{i}\to\mathbf{x}}\tilde{\mathcal{M}}_{\mathrm{Q2}}^{\lambda} & =0\;.\end{aligned}
\label{eq:MQ2_limit}
\end{equation}

\paragraph{Diagram Q3}

We obtain the following expressions for the production amplitudes
for a longitudinally resp. transversally polarized virtual photon:
\begin{equation}
\begin{aligned}\tilde{\mathcal{M}}_{\mathrm{Q3}}^{0} & \!=\!-\frac{p_{0}^{+}\mathbf{q}^{2}-p_{1}^{+}M^{2}}{p_{0}^{+}\mathbf{q}^{2}\!+\!p_{1}^{+}M^{2}}\frac{\alpha_{s}C_{F}}{M}\!\int_{0}^{q^{+}}\!\mathrm{d}\ell_{1}^{+}\frac{4\ell_{1}^{+}(q^{+}\!-\!\ell_{1}^{+})}{q^{+}(p_{0}^{+}\!-\!\ell_{1}^{+})^{2}}\mathcal{A}_{0}(\tilde{Q}^{2})\!\int_{\mathbf{x}}e^{-i\mathbf{k}_{\perp}\cdot\mathbf{x}}\big(U_{\mathbf{x}}\!-\!1\big)\;,
\end{aligned}
\label{eq:MQ3L}
\end{equation}
with
\begin{equation}
\tilde{Q}^{2}\equiv\frac{\ell_{1}^{+}p_{0}^{+}(q^{+}-\ell_{1}^{+})}{p_{1}^{+}(q^{+})^{2}}\mathbf{q}^{2}-i0^{+}\;.
\end{equation}
and:
\begin{equation}
\begin{aligned}\tilde{\mathcal{M}}_{\mathrm{Q3}}^{\lambda} & =0\;.\end{aligned}
\label{eq:MQ3}
\end{equation}
The counterterm for $\tilde{\mathcal{M}}_{\mathrm{Q3}}^{0}$ is given
by:
\begin{equation}
\begin{aligned}\tilde{\mathcal{M}}_{\mathrm{Q3,UV}}^{0} & \!=\!-\frac{p_{0}^{+}\mathbf{q}^{2}\!-\!p_{1}^{+}M^{2}}{p_{0}^{+}\mathbf{q}^{2}\!+\!p_{1}^{+}M^{2}}\frac{\alpha_{s}C_{F}}{M}\!\int_{0}^{q^{+}}\!\!\mathrm{d}\ell_{1}^{+}\frac{4\ell_{1}^{+}(q^{+}\!-\!\ell_{1}^{+})}{q^{+}(p_{0}^{+}\!-\!\ell_{1}^{+})^{2}}\mathcal{A}_{0}(\Delta_{\mathrm{UV}})\!\int_{\mathbf{x}}\!e^{-i\mathbf{k}_{\perp}\cdot\mathbf{x}}\big(\!U_{\mathbf{x}}\!-\!1\!\big)\!\,,
\end{aligned}
\label{eq:MQ3L_UV}
\end{equation}
such that the subtracted amplitude reads:
\begin{equation}
\begin{aligned}\tilde{\mathcal{M}}_{\mathrm{Q3,sub}}^{0} & \!=\!-\frac{p_{0}^{+}\mathbf{q}^{2}\!-\!p_{1}^{+}M^{2}}{p_{0}^{+}\mathbf{q}^{2}\!+\!p_{1}^{+}M^{2}}\frac{\alpha_{s}C_{F}}{4\pi M}\!\int_{0}^{q^{+}}\!\mathrm{d}\ell_{1}^{+}\frac{4\ell_{1}^{+}(q^{+}\!-\!\ell_{1}^{+})}{q^{+}(p_{0}^{+}\!-\!\ell_{1}^{+})^{2}}\ln\frac{\Delta_{\mathrm{UV}}}{\tilde{Q}^{2}}\!\!\int_{\mathbf{x}}\!e^{-i\mathbf{k}_{\perp}\cdot\mathbf{x}}\big(U_{\mathbf{x}}\!-\!1\big)\,.\end{aligned}
\label{eq:MQ3L_sub}
\end{equation}

\subsection{\label{subsec:Inst}Instantaneous $gq\gamma q$ interaction}

\begin{figure}[t]
\begin{centering}
\includegraphics[scale=0.3]{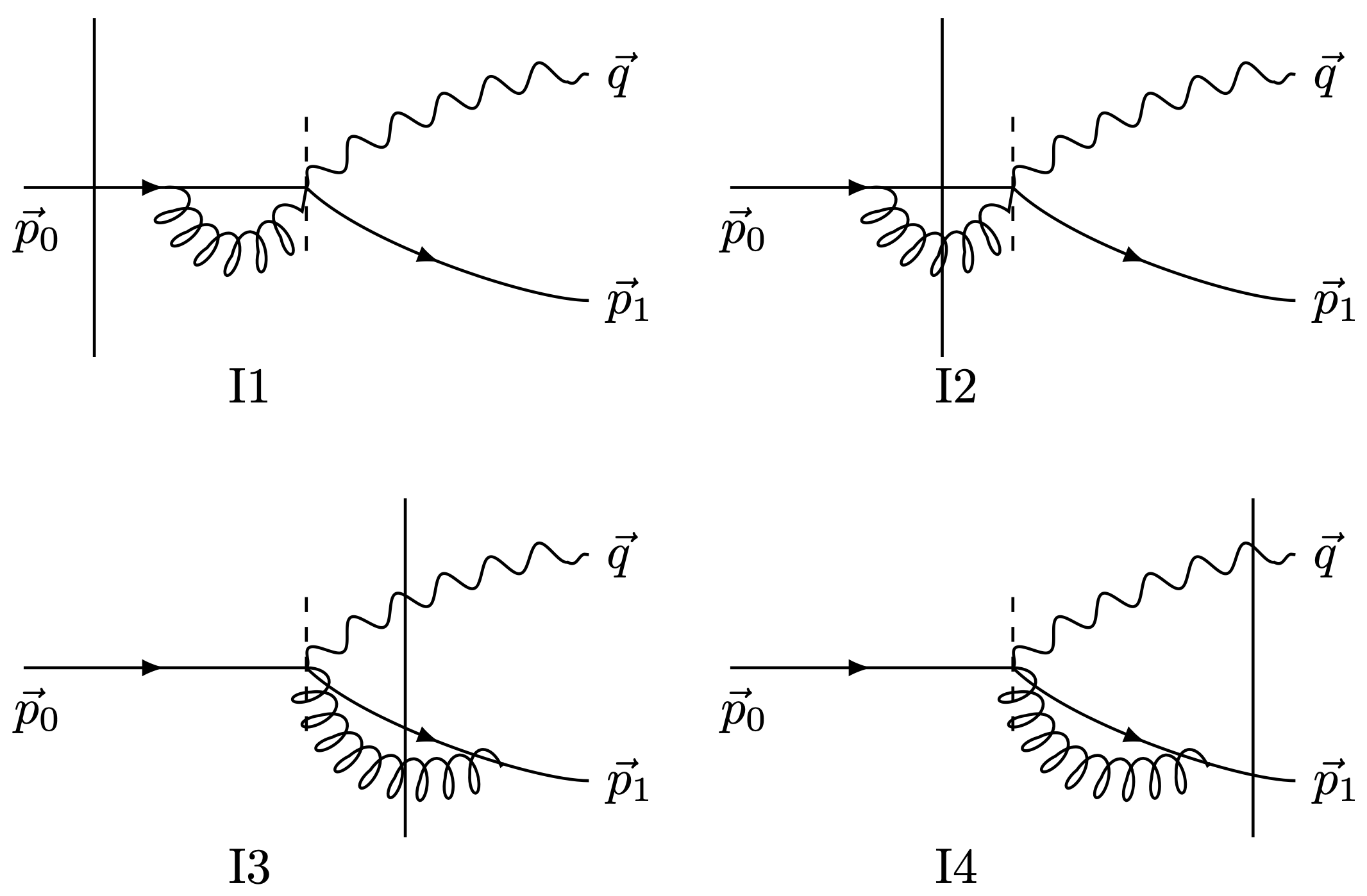}
\par\end{centering}
\caption{\label{fig:I}The four virtual diagrams with an instantaneous $qg\gamma q$-vertex.
Two additional diagrams where the gluon is attached to the asymptotic
incoming or outgoing quark disappear.}
\end{figure}

\paragraph{Diagram I1}

We obtain the following result for the amplitude corresponding to
graph I1 (see figure \ref{fig:I}) in the longitudinal case:
\begin{equation}
\begin{aligned}\tilde{\mathcal{M}}_{\mathrm{I1}}^{0} & \!=\!-\frac{\alpha_{s}C_{F}}{M}\!\!\int_{k_{\mathrm{min}}^{+}}^{p_{0}^{+}}\!\frac{\mathrm{d}k^{+}}{k^{+}}\Big(\frac{k^{+}}{p_{0}^{+}}\Big)^{\!2}\!\frac{q^{+}}{(p_{1}^{+}\!-\!k^{+})}\Big(\frac{2p_{0}^{+}}{k^{+}}\!+\!D\!-\!4\Big)\mathcal{A}_{0}(\Delta_{\mathrm{P}})\!\!\int_{\mathbf{x}}\!e^{-i\mathbf{k}_{\perp}\cdot\mathbf{x}}\big(U_{\mathbf{x}}\!-\!1\big)\!\,.
\end{aligned}
\label{eq:MI1L_final}
\end{equation}
The amplitude disappears when the outgoing virtual photon is transversely
polarized:
\begin{equation}
\begin{aligned}\mathcal{\tilde{M}}_{\mathrm{I1}}^{\lambda} & =0\;.\end{aligned}
\label{eq:MI1}
\end{equation}
Clearly, $\tilde{\mathcal{M}}_{\mathrm{I1}}^{0}$ has a ultraviolet
divergence, contained inside $\mathcal{A}_{0}$. We extract it by
defining the counterterm:
\begin{equation}
\begin{aligned}\tilde{\mathcal{M}}_{\mathrm{I1,UV}}^{0} & =-\frac{\alpha_{s}C_{F}}{M}\int_{k_{\mathrm{min}}^{+}}^{p_{0}^{+}}\frac{\mathrm{d}k^{+}}{k^{+}}\Big(\frac{k^{+}}{p_{0}^{+}}\Big)^{2}\frac{q^{+}}{(p_{1}^{+}-k^{+})}\Big(\frac{2p_{0}^{+}}{k^{+}}+D-4\Big)\\
 & \times\mathcal{A}_{0}(\Delta_{\mathrm{UV}})\int_{\mathbf{x}}e^{-i\mathbf{k}_{\perp}\cdot\mathbf{x}}\big(U_{\mathbf{x}}-1\big)\;,
\end{aligned}
\label{eq:MI1L_UV}
\end{equation}
such that the subtracted amplitude reads:
\begin{equation}
\begin{aligned}\tilde{\mathcal{M}}_{\mathrm{I1,sub}}^{0} & =\frac{\alpha_{s}C_{F}}{4\pi M}\frac{q^{+}}{2p_{0}^{+}}\int_{0}^{p_{0}^{+}}\frac{\mathrm{d}k^{+}}{k^{+}-p_{1}^{+}}\ln\frac{\Delta_{\mathrm{UV}}}{\Delta_{\mathrm{P}}}\int_{\mathbf{x}}e^{-i\mathbf{k}_{\perp}\cdot\mathbf{x}}\big(U_{\mathbf{x}}-1\big)\;.\end{aligned}
\label{eq:MI1L_sub}
\end{equation}

\paragraph{Diagram I2}

We obtain for the amplitude when the virtual photon is longitudinally
polarized:
\begin{equation}
\begin{aligned}\tilde{\mathcal{M}}_{\mathrm{I2}}^{0} & =\frac{\alpha_{s}}{M}\int_{k_{\mathrm{min}}^{+}}^{p_{0}^{+}}\frac{\mathrm{d}k^{+}}{k^{+}}\Big(\frac{k^{+}}{p_{0}^{+}}\Big)^{2}\frac{p_{0}^{+}-k^{+}}{p_{1}^{+}-k^{+}}\int_{\mathbf{x},\mathbf{z}}iA^{\eta^{\prime}}(\mathbf{x}-\mathbf{z})\mathrm{Dirac}^{\eta\eta^{\prime}}\big(\frac{2p_{0}^{+}}{k^{+}}-1\big)\\
 & \times\Bigg[\frac{q^{+}iA^{i}(\mathbf{x}-\mathbf{z},\Delta_{\mathrm{P}})}{p_{0}^{+}-k^{+}}(\delta^{\eta i}+i\sigma^{\eta i})\\
 & -\frac{2p_{1}^{+}-k^{+}}{p_{1}^{+}}\mathbf{P}_{\perp}^{i}\mathcal{K}\big(\mathbf{x}-\mathbf{z},\Delta_{\mathrm{P}}\big)\Big(\delta^{\eta i}+\frac{k^{+}}{2p_{1}^{+}-k^{+}}i\sigma^{\eta i}\Big)\Bigg]\\
 & \times e^{-i\mathbf{k}_{\perp}\cdot\Big(\frac{p_{0}^{+}-k^{+}}{p_{0}^{+}}\mathbf{x}+\frac{k^{+}}{p_{0}^{+}}\mathbf{z}\Big)}\big(t^{a}U_{\mathbf{x}}U_{\mathbf{z}}^{\dagger}t^{a}U_{\mathbf{z}}-C_{F}\big)\;.
\end{aligned}
\label{eq:MI2L_final}
\end{equation}
Likewise, in the transverse case:
\begin{equation}
\begin{aligned}\tilde{\mathcal{M}}_{\mathrm{I2}}^{\lambda} & =\alpha_{s}\int_{0}^{p_{0}^{+}}\frac{\mathrm{d}k^{+}}{k^{+}}\Big(\frac{k^{+}}{p_{0}^{+}}\Big)^{2}(p_{0}^{+}-k^{+}) \int_{\mathbf{x},\mathbf{z}}iA^{\bar{\eta}}(\mathbf{x}-\mathbf{z})\mathcal{K}\big(\mathbf{x}-\mathbf{z},\Delta_{\mathrm{P}}\big)\\
 & \times\Bigg[\Big(\frac{1}{p_{1}^{+}-k^{+}}+\frac{1}{p_{0}^{+}}\Big)\delta^{\eta\lambda}+\Big(\frac{1}{p_{1}^{+}-k^{+}}-\frac{1}{p_{0}^{+}}\Big)i\sigma^{\eta\lambda}\Bigg]\mathrm{Dirac}^{\eta\bar{\eta}}\Big(2\frac{p_{0}^{+}}{k^{+}}-1\Big)\\
 & \times e^{-i\mathbf{k}_{\perp}\cdot\Big(\frac{p_{0}^{+}-k^{+}}{p_{0}^{+}}\mathbf{x}+\frac{k^{+}}{p_{0}^{+}}\mathbf{z}\Big)}\big(t^{a}U_{\mathbf{x}}U_{\mathbf{z}}^{\dagger}t^{a}U_{\mathbf{z}}-C_{F}\big)\;.
\end{aligned}
\label{eq:MI2_final}
\end{equation}
Not surprisingly, the amplitude $\tilde{\mathcal{M}}_{\mathrm{I2}}^{0}$
contains an ultraviolet divergence in the limit $\mathbf{z}\to\mathbf{x}$:
\begin{equation}
\begin{aligned}\lim_{\mathbf{z}\to\mathbf{x}}\tilde{\mathcal{M}}_{\mathrm{I2}}^{0} & =\frac{\alpha_{s}C_{F}}{M}\int_{k_{\mathrm{min}}^{+}}^{p_{0}^{+}}\frac{\mathrm{d}k^{+}}{k^{+}}\Big(\frac{k^{+}}{p_{0}^{+}}\Big)^{2}\frac{q^{+}}{(p_{1}^{+}-k^{+})}\big(\frac{2p_{0}^{+}}{k^{+}}+D-4\big)\\
 & \times\mathcal{A}_{0}(\Delta_{\mathrm{P}})\int_{\mathbf{x}}e^{-i\mathbf{k}_{\perp}\cdot\mathbf{x}}\big(U_{\mathbf{x}}-1\big)\;.
\end{aligned}
\end{equation}
Defining the counterterm:
\begin{equation}
\begin{aligned}\tilde{\mathcal{M}}_{\mathrm{I2,UV}}^{0} & =\frac{\alpha_{s}C_{F}}{M}\int_{k_{\mathrm{min}}^{+}}^{p_{0}^{+}}\frac{\mathrm{d}k^{+}}{k^{+}}\Big(\frac{k^{+}}{p_{0}^{+}}\Big)^{2}\frac{q^{+}}{(p_{1}^{+}-k^{+})}\Big(\frac{2p_{0}^{+}}{k^{+}}+D-4\Big)\\
 & \times\mathcal{A}_{0}(\Delta_{\mathrm{UV}})\int_{\mathbf{x}}e^{-i\mathbf{k}_{\perp}\cdot\mathbf{x}}\big(U_{\mathbf{x}}-1\big) =-\tilde{\mathcal{M}}_{\mathrm{I1,UV}}^{0}\;,
\end{aligned}
\label{eq:MI2L_UV}
\end{equation}
we obtain the following result for the UV-subtracted amplitude:
\begin{equation}
\begin{aligned}\tilde{\mathcal{M}}_{\mathrm{I2}}^{0} & \!=\!\frac{\alpha_{s}}{M}\int_{k_{\mathrm{min}}^{+}}^{p_{0}^{+}}\frac{\mathrm{d}k^{+}}{k^{+}}\frac{k^{+}}{p_{0}^{+}}\Bigg[\frac{k^{+}}{p_{0}^{+}}\frac{p_{0}^{+}-k^{+}}{p_{1}^{+}-k^{+}}\int_{\mathbf{x},\mathbf{z}}iA^{\eta^{\prime}}(\mathbf{x}-\mathbf{z})\mathrm{Dirac}^{\eta\eta^{\prime}}\big(\frac{2p_{0}^{+}}{k^{+}}-1\big)\\
 & \!\times\!\!\Bigg(\!\!\frac{q^{+}iA^{i}(\mathbf{x}\!-\!\mathbf{z},\!\Delta_{\mathrm{P}})}{p_{0}^{+}\!-\!k^{+}}(\delta^{\eta i}\!+\!i\sigma^{\eta i})\!-\!\frac{2p_{1}^{+}\!-\!k^{+}}{p_{1}^{+}}\mathbf{P}_{\perp}^{i}\mathcal{K}\big(\mathbf{x}\!-\!\mathbf{z},\Delta_{\mathrm{P}}\big)\!\Big(\!\delta^{\eta i}\!+\!\frac{k^{+}}{2p_{1}^{+}\!-\!k^{+}}i\sigma^{\eta i}\!\Big)\!\!\Bigg)\\
 & \!\times \!e^{-i\mathbf{k}_{\perp}\cdot\Big(\frac{p_{0}^{+}-k^{+}}{p_{0}^{+}}\mathbf{x}+\frac{k^{+}}{p_{0}^{+}}\mathbf{z}\Big)}\big(t^{a}U_{\mathbf{x}}U_{\mathbf{z}}^{\dagger}t^{a}U_{\mathbf{z}}-C_{F}\big)\\
 &\! -\!\frac{2q^{+}}{(p_{1}^{+}-k^{+})}\mathcal{A}_{0}(\Delta_{\mathrm{UV}})\int_{\mathbf{x}}e^{-i\mathbf{k}_{\perp}\cdot\mathbf{x}}C_{F}\big(U_{\mathbf{x}}-1\big)\Bigg]\;.
\end{aligned}
\label{eq:MI2L_sub_final}
\end{equation}

\paragraph{Diagram I3}

The results for the amplitudes corresponding to the Feynman graph
I3 read:
\begin{equation}
\begin{aligned}\tilde{\mathcal{M}}_{\mathrm{I3}}^{0} & =\frac{\alpha_{s}}{M}\int_{0}^{p_{1}^{+}}\mathrm{d}k^{+}\frac{k^{+}(p_{1}^{+}-k^{+})}{(p_{1}^{+})^{2}}\mathrm{Dirac}^{\eta\bar{\eta}}\big(1-\frac{2p_{1}^{+}}{k^{+}}\big)\int_{\mathbf{x},\mathbf{z}}iA^{\bar{\eta}}(\mathbf{x}-\mathbf{z})\\
 & \times\Bigg[\Big(\frac{2p_{0}^{+}-k^{+}}{p_{1}^{+}(p_{0}^{+}-k^{+})}\mathbf{q}^{i}\mathcal{K}\big(\mathbf{x}-\mathbf{z},\Delta_{\mathrm{P}}\big)+\frac{q^{+}iA^{i}(\mathbf{x}-\mathbf{z},\Delta_{\mathrm{q}})}{(p_{0}^{+}-k^{+})(p_{1}^{+}-k^{+})}\Big)\delta^{\eta i}\\
 & -\Big(\frac{k^{+}}{p_{1}^{+}(p_{0}^{+}-k^{+})}\mathbf{q}^{i}\mathcal{K}\big(\mathbf{x}-\mathbf{z},\Delta_{\mathrm{P}}\big)+\frac{q^{+}iA^{i}(\mathbf{x}-\mathbf{z},\Delta_{\mathrm{q}})}{(p_{0}^{+}-k^{+})(p_{1}^{+}-k^{+})}\Big)i\sigma^{\eta i}\Bigg]\\
 & \times e^{-i\mathbf{k}_{\perp}\cdot\big(\frac{p_{1}^{+}-k^{+}}{p_{1}^{+}}\mathbf{x}+\frac{k^{+}}{p_{1}^{+}}\mathbf{z}\big)}\big(t^{a}U_{\mathbf{x}}U_{\mathbf{z}}^{\dagger}t^{a}U_{\mathbf{z}}-C_{F}\big)\;,
\end{aligned}
\label{eq:MI3L}
\end{equation}
and:
\begin{equation}
\begin{aligned}\tilde{\mathcal{M}}_{\mathrm{I3}}^{\lambda} & =\alpha_{s}\int_{0}^{p_{1}^{+}}\mathrm{d}k^{+}\frac{k^{+}(p_{1}^{+}-k^{+})}{(p_{1}^{+})^{2}}\mathrm{Dirac}^{\eta\bar{\eta}}\Big(1-\frac{2p_{1}^{+}}{k^{+}}\Big)\\
 & \times\Bigg[\Big(\frac{1}{p_{1}^{+}}+\frac{1}{p_{0}^{+}-k^{+}}\Big)\delta^{\eta\lambda}+\Big(\frac{1}{p_{1}^{+}}-\frac{1}{p_{0}^{+}-k^{+}}\Big)i\sigma^{\eta\lambda}\Bigg]\\
 & \times\int_{\mathbf{x},\mathbf{z}}iA^{\bar{\eta}}(\mathbf{x}-\mathbf{z})\mathcal{K}\big(\mathbf{x}-\mathbf{z},\Delta_{\mathrm{P}}\big) e^{-i\mathbf{k}_{\perp}\cdot\big(\frac{p_{1}^{+}-k^{+}}{p_{1}^{+}}\mathbf{x}+\frac{k^{+}}{p_{1}^{+}}\mathbf{z}\big)}\big(t^{c}U_{\mathbf{x}}U_{\mathbf{z}}^{\dagger}t^{c}U_{\mathbf{z}}-C_{F}\big)\;.
\end{aligned}
\label{eq:MI3}
\end{equation}
Amplitude $\tilde{\mathcal{M}}_{\mathrm{I3}}^{0}$ is UV divergent
in the limit $\mathbf{z}\to\mathbf{x}$:
\begin{equation}
\begin{aligned}\lim_{\mathbf{z}\to\mathbf{x}}\tilde{\mathcal{M}}_{\mathrm{I3}}^{0} & =\frac{\alpha_{s}C_{F}}{M}\int_{0}^{p_{1}^{+}}\mathrm{d}k^{+}\frac{k^{+}}{(p_{1}^{+})^{2}}\frac{q^{+}}{(p_{0}^{+}-k^{+})}\big(\frac{2p_{1}^{+}}{k^{+}}+D-4\big)\mathcal{A}_{0}(\Delta_{\mathrm{q}})\\
 & \times e^{-i\mathbf{k}_{\perp}\cdot\mathbf{x}}\big(U_{\mathbf{x}}-1\big)\;.
\end{aligned}
\label{eq:MI3L_limit_UV}
\end{equation}
The counterterm reads:
\begin{equation}
\begin{aligned}\tilde{\mathcal{M}}_{\mathrm{I3,UV}}^{0} & =\frac{\alpha_{s}C_{F}}{M}\int_{0}^{p_{1}^{+}}\mathrm{d}k^{+}\frac{k^{+}}{(p_{1}^{+})^{2}}\frac{q^{+}}{(p_{0}^{+}-k^{+})}\big(\frac{2p_{1}^{+}}{k^{+}}+D-4\big)\\
 & \times\mathcal{A}_{0}(\Delta_{\mathrm{UV}})e^{-i\mathbf{k}_{\perp}\cdot\mathbf{x}}\big(U_{\mathbf{x}}-1\big)\;,
\end{aligned}
\label{eq:MI3L_UV}
\end{equation}
such that the subtracted amplitude becomes:
\begin{equation}
\begin{aligned}\tilde{\mathcal{M}}_{\mathrm{I3,sub}}^{0} & =\frac{\alpha_{s}}{M}\int_{\mathbf{x}}e^{-i\mathbf{k}_{\perp}\cdot\mathbf{x}}\int_{0}^{p_{1}^{+}}\frac{\mathrm{d}k^{+}}{p_{0}^{+}-k^{+}}\Bigg[\int_{\mathbf{z}}\frac{k^{+}(p_{1}^{+}-k^{+})}{(p_{1}^{+})^{2}}iA^{i}(\mathbf{x}-\mathbf{z})\\
 & \times\Bigg(-\frac{2p_{1}^{+}}{k^{+}}\frac{q^{+}iA^{i}(\mathbf{x}-\mathbf{z},\Delta_{\mathrm{q}})}{p_{1}^{+}-k^{+}}\\
 &+\frac{k^{+}}{p_{1}^{+}}\mathbf{q}^{i}\mathcal{K}\big(\mathbf{x}-\mathbf{z},\Delta_{\mathrm{P}}\big)\Big(\frac{2p_{0}^{+}-k^{+}}{k^{+}}\big(1-\frac{2p_{1}^{+}}{k^{+}}\big)-1\Big)\Bigg)\\
 & \times e^{i\frac{k^{+}}{p_{1}^{+}}\mathbf{k}_{\perp}\cdot(\mathbf{x}-\mathbf{z})}\big(t^{a}U_{\mathbf{x}}U_{\mathbf{z}}^{\dagger}t^{a}U_{\mathbf{z}}-C_{F}\big)-\frac{2q^{+}}{p_{1}^{+}}\mathcal{A}_{0}(\Delta_{\mathrm{UV}})C_{F}\big(U_{\mathbf{x}}-1\big)\Bigg]\;.
\end{aligned}
\label{eq:MI3L_sub}
\end{equation}

\paragraph{Diagram I4}

We obtain for the last diagram in figure \ref{fig:I}:
\begin{equation}
\begin{aligned}\tilde{\mathcal{M}}_{\mathrm{I4}}^{0} & =-\frac{\alpha_{s}C_{F}}{M}\int_{0}^{p_{1}^{+}}\mathrm{d}k^{+}\frac{q^{+}k^{+}}{(p_{1}^{+})^{2}(p_{0}^{+}-k^{+})}\big(\frac{2p_{1}^{+}}{k^{+}}+D-4\big)\\
 & \times\mathcal{A}_{0}(\Delta_{\mathrm{q}})\int_{\mathbf{x}}e^{-i\mathbf{k}_{\perp}\cdot\mathbf{x}}\big(U_{\mathbf{x}}-1\big)\;,
\end{aligned}
\label{eq:MI4L_final}
\end{equation}
and:
\begin{equation}
\begin{aligned}\tilde{\mathcal{M}}_{\mathrm{I4}}^{\lambda} & =0\;.\end{aligned}
\label{eq:MI4}
\end{equation}
The counterterm reads:
\begin{equation}
\begin{aligned}\tilde{\mathcal{M}}_{\mathrm{I4,UV}}^{0} & =-\frac{\alpha_{s}C_{F}}{M}\int_{0}^{p_{1}^{+}}\mathrm{d}k^{+}\frac{q^{+}k^{+}}{(p_{1}^{+})^{2}(p_{0}^{+}-k^{+})}\big(\frac{2p_{1}^{+}}{k^{+}}+D-4\big)\\
 & \times\mathcal{A}_{0}(\Delta_{\mathrm{UV}})\int_{\mathbf{x}}e^{-i\mathbf{k}_{\perp}\cdot\mathbf{x}}\big(U_{\mathbf{x}}-1\big) =-\tilde{\mathcal{M}}_{\mathrm{I3,UV}}^{0}\;,
\end{aligned}
\label{eq:MI4L_UV}
\end{equation}
with the subtracted amplitude:
\begin{equation}
\begin{aligned}\tilde{\mathcal{M}}_{\mathrm{I4,sub}}^{0} & =\frac{\alpha_{s}C_{F}}{2\pi M}\int_{0}^{p_{1}^{+}}\mathrm{d}k^{+}\frac{q^{+}}{p_{1}^{+}(k^{+}-p_{0}^{+})}\ln\frac{\Delta_{\mathrm{UV}}}{\Delta_{\mathrm{q}}}\int_{\mathbf{x}}e^{-i\mathbf{k}_{\perp}\cdot\mathbf{x}}\big(U_{\mathbf{x}}-1\big)\;.\end{aligned}
\label{eq:MI4L_sub}
\end{equation}

\subsection{\label{subsec:Z}Field-strength renormalization}

\begin{figure}[t]
\begin{centering}
\includegraphics[scale=0.27]{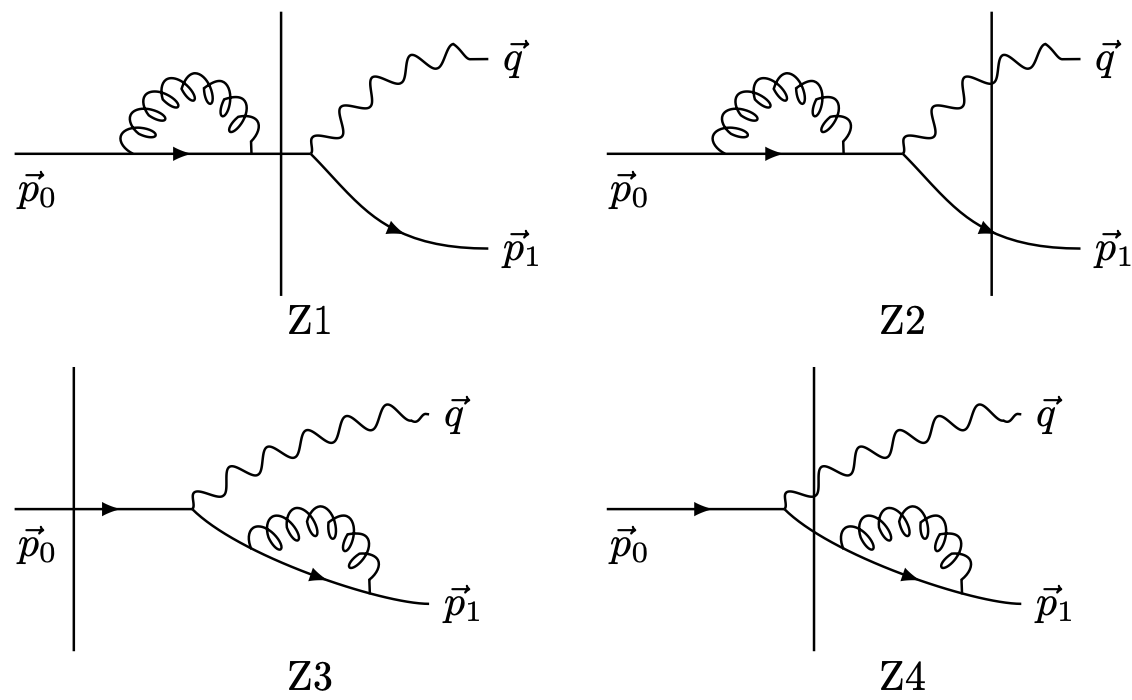}
\par\end{centering}
\caption{\label{fig:Z}The four virtual contributions with a gluon loop on
the asymptotic in- or outgoing quark.}
\end{figure}
The last class of virtual diagrams that we consider are those corresponding
to self-energy corrections on the asymptotic quarks. Although the
corresponding amplitudes vanish in dimensional renormalization, we
will compute them explicitly. Indeed, they consist of two identical
parts with an opposite sign, one corresponding to an UV pole, the
other to an IR one. Since in dimensional renormalization $\epsilon_{\mathrm{UV}}=\epsilon_{\mathrm{IR}}$,
the net result is zero, although in our analysis we will treat the
ultraviolet and infrared parts separately.

It should be noted that, in LCPT, it is not possible to compute these
amplitudes with our usual method, since they feature an energy denominator
$\propto(p_{0}^{-}-p_{0}^{-})^{-1}$. Instead, we make use of the
normalization properties of the Fock states. As demonstrated in appendix
(\ref{subsec:app_Z}), the total contribution of the diagrams in fig.
\ref{fig:Z} can be summarized into the following field-renormalization
constant:
\begin{equation}
\begin{aligned}\mathcal{Z} & =\Bigg[1-\frac{\alpha_{s}C_{F}}{\pi}\Big(\frac{1}{\epsilon_{\mathrm{UV}}}-\frac{1}{\epsilon_{\mathrm{coll}}}\Big)\Big(-\frac{3}{2}+\mathrm{ln}\frac{p_{0}^{+}}{k_{\mathrm{min}}^{+}}+\mathrm{ln}\frac{p_{1}^{+}}{k_{\mathrm{min}}^{+}}\Big)\Bigg]\;,\end{aligned}
\label{eq:fieldrenormalization}
\end{equation}
which appears on the level of the squared amplitude as:
\begin{equation}
\begin{aligned}2\mathcal{M}_{\mathrm{LO}}^{\dagger}\big(\mathcal{M}_{\mathrm{Z1}}+\mathcal{M}_{\mathrm{Z2}}+\mathcal{M}_{\mathrm{Z3}}+\mathcal{M}_{\mathrm{Z4}}\big) & =(\mathcal{Z}-1)\big|\mathcal{M}_{\mathrm{LO}}\big|^{2}\;,\end{aligned}
\label{eq:Z_cross}
\end{equation}
irrespective of the polarization of the emitted virtual photon.

In anticipation of what follows, we can rewrite $\mathcal{Z}$ as
\begin{equation}
\mathcal{Z}=1+\mathcal{Z}_{\mathrm{UV}}+\mathcal{Z}_{\mathrm{UV}}^{\dagger}+\mathcal{Z}_{\mathrm{IS}}+\mathcal{Z}_{\mathrm{IS}}^{\dagger}+\mathcal{Z}_{\mathrm{FS}}+\mathcal{Z}_{\mathrm{FS}}^{\dagger}\;,
\end{equation}
with:
\begin{equation}
\begin{aligned}\mathcal{Z}_{\mathrm{UV}} & =\frac{\alpha_{s}C_{F}}{\pi}\Big(\frac{1}{\epsilon_{\mathrm{UV}}}+\mathrm{ln}\frac{\mu^{2}}{\mu_{\scriptscriptstyle{R}}^{2}}\Big)\Big(\frac{3}{4}-\frac{1}{2}\mathrm{ln}\frac{p_{0}^{+}}{k_{\mathrm{min}}^{+}}-\frac{1}{2}\mathrm{ln}\frac{p_{1}^{+}}{k_{\mathrm{min}}^{+}}\Big)\;,\\
\mathcal{Z}_{\mathrm{IS}} & =\frac{\alpha_{s}C_{F}}{\pi}\Big(\frac{1}{\epsilon_{\mathrm{coll}}}+\mathrm{ln}\frac{\mu^{2}}{\mu_{\scriptscriptstyle{R}}^{2}}\Big)\Big(-\frac{3}{8}+\frac{1}{2}\mathrm{ln}\frac{p_{0}^{+}}{k_{\mathrm{min}}^{+}}\Big)\;,\\
\mathcal{Z}_{\mathrm{FS}} & =\frac{\alpha_{s}C_{F}}{\pi}\Big(\frac{1}{\epsilon_{\mathrm{coll}}}+\mathrm{ln}\frac{\mu^{2}}{\mu_{\scriptscriptstyle{R}}^{2}}\Big)\Big(-\frac{3}{8}+\frac{1}{2}\mathrm{ln}\frac{p_{1}^{+}}{k_{\mathrm{min}}^{+}}\Big)\;.
\end{aligned}
\label{eq:Z}
\end{equation}
Since the scale $\mu_{\scriptscriptstyle{R}}^{2}$ cancels on the cross-section level,
we are free to adjust it at will.

\section{\label{sec:UV}Ultraviolet divergences }

At the perturbative order of our calculation, all short-distance or
ultraviolet divergences are contained in the virtual corrections presented
in the previous section. Because contributions to the running coupling
take place at higher perturbative orders, and since we work in the
massless limit, these singularities have to cancel on the level of
the total virtual amplitude. In this section, we will demonstrate
that this is indeed the case, which is an important check of the calculation. 

Two classes of virtual amplitudes, namely the self-energy corrections
in section \ref{subsec:SE}, and the ones with an instantaneous $gq\gamma q$
interaction (section \ref{subsec:Inst}), are somewhat particular.
Indeed, we have seen that (see eqs. (\ref{eq:MSE2L_UV}), (\ref{eq:MSE2TUV}),
and (\ref{eq:MSE4_UV})):
\begin{equation}
\begin{aligned}\tilde{\mathcal{M}}_{\mathrm{SE2,UV}}^{0,\lambda} & =-\tilde{\mathcal{M}}_{\mathrm{SE1,UV}}^{0,\lambda}\qquad\mathrm{and}\qquad\tilde{\mathcal{M}}_{\mathrm{SE4,UV}}^{0,\lambda}=-\tilde{\mathcal{M}}_{\mathrm{SE3,UV}}^{0,\lambda}\;,\end{aligned}
\end{equation}
and, therefore, all UV divergences already cancel in the sum of amplitudes
belonging to the same class. The same is true for the amplitudes with
an instantaneous $gq\gamma q$ vertex and a longitudinally polarized
photon (eqs. (\ref{eq:MI2L_UV}) and (\ref{eq:MI4L_UV})):
\begin{equation}
\begin{aligned}\tilde{\mathcal{M}}_{\mathrm{I2,UV}}^{0} & =-\tilde{\mathcal{M}}_{\mathrm{I1,UV}}^{0}\qquad\mathrm{and}\qquad\tilde{\mathcal{M}}_{\mathrm{I4,UV}}^{0}=-\tilde{\mathcal{M}}_{\mathrm{I3,UV}}^{0}\;,\end{aligned}
\end{equation}
while in the transverse case, the amplitudes never exhibit UV divergencies
to begin with.

All the ultraviolet divergences we encountered are contained in integrals
$\mathcal{A}_{0}(\Delta)$ (\ref{eq:A0}) which, when evaluated in
dimensional regularization, give: 
\begin{equation}
\begin{aligned}\mathcal{A}_{0}(\Delta) & \equiv\int_{\boldsymbol{\ell}}\frac{1}{\boldsymbol{\ell}^{2}+\Delta}=\frac{1}{4\pi}\Big(\frac{1}{\epsilon_{\mathrm{UV}}}-\gamma_{\scriptscriptstyle{E}}+\ln4\pi+\ln\frac{\mu^{2}}{\Delta}\Big)+\mathcal{O}(\epsilon_{\mathrm{UV}})\;.\end{aligned}
\end{equation}
Following the $\overline{\mathrm{MS}}$ scheme, we will extract the
poles together with the universal constants $-\gamma_{\scriptscriptstyle{E}}+\ln4\pi$,
writing:
\begin{equation}
\begin{aligned}\mathcal{A}_{0}(\Delta) & =\frac{1}{4\pi}\Big(\frac{1}{\tilde{\epsilon}}+\ln\frac{\mu^{2}}{\Delta}\Big)+\mathcal{O}(\epsilon_{\mathrm{UV}})\;,\end{aligned}
\label{eq:A0exp}
\end{equation}
where we defined:
\begin{equation}
\begin{aligned}\frac{1}{\tilde{\epsilon}} & \equiv\frac{1}{\epsilon_{\mathrm{UV}}}-\gamma_{\scriptscriptstyle{E}}+\ln4\pi\;.\end{aligned}
\end{equation}

\subsection{Longitudinal polarization}

\paragraph{Vertex corrections}

We start by collecting the UV-divergent parts of the vertex-correction
amplitudes, which are collected by the counterterms. The counterterms
(\ref{eq:MV1L_UV}) and (\ref{eq:MV2L_UV}) can be combined into:
\begin{equation}
\begin{aligned}\tilde{\mathcal{M}}_{\mathrm{V1,UV}}^{0}+\tilde{\mathcal{M}}_{\mathrm{V2,UV}}^{0} & =\tilde{\mathcal{M}}_{\mathrm{LO1}}^{0}\times-\frac{\alpha_{s}C_{F}}{D-2}\int_{k_{\mathrm{min}}^{+}}^{p_{1}^{+}}\frac{\mathrm{d}k^{+}}{k^{+}}\frac{(k^{+})^{2}}{p_{0}^{+}p_{1}^{+}}\mathrm{Dirac}_{\mathrm{V}}^{j0j}\mathcal{A}_{0}(\Delta_{\mathrm{UV}})\;,\end{aligned}
\label{eq:V12UV}
\end{equation}
while summing the poles from (\ref{eq:MV3L_UV}), and (\ref{eq:MV4L_final})
gives:
\begin{equation}
\begin{aligned}\tilde{\mathcal{M}}_{\mathrm{V3,UV}}^{0}+\tilde{\mathcal{M}}_{\mathrm{V4,UV}}^{0} & =\tilde{\mathcal{M}}_{\mathrm{LO2}}^{0}\times-\frac{\alpha_{s}C_{F}}{D-2}\int_{k_{\mathrm{min}}^{+}}^{p_{1}^{+}}\frac{\mathrm{d}k^{+}}{k^{+}}\frac{(k^{+})^{2}}{p_{0}^{+}p_{1}^{+}}\mathrm{Dirac}_{\mathrm{V}}^{j0j}\mathcal{A}_{0}(\Delta_{\mathrm{UV}})\;.\end{aligned}
\label{eq:V34UV}
\end{equation}
In the above expressions, $\mathrm{Dirac}_{\mathrm{V}}^{j0j}$ is the symmetric part of the Dirac structure (\ref{eq:DiracVL}). We immediately find:
\begin{equation}
\begin{aligned}\tilde{\mathcal{M}}_{\mathrm{V1,UV}}^{0}\!+\!\tilde{\mathcal{M}}_{\mathrm{V2,UV}}^{0} & \!=\!\tilde{\mathcal{M}}_{\mathrm{LO1}}^{0}\frac{\alpha_{s}C_{F}}{\pi}\Bigg[\!\Big(\!\frac{1}{\tilde{\epsilon}}\!+\!\ln\frac{\mu^{2}}{\Delta_{\mathrm{UV}}}\Big)\!\Big(\!-\!\frac{3p_{1}^{+}\!+\!2q^{+}}{4p_{0}^{+}}\!+\!\ln\frac{p_{1}^{+}}{k_{\mathrm{min}}^{+}}\Big)\!-\!\frac{1}{4}\frac{p_{1}^{+}}{p_{0}^{+}}\Bigg]\;,\end{aligned}
\label{eq:V12LUV_final}
\end{equation}
and:
\begin{equation}
\begin{aligned}\tilde{\mathcal{M}}_{\mathrm{V3,UV}}^{0}\!+\!\tilde{\mathcal{M}}_{\mathrm{V4,UV}}^{0} & \!=\!\tilde{\mathcal{M}}_{\mathrm{LO2}}^{0}\frac{\alpha_{s}C_{F}}{\pi}\Bigg[\!\Big(\!\frac{1}{\tilde{\epsilon}}\!+\!\ln\frac{\mu^{2}}{\Delta_{\mathrm{UV}}}\Big)\!\Big(\!-\!\frac{3p_{1}^{+}\!+\!2q^{+}}{4p_{0}^{+}}\!+\!\ln\frac{p_{1}^{+}}{k_{\mathrm{min}}^{+}}\Big)\!-\!\frac{1}{4}\frac{p_{1}^{+}}{p_{0}^{+}}\Bigg]\;.\end{aligned}
\label{eq:V34LUV_final}
\end{equation}

\paragraph{Antiquark vertex corrections}

Likewise, let us collect the UV counterterms for the four amplitudes
with an antiquark vertex. We see that the first three of them: \ref{eq:MA1L_UV}),
(\ref{eq:MA2L_UV}), and (\ref{eq:MA3L_UV}), nicely combine into:
\begin{equation}
\begin{aligned} & \tilde{\mathcal{M}}_{\mathrm{A1,UV}}^{0}+\tilde{\mathcal{M}}_{\mathrm{A2,UV}}^{0}+\tilde{\mathcal{M}}_{\mathrm{A3,UV}}^{0}\\ 
& =\tilde{\mathcal{M}}_{\mathrm{LO1}}^{0}\times-\frac{\alpha_{s}C_{F}}{D-2}\int_{p_{1}^{+}}^{p_{0}^{+}}\frac{\mathrm{d}k^{+}}{k^{+}}\frac{k^{+}(p_{0}^{+}-k^{+})}{p_{0}^{+}q^{+}}\mathrm{Dirac}_{\mathrm{V}}^{j0j}\mathcal{A}_{0}(\Delta_{\mathrm{UV}})\;.\end{aligned}
\label{eq:A123UV}
\end{equation}
The UV-counterterm (\ref{eq:MA4L_UV}) is equal to:
\begin{equation}
\begin{aligned}\tilde{\mathcal{M}}_{\mathrm{A4,UV}}^{0} & =\tilde{\mathcal{M}}_{\mathrm{LO2}}^{0}\times-\frac{\alpha_{s}C_{F}}{D-2}\int_{p_{1}^{+}}^{p_{0}^{+}}\frac{\mathrm{d}k^{+}}{k^{+}}\frac{k^{+}(p_{0}^{+}-k^{+})}{p_{0}^{+}q^{+}}\mathrm{Dirac}_{\mathrm{V}}^{j0j}\mathcal{A}_{0}(\Delta_{\mathrm{UV}})\;.\end{aligned}
\end{equation}
Using expression (\ref{eq:DiracVL}) for the symmetrized part of the
Dirac structure, the plus-momentum integral in the above expressions
can be easily evaluated, yielding:
\begin{equation}
\begin{aligned}\tilde{\mathcal{M}}_{\mathrm{A1+2+3,UV}}^{0} & \!=\!\tilde{\mathcal{M}}_{\mathrm{LO1}}^{0}\frac{\alpha_{s}C_{F}}{\pi}\Bigg[\Big(\frac{1}{\tilde{\epsilon}}\!+\!\ln\frac{\mu^{2}}{\Delta_{\mathrm{UV}}}\Big)\Big(\frac{7}{4}\!+\!\frac{p_{1}^{+}}{4p_{0}^{+}}\!-\!\frac{3p_{1}^{+}\!+\!p_{0}^{+}}{2q^{+}}\ln\frac{p_{0}^{+}}{p_{1}^{+}}\Big)\!-\!\frac{q^{+}}{4p_{0}^{+}}\Bigg]\;,\end{aligned}
\label{eq:MA123L_UV_final}
\end{equation}
and:
\begin{equation}
\begin{aligned}\tilde{\mathcal{M}}_{\mathrm{A4,UV}}^{0} & =\tilde{\mathcal{M}}_{\mathrm{LO2}}^{0}\frac{\alpha_{s}C_{F}}{\pi}\Bigg[\Big(\frac{1}{\tilde{\epsilon}}+\ln\frac{\mu^{2}}{\Delta_{\mathrm{UV}}}\Big)\Big(\frac{7}{4}+\frac{p_{1}^{+}}{4p_{0}^{+}}-\frac{3p_{1}^{+}+p_{0}^{+}}{2q^{+}}\ln\frac{p_{0}^{+}}{p_{1}^{+}}\Big)-\frac{q^{+}}{4p_{0}^{+}}\Bigg]\;.\end{aligned}
\label{eq:MA4L_UV_final}
\end{equation}

\paragraph{Instantaneous four-fermion vertex}

Another category of ultraviolet-singular virtual diagrams are those
with an instantaneous four-fermion vertex. Collecting the counterterms
(\ref{eq:MQ1L_UV}) and (\ref{eq:MQ2L_UV}), we obtain an expression
proportional to the leading-order amplitude $\mathcal{M}_{\mathrm{LO1}}^{0}$
(\ref{eq:MLOL}):
\begin{equation}
\begin{aligned}\tilde{\mathcal{M}}_{\mathrm{Q1,UV}}^{0}+\tilde{\mathcal{M}}_{\mathrm{Q2,UV}}^{0} & =\tilde{\mathcal{M}}_{\mathrm{LO1}}^{0}4\alpha_{s}C_{F}\int_{0}^{q^{+}}\mathrm{d}\ell_{1}^{+}\frac{(q^{+}-\ell_{1}^{+})}{(p_{0}^{+}-\ell_{1}^{+})^{2}}\frac{\ell_{1}^{+}}{q^{+}}\mathcal{A}_{0}(\Delta_{\mathrm{UV}})\;.\end{aligned}
\label{eq:Q12UV}
\end{equation}
Likewise, the counterterm (\ref{eq:MQ3L_UV}) for the amplitude $\tilde{\mathcal{M}}_{\mathrm{Q3}}^{0}$
is:
\begin{equation}
\begin{aligned}\tilde{\mathcal{M}}_{\mathrm{Q3,UV}}^{0} & =\tilde{\mathcal{M}}_{\mathrm{LO2}}^{0}4\alpha_{s}C_{F}\int_{0}^{q^{+}}\mathrm{d}\ell_{1}^{+}\frac{\ell_{1}^{+}(q^{+}-\ell_{1}^{+})}{q^{+}(p_{0}^{+}-\ell_{1}^{+})^{2}}\mathcal{A}_{0}(\Delta_{\mathrm{UV}})\;.\end{aligned}
\label{eq:Q3UV}
\end{equation}
Evaluating the plus-momentum integrals, we obtain:
\begin{equation}
\begin{aligned}\tilde{\mathcal{M}}_{\mathrm{Q1+2,UV}}^{0} & =\tilde{\mathcal{M}}_{\mathrm{LO1}}^{0}\frac{\alpha_{s}C_{F}}{\pi}\Big(\frac{1}{\tilde{\epsilon}}+\ln\frac{\mu^{2}}{\Delta_{\mathrm{UV}}}\Big)\Big(-2+\frac{p_{0}^{+}+p_{1}^{+}}{q^{+}}\ln\frac{p_{0}^{+}}{p_{1}^{+}}\Big)\;,\end{aligned}
\label{eq:Q12UVfinalOK}
\end{equation}
and:
\begin{equation}
\begin{aligned}\tilde{\mathcal{M}}_{\mathrm{Q3,UV}}^{0} & =\tilde{\mathcal{M}}_{\mathrm{LO2}}^{0}\frac{\alpha_{s}C_{F}}{\pi}\Big(\frac{1}{\tilde{\epsilon}}+\ln\frac{\mu^{2}}{\Delta_{\mathrm{UV}}}\Big)\Big(-2+\frac{p_{0}^{+}+p_{1}^{+}}{q^{+}}\ln\frac{p_{0}^{+}}{p_{1}^{+}}\Big)\;.\end{aligned}
\label{eq:Q3UV_final_OK}
\end{equation}

\paragraph{Total UV contribution}

The last set of UV divergences are those stemming from the field-strength
renormalization diagrams (eqs. (\ref{eq:Z_cross}), (\ref{eq:Z})):
\begin{equation}
\begin{aligned}\mathcal{Z}_{\mathrm{UV}}\mathcal{\tilde{M}}_{\mathrm{LO1,2}}^{0} & \!=\!\mathcal{\tilde{M}}_{\mathrm{LO1,2}}^{0}\frac{\alpha_{s}C_{F}}{\pi}\Big(\frac{1}{\tilde{\epsilon}}\!-\!\ln4\pi e^{-\gamma_{\scriptscriptstyle{E}}}\!+\!\mathrm{ln}\frac{\mu^{2}}{\mu_{\scriptscriptstyle{R}}^{2}}\Big)\Big(\frac{3}{4}\!-\!\frac{1}{2}\mathrm{ln}\frac{p_{0}^{+}}{k_{\mathrm{min}}^{+}}\!-\!\frac{1}{2}\mathrm{ln}\frac{p_{1}^{+}}{k_{\mathrm{min}}^{+}}\Big)\;.\end{aligned}
\label{eq:Z_UV}
\end{equation}
Combining them with the results from the previous paragraphs, namely
(\ref{eq:V12LUV_final}), (\ref{eq:MA123L_UV_final}), and (\ref{eq:Q12UVfinalOK}),
we obtain:
\begin{equation}
\begin{aligned} & \tilde{\mathcal{M}}_{\mathrm{V1+2,UV}}^{0}+\tilde{\mathcal{M}}_{\mathrm{A1+2+3,UV}}^{0}+\tilde{\mathcal{M}}_{\mathrm{Q1+2,UV}}^{0}+\mathcal{Z}_{\mathrm{UV}}\mathcal{\tilde{M}}_{\mathrm{LO1}}^{0}\\
 & =\tilde{\mathcal{M}}_{\mathrm{LO1}}^{0}\frac{\alpha_{s}C_{F}}{\pi}\Bigg[\Big(-\frac{3}{4}+\frac{1}{2}\ln\frac{p_{1}^{+}}{k_{\mathrm{min}}^{+}}+\frac{1}{2}\ln\frac{p_{0}^{+}}{k_{\mathrm{min}}^{+}}\Big)\ln\frac{4\pi e^{-\gamma_{\scriptscriptstyle{E}}}\mu_{\scriptscriptstyle{R}}^{2}}{\Delta_{\mathrm{UV}}}-\frac{1}{4}\Bigg]\;.
\end{aligned}
\label{eq:L1UV_final}
\end{equation}
Likewise, combining (\ref{eq:Z_UV}) with eqs. (\ref{eq:V34LUV_final}),
(\ref{eq:MA4L_UV_final}), and (\ref{eq:Q3UV_final_OK}):
\begin{equation}
\begin{aligned} & \tilde{\mathcal{M}}_{\mathrm{V3+4,UV}}^{0}+\tilde{\mathcal{M}}_{\mathrm{A4,UV}}^{0}+\tilde{\mathcal{M}}_{\mathrm{Q3,UV}}^{0}+\mathcal{Z}_{\mathrm{UV}}\mathcal{\tilde{M}}_{\mathrm{LO2}}^{0}\\
 & =\tilde{\mathcal{M}}_{\mathrm{LO2}}^{0}\frac{\alpha_{s}C_{F}}{\pi}\Bigg[\Big(-\frac{3}{4}+\frac{1}{2}\ln\frac{p_{1}^{+}}{k_{\mathrm{min}}^{+}}+\frac{1}{2}\ln\frac{p_{0}^{+}}{k_{\mathrm{min}}^{+}}\Big)\ln\frac{4\pi e^{-\gamma_{\scriptscriptstyle{E}}}\mu_{\scriptscriptstyle{R}}^{2}}{\Delta_{\mathrm{UV}}}-\frac{1}{4}\Bigg]\;.
\end{aligned}
\label{eq:L2UV_final}
\end{equation}
The above two results confirm our earlier claim, namely that in the
longitudinal case all the UV divergences encountered in our calculation
cancel between the different virtual diagrams. Note that there are
still divergences left, in the form of single logarithms depending
on the plus-momentum cutoff $k_{\mathrm{min}}^{+}$. In section \ref{sec:JIMWLK},
we will show how they can be absorbed into the high-energy evolution
of the target gluon density.

\subsection{Transverse polarization}

When the outgoing virtual photon is transversely polarized, the structure
of the UV cancellation is slightly different compared to the previous
subsection. In particular, there are no contributions from the diagrams
with an instantaneous four-fermion vertex. Moreover, since the Dirac
structure is now much more complicated than in the longitudinal case,
we are forced to replace the spinor structures with simpler ones,
obviously in such a way that this procedure still yields the correct
cross section. Hence, although the cancellation of UV divergences
should already take place at the level of the amplitude, as was the
case in the previous section, in practice we will demonstrate this
cancellation in a way closer to the level of the squared amplitude.

The first step is to expand the \textquoteleft even' and \textquoteleft odd'
parts of the Dirac structure $\mathrm{Dirac}_{\mathrm{V}}^{\bar{\eta}\bar{\lambda}\eta^{\prime}}$
(\ref{eq:DiracV}) as follows:
\begin{equation}
\begin{aligned}\mathrm{Dirac}_{\mathrm{V}}^{j\bar{\lambda}j} & =(D-2)\big(1-2\frac{k^{+}-p_{1}^{+}}{q^{+}}\big)\Big(\big(1-2\frac{p_{1}^{+}}{k^{+}}\big)\big(2\frac{p_{0}^{+}}{k^{+}}-1\big)-(D-3)\Big)\delta^{\lambda\bar{\lambda}}\\
 & -(D-2)\Big(\big(1-2\frac{p_{1}^{+}}{k^{+}}\big)\big(2\frac{p_{0}^{+}}{k^{+}}-1\big)-(D-3)+8\frac{D-4}{D-2}\Big)i\sigma^{\lambda\bar{\lambda}}\;,
\end{aligned}
\label{eq:DiracV_UV_even}
\end{equation}
and:
\begin{equation}
\begin{aligned}\epsilon^{\bar{\lambda}\rho}\epsilon^{\bar{\eta}\eta^{\prime}}\mathrm{Dirac}_{\mathrm{V}}^{\bar{\eta}\rho\eta^{\prime}} & =\epsilon^{\bar{\lambda}\rho}\epsilon^{\bar{\eta}\eta^{\prime}}\Bigg[\big(1-2\frac{k^{+}-p_{1}^{+}}{q^{+}}\big)\Big(2\frac{p_{0}^{+}+p_{1}^{+}}{k^{+}}-2-(D-4)\Big)\delta^{\lambda\rho}i\sigma^{\bar{\eta}\eta^{\prime}}\\
 & -\big(1-2\frac{p_{1}^{+}}{k^{+}}\big)\sigma^{\lambda\rho}\sigma^{\bar{\eta}\eta^{\prime}}+\big(2\frac{p_{0}^{+}}{k^{+}}-1\big)\sigma^{\bar{\eta}\eta^{\prime}}\sigma^{\lambda\rho}+i\sigma^{\eta\bar{\eta}}\sigma^{\lambda\rho}\sigma^{\eta\eta^{\prime}}\Bigg]\;.
\end{aligned}
\label{eq:DiracV_UV_odd}
\end{equation}
All the amplitudes we consider here are virtual, hence they will contribute
to the cross section being multiplied with the conjugate of the LO
amplitude, which contains the Dirac structure:
\begin{equation}
\mathrm{Dirac}^{\lambda\lambda^{\prime}}\big(1+\frac{2p_{1}^{+}}{q^{+}}\big)^{\dagger}=\big(1+\frac{2p_{1}^{+}}{q^{+}}\big)\delta^{\lambda\lambda^{\prime}}+i\sigma^{\lambda\lambda^{\prime}}\;.
\end{equation}
After multiplication with the above Dirac structure and evaluating
the trace, the different spinor structures appearing in (\ref{eq:DiracV_UV_even})
and (\ref{eq:DiracV_UV_odd}) all become proportional to $\delta^{\bar{\lambda}\lambda^{\prime}}$:
\begin{equation}
\begin{aligned}\mathrm{Tr}\Big\{\mathcal{P}_{G}\mathrm{Dirac}^{\lambda\lambda^{\prime}\dagger}\big(1+\frac{2p_{1}^{+}}{q^{+}}\big)\delta^{\lambda\bar{\lambda}}\Big\} & =2\big(1+\frac{2p_{1}^{+}}{q^{+}}\big)\delta^{\bar{\lambda}\lambda^{\prime}}\;,\\
\mathrm{Tr}\Big\{\mathcal{P}_{G}\mathrm{Dirac}^{\lambda\lambda^{\prime}\dagger}\big(1+\frac{2p_{1}^{+}}{q^{+}}\big)i\sigma^{\lambda\bar{\lambda}}\Big\} & =-2(D-3)\delta^{\bar{\lambda}\lambda^{\prime}}\;,\\
\epsilon^{\bar{\lambda}\rho}\epsilon^{\bar{\eta}\eta^{\prime}}\mathrm{Tr}\Big\{\mathcal{P}_{G}\mathrm{Dirac}^{\lambda\lambda^{\prime}\dagger}\big(1+\frac{2p_{1}^{+}}{q^{+}}\big)\delta^{\lambda\rho}i\sigma^{\bar{\eta}\eta^{\prime}}\Big\} & =4\delta^{\bar{\lambda}\lambda^{\prime}}\;,\\
\epsilon^{\bar{\lambda}\rho}\epsilon^{\bar{\eta}\eta^{\prime}}\mathrm{Tr}\Big\{\mathcal{P}_{G}\mathrm{Dirac}^{\lambda\lambda^{\prime}\dagger}\big(1+\frac{2p_{1}^{+}}{q^{+}}\big)\sigma^{\lambda\rho}\sigma^{\bar{\eta}\eta^{\prime}}\Big\} & =4\delta^{\bar{\lambda}\lambda^{\prime}}\big(1+\frac{2p_{1}^{+}}{q^{+}}-(D-4)\Big)\;,\\
\epsilon^{\bar{\lambda}\rho}\epsilon^{\bar{\eta}\eta^{\prime}}\mathrm{Tr}\Big\{\mathcal{P}_{G}\mathrm{Dirac}^{\lambda\lambda^{\prime}\dagger}\big(1+\frac{2p_{1}^{+}}{q^{+}}\big)\sigma^{\bar{\eta}\eta^{\prime}}\sigma^{\lambda\rho}\Big\} & =4\delta^{\bar{\lambda}\lambda^{\prime}}\big(1+\frac{2p_{1}^{+}}{q^{+}}+D-4\Big)\;,\\
\epsilon^{\bar{\lambda}\rho}\epsilon^{\bar{\eta}\eta^{\prime}}\mathrm{Tr}\Big\{\mathcal{P}_{G}\mathrm{Dirac}^{\lambda\lambda^{\prime}\dagger}\big(1+\frac{2p_{1}^{+}}{q^{+}}\big)i\sigma^{\eta\bar{\eta}}\sigma^{\lambda\rho}\sigma^{\eta\eta^{\prime}}\Big\} & =4\delta^{\bar{\lambda}\lambda^{\prime}}(D-4)\big(1+\frac{2p_{1}^{+}}{q^{+}}\big)\;,
\end{aligned}
\end{equation}
where we made use of the identities in appendix~\ref{sec:Dirac}.
Therefore, in anticipation of the fact that the virtual amplitudes
we consider will be multiplied with the leading-order one, we can
rescale the Dirac structures in (\ref{eq:DiracV_UV_even}) and (\ref{eq:DiracV_UV_odd})
as follows:
\begin{equation}
\begin{aligned}i\sigma^{\lambda\bar{\lambda}} & \to-(D-3)\frac{q^{+}}{p_{1}^{+}+p_{0}^{+}}\delta^{\bar{\lambda}\lambda}\;,\\
\epsilon^{\bar{\lambda}\rho}\epsilon^{\bar{\eta}\eta^{\prime}}\delta^{\lambda\rho}i\sigma^{\bar{\eta}\eta^{\prime}} & \to2\frac{q^{+}}{p_{1}^{+}+p_{0}^{+}}\delta^{\bar{\lambda}\lambda}\;,\\
\epsilon^{\bar{\lambda}\rho}\epsilon^{\bar{\eta}\eta^{\prime}}\sigma^{\lambda\rho}\sigma^{\bar{\eta}\eta^{\prime}} & \to2\Big(1-(D-4)\frac{q^{+}}{p_{1}^{+}+p_{0}^{+}}\Big)\delta^{\bar{\lambda}\lambda}\;,\\
\epsilon^{\bar{\lambda}\rho}\epsilon^{\bar{\eta}\eta^{\prime}}\sigma^{\bar{\eta}\eta^{\prime}}\sigma^{\lambda\rho} & \to2\Big(1+(D-4)\frac{q^{+}}{p_{1}^{+}+p_{0}^{+}}\Big)\delta^{\bar{\lambda}\lambda}\;,\\
\epsilon^{\bar{\lambda}\rho}\epsilon^{\bar{\eta}\eta^{\prime}}i\sigma^{\eta\bar{\eta}}\sigma^{\lambda\rho}\sigma^{\eta\eta^{\prime}} & \to2(D-4)\delta^{\bar{\lambda}\lambda}\;,
\end{aligned}
\label{eq:Dirac_rescale}
\end{equation}
such that the even and odd parts of $\mathrm{Dirac}_{\mathrm{V}}$
can be replaced by:
\begin{equation}
\begin{aligned}\mathrm{Dirac}_{\mathrm{V}}^{j\bar{\lambda}j} & \to(D-2)\delta^{\bar{\lambda}\lambda}\frac{q^{+}}{p_{1}^{+}+p_{0}^{+}}\Bigg[\Big(\big(1-2\frac{k^{+}-p_{1}^{+}}{q^{+}}\big)\big(1+\frac{2p_{1}^{+}}{q^{+}}\big)+(D-3)\Big)\\
 & \times\Big(\big(1-2\frac{p_{1}^{+}}{k^{+}}\big)\big(2\frac{p_{0}^{+}}{k^{+}}-1\big)-(D-3)\Big)+8\frac{(D-4)(D-3)}{D-2}\Bigg]\;,
\end{aligned}
\label{eq:DiracV_even_rescaled}
\end{equation}
and
\begin{equation}
\begin{aligned}\epsilon^{\bar{\lambda}\rho}\epsilon^{\bar{\eta}\eta^{\prime}}\mathrm{Dirac}_{\mathrm{V}}^{\bar{\eta}\rho\eta^{\prime}} & \to\frac{8q^{+}}{p_{1}^{+}+p_{0}^{+}}\delta^{\bar{\lambda}\lambda}\Bigg[\Big(1+\frac{2p_{1}^{+}-k^{+}}{q^{+}}\Big)\Big(\frac{p_{0}^{+}+p_{1}^{+}}{k^{+}}-1\Big)\\
 & +\frac{D-4}{2}\Big(\frac{q^{+}}{k^{+}}+\frac{k^{+}}{q^{+}}\Big)\Bigg]\;.
\end{aligned}
\label{eq:DiracV_odd_rescaled}
\end{equation}
Likewise, the leading-order Dirac structure can be replaced by:
\begin{align}
\mathrm{Dirac}^{\lambda\bar{\lambda}}\big(1+\frac{2p_{1}^{+}}{q^{+}}\big) & \to\Big(\big(1+\frac{2p_{1}^{+}}{q^{+}}\big)+(D-3)\frac{q^{+}}{p_{1}^{+}+p_{0}^{+}}\Big)\delta^{\bar{\lambda}\lambda}\;.
\end{align}
In the following paragraphs, we will use an underline to denote the
(reduced) amplitudes in which the rescaling procedure (\ref{eq:Dirac_rescale})
has been performed. For instance, the leading-order amplitudes (\ref{eq:MLOT})
become:
\begin{equation}
\begin{aligned}\underline{\mathcal{\tilde{M}}}_{\mathrm{LO1}}^{\lambda^{\prime}} & =\frac{-q^{+}\mathbf{P}_{\perp}^{\lambda^{\prime}}}{p_{0}^{+}\mathbf{P}_{\perp}^{2}+p_{1}^{+}M^{2}}\frac{q^{+}}{p_{1}^{+}+p_{0}^{+}}\Big(\big(\frac{p_{1}^{+}+p_{0}^{+}}{q^{+}}\big)^{2}+D-3\Big)\int_{\mathbf{x}}e^{-i\mathbf{k}_{\perp}\cdot\mathbf{x}}\big[U_{\mathbf{x}}-1\big]\;,\\
\underline{\mathcal{\tilde{M}}}_{\mathrm{LO2}}^{\lambda^{\prime}} & =\frac{-q^{+}\mathbf{q}^{\lambda^{\prime}}}{p_{0}^{+}\mathbf{q}^{2}+p_{1}^{+}M^{2}}\frac{q^{+}}{p_{1}^{+}+p_{0}^{+}}\Big(\big(\frac{p_{1}^{+}+p_{0}^{+}}{q^{+}}\big)^{2}+D-3\Big)\int_{\mathbf{x}}e^{-i\mathbf{k}_{\perp}\cdot\mathbf{x}}\big[U_{\mathbf{x}}-1\big]\;.
\end{aligned}
\end{equation}

\paragraph{Vertex corrections}

Let us first study the part of (\ref{eq:MV1T_UV}) that contains the
UV divergence. Applying the rescaling procedure $\tilde{\mathcal{M}}_{\mathrm{V1,UV}}^{\lambda}\to\underline{\tilde{\mathcal{M}}}_{\mathrm{V1,UV}}^{\lambda}$
and expanding $\mathcal{A}_{0}(\Delta_{\mathrm{P}})$ (\ref{eq:DeltaP}), we
get:
\begin{equation}
\begin{aligned}\underline{\tilde{\mathcal{M}}}_{\mathrm{V1,UV}}^{\lambda} & =\underline{\mathcal{\tilde{M}}}_{\mathrm{LO1}}^{\lambda}\times-\frac{\alpha_{s}C_{F}}{\pi}\frac{1}{\big(\frac{p_{1}^{+}+p_{0}^{+}}{q^{+}}\big)^{2}\!+\!D\!-\!3}\int_{k_{\mathrm{min}}^{+}}^{p_{1}^{+}}\mathrm{d}k^{+}\frac{q^{+}(k^{+})^{2}}{4p_{1}^{+}p_{0}^{+}(p_{1}^{+}\!-\!k^{+})(p_{0}^{+}\!-\!k^{+})}\\
 & \times\Bigg\{\Big(\frac{1}{2}+p_{0}^{+}\frac{p_{1}^{+}-k^{+}}{k^{+}q^{+}}\Big)\Bigg[\Big(\big(1-2\frac{k^{+}-p_{1}^{+}}{q^{+}}\big)\big(1+\frac{2p_{1}^{+}}{q^{+}}\big)+(D-3)\Big)\\
 & \times\Big(\big(1-2\frac{p_{1}^{+}}{k^{+}}\big)\big(2\frac{p_{0}^{+}}{k^{+}}-1\big)-(D-3)\Big)+8\frac{(D-4)(D-3)}{D-2}\Bigg]\\
 & +\frac{4}{(D\!-\!3)(D\!-\!2)}\Bigg[\!\Big(1\!+\!\frac{2p_{1}^{+}\!-\!k^{+}}{q^{+}}\Big)\Big(\frac{p_{0}^{+}\!+\!p_{1}^{+}}{k^{+}}\!-\!1\Big)\!+\!\frac{D-4}{2}\Big(\frac{q^{+}}{k^{+}}\!+\!\frac{k^{+}}{q^{+}}\Big)\!\Bigg]\!\Bigg\}\\
 & \times\Big(\frac{1}{\tilde{\epsilon}}+\ln\frac{\mu^{2}}{\Delta_{\mathrm{UV}}}\Big)
\end{aligned}
\label{eq:V1T_UV_rescaled}
\end{equation}
Expanding around $\epsilon_{\mathrm{UV}}=0$, we obtain, introducing
for simplicity $x=k^{+}/p_{1}^{+}$, $a=p_{0}^{+}/p_{1}^{+}$, and
$b=k_{\mathrm{min}}^{+}/p_{1}^{+}$:
\begin{equation}
\begin{aligned}\underline{\tilde{\mathcal{M}}}_{\mathrm{V1,UV}}^{\lambda} & =\underline{\mathcal{\tilde{M}}}_{\mathrm{LO1}}^{\lambda}\frac{\alpha_{s}C_{F}}{\pi}\Bigg\{\frac{3}{4}\frac{a(a-1)}{a^{2}+1}\int_{0}^{1}\frac{\mathrm{d}x}{x-1}\\
 & +\Big(\frac{1}{\tilde{\epsilon}}+\ln\frac{\mu^{2}}{\Delta_{\mathrm{UV}}}\Big)\Big(-\frac{1+2a}{2(1+a^{2})}-\ln b\Big)\\
 & +\frac{6a^{5}-17a^{4}+14a^{3}-6a^{2}-8a+3}{8a(a^{2}+1)^{2}}-\frac{3}{4}\frac{a-1}{a^{2}+1}\ln\frac{a-1}{a}\Bigg\}\;.
\end{aligned}
\label{eq:V1T_UV_final_OK}
\end{equation}
Following the same procedure for (\ref{eq:MV4_final}), we get:
\begin{equation}
\begin{aligned}\underline{\tilde{\mathcal{M}}}_{\mathrm{V4,UV}}^{\lambda} & =\underline{\mathcal{\tilde{M}}}_{\mathrm{LO2}}^{\lambda}\frac{\alpha_{s}C_{F}}{\pi}\Bigg\{\frac{3}{4}\frac{a(a\!-\!1)}{a^{2}\!+\!1}\int_{0}^{1}\frac{\mathrm{d}x}{x\!-\!1}\!+\!\Big(\frac{1}{\tilde{\epsilon}}\!+\!\ln\frac{\mu^{2}}{\Delta_{\mathrm{UV}}}\Big)\!\Big(\!-\frac{1\!+\!2a}{2(1\!+\!a^{2})}\!-\!\ln b\Big)\\
 & +\frac{6a^{5}-17a^{4}+14a^{3}-6a^{2}-8a+3}{8a(a^{2}+1)^{2}}-\frac{3}{4}\frac{a-1}{a^{2}+1}\ln\frac{a-1}{a}\Bigg\}\;.
\end{aligned}
\label{eq:MV4_final_UV}
\end{equation}

\paragraph{Antiquark vertex diagrams}

The counterterm $\tilde{\mathcal{M}}_{\mathrm{A1,UV}}^{\lambda}$
(\ref{eq:MA1T_UV}) reads:
\begin{equation}
\begin{aligned}\tilde{\mathcal{M}}_{\mathrm{A1,UV}}^{\lambda} & =\frac{\mathbf{P}_{\perp}^{\bar{\lambda}}}{p_{0}^{+}\mathbf{P}_{\perp}^{2}+p_{1}^{+}M^{2}}\frac{\alpha_{s}C_{F}}{D-2}\int_{p_{1}^{+}}^{p_{0}^{+}}\frac{\mathrm{d}k^{+}}{k^{+}}\frac{k^{+}}{2p_{1}^{+}}\frac{k^{+}p_{1}^{+}q^{+}}{p_{0}^{+}(p_{1}^{+}-k^{+})}\\
 & \times\Bigg\{\mathrm{Dirac}_{\mathrm{V}}^{j\bar{\lambda}j}+\frac{1}{D-3}\epsilon^{\bar{\lambda}\rho}\epsilon^{ij}\mathrm{Dirac}_{\mathrm{V}}^{i\rho j}\Bigg\}\mathcal{A}_{0}(\Delta_{\mathrm{UV}})\int_{\mathbf{x}}e^{-i\mathbf{k}_{\perp}\cdot\mathbf{x}}\big(U_{\mathbf{x}}-1\big)\;,
\end{aligned}
\label{eq:MA1_UV_0}
\end{equation}
or, after rescaling in the sense of eq. (\ref{eq:Dirac_rescale}):
\begin{equation}
\begin{aligned}\tilde{\mathcal{\underline{M}}}_{\mathrm{A1,UV}}^{\lambda} & =\underline{\mathcal{\tilde{M}}}_{\mathrm{LO1}}^{\lambda}\times-\frac{\alpha_{s}C_{F}}{\pi}\frac{1}{\big(\frac{p_{1}^{+}+p_{0}^{+}}{q^{+}}\big)^{2}+D-3}\int_{p_{1}^{+}}^{p_{0}^{+}}\frac{\mathrm{d}k^{+}}{k^{+}}\frac{k^{+}}{8p_{1}^{+}}\frac{k^{+}p_{1}^{+}}{p_{0}^{+}(p_{1}^{+}-k^{+})}\\
 & \times\Big(\frac{1}{\tilde{\epsilon}}+\ln\frac{\mu^{2}}{\Delta_{\mathrm{UV}}}\Big)\Bigg\{\Bigg[\Big(\big(1-2\frac{k^{+}-p_{1}^{+}}{q^{+}}\big)\big(1+\frac{2p_{1}^{+}}{q^{+}}\big)+(D-3)\Big)\\
 & \times\Big(\big(1-2\frac{p_{1}^{+}}{k^{+}}\big)\big(2\frac{p_{0}^{+}}{k^{+}}-1\big)-(D-3)\Big)+8\frac{(D-4)(D-3)}{D-2}\Bigg]\\
 & +\frac{8}{(D\!-\!3)(D\!-\!2)}\Bigg[\Big(1\!+\!\frac{2p_{1}^{+}\!-\!k^{+}}{q^{+}}\Big)\Big(\frac{p_{0}^{+}\!+\!p_{1}^{+}}{k^{+}}\!-\!1\Big)\!+\!\frac{D-4}{2}\Big(\frac{q^{+}}{k^{+}}\!+\!\frac{k^{+}}{q^{+}}\Big)\!\Bigg]\!\Bigg\}\;.
\end{aligned}
\label{eq:MA1_UV_rescaling}
\end{equation}
Expanding around $\epsilon_{\mathrm{UV}}=0$, and reintroducing $x=k^{+}/p_{1}^{+}$,
$a=p_{0}^{+}/p_{1}^{+}$:
\begin{equation}
\begin{aligned}\tilde{\mathcal{\underline{M}}}_{\mathrm{A1,UV}}^{\lambda} & =\underline{\mathcal{\tilde{M}}}_{\mathrm{LO1}}^{\lambda}\frac{\alpha_{s}C_{F}}{\pi}\Bigg\{\Big(\frac{1}{\tilde{\epsilon}}+\ln\frac{\mu^{2}}{\Delta_{\mathrm{UV}}}\Big)\Big(-\frac{3}{4}+\frac{1+2a}{2(1+a^{2})}+\frac{1}{2}\ln a\Big)\\
 & +\frac{3}{4}\frac{a(a-1)}{a^{2}+1}\int_{1}^{a}\frac{\mathrm{d}x}{x-1}+(a-1)\frac{3+6a-18a^{2}+18a^{3}-17a^{4}}{8a(a^{2}+1)^{2}}\Bigg\}\;.
\end{aligned}
\label{eq:A1T_UV_final_OK}
\end{equation}
Likewise, the UV divergent part of (\ref{eq:MA4_final}) becomes:
\begin{equation}
\begin{aligned}\tilde{\mathcal{\underline{M}}}_{\mathrm{A4,UV}}^{\lambda} & =\underline{\mathcal{\tilde{M}}}_{\mathrm{LO2}}^{\lambda}\times-\alpha_{s}C_{F}\frac{1}{\big(\frac{p_{1}^{+}+p_{0}^{+}}{q^{+}}\big)^{2}+D-3}\int_{p_{1}^{+}}^{p_{0}^{+}}\frac{\mathrm{d}k^{+}}{k^{+}}\frac{(k^{+})^{2}}{2p_{0}^{+}(p_{1}^{+}-k^{+})}\\
 & \times\Big(\frac{1}{\tilde{\epsilon}}+\ln\frac{\mu^{2}}{\Delta_{\mathrm{UV}}}\Big)\Bigg\{\Bigg[\Big(\big(1-2\frac{k^{+}-p_{1}^{+}}{q^{+}}\big)\big(1+\frac{2p_{1}^{+}}{q^{+}}\big)+(D-3)\Big)\\
 & \times\Big(\big(1-2\frac{p_{1}^{+}}{k^{+}}\big)\big(2\frac{p_{0}^{+}}{k^{+}}-1\big)-(D-3)\Big)+8\frac{(D-4)(D-3)}{D-2}\Bigg]\\
 & +\frac{8}{(D\!-\!2)(D\!-\!3)}\Bigg[\!\Big(1\!+\!\frac{2p_{1}^{+}\!-\!k^{+}}{q^{+}}\Big)\Big(\frac{p_{0}^{+}\!+\!p_{1}^{+}}{k^{+}}\!-\!1\Big)\!+\!\frac{D\!-\!4}{2}\Big(\frac{q^{+}}{k^{+}}\!+\!\frac{k^{+}}{q^{+}}\Big)\!\Bigg]\!\Bigg\}\;.
\end{aligned}
\label{eq:MA4_UV_rescaled}
\end{equation}
Or, after performing the integrations over the gluon plus-momentum:
\begin{equation}
\begin{aligned}\tilde{\mathcal{\underline{M}}}_{\mathrm{A4,UV}}^{\lambda} & =\underline{\mathcal{\tilde{M}}}_{\mathrm{LO2}}^{\lambda}\frac{\alpha_{s}C_{F}}{\pi}\Bigg\{\Big(\frac{1}{\tilde{\epsilon}}+\ln\frac{\mu^{2}}{\Delta_{\mathrm{UV}}}\Big)\Big(-\frac{3}{4}+\frac{1+2a}{2(1+a^{2})}+\frac{1}{2}\ln a\Big)\\
 & +\frac{3}{4}\frac{a(a-1)}{a^{2}+1}\int_{1}^{a}\frac{\mathrm{d}x}{x-1}+(a-1)\frac{3+6a-18a^{2}+18a^{3}-17a^{4}}{8a(a^{2}+1)^{2}}\Bigg\}\;.
\end{aligned}
\label{eq:A4T_UV_final_OK}
\end{equation}

\paragraph{Total UV contributions}

In order to add the results (\ref{eq:V1T_UV_final_OK}) and (\ref{eq:A1T_UV_final_OK}),
some closer attention is needed to the two integrals we left unevaluated.
Indeed, there is a singularity for $x\to1$ at the end- resp. starting
point of the leftover integration in $\underline{\tilde{\mathcal{M}}}_{\mathrm{V1,UV}}^{\lambda}$
and $\tilde{\mathcal{\underline{M}}}_{\mathrm{A1,UV}}^{\lambda}$.
Introducing an infinitesimal regulator $i0^{+}$,\footnote{Note that, whenever necessary, we consistently assign a positive infinitesimal
imaginary part to the gluon plus-momentum.} these integrals can be added and evaluated using the Sokhotski--Plemelj
theorem:
\begin{equation}
\begin{aligned}\int_{0}^{1}\frac{\mathrm{d}x}{x-1+i0^{+}}+\int_{1}^{a}\frac{\mathrm{d}x}{x-1+i0^{+}} & =\int_{-1}^{a-1}\frac{\mathrm{d}y}{y+i0^{+}}=\mathcal{P}\int_{-1}^{a-1}\frac{\mathrm{d}y}{y}-i\pi\;,\\
 & =\ln(a-1)-i\pi\;.
\end{aligned}
\end{equation}
The same holds for the integrals in (\ref{eq:MV4_final_UV}) and (\ref{eq:A4T_UV_final_OK}).

With the above manipulation, we are finally in a position to add (\ref{eq:V1T_UV_final_OK})
and (\ref{eq:A1T_UV_final_OK}) to the contribution from the quark
field-strength renormalization (\ref{eq:Z}), yielding:
\begin{equation}
\begin{aligned} & \underline{\tilde{\mathcal{M}}}_{\mathrm{V1,UV}}^{\lambda}+\tilde{\mathcal{\underline{M}}}_{\mathrm{A1,UV}}^{\lambda}+\mathcal{Z}_{\mathrm{UV}}\mathcal{\tilde{M}}_{\mathrm{LO1}}^{\lambda} \\
& =\underline{\mathcal{\tilde{M}}}_{\mathrm{LO1}}^{\lambda}\frac{\alpha_{s}C_{F}}{\pi}\Bigg\{\frac{3}{4}\frac{p_{0}^{+}q^{+}}{(p_{0}^{+})^{2}+(p_{1}^{+})^{2}}\Big(-i\pi+\ln\frac{q^{+}}{p_{1}^{+}}\Big)\\
 & +\mathrm{ln}\frac{\Delta_{\mathrm{UV}}}{4\pi e^{-\gamma_{\scriptscriptstyle{E}}}\mu_{\scriptscriptstyle{R}}^{2}}\Big(\frac{3}{4}-\frac{1}{2}\mathrm{ln}\frac{p_{0}^{+}}{k_{\mathrm{min}}^{+}}-\frac{1}{2}\mathrm{ln}\frac{p_{1}^{+}}{k_{\mathrm{min}}^{+}}\Big)\\
 & -\frac{11}{8}+\frac{9}{4}\frac{p_{1}^{+}p_{0}^{+}}{(p_{0}^{+})^{2}+(p_{1}^{+})^{2}}-\frac{3}{4}\frac{p_{1}^{+}q^{+}}{(p_{0}^{+})^{2}+(p_{1}^{+})^{2}}\ln\frac{q^{+}}{p_{0}^{+}}\Bigg\}\;,
\end{aligned}
\label{eq:1T_UV_final}
\end{equation}
and similarly for (\ref{eq:MV4_final_UV}) and (\ref{eq:A4T_UV_final_OK}):
\begin{equation}
\begin{aligned} & \underline{\tilde{\mathcal{M}}}_{\mathrm{V4,UV}}^{\lambda}+\tilde{\mathcal{\underline{M}}}_{\mathrm{A4,UV}}^{\lambda}+\mathcal{Z}_{\mathrm{UV}}\mathcal{\tilde{M}}_{\mathrm{LO2}}^{\lambda} \\
& =\underline{\mathcal{\tilde{M}}}_{\mathrm{LO2}}^{\lambda}\frac{\alpha_{s}C_{F}}{\pi}\Bigg\{\frac{3}{4}\frac{p_{0}^{+}q^{+}}{(p_{0}^{+})^{2}+(p_{1}^{+})^{2}}\Big(-i\pi+\ln\frac{q^{+}}{p_{1}^{+}}\Big)\\
 & +\mathrm{ln}\frac{\Delta_{\mathrm{UV}}}{4\pi e^{-\gamma_{\scriptscriptstyle{E}}}\mu_{\scriptscriptstyle{R}}^{2}}\Big(\frac{3}{4}-\frac{1}{2}\mathrm{ln}\frac{p_{0}^{+}}{k_{\mathrm{min}}^{+}}-\frac{1}{2}\mathrm{ln}\frac{p_{1}^{+}}{k_{\mathrm{min}}^{+}}\Big)\\
 & -\frac{11}{8}+\frac{9}{4}\frac{p_{1}^{+}p_{0}^{+}}{(p_{0}^{+})^{2}+(p_{1}^{+})^{2}}-\frac{3}{4}\frac{p_{1}^{+}q^{+}}{(p_{0}^{+})^{2}+(p_{1}^{+})^{2}}\ln\frac{q^{+}}{p_{0}^{+}}\Bigg\}\;.
\end{aligned}
\label{eq:2T_UV_final}
\end{equation}

Hence, we have proven that, also in the case of a transversely polarized
photon, the total virtual contribution to the cross section is free
from UV divergences. Just like in the longitudinal case, however,
there are rapidity divergences left, which will be treated in section
\ref{sec:JIMWLK}.

Note that, in this section, all our expressions are proportional to
one of the two leading-order amplitudes. We have chosen to keep these
amplitudes defined in $D$ dimensions. However, we have checked that,
when factorizing the LO amplitudes out in $D=4$ dimensions instead,
the extra finite terms that appear all compensate each other. One,
therefore, obtains exactly the same result irrespective of the precise
procedure followed, which is an interesting nontrivial check of our
calculation. 

\subsection{Contribution to the cross section due to UV counterterms}

Collecting our results (\ref{eq:L1UV_final}) and (\ref{eq:L2UV_final})
for the sum of the UV counterterms, and applying the definition (\ref{eq:parton2hadron}),
we eventually obtain the following contribution to the cross section
in case of a longitudinally polarized photon:
\begin{equation}
\begin{aligned}\mathrm{d}\sigma_{\mathrm{UV}}^{\mathrm{L}} & =\mathrm{d}\sigma_{\mathrm{LO}}^{\mathrm{L}}\frac{\alpha_{s}C_{F}}{\pi}\Bigg[\Big(-\frac{3}{2}+\ln\frac{p_{1}^{+}}{k_{\mathrm{min}}^{+}}+\ln\frac{p_{0}^{+}}{k_{\mathrm{min}}^{+}}\Big)\ln\frac{4\pi e^{-\gamma_{\scriptscriptstyle{E}}}\mu_{\scriptscriptstyle{R}}^{2}}{\Delta_{\mathrm{UV}}}-\frac{1}{2}\Bigg]\;.\end{aligned}
\label{eq:sigmaL_UV}
\end{equation}
Similarly, (\ref{eq:1T_UV_final}) and (\ref{eq:2T_UV_final}) can
be combined into:
\begin{equation}
\begin{aligned}\mathrm{d}\sigma_{\mathrm{UV}}^{\mathrm{T}} & =\mathrm{d}\sigma_{\mathrm{LO}}^{\mathrm{T}}\frac{\alpha_{s}C_{F}}{\pi}\Bigg[\Big(-\frac{3}{2}+\ln\frac{p_{1}^{+}}{k_{\mathrm{min}}^{+}}+\ln\frac{p_{0}^{+}}{k_{\mathrm{min}}^{+}}\Big)\ln\frac{4\pi e^{-\gamma_{\scriptscriptstyle{E}}}\mu_{\scriptscriptstyle{R}}^{2}}{\Delta_{\mathrm{UV}}}\\
 & +\frac{3}{2}\frac{p_{0}^{+}q^{+}}{(p_{0}^{+})^{2}+(p_{1}^{+})^{2}}\Big(-i\pi+\ln\frac{q^{+}}{p_{1}^{+}}+3\frac{p_{1}^{+}}{q^{+}}-\frac{p_{1}^{+}}{p_{0}^{+}}\ln\frac{q^{+}}{p_{0}^{+}}\Big)-\frac{11}{4}\Bigg]\;.
\end{aligned}
\label{eq:sigmaT_UV}
\end{equation}

\section{\label{sec:real}Real next-to-leading order corrections}

In this section, we present the amplitudes for the real NLO corrections.
One should be very careful to note that, due to momentum conservation,
the plus-momentum of the incoming quark is equal to
\begin{equation}
p_{0\scriptscriptstyle{R}}^{+}=p_{1}^{+}+q^{+}+p_{3}^{+}\;.
\end{equation}
We denote this quantity with an additional index $R$, to avoid confusion
with the same momentum component but in leading-order and virtual
amplitudes, where $p_{0}^{+}=p_{1}^{+}+q^{+}$. 

\subsection{Initial-state radiation}

\begin{figure}[t]
\begin{centering}
\includegraphics[scale=0.35]{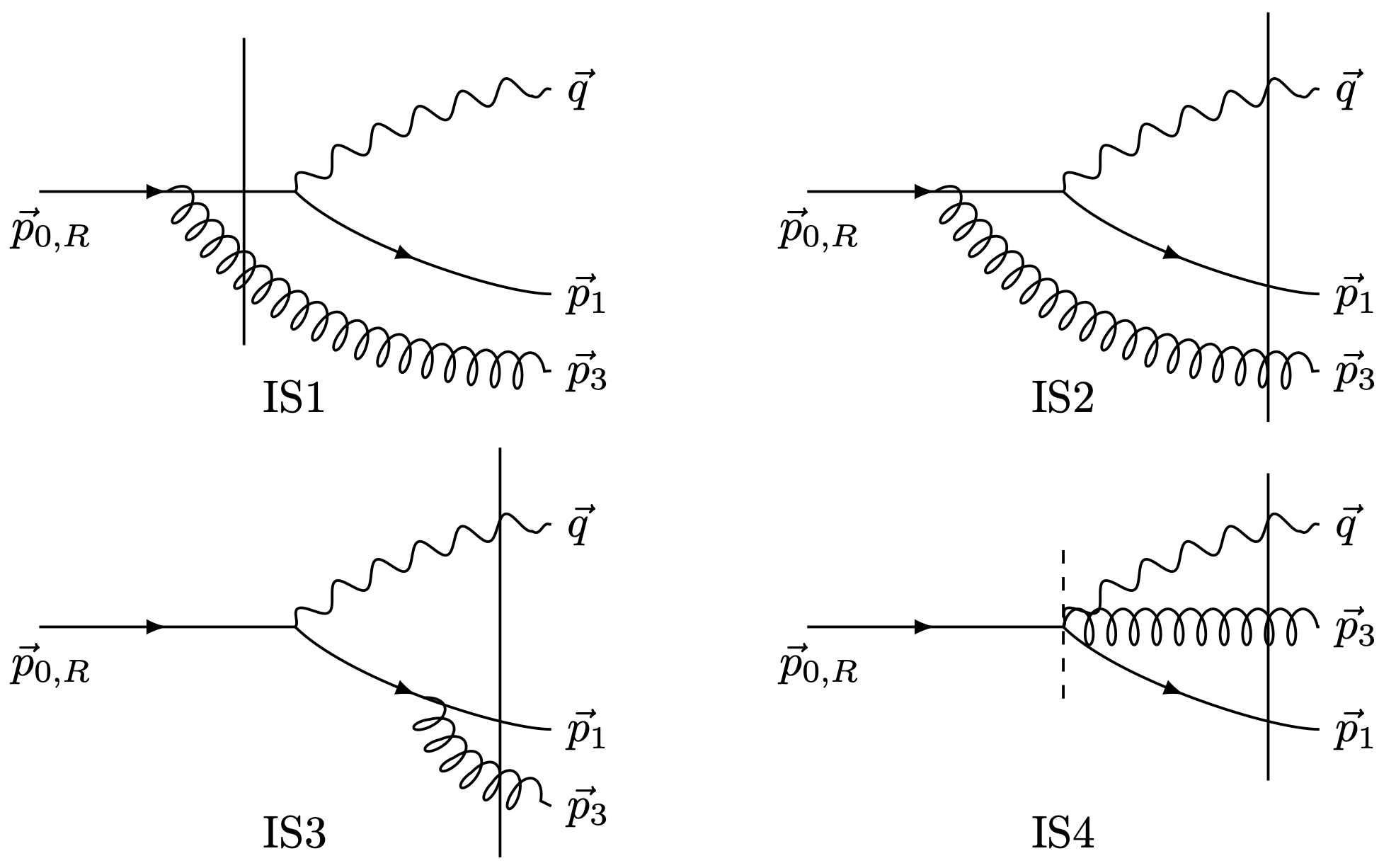}
\par\end{centering}
\caption{\label{fig:ISR}The four diagrams corresponding to real gluon emission
in the initial state. }
\end{figure}

\paragraph{Diagram IS1}

The reduced amplitudes corresponding to graph ISR1 in figure (\ref{fig:ISR})
read:
\begin{equation}
\begin{aligned}\tilde{\mathcal{M}}_{\mathrm{IS1}}^{0\eta} & =-\frac{p_{0}^{+}\mathbf{P}_{\perp}^{2}-p_{1}^{+}M^{2}}{p_{0}^{+}\mathbf{P}_{\perp}^{2}+p_{1}^{+}M^{2}}\frac{1}{M}\frac{p_{3}^{+}}{p_{0\scriptscriptstyle{R}}^{+}}\mathrm{Dirac}^{\eta\bar{\eta}}\big(1+\frac{2p_{0}^{+}}{p_{3}^{+}}\big)\\
 & \times\int_{\mathbf{x},\mathbf{z}}iA^{\bar{\eta}}(\mathbf{x}-\mathbf{z})e^{-i\mathbf{k}_{\perp}\cdot\mathbf{x}}e^{-i\mathbf{p}_{3}\cdot\mathbf{z}}\big(U_{\mathbf{x}}U_{\mathbf{z}}^{\dagger}t^{c}U_{\mathbf{z}}-t^{c}\big)\;,
\end{aligned}
\label{eq:MIS1L}
\end{equation}
and: 

\begin{equation}
\begin{aligned}\tilde{\mathcal{M}}_{\mathrm{IS1}}^{\lambda\eta} & =\frac{q^{+}\mathbf{P}_{\perp}^{\bar{\lambda}}}{p_{0}^{+}\mathbf{P}_{\perp}^{2}+p_{1}^{+}M^{2}}\frac{p_{3}^{+}}{p_{0\scriptscriptstyle{R}}^{+}}\mathrm{Dirac}^{\lambda\bar{\lambda}}\big(1+\frac{2p_{1}^{+}}{q^{+}}\big)\mathrm{Dirac}^{\eta\bar{\eta}}\big(1+\frac{2p_{0}^{+}}{p_{3}^{+}}\big)\\
 & \times\int_{\mathbf{x},\mathbf{z}}iA^{\bar{\eta}}(\mathbf{x}-\mathbf{z})e^{-i\mathbf{k}_{\perp}\cdot\mathbf{x}}e^{-i\mathbf{p}_{3}\cdot\mathbf{z}}\big(U_{\mathbf{x}}U_{\mathbf{z}}^{\dagger}t^{c}U_{\mathbf{z}}-t^{c}\big)\;.
\end{aligned}
\label{eq:MIS1}
\end{equation}

\paragraph{Diagram IS2}

The amplitude for diagram IS2, when the photon is longitudinally polarized,
reads:
\begin{equation}
\begin{aligned}\tilde{\mathcal{M}}_{\mathrm{IS2}}^{0\eta} & =\frac{1}{M}\mathrm{Dirac}^{\eta\bar{\eta}}\big(1+\frac{2p_{0}^{+}}{p_{3}^{+}}\big)\frac{p_{3}^{+}}{p_{0\scriptscriptstyle{R}}^{+}}\\
 & \times\int_{\boldsymbol{\ell}}\frac{\boldsymbol{\ell}^{\bar{\eta}}}{\boldsymbol{\ell}^{2}}\frac{p_{3}^{+}\big(q^{+}\boldsymbol{\ell}+p_{0}^{+}\mathbf{q}\big)^{2}-p_{1}^{+}p_{3}^{+}p_{0}^{+}M^{2}}{p_{3}^{+}\big(q^{+}\boldsymbol{\ell}+p_{0}^{+}\mathbf{q}\big)^{2}+p_{1}^{+}q^{+}p_{0R}^{+}\boldsymbol{\ell}^{2}+p_{1}^{+}p_{3}^{+}p_{0}^{+}M^{2}}\\
 & \times\int_{\mathbf{x},\mathbf{z}}e^{-i\boldsymbol{\ell}\cdot(\mathbf{\mathbf{x}-\mathbf{z}})}e^{-i\mathbf{k}_{\perp}\cdot\mathbf{x}}e^{-i\mathbf{p}_{3}\cdot\mathbf{z}}\big(U_{\mathbf{x}}U_{\mathbf{z}}^{\dagger}t^{c}U_{\mathbf{z}}-t^{c}\big)\;.
\end{aligned}
\label{eq:MIS2L}
\end{equation}
In the transverse case, we obtain:
\begin{equation}
\begin{aligned}\tilde{\mathcal{M}}_{\mathrm{IS2}}^{\lambda\eta} & =\frac{q^{+}(p_{3}^{+})^{2}}{p_{0\scriptscriptstyle{R}}^{+}}\mathrm{Dirac}^{\lambda\bar{\lambda}}\big(1+\frac{2p_{1}^{+}}{q^{+}}\big)\mathrm{Dirac}^{\eta\bar{\eta}}\big(1+\frac{2p_{0}^{+}}{p_{3}^{+}}\big)\\
 & \times\int_{\mathbf{x},\mathbf{z}}\int_{\boldsymbol{\ell}}e^{-i\boldsymbol{\ell}\cdot(\mathbf{\mathbf{x}-\mathbf{z}})}\frac{\boldsymbol{\ell}^{\bar{\eta}}}{\boldsymbol{\ell}^{2}}\frac{q^{+}\boldsymbol{\ell}^{\bar{\lambda}}+p_{0}^{+}\mathbf{q}^{\bar{\lambda}}}{p_{3}^{+}\big(q^{+}\boldsymbol{\ell}+p_{0}^{+}\mathbf{q}\big)^{2}+p_{1}^{+}q^{+}p_{0R}^{+}\boldsymbol{\ell}^{2}+p_{1}^{+}p_{3}^{+}p_{0}^{+}M^{2}}\\
 & \times e^{-i\mathbf{k}_{\perp}\cdot\mathbf{x}}e^{-i\mathbf{p}_{3}\cdot\mathbf{z}}\big(U_{\mathbf{x}}U_{\mathbf{z}}^{\dagger}t^{c}U_{\mathbf{z}}-t^{c}\big)\;.
\end{aligned}
\label{eq:MIS2}
\end{equation}
In the above expressions, we have used the definition:
\begin{equation}
\begin{aligned}\Delta_{\mathrm{IS}} & \equiv\frac{p_{1}^{+}p_{3}^{+}}{q^{+}(p_{1}^{+}+p_{3}^{+})^{2}}\big(p_{0\scriptscriptstyle{R}}^{+}\mathbf{q}^{2}+(p_{1}^{+}+p_{3}^{+})M^{2}\big)\;.\end{aligned}
\label{eq:deltaISR}
\end{equation}

\paragraph{Diagram IS3}

The amplitudes for the production of a longitudinally or transversely
polarized virtual photon read, respectively:
\begin{equation}
\begin{aligned}\tilde{\mathcal{M}}_{\mathrm{IS3}}^{0\eta} & =\frac{p_{0\scriptscriptstyle{R}}^{+}\mathbf{q}^{2}-(p_{1}^{+}+p_{3}^{+})M^{2}}{p_{0\scriptscriptstyle{R}}^{+}\mathbf{q}^{2}+(p_{1}^{+}+p_{3}^{+})M^{2}}\frac{1}{M}\frac{p_{3}^{+}}{p_{1}^{+}+p_{3}^{+}}\mathrm{Dirac}^{\eta\bar{\eta}}\big(1+\frac{2p_{1}^{+}}{p_{3}^{+}}\big)\\
 & \times\int_{\mathbf{x},\mathbf{z}}iA^{\bar{\eta}}(\mathbf{x}-\mathbf{z},\Delta_{\mathrm{IS}})e^{i\frac{p_{3}^{+}}{p_{1}^{+}+p_{3}^{+}}\mathbf{q}\cdot(\mathbf{x}-\mathbf{z})}e^{-i\mathbf{k}_{\perp}\cdot\mathbf{x}}e^{-i\mathbf{p}_{3}\cdot\mathbf{z}}\big(U_{\mathbf{x}}U_{\mathbf{z}}^{\dagger}t^{c}U_{\mathbf{z}}-t^{c}\big)\;,
\end{aligned}
\label{eq:MIS3L}
\end{equation}
and:
\begin{equation}
\begin{aligned}\tilde{\mathcal{M}}_{\mathrm{IS3}}^{\lambda\eta} & =\frac{q^{+}\mathbf{q}^{\bar{\lambda}}}{p_{0\scriptscriptstyle{R}}^{+}\mathbf{q}^{2}+(p_{1}^{+}+p_{3}^{+})M^{2}}\frac{p_{3}^{+}}{p_{1}^{+}+p_{3}^{+}}\mathrm{Dirac}^{\eta\bar{\eta}}\big(1+\frac{2p_{1}^{+}}{p_{3}^{+}}\big)\mathrm{Dirac}^{\lambda\bar{\lambda}}\big(1+2\frac{p_{1}^{+}+p_{3}^{+}}{q^{+}}\big)\\
 & \times\int_{\mathbf{x},\mathbf{z}}iA^{\bar{\eta}}(\mathbf{x}-\mathbf{z},\Delta_{\mathrm{IS}})e^{i\frac{p_{3}^{+}}{p_{1}^{+}+p_{3}^{+}}\mathbf{q}\cdot(\mathbf{x}-\mathbf{z})}e^{-i\mathbf{k}_{\perp}\cdot\mathbf{x}}e^{-i\mathbf{p}_{3}\cdot\mathbf{z}}\big(U_{\mathbf{x}}U_{\mathbf{z}}^{\dagger}t^{c}U_{\mathbf{z}}-t^{c}\big)\;.
\end{aligned}
\label{eq:MIS3}
\end{equation}

\paragraph{Diagram IS4}

Finally, for the instantaneous $q\to\gamma^{*}gq$ splitting, we obtain
the following result for the longitudinally resp. transversely polarized
amplitudes:
\begin{equation}
\begin{aligned}\tilde{\mathcal{M}}_{\mathrm{IS4}}^{0\eta} & =\frac{-p_{1}^{+}p_{3}^{+}}{(p_{1}^{+}+p_{3}^{+})M}\int_{\mathbf{x},\mathbf{z}}\Bigg[\frac{q^{+}}{p_{1}^{+}p_{0}^{+}}iA^{i}\big(\mathbf{x}-\mathbf{z},\Delta_{\mathrm{IS}})\big(\delta^{i\eta}+i\sigma^{i\eta}\big)\\
 & +\frac{2p_{0}^{+}+p_{3}^{+}}{p_{0}^{+}(p_{1}^{+}+p_{3}^{+})}\mathbf{q}^{i}\mathcal{K}\big(\mathbf{x}-\mathbf{z},\Delta_{\mathrm{IS}})\big(\delta^{i\eta}+\frac{p_{3}^{+}}{2p_{0}^{+}+p_{3}^{+}}i\sigma^{i\eta}\big)\Bigg]\\
 & \times e^{i\frac{p_{3}^{+}}{p_{1}^{+}+p_{3}^{+}}\mathbf{q}\cdot(\mathbf{x}-\mathbf{z})}e^{-i\mathbf{k}_{\perp}\cdot\mathbf{x}}e^{-i\mathbf{p}_{3}\cdot\mathbf{z}}\big(U_{\mathbf{x}}U_{\mathbf{z}}^{\dagger}t^{c}U_{\mathbf{z}}-t^{c}\big)\;,
\end{aligned}
\label{eq:MIS4L_final}
\end{equation}
and:
\begin{equation}
\begin{aligned}\tilde{\mathcal{M}}_{\mathrm{IS4}}^{\lambda\eta} & =\frac{-p_{1}^{+}p_{3}^{+}}{p_{1}^{+}+p_{3}^{+}}\Bigg[\Big(\frac{1}{p_{1}^{+}+p_{3}^{+}}+\frac{1}{p_{0}^{+}}\Big)\delta^{\eta\lambda}+\Big(\frac{1}{p_{1}^{+}+p_{3}^{+}}-\frac{1}{p_{0}^{+}}\Big)i\sigma^{\eta\lambda}\Bigg]\\
 & \times\int_{\mathbf{x},\mathbf{z}}\mathcal{K}\big(\mathbf{x}-\mathbf{z},\Delta_{\mathrm{IS}}\big)e^{i\frac{p_{3}^{+}}{p_{1}^{+}+p_{3}^{+}}\mathbf{q}\cdot(\mathbf{x}-\mathbf{z})}e^{-i\mathbf{k}_{\perp}\cdot\mathbf{x}}e^{-i\mathbf{p}_{3}\cdot\mathbf{z}}\big(U_{\mathbf{x}}U_{\mathbf{z}}^{\dagger}t^{c}U_{\mathbf{z}}-t^{c}\big)\;.
\end{aligned}
\label{eq:MIS4}
\end{equation}

\subsection{Final-state radiation}

\begin{figure}[t]
\begin{centering}
\includegraphics[scale=0.3]{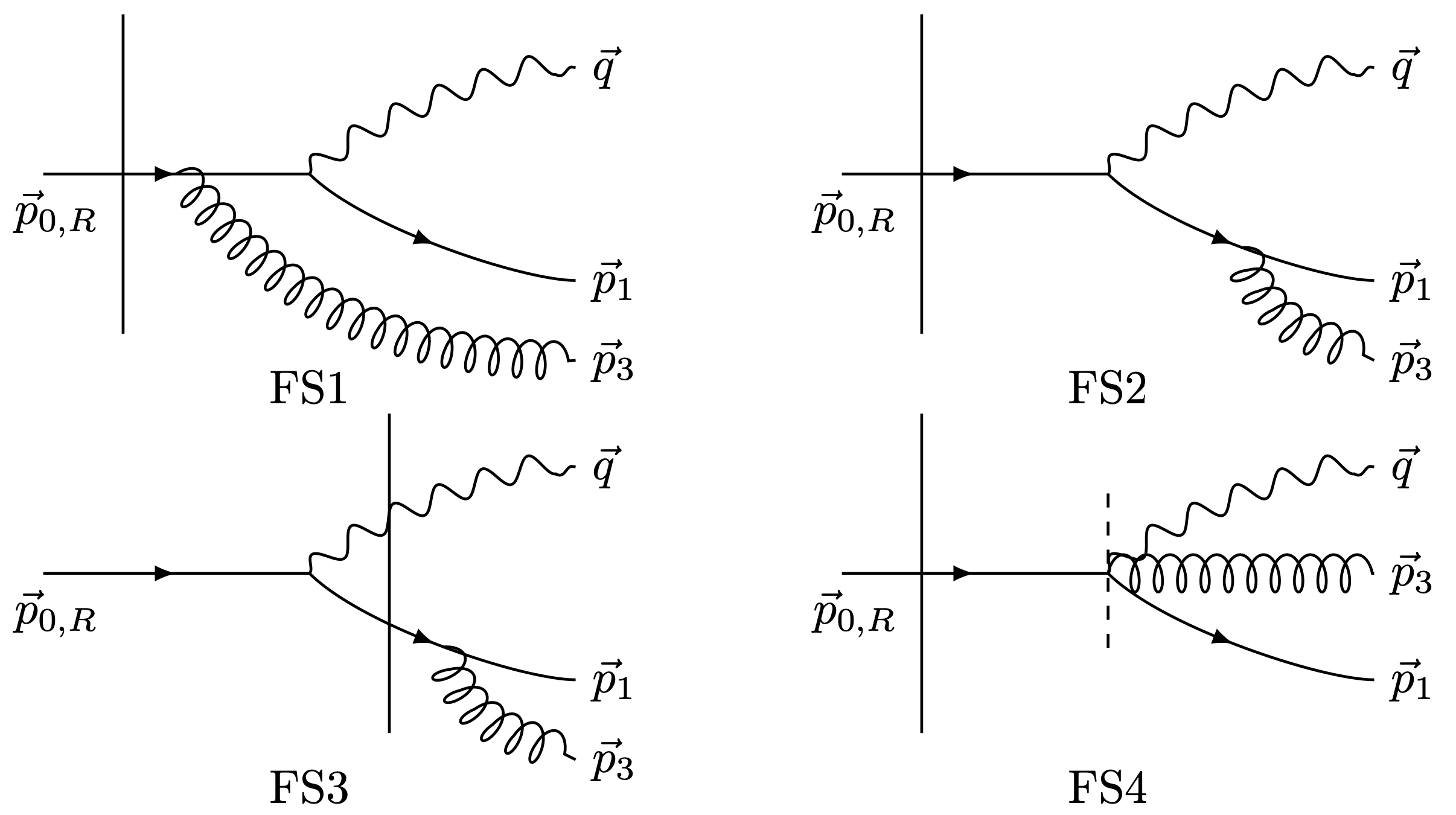}
\par\end{centering}
\caption{\label{fig:FS}The four diagrams corresponding to real gluon emission
in the final state.}
\end{figure}

\paragraph{Diagram FS1}

The amplitude corresponding to the first final-state radiation diagram
in figure (\ref{fig:FS}) gives, in the longitudinally polarized case:
\begin{equation}
\begin{aligned}\tilde{\mathcal{M}}_{\mathrm{FS1}}^{0\eta} & =\frac{p_{0}^{+}\mathbf{P}_{\perp}^{2}-p_{1}^{+}M^{2}}{p_{0}^{+}\mathbf{P}_{\perp}^{2}+p_{1}^{+}M^{2}}\frac{1}{M}\frac{p_{3}^{+}}{p_{0}^{+}}\mathrm{Dirac}^{\eta\bar{\eta}}\big(1+\frac{2p_{0}^{+}}{p_{3}^{+}}\big)\\
 & \times\frac{\big(\mathbf{p}_{3}-\frac{p_{3}^{+}}{p_{0}^{+}}\mathbf{k}_{\perp}\big)^{\bar{\eta}}}{\big(\mathbf{p}_{3}-\frac{p_{3}^{+}}{p_{0}^{+}}\mathbf{k}_{\perp}\big)^{2}+\Delta_{\mathrm{FS}}}\int_{\mathbf{x}}e^{-i\mathbf{x}\cdot(\mathbf{q}+\mathbf{p}_{1}+\mathbf{p}_{3})}t^{c}\big(U_{\mathbf{x}}-1\big)\;,
\end{aligned}
\label{eq:MFS1L}
\end{equation}
and when the photon is transversely polarized:
\begin{equation}
\begin{aligned}\tilde{\mathcal{M}}_{\mathrm{FS1}}^{\lambda\eta} & =-\frac{q^{+}\mathbf{P}_{\perp}^{\bar{\lambda}}}{p_{0}^{+}\mathbf{P}_{\perp}^{2}+p_{1}^{+}M^{2}}\frac{p_{3}^{+}}{p_{0}^{+}}\mathrm{Dirac}^{\lambda\bar{\lambda}}\big(1+\frac{2p_{1}^{+}}{q^{+}}\big)\mathrm{Dirac}^{\eta\bar{\eta}}\big(1+\frac{2p_{0}^{+}}{p_{3}^{+}}\big)\\
 & \times\frac{\big(\mathbf{p}_{3}-\frac{p_{3}^{+}}{p_{0}^{+}}\mathbf{k}_{\perp}\big)^{\bar{\eta}}}{\big(\mathbf{p}_{3}-\frac{p_{3}^{+}}{p_{0}^{+}}\mathbf{k}_{\perp}\big)^{2}+\Delta_{\mathrm{FS}}}\int_{\mathbf{x}}e^{-i\mathbf{x}\cdot(\mathbf{q}+\mathbf{p}_{1}+\mathbf{p}_{3})}t^{c}\big(U_{\mathbf{x}}-1\big)\;.
\end{aligned}
\label{eq:MFS1}
\end{equation}
In the above, we have introduced:
\begin{equation}
\Delta_{\mathrm{FS}}\equiv\frac{p_{3}^{+}p_{0\scriptscriptstyle{R}}^{+}}{p_{1}^{+}q^{+}p_{0}^{+}}\big(p_{0}^{+}\mathbf{P}_{\perp}^{2}+p_{1}^{+}M^{2}\big)\;.\label{eq:DeltaFS}
\end{equation}
For later purposes (for instance the high-energy resummation in section
(\ref{sec:JIMWLK})), it can be convenient to go to transverse coordinate
space, writing (obtained by inverting (\ref{eq:WWfieldmass})):
\begin{equation}
\begin{aligned}\frac{\big(\mathbf{p}_{3}-\frac{p_{3}^{+}}{p_{0}^{+}}\mathbf{k}_{\perp}\big)^{\bar{\eta}}}{\big(\mathbf{p}_{3}-\frac{p_{3}^{+}}{p_{0}^{+}}\mathbf{k}_{\perp}\big)^{2}+\Delta_{\mathrm{FS}}}e^{-i\mathbf{p}_{3}\cdot\mathbf{x}} & =\frac{\big(\mathbf{p}_{3}-\frac{p_{3}^{+}}{p_{0}^{+}}\mathbf{k}_{\perp}\big)^{\bar{\eta}}}{\big(\mathbf{p}_{3}-\frac{p_{3}^{+}}{p_{0}^{+}}\mathbf{k}_{\perp}\big)^{2}+\Delta_{\mathrm{FS}}}e^{-i\big(\mathbf{p}_{3}-\frac{p_{3}^{+}}{p_{0}^{+}}\mathbf{k}_{\perp}\big)\cdot\mathbf{x}}e^{-i\frac{p_{3}^{+}}{p_{0}^{+}}\mathbf{k}_{\perp}\cdot\mathbf{x}}\;,\\
 & =e^{-i\frac{p_{3}^{+}}{p_{0}^{+}}\mathbf{k}_{\perp}\cdot\mathbf{x}}\int_{\mathbf{z}}e^{-i\big(\mathbf{p}_{3}-\frac{p_{3}^{+}}{p_{0}^{+}}\mathbf{k}_{\perp}\big)\cdot\mathbf{z}}iA^{\bar{\eta}}(\mathbf{x}-\mathbf{z},\Delta_{\mathrm{FS}})
\end{aligned}
\label{eq:FS1coordspace}
\end{equation}
such that:
\begin{equation}
\begin{aligned}\tilde{\mathcal{M}}_{\mathrm{FS1}}^{0\eta} & =\frac{p_{0}^{+}\mathbf{P}_{\perp}^{2}-p_{1}^{+}M^{2}}{p_{0}^{+}\mathbf{P}_{\perp}^{2}+p_{1}^{+}M^{2}}\frac{1}{M}\frac{p_{3}^{+}}{p_{0}^{+}}\mathrm{Dirac}^{\eta\bar{\eta}}\big(1+\frac{2p_{0}^{+}}{p_{3}^{+}}\big)\\
 & \times\int_{\mathbf{x},\mathbf{z}}e^{-i\big(\mathbf{p}_{3}-\frac{p_{3}^{+}}{p_{0}^{+}}\mathbf{k}_{\perp}\big)\cdot\mathbf{z}}e^{-i\frac{p_{3}^{+}}{p_{0}^{+}}\mathbf{k}_{\perp}\cdot\mathbf{x}}e^{-i\mathbf{k}_{\perp}\cdot\mathbf{x}}iA^{\bar{\eta}}(\mathbf{x}-\mathbf{z},\Delta_{\mathrm{FS}})t^{c}\big(U_{\mathbf{x}}-1\big)\;,
\end{aligned}
\label{eq:MFS1L_coord}
\end{equation}
and:
\begin{equation}
\begin{aligned}\tilde{\mathcal{M}}_{\mathrm{FS1}}^{\lambda\eta} & =-\frac{q^{+}\mathbf{P}_{\perp}^{\bar{\lambda}}}{p_{0}^{+}\mathbf{P}_{\perp}^{2}+p_{1}^{+}M^{2}}\frac{p_{3}^{+}}{p_{0}^{+}}\mathrm{Dirac}^{\lambda\bar{\lambda}}\big(1+\frac{2p_{1}^{+}}{q^{+}}\big)\mathrm{Dirac}^{\eta\bar{\eta}}\big(1+\frac{2p_{0}^{+}}{p_{3}^{+}}\big)\\
 & \times\int_{\mathbf{x},\mathbf{z}}e^{-i\big(\mathbf{p}_{3}-\frac{p_{3}^{+}}{p_{0}^{+}}\mathbf{k}_{\perp}\big)\cdot\mathbf{z}}e^{-i\frac{p_{3}^{+}}{p_{0}^{+}}\mathbf{k}_{\perp}\cdot\mathbf{x}}e^{-i\mathbf{k}_{\perp}\cdot\mathbf{x}}iA^{\bar{\eta}}(\mathbf{x}-\mathbf{z},\Delta_{\mathrm{FS}})t^{c}\big(U_{\mathbf{x}}-1\big)\;.
\end{aligned}
\label{eq:MFS1T_coord}
\end{equation}

\paragraph{Diagram FS2}

The amplitudes read:
\begin{equation}
\begin{aligned}\tilde{\mathcal{M}}_{\mathrm{FS2}}^{0\eta} & =\frac{1}{M}\frac{p_{3}^{+}}{p_{1}^{+}}\mathrm{Dirac}^{\eta\bar{\eta}}\big(1+\frac{2p_{1}^{+}}{p_{3}^{+}}\big)\\
 & \times\frac{\mathbf{p}_{3}^{\bar{\eta}}-\frac{p_{3}^{+}}{p_{1}^{+}}\mathbf{p}_{1}^{\bar{\eta}}}{\big(\mathbf{p}_{3}\!-\!\frac{p_{3}^{+}}{p_{1}^{+}}\mathbf{p}_{1}\big)^{2}}\frac{\big(\mathbf{p}_{3}\!+\!\mathbf{p}_{1}\!-\!\frac{p_{1}^{+}\!+p_{3}^{+}}{q^{+}}\mathbf{q}\big)^{2}\!-\!\frac{p_{0\scriptscriptstyle{R}}^{+}(p_{1}^{+}\!+p_{3}^{+})}{(q^{+})^{2}}M^{2}}{\frac{p_{1}^{+}p_{0\scriptscriptstyle{R}}^{+}}{q^{+}p_{3}^{+}}\big(\mathbf{p}_{3}\!-\!\frac{p_{3}^{+}}{p_{1}^{+}}\mathbf{p}_{1}\big)^{2}\!+\!\big(\mathbf{p}_{3}\!+\!\mathbf{p}_{1}\!-\!\frac{p_{1}^{+}\!+p_{3}^{+}}{q^{+}}\mathbf{q}\big)^{2}\!+\!\frac{p_{0\scriptscriptstyle{R}}^{+}(p_{1}^{+}\!+p_{3}^{+})}{(q^{+})^{2}}M^{2}}\\
 & \times\int_{\mathbf{x}}e^{-i\mathbf{x}\cdot(\mathbf{q}+\mathbf{p}_{1}+\mathbf{p}_{3})}t^{c}\big(U_{\mathbf{x}}-1\big)\;,
\end{aligned}
\label{eq:MFS2L}
\end{equation}
and: 
\begin{equation}
\begin{aligned}\tilde{\mathcal{M}}_{\mathrm{FS2}}^{\lambda\eta} & =-\frac{p_{3}^{+}}{p_{1}^{+}}\mathrm{Dirac}^{\eta\bar{\eta}}\big(1+\frac{2p_{1}^{+}}{p_{3}^{+}}\big)\mathrm{Dirac}^{\lambda\bar{\lambda}}\big(1+\frac{2(p_{1}^{+}+p_{3}^{+})}{q^{+}}\big)\\
 & \times\frac{\mathbf{p}_{3}^{\bar{\eta}}\!-\!\frac{p_{3}^{+}}{p_{1}^{+}}\mathbf{p}_{1}^{\bar{\eta}}}{\big(\mathbf{p}_{3}\!-\!\frac{p_{3}^{+}}{p_{1}^{+}}\mathbf{p}_{1}\big)^{2}}\frac{\mathbf{p}_{3}^{\bar{\lambda}}\!+\!\mathbf{p}_{1}^{\bar{\lambda}}\!-\!\frac{p_{1}^{+}\!+p_{3}^{+}}{q^{+}}\mathbf{q}^{\bar{\lambda}}}{\frac{p_{1}^{+}p_{0\scriptscriptstyle{R}}^{+}}{q^{+}p_{3}^{+}}\big(\mathbf{p}_{3}\!-\!\frac{p_{3}^{+}}{p_{1}^{+}}\mathbf{p}_{1}\big)^{2}\!+\!\big(\mathbf{p}_{3}\!+\!\mathbf{p}_{1}\!-\!\frac{p_{1}^{+}\!+p_{3}^{+}}{q^{+}}\mathbf{q}\big)^{2}\!+\!\frac{p_{0\scriptscriptstyle{R}}^{+}(p_{1}^{+}\!+p_{3}^{+})}{(q^{+})^{2}}M^{2}}\\
 & \times\int_{\mathbf{x}}e^{-i\cdot(\mathbf{q}+\mathbf{p}_{1}+\mathbf{p}_{3})\cdot\mathbf{x}}t^{c}\big(U_{\mathbf{x}}-1\big)\;.
\end{aligned}
\label{eq:MFS2}
\end{equation}

\paragraph{Diagram FS3}
The amplitudes corresponding to Feynman graph FS3 (fig. (\ref{fig:FS}))
for the emission of a longitudinally or transversely polarized photon
read, respectively:
\begin{equation}
\begin{aligned}\tilde{\mathcal{M}}_{\mathrm{FS3}}^{0\eta} & =-\frac{1}{M}\frac{p_{0\scriptscriptstyle{R}}^{+}\mathbf{q}^{2}-(p_{1}^{+}+p_{3}^{+})M^{2}}{p_{0\scriptscriptstyle{R}}^{+}\mathbf{q}^{2}+(p_{1}^{+}+p_{3}^{+})M^{2}}\frac{p_{3}^{+}}{p_{1}^{+}}\mathrm{Dirac}^{\eta\bar{\eta}}\big(1+\frac{2p_{1}^{+}}{p_{3}^{+}}\big)\\
 & \times\frac{\mathbf{p}_{3}^{\bar{\eta}}-\frac{p_{3}^{+}}{p_{1}^{+}}\mathbf{p}_{1}^{\bar{\eta}}}{\big(\mathbf{p}_{3}-\frac{p_{3}^{+}}{p_{1}^{+}}\mathbf{p}_{1}\big)^{2}}\int_{\mathbf{x}}e^{-i\mathbf{x}\cdot(\mathbf{q}+\mathbf{p}_{1}+\mathbf{p}_{3})}t^{c}\big(U_{\mathbf{x}}-1\big)\;,
\end{aligned}
\label{eq:MFS3L}
\end{equation}
and:
\begin{align}
\tilde{\mathcal{M}}_{\mathrm{FS3}}^{\lambda\eta} & =\frac{-q^{+}\mathbf{q}^{\bar{\lambda}}}{p_{0\scriptscriptstyle{R}}^{+}\mathbf{q}^{2}+(p_{1}^{+}+p_{3}^{+})M^{2}}\frac{\mathbf{p}_{3}^{\bar{\eta}}-\frac{p_{3}^{+}}{p_{1}^{+}}\mathbf{p}_{1}^{\bar{\eta}}}{\big(\mathbf{p}_{3}-\frac{p_{3}^{+}}{p_{1}^{+}}\mathbf{p}_{1}\big)^{2}}\label{eq:MFS3}\\
 & \times\frac{p_{3}^{+}}{p_{1}^{+}}\mathrm{Dirac}^{\eta\bar{\eta}}\big(1+\frac{2p_{1}^{+}}{p_{3}^{+}}\big)\mathrm{Dirac}^{\lambda\bar{\lambda}}\big(1+\frac{2(p_{1}^{+}+p_{3}^{+})}{q^{+}}\big)\nonumber \\
 & \times\int_{\mathbf{x}}e^{-i\mathbf{x}\cdot(\mathbf{q}+\mathbf{p}_{1}+\mathbf{p}_{3})}t^{c}\big(U_{\mathbf{x}}-1\big)\;.\nonumber 
\end{align}
With the help of the relation:
\begin{equation}
\begin{aligned}\frac{\mathbf{p}_{3}^{\bar{\eta}}-\frac{p_{3}^{+}}{p_{1}^{+}}\mathbf{p}_{1}^{\bar{\eta}}}{\big(\mathbf{p}_{3}-\frac{p_{3}^{+}}{p_{1}^{+}}\mathbf{p}_{1}\big)^{2}}e^{-i\mathbf{x}\cdot\mathbf{p}_{3}} & =e^{-i\frac{p_{3}^{+}}{p_{1}^{+}}\mathbf{p}_{1}\cdot\mathbf{x}}\int_{\mathbf{z}}e^{-i\big(\mathbf{p}_{3}-\frac{p_{3}^{+}}{p_{1}^{+}}\mathbf{p}_{1}\big)\cdot\mathbf{z}}iA^{\bar{\eta}}(\mathbf{x}-\mathbf{z})\;,\end{aligned}
\end{equation}
the above amplitudes can also be cast in coordinate space, yielding:
\begin{equation}
\begin{aligned}\tilde{\mathcal{M}}_{\mathrm{FS3}}^{0\eta} & =-\frac{1}{M}\frac{p_{0\scriptscriptstyle{R}}^{+}\mathbf{q}^{2}-(p_{1}^{+}+p_{3}^{+})M^{2}}{p_{0\scriptscriptstyle{R}}^{+}\mathbf{q}^{2}+(p_{1}^{+}+p_{3}^{+})M^{2}}\frac{p_{3}^{+}}{p_{1}^{+}}\mathrm{Dirac}^{\eta\bar{\eta}}\big(1+\frac{2p_{1}^{+}}{p_{3}^{+}}\big)\\
 & \times\int_{\mathbf{x},\mathbf{z}}e^{-i\big(\mathbf{p}_{3}-\frac{p_{3}^{+}}{p_{1}^{+}}\mathbf{p}_{1}\big)\cdot\mathbf{z}}e^{-i\frac{p_{3}^{+}}{p_{1}^{+}}\mathbf{p}_{1}\cdot\mathbf{x}}e^{-i\mathbf{k}_{\perp}\cdot\mathbf{x}}iA^{\bar{\eta}}(\mathbf{x}-\mathbf{z})t^{c}\big(U_{\mathbf{x}}-1\big)\;,
\end{aligned}
\label{eq:MFS3L_coord}
\end{equation}
and:
\begin{align}
\tilde{\mathcal{M}}_{\mathrm{FS3}}^{\lambda\eta} & =\frac{-q^{+}\mathbf{q}^{\bar{\lambda}}}{p_{0\scriptscriptstyle{R}}^{+}\mathbf{q}^{2}+(p_{1}^{+}+p_{3}^{+})M^{2}}\frac{p_{3}^{+}}{p_{1}^{+}}\mathrm{Dirac}^{\eta\bar{\eta}}\big(1+\frac{2p_{1}^{+}}{p_{3}^{+}}\big)\mathrm{Dirac}^{\lambda\bar{\lambda}}\big(1+\frac{2(p_{1}^{+}+p_{3}^{+})}{q^{+}}\big)\label{eq:MFS3T_coord}\\
 & \times\int_{\mathbf{x},\mathbf{z}}e^{-i\big(\mathbf{p}_{3}-\frac{p_{3}^{+}}{p_{1}^{+}}\mathbf{p}_{1}\big)\cdot\mathbf{z}}e^{-i\frac{p_{3}^{+}}{p_{1}^{+}}\mathbf{p}_{1}\cdot\mathbf{x}}e^{-i\mathbf{k}_{\perp}\cdot\mathbf{x}}iA^{\bar{\eta}}(\mathbf{x}-\mathbf{z})t^{c}\big(U_{\mathbf{x}}-1\big)\;.\nonumber 
\end{align}

\paragraph{Diagram FS4}
The last diagrams we consider correspond to the instantaneous $q\to q\gamma^{*}g$
splitting after the initial quark has interacted with the shockwave.
In the longitudinal case, we find:
\begin{equation}
\begin{aligned}\tilde{\mathcal{M}}_{\mathrm{FS4}}^{0\eta} & \!=\!-\frac{1}{M}\frac{p_{3}^{+}p_{0\scriptscriptstyle{R}}^{+}}{p_{0}^{+}}\frac{1}{\big(\mathbf{p}_{3}\!-\!\frac{p_{3}^{+}}{p_{0}^{+}}\mathbf{k}_{\perp}\big)^{2}\!+\!\Delta_{\mathrm{FS}}}\Bigg[\frac{q^{+}}{p_{0\scriptscriptstyle{R}}^{+}(p_{1}^{+}\!+\!p_{3}^{+})}\big(\mathbf{p}_{3}\!-\!\frac{p_{3}^{+}}{p_{0}^{+}}\mathbf{k}_{\perp}\big)^{\bar{\eta}}(\delta^{\bar{\eta}\eta}\!-\!i\sigma^{\bar{\eta}\eta})\\
 & \!+\!\frac{2p_{1}^{+}\!+\!p_{3}^{+}}{p_{1}^{+}(p_{1}^{+}\!+\!p_{3}^{+})}\mathbf{P}_{\perp}^{\bar{\eta}}\Big(\delta^{\bar{\eta}\eta}\!+\!\frac{p_{3}^{+}}{2p_{1}^{+}+p_{3}^{+}}i\sigma^{\bar{\eta}\eta}\Big)\Bigg]\!\int_{\mathbf{x}}e^{-i\mathbf{x}\cdot(\mathbf{q}+\mathbf{p}_{1}+\mathbf{p}_{3})}t^{c}\big(U_{\mathbf{x}}\!-\!1\big)\;,
\end{aligned}
\label{eq:FS4L}
\end{equation}
while we obtain for the emission of a transversely polarized virtual photon:
\begin{equation}
\begin{aligned}\tilde{\mathcal{M}}_{\mathrm{FS4}}^{\lambda\eta} & =\frac{p_{3}^{+}p_{0\scriptscriptstyle{R}}^{+}}{p_{0}^{+}}\frac{1}{\big(\mathbf{p}_{3}-\frac{p_{3}^{+}}{p_{0}^{+}}\mathbf{k}_{\perp}\big)^{2}+\Delta_{\mathrm{FS}}}\Bigg[\Big(\frac{1}{p_{1}^{+}+p_{3}^{+}}+\frac{1}{p_{0}^{+}}\Big)\delta^{\eta\lambda}\\
&+i\sigma^{\eta\lambda}\Big(\frac{1}{p_{1}^{+}+p_{3}^{+}}-\frac{1}{p_{0}^{+}}\Big)\Bigg]\int_{\mathbf{x}}e^{-i\mathbf{x}\cdot(\mathbf{q}+\mathbf{p}_{1}+\mathbf{p}_{3})}t^{c}\big(U_{\mathbf{x}}-1\big)\;.
\end{aligned}
\label{eq:FS4}
\end{equation}
For later purposes, it can be useful to write:
\begin{equation}
\begin{aligned}\frac{1}{\big(\mathbf{p}_{3}-\frac{p_{3}^{+}}{p_{0}^{+}}\mathbf{k}_{\perp}\big)^{2}+\Delta_{\mathrm{FS}}}e^{-i\mathbf{p}_{3}\cdot\mathbf{x}} & =\frac{1}{\big(\mathbf{p}_{3}-\frac{p_{3}^{+}}{p_{0}^{+}}\mathbf{k}_{\perp}\big)^{2}+\Delta_{\mathrm{FS}}}e^{-i\big(\mathbf{p}_{3}-\frac{p_{3}^{+}}{p_{0}^{+}}\mathbf{k}_{\perp}\big)\cdot\mathbf{x}}e^{-i\frac{p_{3}^{+}}{p_{0}^{+}}\mathbf{k}_{\perp}\cdot\mathbf{x}}\;,\\
 & =e^{-i\frac{p_{3}^{+}}{p_{0}^{+}}\mathbf{k}_{\perp}\cdot\mathbf{x}}\int_{\mathbf{z}}e^{-i\big(\mathbf{p}_{3}-\frac{p_{3}^{+}}{p_{0}^{+}}\mathbf{k}_{\perp}\big)\cdot\mathbf{z}}\mathcal{K}(\mathbf{x}-\mathbf{z},\Delta_{\mathrm{FS}})\;,
\end{aligned}
\label{eq:FS4_coordspace}
\end{equation}
such that:
\begin{equation}
\begin{aligned}\tilde{\mathcal{M}}_{\mathrm{FS4}}^{0\eta} & =-\frac{1}{M}\frac{p_{3}^{+}p_{0\scriptscriptstyle{R}}^{+}}{p_{0}^{+}}\Bigg[\frac{q^{+}}{p_{0\scriptscriptstyle{R}}^{+}(p_{1}^{+}+p_{3}^{+})}\big(\mathbf{p}_{3}-\frac{p_{3}^{+}}{p_{0}^{+}}\mathbf{k}_{\perp}\big)^{\bar{\eta}}(\delta^{\bar{\eta}\eta}-i\sigma^{\bar{\eta}\eta})\\
 & +\frac{2p_{1}^{+}+p_{3}^{+}}{p_{1}^{+}(p_{1}^{+}+p_{3}^{+})}\mathbf{P}_{\perp}^{\bar{\eta}}\Big(\delta^{\bar{\eta}\eta}+\frac{p_{3}^{+}}{2p_{1}^{+}+p_{3}^{+}}i\sigma^{\bar{\eta}\eta}\Big)\Bigg]\\
 & \times e^{-i\frac{p_{3}^{+}}{p_{0}^{+}}\mathbf{k}_{\perp}\cdot\mathbf{x}}\int_{\mathbf{x},\mathbf{z}}e^{-i\big(\mathbf{p}_{3}-\frac{p_{3}^{+}}{p_{0}^{+}}\mathbf{k}_{\perp}\big)\cdot\mathbf{z}}\mathcal{K}(\mathbf{x}-\mathbf{z},\Delta_{\mathrm{FS}})\int_{\mathbf{x}}e^{-i\mathbf{k}_{\perp}\cdot\mathbf{x}}t^{c}\big(U_{\mathbf{x}}-1\big)\;,
\end{aligned}
\label{eq:FS4L_coord}
\end{equation}
and:
\begin{equation}
\begin{aligned}\tilde{\mathcal{M}}_{\mathrm{FS4}}^{\lambda\eta} & =\frac{p_{3}^{+}p_{0\scriptscriptstyle{R}}^{+}}{p_{0}^{+}}\Bigg[\Big(\frac{1}{p_{1}^{+}+p_{3}^{+}}+\frac{1}{p_{0}^{+}}\Big)\delta^{\eta\lambda}+i\sigma^{\eta\lambda}\Big(\frac{1}{p_{1}^{+}+p_{3}^{+}}-\frac{1}{p_{0}^{+}}\Big)\Bigg]\\
 & \times e^{-i\frac{p_{3}^{+}}{p_{0}^{+}}\mathbf{k}_{\perp}\cdot\mathbf{x}}\int_{\mathbf{x},\mathbf{z}}e^{-i\big(\mathbf{p}_{3}-\frac{p_{3}^{+}}{p_{0}^{+}}\mathbf{k}_{\perp}\big)\cdot\mathbf{z}}\mathcal{K}(\mathbf{x}-\mathbf{z},\Delta_{\mathrm{FS}})\int_{\mathbf{x}}e^{-i\mathbf{k}_{\perp}\cdot\mathbf{x}}t^{c}\big(U_{\mathbf{x}}-1\big)\;.
\end{aligned}
\label{eq:FS4_coord}
\end{equation}

\section{\label{sec:DGLAP}Infrared safety in the initial state}

In this section, we show that two of the initial-state radiation diagrams
(figure \ref{fig:ISR}), namely $\mathrm{IS1}$ and $\mathrm{IS2}$,
lead to infrared divergences on the level of the cross section, i.e.
after squaring and integrating over the outgoing gluon. These divergences
stem from the kinematic region in which the gluon, radiated from the
incoming quark, is collinear to its emitter. First, we will evaluate
the divergent integrals over the transverse momentum of the outgoing
gluon $\mathbf{p}_{3}$ using dimensional regularization. The collinear
singularities will be parameterized by the poles $\epsilon_{\mathrm{coll}}$.
We shall then demonstrate how these poles can be absorbed into the
DGLAP evolution equations of the incoming quark.

\subsection{Collinear divergences from initial-state radiation}

Real corrections to the NLO cross section are obtained by computing
the $2\to3$ partonic amplitudes, multiplying them with their complex
conjugate, and integrating the additional parton out (see below, subsection
\ref{subsec:parton2hadron}). When considering the initial-state radiation
amplitudes $\mathcal{M}_{\mathrm{IS1}}^{0\eta}$ (\ref{eq:MIS1L})
and $\mathcal{M}_{\mathrm{IS2}}^{0\eta}$ (\ref{eq:MIS2L}), we obtain
after promoting the reduced amplitudes to the full ones according
to formula (\ref{eq:reduceddef}), multiplying them with their complex
conjugate, and integrating over the gluon transverse momentum:
\begin{equation}
\begin{aligned}\int_{\mathbf{p}_{3}}\Big|\mathcal{M}_{\mathrm{IS1}}^{0\eta}+\mathcal{M}_{\mathrm{IS2}}^{0\eta}\Big|^{2} & =\frac{g_{\mathrm{em}}^{2}g_{s}^{2}N_{c}C_{F}}{M^{2}}\Big(\frac{p_{3}^{+}}{p_{0\scriptscriptstyle{R}}^{+}}\Big)^{2}8p_{1}^{+}p_{0\scriptscriptstyle{R}}^{+}\Big(\big(1+\frac{2p_{0}^{+}}{p_{3}^{+}}\big)^{2}+D-3\Big)\\
 & \times\int_{\mathbf{x},\mathbf{x}^{\prime}}\int_{\boldsymbol{\ell}}\frac{1}{\boldsymbol{\ell}^{2}}e^{-i\boldsymbol{\ell}\cdot(\mathbf{\mathbf{x}}-\mathbf{x}^{\prime})}\Big(-\frac{p_{0}^{+}\mathbf{P}_{\perp}^{2}-p_{1}^{+}M^{2}}{p_{0}^{+}\mathbf{P}_{\perp}^{2}+p_{1}^{+}M^{2}} \\
 &+\frac{p_{3}^{+}\big(q^{+}\boldsymbol{\ell}+p_{0}^{+}\mathbf{q}\big)^{2}-p_{0}^{+}p_{1}^{+}p_{3}^{+}M^{2}}{p_{3}^{+}\big(q^{+}\boldsymbol{\ell}+p_{0}^{+}\mathbf{q}\big)^{2}+p_{1}^{+}q^{+}p_{0R}^{+}\boldsymbol{\ell}^{2}+p_{0}^{+}p_{1}^{+}p_{3}^{+}M^{2}}\Big)^{2}\\
 & \times e^{-i\mathbf{k}_{\perp}\cdot(\mathbf{x}-\mathbf{x}^{\prime})}\big(s_{\mathbf{x}^{\prime}\mathbf{x}}+1\big)\;.
\end{aligned}
\label{eq:IScollL_0}
\end{equation}
Power counting teaches us that the integral over the gluon transverse
momentum with respect to its emitter contains a collinear divergence:\footnote{Ultraviolet divergences $\boldsymbol{\ell}^{2}\to\infty$ are cut
off by the phase.}
\begin{equation}
\int_{\boldsymbol{\ell}}\frac{1}{\boldsymbol{\ell}^{2}}e^{-i\boldsymbol{\ell}\cdot(\mathbf{\mathbf{x}}-\mathbf{x}^{\prime})}=-\frac{1}{4\pi}\Big(\frac{1}{\epsilon_{\mathrm{\mathrm{coll}}}}+\gamma_{\scriptscriptstyle{E}}+\mathrm{ln}\big(\mu^{2}\pi(\mathbf{x}-\mathbf{x}^{\prime})^{2}\big)\Big)+\mathcal{O}(\epsilon_{\mathrm{coll}})\;,
\end{equation}
which is easily proven with the help of the Schwinger trick (\ref{eq:Schwinger}).
Therefore, in the collinear limit $\boldsymbol{\ell}\to0$, the third
line in formula (\ref{eq:IScollL_0}) can be approximated by:
\begin{equation}
\frac{p_{3}^{+}\big(q^{+}\boldsymbol{\ell}+p_{0}^{+}\mathbf{q}\big)^{2}-p_{0}^{+}p_{1}^{+}p_{3}^{+}M^{2}}{p_{3}^{+}\big(q^{+}\boldsymbol{\ell}+p_{0}^{+}\mathbf{q}\big)^{2}+p_{1}^{+}q^{+}p_{0R}^{+}\boldsymbol{\ell}^{2}+p_{1}^{+}p_{3}^{+}p_{0}^{+}M^{2}}\to\frac{p_{0}^{+}\mathbf{q}^{2}-p_{1}^{+}M^{2}}{p_{0}^{+}\mathbf{q}^{2}+p_{1}^{+}M^{2}}\;.\label{eq:IS_coll_approx}
\end{equation}
With these simplifications, the collinearly divergent part of eq.
(\ref{eq:IScollL_0}) can be rewritten as:\footnote{Of course, the above notation is somewhat symbolic, as the square
of the leading order part $|\mathcal{M}_{\mathrm{LO}}|^{2}$ still
contains the integrals over $\mathbf{x}$ and $\mathbf{x}^{\prime}$,
which also appear in the phase $e^{-i\mathbf{k}\cdot(\mathbf{\mathbf{x}}-\mathbf{x}^{\prime})}$.}
\begin{equation}
\begin{aligned}\int_{\mathbf{p}_{3}}\Big|\mathcal{M}_{\mathrm{IS1}}^{(0,\lambda)\eta}+\mathcal{M}_{\mathrm{IS2}}^{(0,\lambda)\eta}\Big|^{2} & \overset{\mathrm{coll.}}{=}|\mathcal{M}_{\mathrm{LO}}^{0,\lambda}|^{2}\times-\alpha_{s}C_{F}\Big(\frac{1}{\epsilon_{\mathrm{coll}}}+\gamma_{\scriptscriptstyle{E}}+\mathrm{ln}\big(\mu^{2}\pi(\mathbf{x}-\mathbf{x}^{\prime})^{2}\big)\Big)\\
 & \times\frac{(p_{3}^{+})^{2}}{p_{0}^{+}p_{0\scriptscriptstyle{R}}^{+}}\Big(\big(1+\frac{2p_{0}^{+}}{p_{3}^{+}}\big)^{2}+D-3\Big)\;.
\end{aligned}
\label{eq:IScoll_final}
\end{equation}
It is easily verified that the above result also holds in the case
of a transversely polarized virtual photon, hence the index $(0,\lambda)$.
This implies that we have extracted a universal collinear factor,
which is independent of the details of the leading-order process.
It is, therefore, natural to associate it with quantum fluctuations
of the incoming quark before it participates in the hard process.
Indeed, in the next section we demonstrate that the collinear pole
can be absorbed into the DGLAP evolution of the quark PDF.

\subsection{\label{subsec:parton2hadron}From parton to hadron level}

On the partonic level, the amplitudes for the real NLO corrections
are transformed into contributions to the cross section using the
following formula:
\begin{equation}
\begin{aligned}\mathrm{d}\hat{\sigma}_{\mathrm{real}} & =\frac{1}{2p_{0\scriptscriptstyle{R}}^{+}}\mathrm{PS}(\vec{p}_{1},\vec{q},\vec{p}_{3})2\pi\delta(p_{0\scriptscriptstyle{R}}^{+}-p_{1}^{+}-q^{+}-p_{3}^{+})\frac{1}{D-2}\big|\mathcal{M}_{\mathrm{real}}\big|^{2}\;,\end{aligned}
\label{eq:crosssectiondef_real}
\end{equation}
where we used the short-hand notation:
\begin{equation}
\mathrm{PS}(\vec{k})=\frac{\mathrm{d}k^{+}\mathrm{d}^{D-2}\mathbf{k}}{(2\pi)^{D-3}2k^{+}}\;,\label{eq:PSdef}
\end{equation}
for the phase-space integration measure. As discussed in section \ref{sec:LO},
eq. (\ref{eq:parton2hadron}), the partonic cross section needs to
be transformed into a hadronic one by convolving with the leading-order
quark PDF $f_{q}^{(0)}$, and taking the CGC average:
\begin{equation}
\begin{aligned}\mathrm{d}\sigma_{\mathrm{real}} & =\int\frac{\mathrm{d}p_{0\scriptscriptstyle{R}}^{+}}{p_{p}^{+}}f_{q}^{(0)}\big(\frac{x_{p}}{\xi}\big)\langle\mathrm{d}\hat{\sigma}_{\mathrm{real}}\rangle\;.\end{aligned}
\label{eq:hadronic_real}
\end{equation}
Finally, the radiated gluon is integrated out in order to obtain the
real radiative corrections to the full NLO cross section:
\begin{equation}
\begin{aligned}\mathrm{d}\sigma_{\mathrm{NLO}} & =\mathrm{d}\sigma_{\mathrm{LO}}+\mathrm{d}\sigma_{\mathrm{virtual}}+\int\mathrm{d}^{D-1}\vec{p}_{3}\frac{\mathrm{d}\sigma_{\mathrm{real}}}{\mathrm{d}^{D-1}p_{3}}\;.\end{aligned}
\label{eq:NLOdef}
\end{equation}
We shall now apply the procedure above to transform eq. (\ref{eq:IScoll_final})
into a contribution to the NLO cross section. Promoting to the partonic
cross section, performing the convolution with the PDF, and averaging
over the semiclassical background fields, we find:\footnote{Note that we do not distinguish anymore between the polarization states
of the virtual photon, since the collinear initial-state radiation
is insensitive to them.}
\begin{equation}
\begin{aligned}\int\mathrm{d}^{D\!-\!1}\vec{p}_{3}\frac{\mathrm{d}\sigma_{\mathrm{IS1,2}}}{\mathrm{d}^{D\!-\!1}p_{3}} & \!\overset{\mathrm{coll.}}{=}\!\!\int\frac{\mathrm{d}p_{0\scriptscriptstyle{R}}^{+}}{p_{p}^{+}}f_{q}^{(0)}\big(\frac{x_{p}}{\xi}\big)\!\int_{k_{\mathrm{min}}^{+}}\frac{\mathrm{d}p_{3}^{+}}{(2\pi)^{D\!-\!3}2p_{3}^{+}}\Big(\frac{p_{3}^{+}}{p_{0\scriptscriptstyle{R}}^{+}}\Big)^{\!2}\Big(\big(1\!+\!\frac{2p_{0}^{+}}{p_{3}^{+}}\big)^{\!2}\!+\!D\!-\!3\Big)\\
 & \times\frac{1}{2p_{0}^{+}}\int\mathrm{PS}(\vec{p}_{1},\vec{q})2\pi\delta(p_{0\scriptscriptstyle{R}}^{+}-p_{1}^{+}-q^{+}-p_{3}^{+})\frac{1}{D-2}\big\langle\big|\mathcal{M}_{\mathrm{LO}}\big|^{2}\big\rangle\\
 & \times-\alpha_{s}C_{F}\Big(\frac{1}{\epsilon_{\mathrm{coll}}}+\gamma_{\scriptscriptstyle{E}}+\mathrm{ln}\big(\mu^{2}\pi(\mathbf{x}-\mathbf{x}^{\prime})^{2}\big)\Big)\;.
\end{aligned}
\end{equation}
After a change of variables in the integral over gluon plus-momentum,
introducing $\xi=p_{0}^{+}/p_{0\scriptscriptstyle{R}}^{+}$ in function of which $p_{3}^{+}=p_{0}^{+}\frac{1-\xi}{\xi}$
with $\mathrm{d}\xi/\xi=\mathrm{d}p_{3}^{+}/p_{0\scriptscriptstyle{R}}^{+}$, and remembering
the leading-order definition of projectile quark momentum fraction
$x_{p}=p_{0}^{+}/p_{p}^{+}$, we finally obtain:
\begin{equation}
\begin{aligned}\int\mathrm{d}^{D-1}\vec{p}_{3}\frac{\mathrm{d}\sigma_{\mathrm{IS1,2}}}{\mathrm{d}^{D-1}p_{3}} & \overset{\mathrm{coll.}}{=}\int_{x_{p}}^{p_{0}^{+}/(p_{0}^{+}+k_{\mathrm{min}}^{+})}\mathrm{d}\xi\frac{1+\xi^{2}}{1-\xi}\frac{x_{p}}{\xi}f_{q}^{(0)}\big(\frac{x_{p}}{\xi}\big)\\
 & \times\frac{1}{2(p_{0}^{+})^{2}}\int\mathrm{PS}(\vec{p}_{1},\vec{q})\frac{1}{2}\big\langle\big|\mathcal{M}_{\mathrm{LO}}\big|^{2}\big\rangle\\
 & \times-\alpha_{s}C_{F}\Big(\frac{1}{\epsilon_{\mathrm{coll}}}+\gamma_{\scriptscriptstyle{E}}+\mathrm{ln}\big(\mu^{2}\pi(\mathbf{x}-\mathbf{x}^{\prime})^{2}\big)\Big)\;.
\end{aligned}
\label{eq:sigmaIS12coll_0}
\end{equation}
The upper bound $p_{0}^{+}/(p_{0}^{+}+k_{\mathrm{min}}^{+})>\xi$
in the integral stems from the rapidity cutoff $p_{3}^{+}>k_{\mathrm{min}}^{+}$,
while the requirement $\xi>x_{p}$ comes from the domain of the PDF.
To cast it in a form in which one can recognize DGLAP, let us start
by splitting up the integration:
\begin{equation}
\begin{aligned} & \int_{x_{p}}^{p_{0}^{+}/(p_{0}^{+}+k_{\mathrm{min}}^{+})}\mathrm{d}\xi\frac{1+\xi^{2}}{1-\xi}\frac{x_{p}}{\xi}f_{q}^{(0)}\big(\frac{x_{p}}{\xi}\big)\\
 & =\Big(\int_{x_{p}}^{1}\mathrm{d}\xi+\int_{1}^{p_{0}^{+}/(p_{0}^{+}+k_{\mathrm{min}}^{+})}\mathrm{d}\xi\Big)\frac{1+\xi^{2}}{1-\xi}\frac{x_{p}}{\xi}f_{q}^{(0)}\big(\frac{x_{p}}{\xi}\big)\;.
\end{aligned}
\label{eq:dirtytrick}
\end{equation}
The next step is to make use of the definition of the plus-distribution:
\begin{equation}
\begin{aligned}\int_{z}^{1}\mathrm{d}\xi\frac{f(\xi)}{(1-\xi)_{+}} & \equiv\int_{z}^{1}\mathrm{d}\xi\frac{f(\xi)-f(1)}{1-\xi}-f(1)\int_{0}^{z}\frac{\mathrm{d}\xi}{1-\xi}\;,\end{aligned}
\label{eq:plusdef}
\end{equation}
using which the first integration in (\ref{eq:dirtytrick}) can be
written as:
\begin{equation}
\begin{aligned}\int_{x_{p}}^{1}\mathrm{d}\xi\frac{1+\xi^{2}}{1-\xi}\frac{x_{p}}{\xi}f_{q}^{(0)}\big(\frac{x_{p}}{\xi}\big) & =\int_{x_{p}}^{1}\mathrm{d}\xi\frac{1+\xi^{2}}{(1-\xi)_{+}}\frac{x_{p}}{\xi}f_{q}^{(0)}\big(\frac{x_{p}}{\xi}\big)+x_{p}f_{q}^{(0)}(x_{p})\int_{0}^{1}\mathrm{d\xi}\frac{2}{1-\xi}\;.\end{aligned}
\label{eq:DGLAPtrick1}
\end{equation}
The second integral in (\ref{eq:dirtytrick}) is the one that contains
the divergence for $p_{3}^{+}\to0$, or equivalently, $\xi\to1$.
In this limit, one can write: 
\begin{equation}
\begin{aligned}\int_{1}^{p_{0}^{+}/(p_{0}^{+}+k_{\mathrm{min}}^{+})}\mathrm{d}\xi\frac{1+\xi^{2}}{1-\xi}\frac{x_{p}}{\xi}f_{q}^{(0)}\big(\frac{x_{p}}{\xi}\big) & \simeq x_{p}f_{q}^{(0)}(x_{p})\int_{1}^{p_{0}^{+}/(p_{0}^{+}+k_{\mathrm{min}}^{+})}\mathrm{d}\xi\frac{2}{1-\xi}\;.\end{aligned}
\label{eq:DGLAPtrick2}
\end{equation}
Combining (\ref{eq:DGLAPtrick1}) and (\ref{eq:DGLAPtrick2}), eq.
(\ref{eq:dirtytrick}) eventually becomes:
\begin{equation}
\begin{aligned} & \int_{x_{p}}^{p_{0}^{+}/(p_{0}^{+}+k_{\mathrm{min}}^{+})}\mathrm{d}\xi\frac{1+\xi^{2}}{1-\xi}\frac{x_{p}}{\xi}f_{q}^{(0)}\big(\frac{x_{p}}{\xi}\big)\\
 & =\int_{x_{p}}^{1}\mathrm{d}\xi\frac{1+\xi^{2}}{(1-\xi)_{+}}\frac{x_{p}}{\xi}f_{q}^{(0)}\big(\frac{x_{p}}{\xi}\big)+x_{p}f_{q}^{(0)}(x_{p})\int_{0}^{p_{0}^{+}/(p_{0}^{+}+k_{\mathrm{min}}^{+})}\mathrm{d\xi}\frac{2}{1-\xi}\;,\\
 & =\int_{x_{p}}^{1}\mathrm{d}\xi\Big(\frac{1+\xi^{2}}{(1-\xi)_{+}}+\frac{3}{2}\delta(1-\xi)\Big)\frac{x_{p}}{\xi}f_{q}^{(0)}\big(\frac{x_{p}}{\xi}\big)+\Big(-\frac{3}{2}+2\ln\frac{p_{0}^{+}}{k_{\mathrm{min}}^{+}}\Big)x_{p}f_{q}^{(0)}(x_{p})\;.
\end{aligned}
\end{equation}
In the last line of the above formula, one can recognize the Altarelli-Parisi
$q\to q$ splitting function:
\begin{equation}
P_{qq}^{(0)}(\xi)=C_{F}\Big(\frac{1+\xi^{2}}{(1-\xi)_{+}}+\frac{3}{2}\delta(1-\xi)\Big)\;.\label{eq:splittingfunction}
\end{equation}
We, therefore, have shown that eq. (\ref{eq:sigmaIS12coll_0}) can
be rewritten as:
\begin{equation}
\begin{aligned}\int\mathrm{d}^{D-1}\vec{p}_{3}\frac{\mathrm{d}\sigma_{\mathrm{IS1,2}}}{\mathrm{d}^{D-1}p_{3}} &\! \overset{\mathrm{coll.}}{=}\!\!\Bigg[\!\int_{x_{p}}^{1}\mathrm{d}\xi P_{qq}^{(0)}(\xi)\frac{x_{p}}{\xi}f_{q}^{(0)}\big(\frac{x_{p}}{\xi}\big)\!+\!C_{F}\Big(\!-\!\frac{3}{2}\!+\!2\ln\frac{p_{0}^{+}}{k_{\mathrm{min}}^{+}}\Big)x_{p}f_{q}^{(0)}(x_{p})\Bigg]\\
 & \times\frac{1}{2(p_{0}^{+})^{2}}\int\mathrm{PS}(\vec{p}_{1},\vec{q})\frac{1}{2}\big\langle\big|\mathcal{M}_{\mathrm{LO}}\big|^{2}\big\rangle\\
 & \times-\alpha_{s}\Big(\frac{1}{\epsilon_{\mathrm{coll}}}+\gamma_{\scriptscriptstyle{E}}+\mathrm{ln}\big(\mu^{2}\pi(\mathbf{x}-\mathbf{x}^{\prime})^{2}\big)\Big)\;.
\end{aligned}
\label{eq:IS2DGLAP}
\end{equation}
However, the above pole, stemming from real gluon radiation in the
initial-state, is not the first collinear singularity we have encountered
in this work. Remember that in subsection \ref{subsec:Z}, we discovered
collinear poles stemming from the quark field-strength renormalization:
\begin{equation}
\begin{aligned}\big(\mathcal{Z}_{\mathrm{IS}}+\mathcal{Z}_{\mathrm{IS}}^{\dagger}\big)\mathrm{d}\sigma_{\mathrm{LO}} & =\alpha_{s}C_{F}\Big(\frac{1}{\epsilon_{\mathrm{coll}}}+\mathrm{ln}\frac{\mu^{2}}{\mu_{\scriptscriptstyle{R}}^{2}}\Big)\Big(-\frac{3}{2}+2\mathrm{ln}\frac{p_{0}^{+}}{k_{\mathrm{min}}^{+}}\Big)\\
 & \times x_{p}f_{q}^{(0)}(x_{p})\frac{1}{2(p_{0}^{+})^{2}}\int\mathrm{PS}(\vec{p}_{1},\vec{q})\frac{1}{2}\big\langle\big|\mathcal{M}_{\mathrm{LO}}\big|^{2}\big\rangle\;.
\end{aligned}
\label{eq:ZIS2DGLAP}
\end{equation}
Adding (\ref{eq:IS2DGLAP}) and (\ref{eq:ZIS2DGLAP}), one obtains:
\begin{equation}
\begin{aligned} & \int\mathrm{d}^{D-1}\vec{p}_{3}\frac{\mathrm{d}\sigma_{\mathrm{IS1,2}}}{\mathrm{d}^{D-1}p_{3}}\Big|_{\mathrm{coll.}}+\big(\mathcal{Z}_{\mathrm{IS}}+\mathcal{Z}_{\mathrm{IS}}^{\dagger}\big)\mathrm{d}\sigma_{\mathrm{LO}}\\
 & =\frac{1}{2(p_{0}^{+})^{2}}\int\mathrm{PS}(\vec{p}_{1},\vec{q})\frac{1}{2}\big\langle\big|\mathcal{M}_{\mathrm{LO}}\big|^{2}\big\rangle\\
 & \times\Bigg[-\alpha_{s}\Big(\frac{1}{\epsilon_{\mathrm{coll}}}+\gamma_{\scriptscriptstyle{E}}+\mathrm{ln}\big(\mu^{2}\pi(\mathbf{x}-\mathbf{x}^{\prime})^{2}\big)\Big)\int_{x_{p}}^{1}\mathrm{d}\xi P_{qq}^{(0)}(\xi)\frac{x_{p}}{\xi}f_{q}^{(0)}\big(\frac{x_{p}}{\xi}\big)\\
 & +\alpha_{s}C_{F}\Big(\mathrm{ln}\frac{e^{-\gamma_{\scriptscriptstyle{E}}}}{\mu_{\scriptscriptstyle{R}}^{2}\pi(\mathbf{x}-\mathbf{x}^{\prime})^{2}}\Big)\Big(-\frac{3}{2}+2\mathrm{ln}\frac{p_{0}^{+}}{k_{\mathrm{min}}^{+}}\Big)x_{p}f_{q}^{(0)}(x_{p})\Bigg]\;.
\end{aligned}
\label{eq:IS+Z_DGLAP}
\end{equation}
In the above equation, the only collinear singularity left is proportional
to the leading-order squared amplitude times the quark PDF convolved
with the Altarelli-Parisi splitting function. The final step is to
notice that the ($\overline{\mathrm{MS}}$-) quark PDF at NLO: $f_{q}^{(1)}$,
is related to the leading order one $f_{q}^{(0)}$ as follows:\footnote{In principle, eq. (\ref{eq:PDFMSbar}) contains an additional term
proportional to the gluon PDF. The corresponding collinear pole should
cancel with initial-state radiation in the $g+A\to\gamma^{*}+q+\bar{q}$
contribution to the cross section. The analysis of the gluon channel
is left for future work, see also the discussion around fig. \ref{fig:gluonchannel}.}
\begin{equation}
\begin{aligned} x_{p}f_{q}^{(1)}(x_{p},\mu^{2})  &=x_{p}f_{q}^{(0)}(x_{p})\\
&-\Big(\frac{1}{\epsilon_{\mathrm{coll}}}-\gamma_{\scriptscriptstyle{E}}+\ln4\pi\Big)\frac{\alpha_{s}}{2\pi}\int_{x_{p}}^{1}\frac{\mathrm{d}\xi}{\xi}P_{qq}^{(0)}(\xi)x_{p}f_{q}^{(0)}\big(\frac{x_{p}}{\xi}\big)+\mathcal{O}(\alpha_{s}^{2})\;.\end{aligned}
\label{eq:PDFMSbar}
\end{equation}
Therefore, we can rewrite the LO cross section as:
\begin{equation}
\begin{aligned}\mathrm{d}\sigma_{\mathrm{LO}} & =\mathrm{d}\sigma_{\mathrm{LO}+\mathrm{DGLAP}}+\Big(\frac{1}{\epsilon_{\mathrm{coll}}}-\gamma_{\scriptscriptstyle{E}}+\ln4\pi\Big)\frac{\alpha_{s}}{2\pi}\int_{x_{p}}^{1}\frac{\mathrm{d}\xi}{\xi}P_{qq}^{(0)}(\xi)x_{p}f_{q}^{(0)}\big(\frac{x_{p}}{\xi}\big)\;\\
 & \times\frac{2\pi}{2(p_{0}^{+})^{2}}\int\mathrm{PS}(\vec{p}_{1},\vec{q})\frac{1}{D-2}\big\langle\big|\mathcal{M}_{\mathrm{LO}}\big|^{2}\big\rangle\;,
\end{aligned}
\label{eq:LO+DGLAP}
\end{equation}
where we defined $\mathrm{d}\sigma_{\mathrm{LO}+\mathrm{DGLAP}}$
as the leading-order cross section convolved with the NLO quark PDF:
\begin{equation}
\mathrm{d}\sigma_{\mathrm{LO}+\mathrm{DGLAP}}=x_{p}f_{q}^{(1)}\big(x_{p},\mu^{2}\big)\frac{2\pi}{2(p_{0}^{+})^{2}}\int\mathrm{PS}(\vec{p}_{1},\vec{q})\frac{1}{D-2}\big\langle\big|\mathcal{M}_{\mathrm{LO}}\big|^{2}\big\rangle\;.\label{eq:sigma_LO+DGLAP}
\end{equation}
Adding (\ref{eq:IS+Z_DGLAP}) and (\ref{eq:LO+DGLAP}), we finally
obtain:
\begin{equation}
\begin{aligned}\mathrm{d}\sigma_{\mathrm{LO}}+\int\mathrm{d}^{D-1}\vec{p}_{3}\frac{\mathrm{d}\sigma_{\mathrm{IS1,2}}}{\mathrm{d}^{D-1}p_{3}}\Big|_{\mathrm{coll.}}+\big(\mathcal{Z}_{\mathrm{IS}}+\mathcal{Z}_{\mathrm{IS}}^{\dagger}\big)\mathrm{d}\sigma_{\mathrm{LO}}=\mathrm{d}\sigma_{\mathrm{LO}+\mathrm{DGLAP}}+ & \mathrm{d}\sigma_{\mathrm{IS}}\;,\end{aligned}
\label{eq:DGLAP_final}
\end{equation}
with:
\begin{equation}
\begin{aligned}\mathrm{d}\sigma_{\mathrm{IS}} & \equiv\alpha_{s}\Bigg[\ln\frac{c_{0}^{2}}{\mu^{2}(\mathbf{x}-\mathbf{x}^{\prime})^{2}}\int_{x_{p}}^{1}\mathrm{d}\xi P_{qq}^{(0)}(\xi)\frac{x_{p}}{\xi}f_{q}^{(0)}\big(\frac{x_{p}}{\xi}\big)\\
 & +C_{F}\Big(\mathrm{ln}\frac{e^{-\gamma_{\scriptscriptstyle{E}}}}{\mu_{\scriptscriptstyle{R}}^{2}\pi(\mathbf{x}-\mathbf{x}^{\prime})^{2}}\Big)\Big(-\frac{3}{2}+2\mathrm{ln}\frac{p_{0}^{+}}{k_{\mathrm{min}}^{+}}\Big)x_{p}f_{q}^{(0)}(x_{p})\Bigg] \\
& \times\frac{1}{2(p_{0}^{+})^{2}}\int\mathrm{PS}(\vec{p}_{1},\vec{q})\frac{1}{2}\big\langle\big|\mathcal{M}_{\mathrm{LO}}\big|^{2}\big\rangle\;.
\end{aligned}
\label{eq:sigma_IS}
\end{equation}
The above result proves that, at least at NLO accuracy, the hybrid
dilute-dense factorization Ansatz employed in this work holds, as
a collinear PDF is sufficient to absorb all collinear singularities
stemming from initial-state radiation.

\section{\label{sec:jet}Infrared safety in the final state}

\subsection{Jet algorithm}

For the leading-order and virtual contributions to the cross section, the outgoing
quark can be trivially related to the jet which it will initiate.
Indeed, it is sufficient to assume that the jet will be centered around
the outgoing quark, such that $\vec{p}_{\mathrm{jet}}=\vec{p}_{1}$.
However, for the real NLO corrections, this is not true anymore. Instead,
we need to specify what we mean by a jet, using a specific jet algorithm.
This algorithm allows one to distinguish the case in which the quark
and the gluon are grouped inside the same jet, when the reaction has
a similar topology as the leading-order one, from the situation where
both the quark and the gluon each initiate a separate jet. Since our
main purpose here is to prove final-state infrared safety, we can
limit ourselves to the following straightforward algorithm: the outgoing
quark and gluon are grouped inside the same jet when their momenta
satisfy the condition:
\begin{equation}
\begin{aligned}\frac{p_{1}^{+}+p_{3}^{+}}{|\mathbf{p}_{1}+\mathbf{p}_{3}|}\Big|\frac{\mathbf{p}_{1}}{p_{1}^{+}}-\frac{\mathbf{p}_{3}}{p_{3}^{+}}\Big| & <R\;.\end{aligned}
\label{eq:jetalgorithm}
\end{equation}
The parameter $R$ is known as the jet radius parameter and is bounded
by $0<R<1$. Moreover, for our purposes it is sufficient to work in
the so-called narrow-jet limit $R\to0$, suppressing all positive
powers of $R$ in the calculation.\footnote{We note that, in this limit, our jet definition becomes equivalent to the Cambridge/Aachen algorithm~\cite{Dokshitzer:1997in,Wobisch:1998wt}}  In practice, this means that there
is only a single configuration where the quark and gluon are part
of the same jet, namely when the gluon is radiated collinearly to
the quark in the final state. Note that, in this work, we always consider
the case where the observed jet is initiated by the quark and the
gluon is integrated out, whether the latter forms its own jet or belongs
to the quark one. For later phenomenological applications, our intermediate
$q+A\to\gamma^{*}+q+g$ results can be directly used to calculate
the contribution to the cross section where the observed jet stems
from the gluon, while the quark is either part of this same jet or stays undetected. Likewise, if desired, it always remains possible to
employ a more sophisticated jet algorithm.

Let us start by adding and squaring the amplitudes corresponding to
Feynman diagrams $\mathrm{FS2}$ (\ref{eq:MFS2}) and $\mathrm{FS3}$
(\ref{eq:MFS3}), starting with the transverse case:
\begin{equation}
\begin{aligned}|\mathcal{M}_{\mathrm{FS2+3}}^{\lambda}|^{2} & \!=\!g_{\mathrm{em}}^{2}g_{s}^{2}C_{F}N_{c}\Big(\frac{p_{3}^{+}}{p_{1}^{+}}\Big)^{\!2}8p_{1}^{+}p_{0\scriptscriptstyle{R}}^{+}\!\Big(\!\big(1\!+\!\frac{2p_{1}^{+}}{p_{3}^{+}}\!\big)^{\!2}\!+\!D\!-\!3\!\Big)\!\Big(\!\big(1\!+\!2\frac{p_{1}^{+}\!+\!p_{3}^{+}}{q^{+}}\big)^{\!2}\!+\!D\!-\!3\!\Big)\\
 & \times\frac{1}{\big(\mathbf{p}_{3}-\frac{p_{3}^{+}}{p_{1}^{+}}\mathbf{p}_{1}\big)^{2}}\Bigg[\frac{p_{3}^{+}}{p_{0}^{+}}\frac{\mathbf{q}-\frac{q^{+}}{p_{1}^{+}+p_{3}^{+}}(\mathbf{p}_{1}+\mathbf{p}_{3})}{\big(\mathbf{p}_{3}-\frac{p_{3}^{+}}{p_{0}^{+}}\mathbf{k}_{\perp}\big)^{2}+\Delta_{\mathrm{FS}}}-\frac{q^{+}\mathbf{q}}{p_{0\scriptscriptstyle{R}}^{+}\mathbf{q}^{2}+(p_{1}^{+}+p_{3}^{+})M^{2}}\Bigg]^{2}\\
 & \times\int_{\mathbf{x},\mathbf{x}^{\prime}}e^{-i(\mathbf{q}+\mathbf{p}_{1}+\mathbf{p}_{3})\cdot(\mathbf{x}-\mathbf{x}^{\prime})}\big(s_{\mathbf{x}\mathbf{x}^{\prime}}+1\big)\;,
\end{aligned}
\label{eq:MFS2+3T}
\end{equation}
where the spinor trace was evaluated with the help of the relations
in sec.~\ref{sec:Dirac}. Similarly, in the longitudinal case:
\begin{equation}
\begin{aligned}\big|\mathcal{M}_{\mathrm{FS2+3}}^{0}\big|^{2} & =g_{\mathrm{em}}^{2}g_{s}^{2}C_{F}N_{c}\Big(\frac{p_{3}^{+}}{p_{1}^{+}}\Big)^{2}8p_{1}^{+}p_{0\scriptscriptstyle{R}}^{+}\Big(\big(1+\frac{2p_{1}^{+}}{p_{3}^{+}}\big)^{2}+D-3\Big)\frac{1}{M^{2}}\frac{1}{\big(\mathbf{p}_{3}-\frac{p_{3}^{+}}{p_{1}^{+}}\mathbf{p}_{1}\big)^{2}}\\
 & \times\Bigg[\frac{p_{0\scriptscriptstyle{R}}^{+}p_{3}^{+}}{q^{+}p_{0}^{+}}\frac{\frac{\big((p_{1}^{+}+p_{3}^{+})\mathbf{q}-q^{+}(\mathbf{p}_{1}+\mathbf{p}_{3})\big)^{2}}{p_{0\scriptscriptstyle{R}}^{+}(p_{1}^{+}+p_{3}^{+})}-M^{2}}{\big(\mathbf{p}_{3}-\frac{p_{3}^{+}}{p_{0}^{+}}\mathbf{k}_{\perp}\big)^{2}+\Delta_{\mathrm{FS}}}-\frac{p_{0\scriptscriptstyle{R}}^{+}\mathbf{q}^{2}-(p_{1}^{+}+p_{3}^{+})M^{2}}{p_{0\scriptscriptstyle{R}}^{+}\mathbf{q}^{2}+(p_{1}^{+}+p_{3}^{+})M^{2}}\Bigg]^{2}\\
 & \times\int_{\mathbf{x},\mathbf{x}^{\prime}}e^{-i(\mathbf{q}+\mathbf{p}_{1}+\mathbf{p}_{3})\cdot(\mathbf{x}-\mathbf{x}^{\prime})}\big(s_{\mathbf{x}\mathbf{x}^{\prime}}+1\big)\;.
\end{aligned}
\label{eq:MFS2+3L}
\end{equation}
From the structure of the above expressions, it is clear that both
of them contain a divergence in the limit $p_{1}^{+}\mathbf{p}_{3}-p_{3}^{+}\mathbf{p}_{1}\to0$,
i.e. when the gluon is radiated collinearly to its parent quark. As
discussed in the previous paragraph, this implies that eqs. (\ref{eq:MFS2+3T})
and (\ref{eq:MFS2+3L}) will contribute to the scenario in which the
jet algorithm groups the quark and gluon inside the same jet.

\subsection{Gluon inside the jet}

Let us first consider the case when the quark and the gluon are grouped
inside the same jet. Integrating over the gluon momentum, the jet
algorithm (\ref{eq:jetalgorithm}) is then imposed by adding the following
step function to the integrand:
\begin{equation}
\begin{aligned}\theta_{\mathrm{in}}(\vec{p}_{1},\vec{p}_{3}) & =\theta\Bigg((\mathbf{p}_{1}+\mathbf{p}_{3})^{2}R^{2}-(p_{1}^{+}+p_{3}^{+})^{2}\Big(\frac{\mathbf{p}_{1}}{p_{1}^{+}}-\frac{\mathbf{p}_{3}}{p_{3}^{+}}\Big)^{2}\Bigg)\;.\end{aligned}
\end{equation}
Moreover, the quark momentum $\vec{p}_{1}$ is not the correct physical
parameter anymore, but rather we should work with the jet momentum
$\vec{p}_{j}\equiv\vec{p}_{1}+\vec{p}_{3}$. With the help of the
intermediate relations
\begin{equation}
\begin{aligned}\frac{1}{\big(\mathbf{p}_{3}-\frac{p_{3}^{+}}{k_{\perp}^{+}}\mathbf{k}_{\perp}\big)^{2}+\Delta_{\mathrm{FS}}} & \overset{\vec{p}_{1}\to\vec{p}_{j}-\vec{p}_{3}}{=}\frac{(p_{j}^{+}-p_{3}^{+})(q^{+}+p_{j}^{+}-p_{3}^{+})}{(p_{j}^{+}+q^{+})p_{j}^{+}}\\
 & \times\frac{1}{\big(\mathbf{p}_{3}-\frac{p_{3}^{+}}{p_{j}^{+}}\mathbf{p}_{j}\big)^{2}+\frac{p_{3}^{+}(p_{j}^{+}-p_{3}^{+})}{(p_{j}^{+})^{2}q^{+}}\big((p_{j}^{+}+q^{+})\mathbf{P}_{j\perp}^{2}+p_{j}^{+}M^{2}\big)}\;,
\end{aligned}
\end{equation}
and
\begin{equation}
\begin{aligned}\big(\mathbf{p}_{3}-\frac{p_{3}^{+}}{p_{1}^{+}}\mathbf{p}_{1}\big)^{2} & \overset{\vec{p}_{1}\to\vec{p}_{j}-\vec{p}_{3}}{=}\Big(\frac{p_{j}^{+}}{p_{j}^{+}-p_{3}^{+}}\Big)^{2}\big(\mathbf{p}_{3}-\frac{p_{3}^{+}}{p_{j}^{+}}\mathbf{p}_{j}\big)^{2}\;,\end{aligned}
\end{equation}
one then obtains, writing $\boldsymbol{\ell}=\mathbf{p}_{3}-\frac{p_{3}^{+}}{p_{j}^{+}}\mathbf{p}_{j}$:
\begin{equation}
\begin{aligned} & \int_{\mathbf{p}_{3}}|\mathcal{M}_{\mathrm{FS2+3}}^{\mathrm{\lambda}}|^{2}\theta_{\mathrm{in}}(\vec{p}_{1},\vec{p}_{3}) \\
& \!\!\overset{\vec{p}_{1}\to\vec{p}_{j}-\vec{p}_{3}}{=}\!\!g_{\mathrm{em}}^{2}g_{s}^{2}C_{F}N_{c}\Big(\frac{p_{3}^{+}}{p_{j}^{+}}\!\Big)^{\!2}\!8(p_{j}^{+}\!-\!p_{3}^{+})(p_{j}^{+}\!+\!q^{+})\!\Big(\!\big(2\frac{p_{j}^{+}}{p_{3}^{+}}\!-\!1\!\big)^{2}\!+\!D\!-\!3\!\Big)\!\Big(\!\big(\!1\!+\!2\frac{p_{j}^{+}}{q^{+}}\!\big)^{2}\!+\!D\!-\!3\!\Big)\\
 & \times\int_{\boldsymbol{\ell}}\frac{1}{\boldsymbol{\ell}^{2}}\Bigg[\frac{(p_{j}^{+}-p_{3}^{+})p_{3}^{+}}{(p_{j}^{+})^{2}}\frac{\mathbf{P}_{j\perp}}{\boldsymbol{\ell}^{2}+\frac{p_{3}^{+}(p_{j}^{+}-p_{3}^{+})}{(p_{j}^{+})^{2}q^{+}}\big((p_{j}^{+}+q^{+})\mathbf{P}_{j\perp}^{2}+p_{j}^{+}M^{2}\big)}+\frac{q^{+}\mathbf{q}}{p_{0}^{+}\mathbf{q}^{2}+p_{j}^{+}M^{2}}\Bigg]^{2}\\
 & \times\theta\Bigg(\frac{(p_{3}^{+})^{2}(p_{j}^{+}-p_{3}^{+})^{2}}{(p_{j}^{+})^{4}}\mathbf{p}_{j}^{2}R^{2}-\boldsymbol{\ell}^{2}\Bigg)\int_{\mathbf{x},\mathbf{x}^{\prime}}e^{-i(\mathbf{q}+\mathbf{p}_{j})\cdot(\mathbf{x}-\mathbf{x}^{\prime})}\big(s_{\mathbf{x}\mathbf{x}^{\prime}}+1\big)\;.
\end{aligned}
\label{eq:MFS2+3T_injet}
\end{equation}
In the above expression, we changed the integration variable from
$\mathbf{p}_{3}$ to $\boldsymbol{\ell}\equiv\mathbf{p}_{3}-\frac{p_{3}^{+}}{p_{j}^{+}}\mathbf{p}_{j}$,
and introduced the analogues of the momentum combinations (\ref{eq:defPkperp}),
now with respect to the jet momentum:
\begin{equation}
\begin{aligned}\mathbf{P}_{j\perp} & \equiv\frac{q^{+}\mathbf{p}_{j}-p_{j}^{+}\mathbf{q}}{q^{+}+p_{j}^{+}}\qquad\mathrm{and}\qquad
\mathbf{k}_{j\perp} & \equiv\mathbf{p}_{j}+\mathbf{q}\;.
\end{aligned}
\label{eq:defPkjet}
\end{equation}
Similarly, in the longitudinally polarized case, we have:
\begin{equation}
\begin{aligned}&\int_{\mathbf{p}_{3}}\big|\mathcal{M}_{\mathrm{FS2+3}}^{0}\big|^{2}  \theta_{\mathrm{in}}(\vec{p}_{1},\vec{p}_{3})\\
&\overset{\vec{p}_{1}\to\vec{p}_{j}-\vec{p}_{3}}{=}\frac{g_{\mathrm{em}}^{2}g_{s}^{2}}{M^{2}}C_{F}N_{c}8(p_{j}^{+}-p_{3}^{+})(p_{j}^{+}+q^{+})\Big(\big(1+\frac{2(p_{j}^{+}-p_{3}^{+})}{p_{3}^{+}}\big)^{2}+D-3\Big)\Big(\frac{p_{3}^{+}}{p_{j}^{+}}\Big)^{2}\\
 & \times\int_{\boldsymbol{\ell}}\frac{1}{\boldsymbol{\ell}^{2}}\Bigg[\frac{p_{3}^{+}}{q^{+}}\frac{(p_{j}^{+}-p_{3}^{+})}{(p_{j}^{+})^{2}}\frac{p_{0}^{+}\mathbf{P}_{j\perp}^{2}-p_{j}^{+}M^{2}}{\boldsymbol{\ell}^{2}+\frac{p_{3}^{+}(p_{j}^{+}-p_{3}^{+})}{(p_{j}^{+})^{2}q^{+}}\big(p_{0}^{+}\mathbf{P}_{j\perp}^{2}+p_{j}^{+}M^{2}\big)}-\frac{p_{0}^{+}\mathbf{q}^{2}-p_{j}^{+}M^{2}}{p_{0}^{+}\mathbf{q}^{2}+p_{j}^{+}M^{2}}\Bigg]^{2}\\
 & \times\theta\Bigg(\frac{(p_{3}^{+})^{2}(p_{j}^{+}-p_{3}^{+})^{2}}{(p_{j}^{+})^{4}}\mathbf{p}_{j}^{2}R^{2}-\boldsymbol{\ell}^{2}\Bigg)\int_{\mathbf{x},\mathbf{x}^{\prime}}e^{-i(\mathbf{q}+\mathbf{p}_{j})\cdot(\mathbf{x}-\mathbf{x}^{\prime})}\big(s_{\mathbf{x}\mathbf{x}^{\prime}}+1\big)\;.
\end{aligned}
\label{eq:MFS2+3L_injet}
\end{equation}
The transverse integrals in eq. (\ref{eq:MFS2+3T_injet}) can be evaluated
with the help of the following identities, which are easily proven
in dimensional regularization:
\begin{equation}
\begin{aligned}\mu^{4-D}\int\frac{\mathrm{d}^{D-2}\boldsymbol{\ell}}{(2\pi)^{D-2}}\frac{\theta(\mathbf{b}-\boldsymbol{\ell})}{\boldsymbol{\ell}^{2}} & =-\frac{1}{4\pi}\Big[\frac{1}{\epsilon_{\mathrm{coll}}}+\ln(\frac{4\pi e^{-\gamma_{\scriptscriptstyle{E}}}\mu^{2}}{b^{2}})\Big]+\mathcal{O}(b,\epsilon_{\mathrm{coll}})\;,\\
\mu^{4-D}\int\frac{\mathrm{d}^{D-2}\boldsymbol{\ell}}{(2\pi)^{D-2}}\frac{\theta(\mathbf{b}-\boldsymbol{\ell})}{\boldsymbol{\ell}^{2}(\boldsymbol{\ell}^{2}+m)} & =-\frac{1}{4\pi m}\Big[\frac{1}{\epsilon_{\mathrm{coll}}}+\ln(\frac{4\pi e^{-\gamma_{\scriptscriptstyle{E}}}\mu^{2}}{b^{2}})\Big]+\mathcal{O}(b,\epsilon_{\mathrm{coll}})\;,\\
\mu^{4-D}\int\frac{\mathrm{d}^{D-2}\boldsymbol{\ell}}{(2\pi)^{D-2}}\frac{\theta(\mathbf{b}-\boldsymbol{\ell})}{\boldsymbol{\ell}^{2}(\boldsymbol{\ell}^{2}+m)^{2}} & =-\frac{1}{4\pi m^{2}}\Big[\frac{1}{\epsilon_{\mathrm{coll}}}+\ln(\frac{4\pi e^{-\gamma_{\scriptscriptstyle{E}}}\mu^{2}}{b^{2}})\Big]+\mathcal{O}(b,\epsilon_{\mathrm{coll}})\;,
\end{aligned}
\label{eq:injetintegrals}
\end{equation}
where we should note remind the reader that we work in the narrow-cone
approximation $R\to0$, keeping only negative power-like or logarithmic
dependencies on $R$. We also expand $D-3=1-2\epsilon_{\mathrm{coll}}$
from the Dirac structure $\big(2\frac{p_{j}^{+}}{p_{3}^{+}}-1\big)^{2}+D-3$.
In both the transverse (\ref{eq:MFS2+3T_injet}) and longitudinally
(\ref{eq:MFS2+3L_injet}) polarized case, we obtain a result proportional
to the corresponding leading-order amplitude: 
\begin{equation}
\begin{aligned} & \int_{\mathbf{p}_{3}}|\mathcal{M}_{\mathrm{FS2+3}}^{0,\lambda}|^{2}\theta_{\mathrm{in}}(\vec{p}_{1},\vec{p}_{3})\\
 &=\big|\mathcal{M}_{\mathrm{LO}}^{0,\lambda}\big|^{2}\times-\alpha_{s}C_{F}\Bigg[\frac{1}{\epsilon_{\mathrm{coll}}}+\ln\big(\frac{4\pi e^{-\gamma_{\scriptscriptstyle{E}}}\mu^{2}(p_{j}^{+})^{4}}{(p_{3}^{+})^{2}(p_{j}^{+}-p_{3}^{+})^{2}\mathbf{p}_{j}^{2}R^{2}}\big)-\frac{(p_{3}^{+})^{2}}{(p_{j}^{+})^{2}+(p_{j}^{+}-p_{3}^{+})^{2}}\Bigg]\\
 & \times\Big(\frac{p_{3}^{+}}{p_{j}^{+}}\Big)^{2}\frac{(p_{j}^{+}-p_{3}^{+})}{p_{j}^{+}}\Big(\big(2\frac{p_{j}^{+}}{p_{3}^{+}}-1\big)^{2}+1\Big)+\mathcal{O}(\epsilon_{\mathrm{coll}})\;.
\end{aligned}
\end{equation}
Note that, in the transverse case, the structure $\big(1\!+\!2\frac{p_{j}^{+}}{q^{+}}\big)^{2}\!+\!D\!-\!3$
is absorbed into $\big|\mathcal{M}_{\mathrm{LO1+2}}^{\mathrm{T}}\big|^{2}$,
which hence stays $D$-dimensional.

The next step is to integrate over the gluon plus-momentum. However,
due to the shift $\vec{p}_{1}\to\vec{p}_{j}-\vec{p}_{3}$ dictated
by the jet algorithm, an additional dependence on $p_{3}^{+}$ is
introduced on the level of the cross section:
\begin{equation}
\begin{aligned}\mathrm{d}\sigma_{\mathrm{in}}^{\mathrm{T},\mathrm{L}} & =\int\frac{\mathrm{d}p_{3}^{+}\mathrm{\mathrm{d}}^{D-2}\mathbf{p}_{3}}{(2\pi)^{D-1}2p_{3}^{+}}\frac{1}{2p_{0\scriptscriptstyle{R}}^{+}}\frac{\mathrm{d}p_{1}^{+}\mathrm{\mathrm{d}}^{D-2}\mathbf{p}_{1}}{(2\pi)^{D-1}2p_{1}^{+}}\frac{\mathrm{d}q^{+}\mathrm{\mathrm{d}}^{2}\mathbf{q}}{(2\pi)^{D-1}2q^{+}}\\
 & \times2\pi\delta(p_{0\scriptscriptstyle{R}}^{+}-p_{1}^{+}-q^{+}-p_{3}^{+})\frac{1}{D-2}|\mathcal{M}_{\mathrm{FS2+3}}^{0,\lambda}|^{2}\theta_{\mathrm{in}}(\vec{p}_{1},\vec{p}_{3})\;,\\
 & \overset{\vec{p}_{1}\to\vec{p}_{j}-\vec{p}_{3}}{=}\int\frac{\mathrm{d}p_{3}^{+}\mathrm{\mathrm{d}}^{D-2}\mathbf{p}_{3}}{(2\pi)^{D-1}2p_{3}^{+}}\frac{1}{2p_{0}^{+}}\frac{\mathrm{d}p_{j}^{+}\mathrm{\mathrm{d}}^{D-2}\mathbf{p}_{j}}{(2\pi)^{D-1}2(p_{j}^{+}-p_{3}^{+})}\frac{\mathrm{d}q^{+}\mathrm{\mathrm{d}}^{2}\mathbf{q}}{(2\pi)^{D-1}2q^{+}}\\
 & \times2\pi\delta(p_{0}^{+}-p_{j}^{+}-q^{+})\frac{1}{D-2}|\mathcal{M}_{\mathrm{FS2+3}}^{0,\lambda}|^{2}\theta_{\mathrm{in}}(\vec{p}_{j}-\vec{p}_{3},\vec{p}_{3})\;,\\
 & =\frac{1}{2p_{0}^{+}}\frac{\mathrm{d}p_{j}^{+}\mathrm{\mathrm{d}}^{D-2}\mathbf{p}_{j}}{(2\pi)^{D-1}2p_{j}^{+}}\frac{\mathrm{d}q^{+}\mathrm{\mathrm{d}}^{2}\mathbf{q}}{(2\pi)^{D-1}2q^{+}}2\pi\delta(p_{0}^{+}-p_{j}^{+}-q^{+})\frac{1}{D-2}\\
 & \times\int\frac{\mathrm{d}p_{3}^{+}}{(2\pi)^{D-3}2p_{3}^{+}}\frac{p_{j}^{+}}{(p_{j}^{+}-p_{3}^{+})}\int_{\mathbf{p}_{3}}|\mathcal{M}_{\mathrm{FS2+3}}^{0,\lambda}|^{2}\theta_{\mathrm{in}}(\vec{p}_{j}-\vec{p}_{3},\vec{p}_{3})\;.
\end{aligned}
\end{equation}
Evaluating the plus-momentum integral, we finally obtain:
\begin{equation}
\begin{aligned}\mathrm{d}\sigma_{\mathrm{in}}^{\mathrm{T,L}} & \!=\!\mathrm{d}\sigma_{\mathrm{LO}}^{\mathrm{T},\mathrm{L}}\,\frac{\alpha_{s}C_{F}}{\pi}\!\Bigg[\!\Big(\frac{1}{\epsilon_{\mathrm{coll}}}\!+\!\ln\big(\frac{4\pi e^{-\gamma_{\scriptscriptstyle{E}}}\mu^{2}}{\mathbf{p}_{j}^{2}R^{2}}\big)\!\Big)\!\Big(\frac{3}{4}\!-\!\ln\frac{p_{j}^{+}}{k_{\mathrm{min}}^{+}}\Big)\!+\!\frac{13}{4}\!-\!\frac{\pi^{2}}{3}\!-\!\ln^{2}\!\frac{p_{j}^{+}}{k_{\mathrm{min}}^{+}}\Bigg]\;,\end{aligned}
\label{eq:FS2+3_in}
\end{equation}
where $\mathrm{d}\sigma_{\mathrm{LO}}^{\mathrm{T},\mathrm{L}}$ is
the leading-order amplitude in terms of the jet instead of the quark.
Adding the above result to the \textquoteleft final-state' part of
the virtual field-strength renormalization corrections (\ref{eq:Z}):
\begin{equation}
\begin{aligned}\mathrm{d}\sigma_{\mathrm{\mathcal{Z}}_{\mathrm{FS}}} & =\mathrm{d}\sigma_{\mathrm{LO}}\frac{\alpha_{s}C_{F}}{\pi}\Big(\frac{1}{\epsilon_{\mathrm{coll}}}+\mathrm{ln}\frac{\mu^{2}}{\mu_{\scriptscriptstyle{R}}^{2}}\Big)\Big(-\frac{3}{4}+\mathrm{ln}\frac{p_{j}^{+}}{k_{\mathrm{min}}^{+}}\Big)\;,\end{aligned}
\end{equation}
where we can identify the quark momentum with the jet one, we obtain:
\begin{equation}
\begin{aligned}\mathrm{d}\sigma_{\mathrm{in}}\!+\!\mathrm{d}\sigma_{\mathrm{\mathcal{Z}}_{\mathrm{FS}}} & \!=\!\mathrm{d}\sigma_{\mathrm{LO}}\!\times\!\frac{\alpha_{s}C_{F}}{\pi}\Bigg[\ln\big(\frac{4\pi e^{-\gamma_{\scriptscriptstyle{E}}}\mu_{\scriptscriptstyle{R}}^{2}}{\mathbf{p}_{j}^{2}R^{2}}\big)\Big(\frac{3}{4}-\ln\!\frac{p_{j}^{+}}{k_{\mathrm{min}}^{+}}\Big)\!+\!\frac{13}{4}\!-\!\frac{\pi^{2}}{3}\!-\!\ln^{2}\!\frac{p_{j}^{+}}{k_{\mathrm{min}}^{+}}\!\Bigg]\;.\end{aligned}
\label{eq:jet+ZFS}
\end{equation}
We omitted the labels indicating the photon polarization since the
above result does not depend on it. The above expression is an important
result, demonstrating the cancellation of final-state collinear divergencies
in our cross section. There is, however, one loose end in the form
of a double logarithm in the rapidity cutoff $k_{\mathrm{min}}^{+}$.
Since high-energy resummation only involves single large logarithms,
this term is unphysical and needs to cancel in the final cross section.
In the next subsection, we will show that the unphysical double logarithm
cancels when adding the contribution of a soft gluon just outside
the jet.

\subsection{Gluon outside the jet}

When the gluon and its parent quark each form a distinct jet, the
quark momentum can be directly identified with the one of the jet:
$\vec{p}_{j}=\vec{p}_{1}$. The jet function becomes, introducing
$\boldsymbol{\ell}\equiv\mathbf{p}_{3}-\frac{p_{3}^{+}}{p_{j}^{+}}\mathbf{p}_{j}$:
\begin{equation}
\begin{aligned}1-\theta_{\mathrm{in}}(\vec{p}_{1}\to\vec{p}_{j},\vec{p}_{3}) & =\theta\Big((p_{j}^{+}+p_{3}^{+})^{2}\Big(\frac{\mathbf{p}_{j}}{p_{j}^{+}}-\frac{\mathbf{p}_{3}}{p_{3}^{+}}\Big)^{2}-(\mathbf{p}_{j}+\mathbf{p}_{3})^{2}R^{2}\Big)\;,\\
 & =\theta\Big(\boldsymbol{\ell}^{2}-\Big(\frac{p_{3}^{+}}{p_{j}^{+}+p_{3}^{+}}\Big)^{2}\big(\boldsymbol{\ell}+\frac{p_{3}^{+}+p_{j}^{+}}{p_{j}^{+}}\mathbf{p}_{j}\big)^{2}R^{2}\Big)\;,\\
 & \approx\theta\Big(\boldsymbol{\ell}^{2}-\Big(\frac{p_{3}^{+}}{p_{j}^{+}}\Big)^{2}\mathbf{p}_{j}^{2}R^{2}\Big)\;,
\end{aligned}
\end{equation}
where the last equality holds in the narrow-jet limit $R\to0$. Integrating
over the gluon momentum in (\ref{eq:MFS2+3T}) then gives:
\begin{equation}
\begin{aligned} & \int\mathrm{PS}(\vec{p}_{3})|\mathcal{M}_{\mathrm{FS2+3}}^{\lambda\eta}|^{2}\big(1-\mathrm{\theta}_{\mathrm{in}}(\vec{p}_{1},\vec{p}_{3})\big)\\
 & =g_{\mathrm{em}}^{2}\alpha_{s}C_{F}N_{c}\int\frac{\mathrm{d}p_{3}^{+}}{p_{3}^{+}}\int_{\boldsymbol{\ell}}\frac{1}{\boldsymbol{\ell}^{2}}\Bigg[\frac{p_{3}^{+}}{p_{0}^{+}}\frac{q^{+}}{p_{j}^{+}+p_{3}^{+}}\frac{\boldsymbol{\ell}+\frac{p_{0}^{+}}{p_{j}^{+}}\frac{p_{j}^{+}+p_{3}^{+}}{q^{+}}\mathbf{P}_{\perp}}{\big(\boldsymbol{\ell}+\frac{p_{3}^{+}}{p_{j}^{+}}\mathbf{P}_{\perp}\big)^{2}+\Delta_{\mathrm{FS}}}+\frac{q^{+}\mathbf{q}}{p_{0\scriptscriptstyle{R}}^{+}\mathbf{q}^{2}+(p_{j}^{+}+p_{3}^{+})M^{2}}\Bigg]^{2}\\
 & \times\theta\Big(\boldsymbol{\ell}^{2}-\Big(\frac{p_{3}^{+}}{p_{j}^{+}}\Big)^{2}\mathbf{p}_{j}^{2}R^{2}\Big)\\
 & \times\Big(\frac{p_{3}^{+}}{p_{j}^{+}}\Big)^{2}8p_{j}^{+}p_{0\scriptscriptstyle{R}}^{+}\Big(\big(1+\frac{2p_{j}^{+}}{p_{3}^{+}}\big)^{2}+1\Big)\Big(\big(1+2\frac{p_{j}^{+}+p_{3}^{+}}{q^{+}}\big)^{2}+1\Big)\\
 & \times\int_{\mathbf{x},\mathbf{x}^{\prime}}e^{-i(\boldsymbol{\ell}+\frac{p_{3}^{+}}{p_{j}^{+}}\mathbf{P}_{\perp}+\frac{p_{0\scriptscriptstyle{R}}^{+}}{p_{0}^{+}}\mathbf{k}_{\perp})\cdot(\mathbf{x}-\mathbf{x}^{\prime})}\big(s_{\mathbf{x}\mathbf{x}^{\prime}}+1\big)\;.
\end{aligned}
\label{eq:MFS2+3_out}
\end{equation}
From experience \cite{Taels:2022tza}, we expect that a similar unphysical
double logarithm as the one in eq. (\ref{eq:jet+ZFS}) will appear
from the phase-space integration over the gluon, in the kinematical
region where the latter is just outside the jet with small plus-momentum,
i.e. when simultaneously:
\begin{equation}
\begin{aligned}\boldsymbol{\ell}=\mathbf{p}_{3}-\frac{p_{3}^{+}}{p_{j}^{+}}\mathbf{p}_{j} & \to0\quad\mathrm{and}\quad p_{3}^{+}\to0\;.\end{aligned}
\end{equation}
In this regime, eq. (\ref{eq:MFS2+3_out}) can be approximated by:
\begin{equation}
\begin{aligned} & \lim_{\mathrm{soft}}\int\mathrm{PS}(\vec{p}_{3})|\mathcal{M}_{\mathrm{FS2+3}}|^{2}\big(1-\mathrm{\theta}_{\mathrm{in}}(\vec{p}_{1},\vec{p}_{3})\big)\\
 & =|\mathcal{M}_{\mathrm{LO}}|^{2}4\alpha_{s}C_{F}\int_{k_{\mathrm{min}}^{+}}^{+\infty}\frac{\mathrm{d}p_{3}^{+}}{p_{3}^{+}}\int_{\boldsymbol{\ell}}\frac{e^{-i\boldsymbol{\ell}\cdot(\mathbf{x}-\mathbf{x}^{\prime})}}{\boldsymbol{\ell}^{2}}\theta\Big(\boldsymbol{\ell}^{2}-\Big(\frac{p_{3}^{+}}{p_{j}^{+}}\Big)^{2}\mathbf{p}_{j}^{2}R^{2}\Big)\;,\\
 & =|\mathcal{M}_{\mathrm{LO}}|^{2}\frac{\alpha_{s}C_{F}}{\pi}\Big(2\ln\frac{p_{j}^{+}}{k_{\mathrm{min}}^{+}}\ln\frac{c_{0}}{R|\mathbf{p}_{j}||\mathbf{x}-\mathbf{x}^{\prime}|}+\ln^{2}\frac{p_{j}^{+}}{k_{\mathrm{min}}^{+}}\Big)\;,
\end{aligned}
\label{eq:outsoftalmost}
\end{equation}
where we used the identity (with $c_{0}=2e^{-\gamma_{\scriptscriptstyle{E}}}$):
\begin{equation}
\begin{aligned}\int_{\boldsymbol{\ell}}\frac{e^{-i\boldsymbol{\ell}\cdot\mathbf{x}}}{\boldsymbol{\ell}^{2}}\theta(\boldsymbol{\ell}^{2}-\mathbf{b}^{2}) & =\frac{1}{2\pi}\int_{b}^{\infty}\frac{\mathrm{d}\ell}{\ell}J_{0}\big(\ell|\mathbf{x}|\big)=\frac{1}{2\pi}\ln\frac{c_{0}}{b|\mathbf{x}|}\;.\end{aligned}
\end{equation}
and where we suppressed the polarization labels since the same relation
holds in the longitudinal case. Promoting the result (\ref{eq:outsoftalmost})
to the cross-section level, and adding it to eq. (\ref{eq:jet+ZFS}),
we end up with:
\begin{equation}
\begin{aligned}\mathrm{d}\sigma_{\mathrm{jet}} & \equiv\mathrm{d}\sigma_{\mathrm{in}}+\mathrm{d}\sigma_{\mathrm{\mathcal{Z}}_{\mathrm{FS}}}+\mathrm{d}\sigma_{\mathrm{out,soft}}\\
 & =\mathrm{d}\sigma_{\mathrm{LO}}\times\frac{\alpha_{s}C_{F}}{\pi}\Bigg[\frac{3}{4}\ln\big(\frac{4\pi e^{-\gamma_{\scriptscriptstyle{E}}}\mu_{\scriptscriptstyle{R}}^{2}}{\mathbf{p}_{j}^{2}R^{2}}\big)+\frac{13}{4}-\frac{\pi^{2}}{3}+\ln\frac{p_{j}^{+}}{k_{\mathrm{min}}^{+}}\ln\frac{c_{0}}{2\pi\mu_{\scriptscriptstyle{R}}^{2}(\mathbf{x}-\mathbf{x}^{\prime})^{2}}\Bigg]\;,
\end{aligned}
\label{eq:jet_final}
\end{equation}
in which the pathological double logarithm has cancelled.

\section{\label{sec:JIMWLK}High-energy resummation}

In sections \ref{sec:UV}, \ref{sec:DGLAP}, and \ref{sec:jet}, we have demonstrated that the different ultraviolet- and collinear divergences
encountered in our calculation all cancel or, in the case of initial-state
collinear poles, can be absorbed into the DGLAP evolution of the incoming
quark. We are now in a position to treat the remaining high-energy
or rapidity divergences, which either stem from the $k_{3}^{+}\rightarrow0$
limit in the gluon loop for virtual diagrams, either from the $p_{3}^{+}\rightarrow0$
region in the gluon phase-space integration of real NLO contributions
in the cross section. In both cases, they are regularized with a cutoff
$k_{\mathrm{min}}^{+}$ and come in the form of single large logarithms.
To be more precise, let us formally separate the integral over gluon
plus-momentum in the (fixed-order) NLO cross section as follows: 
\begin{equation}
\begin{aligned}\mathrm{d}\sigma_{\mathrm{NLO}} & =\int_{k_{\mathrm{min}}^{+}}^{+\infty}\frac{\mathrm{d}p_{3}^{+}}{p_{3}^{+}}\mathrm{d}\tilde{\sigma}_{\mathrm{NLO}}\;.\end{aligned}
\end{equation}
Introducing the high-energy factorization scale $k_{f}^{+}$, we can rewrite the equation above:
\begin{equation}
\begin{aligned}\mathrm{d}\sigma_{\mathrm{NLO}} & \!=\!\!\int_{k_{\mathrm{min}}^{+}}^{k_{f}^{+}}\!\frac{\mathrm{d}p_{3}^{+}}{p_{3}^{+}}\hat{H}_{\mathrm{JIMWLK}}\mathrm{d}\sigma_{\mathrm{LO}}\!+\!\!\int_{0}^{+\infty}\!\frac{\mathrm{d}p_{3}^{+}}{p_{3}^{+}}\Big(\mathrm{d}\tilde{\sigma}_{\mathrm{NLO}}\!-\!\theta(k_{f}^{+}\!-\!p_{3}^{+})\hat{H}_{\mathrm{JIMWLK}}\mathrm{d}\sigma_{\mathrm{LO}}\Big)\end{aligned}
\label{eq:JIMWLKsubtraction}
\end{equation}
where $\hat{H}_{\mathrm{JIMWLK}}$ is the JIMWLK Hamiltonian, acting
on the target averages in the leading-order cross section. We will
show in the following subsections that the above subtraction method
works, namely that all the rapidity-divergent contributions in our
NLO cross section can be combined into the first term of the above
equation, such that the second term is a completely finite rapidity-subtracted
cross section. Explicitly, the action of the JIMWLK Hamiltonian on
the dipole operator is:
\begin{equation}
\begin{aligned}\hat{H}_{\mathrm{JIMWLK}}\big\langle s_{\mathbf{x}\mathbf{x}^{\prime}}+1\big\rangle & =-\frac{\alpha_{s}N_{c}}{2\pi^{2}}\int_{\mathbf{z}}\frac{(\mathbf{x}-\mathbf{x}^{\prime})^{2}}{(\mathbf{x}-\mathbf{z})^{2}(\mathbf{x}^{\prime}-\mathbf{z})^{2}}\big\langle s_{\mathbf{x^{\prime}}\mathbf{x}}-s_{\mathbf{x}^{\prime}\mathbf{z}}s_{\mathbf{z}\mathbf{x}}\big\rangle\;.\end{aligned}
\label{eq:dipoleJIMWLK}
\end{equation}

Moreover, at fixed order, the target average of the Wilson-line structure
in the leading-order cross section (\ref{eq:crossLO}) does not contain
any evolution, hence we label it with a zero:
\begin{equation}
\big\langle s_{\mathbf{x}\mathbf{x}^{\prime}}+1\big\rangle_{0}\;.
\end{equation}
However, with JIMWLK we can resum the high-energy leading logarithms
$Y_{f}^{+}\equiv\ln(k_{f}^{+}/k_{\mathrm{min}}^{+})$ as follows:
\begin{equation}
\begin{aligned}\partial_{Y_{f}^{+}}\big\langle s_{\mathbf{x}\mathbf{x}^{\prime}}+1\big\rangle_{Y_{f}^{+}} & =\Big\langle\hat{H}_{\mathrm{JIMWLK}}\big(s_{\mathbf{x}\mathbf{x}^{\prime}}+1\big)\Big\rangle_{Y_{f}^{+}}\;.\end{aligned}
\label{eq:JIMWLK_eq}
\end{equation}
Integrating the above equation, we obtain:
\begin{equation}
\begin{aligned}\big\langle s_{\mathbf{x}\mathbf{x}^{\prime}}+1\big\rangle_{Y_{f}^{+}} & =\big\langle s_{\mathbf{x}\mathbf{x}^{\prime}}+1\big\rangle_{0}+\int_{0}^{Y_{f}^{+}}\mathrm{d}Y^{+}\Big\langle\hat{H}_{\mathrm{JIMWLK}}\big(s_{\mathbf{x}\mathbf{x}^{\prime}}+1\big)\Big\rangle_{Y^{+}}\;,\\
 & =\big\langle s_{\mathbf{x}\mathbf{x}^{\prime}}+1\big\rangle_{0}+Y_{f}^{+}\Big\langle\hat{H}_{\mathrm{JIMWLK}}\big(s_{\mathbf{x}\mathbf{x}^{\prime}}+1\big)\Big\rangle+{\cal O}(\alpha_{s}^{2})\;.
\end{aligned}
\label{eq:JIMWLK_eq_int}
\end{equation}
In the last line, we have performed a fixed-order expansion, where
we should note that the JIMWLK Hamiltonian is of order $\alpha_{s}$
and that the dependence of the target average on the rapidity scale
is not specified since it is a higher-order effect. We can, therefore,
write the analogue of eq. (\ref{eq:LO+DGLAP}):
\begin{equation}
\begin{aligned}\mathrm{d}\sigma_{\mathrm{LO}} & =\mathrm{d}\sigma_{\mathrm{LO+JIMWLK}}-\ln\frac{k_{f}^{+}}{k_{\mathrm{min}}^{+}}\hat{H}_{\mathrm{JIMWLK}}\mathrm{d}\sigma_{\mathrm{LO}}\;.\end{aligned}
\label{eq:LO_JIMWLK}
\end{equation}
Combining eqs. (\ref{eq:LO_JIMWLK}) and (\ref{eq:JIMWLKsubtraction}),
one then obtains the final result:
\begin{equation}
\begin{aligned}\mathrm{d}\sigma_{\mathrm{LO}}\!+\!\mathrm{d}\sigma_{\mathrm{NLO}} & \!=\!\mathrm{d}\sigma_{\mathrm{LO+JIMWLK}}\!+\!\int_{0}^{+\infty}\frac{\mathrm{d}p_{3}^{+}}{p_{3}^{+}}\Big(\mathrm{d}\tilde{\sigma}_{\mathrm{NLO}}\!-\!\theta(k_{f}^{+}\!-\!p_{3}^{+})\hat{H}_{\mathrm{JIMWLK}}\mathrm{d}\sigma_{\mathrm{LO}}\Big)\;.\end{aligned}
\label{eq:LO+NLO_finite}
\end{equation}

\subsection{\label{subsec:jimwlk_leftover}Leftovers from UV and collinear subtractions}

In section \ref{sec:UV}, we have demonstrated how the sum of the
different UV counterterms in the virtual amplitudes yields a result
which is free from any UV poles. There are, however, still rapidity
divergences left, regularized by the cutoff $k_{\mathrm{min}}^{+}$.
On the level of the cross section, we have for the longitudinally
polarized photon production (\ref{eq:sigmaL_UV}):
\begin{equation}
\begin{aligned}\mathrm{d}\sigma_{\mathrm{UV}}^{\mathrm{L}} & =\mathrm{d}\sigma_{\mathrm{LO}}^{\mathrm{L}}\times\frac{\alpha_{s}C_{F}}{\pi}\Bigg[\Big(-\frac{3}{2}+\ln\frac{p_{1}^{+}}{k_{\mathrm{min}}^{+}}+\ln\frac{p_{0}^{+}}{k_{\mathrm{min}}^{+}}\Big)\ln\frac{4\pi e^{-\gamma_{\scriptscriptstyle{E}}}\mu_{\scriptscriptstyle{R}}^{2}}{\Delta_{\mathrm{UV}}}-\frac{1}{2}\Bigg]\;.\end{aligned}
\end{equation}
Following the usual high-energy resummation procedure, we will extract
from above cross section the terms enhanced by a large rapidity or
high-energy logarithm, and cut the (implicit) integration over the
gluon plus-momentum off at the rapidity factorization scale $k_{f}^{+}$:
\begin{equation}
\lim_{k^{+}\to0}\begin{aligned}\mathrm{d}\sigma_{\mathrm{UV}}^{\mathrm{L}} & =\mathrm{d}\sigma_{\mathrm{LO}}^{\mathrm{L}}\times\ln\frac{k_{f}^{+}}{k_{\mathrm{min}}^{+}}\times\frac{2\alpha_{s}C_{F}}{\pi}\ln\frac{4\pi e^{-\gamma_{\scriptscriptstyle{E}}}\mu_{\scriptscriptstyle{R}}^{2}}{\Delta_{\mathrm{UV}}}\;.\end{aligned}
\label{eq:UV_L_JIMWLK}
\end{equation}
We obtain the same result when the photon is transversally polarized
(\ref{eq:sigmaT_UV}):
\begin{equation}
\begin{aligned}\lim_{k^{+}\to0}\mathrm{d}\sigma_{\mathrm{UV}}^{\mathrm{T}} & =\mathrm{d}\sigma_{\mathrm{LO}}^{\mathrm{T}}\times\ln\frac{k_{f}^{+}}{k_{\mathrm{min}}^{+}}\times\frac{2\alpha_{s}C_{F}}{\pi}\ln\frac{4\pi e^{-\gamma_{\scriptscriptstyle{E}}}\mu_{\scriptscriptstyle{R}}^{2}}{\Delta_{\mathrm{UV}}}\;.\end{aligned}
\label{eq:UV_T_JIMWLK}
\end{equation}
Let us now turn to the real NLO corrections that contributed to the
DGLAP evolution of the incoming quark (\ref{eq:DGLAP_final}). Keeping
only the high-energy logarithms, we find:
\begin{equation}
\begin{aligned}\lim_{k^{+}\to0}\mathrm{d}\sigma_{\mathrm{IS}} & =\mathrm{d}\sigma_{\mathrm{LO}}\times\mathrm{ln}\frac{k_{f}^{+}}{k_{\mathrm{min}}^{+}}\times\frac{\alpha_{s}C_{F}}{\pi}\mathrm{ln}\frac{e^{-\gamma_{\scriptscriptstyle{E}}}}{\mu_{\scriptscriptstyle{R}}^{2}\pi(\mathbf{x}-\mathbf{x}^{\prime})^{2}}\;.\end{aligned}
\label{eq:DGLAP_JIMWLK}
\end{equation}
irrespective of the polarization of the photon.

Finally, there are leftover rapidity logarithms after the cancellation
of collinear divergences in the final state by the jet algorithm (\ref{eq:jet_final}),
namely:
\begin{equation}
\begin{aligned}\lim_{k^{+}\to0}\mathrm{d}\sigma_{\mathrm{jet}} & =\mathrm{d}\sigma_{\mathrm{LO}}\times\mathrm{ln}\frac{k_{f}^{+}}{k_{\mathrm{min}}^{+}}\times\frac{\alpha_{s}C_{F}}{\pi}\ln\frac{c_{0}}{2\pi\mu_{\scriptscriptstyle{R}}^{2}(\mathbf{x}-\mathbf{x}^{\prime})^{2}}\;.\end{aligned}
\label{eq:jet_JIMWLK}
\end{equation}
Adding (\ref{eq:UV_L_JIMWLK}) or (\ref{eq:UV_T_JIMWLK}) to (\ref{eq:DGLAP_JIMWLK})
and (\ref{eq:jet_JIMWLK}), we obtain both for transverse and longitudinally
polarized photons:
\begin{equation}
\begin{aligned}\lim_{k^{+}\to0}\Big(\mathrm{d}\sigma_{\mathrm{UV}}+\mathrm{d}\sigma_{\mathrm{IS}}+\mathrm{d}\sigma_{\mathrm{jet}}\Big) & =\mathrm{d}\sigma_{\mathrm{LO}}\times\ln\frac{k_{f}^{+}}{k_{\mathrm{min}}^{+}}\times\frac{2\alpha_{s}C_{F}}{\pi}\ln\frac{c_{0}^{2}}{\Delta_{\mathrm{UV}}(\mathbf{x}-\mathbf{x}^{\prime})^{2}}\;.\end{aligned}
\label{eq:UVDGLAPJET_JIMWLK}
\end{equation}
It is worth remarking that, in the above result, the arbitrary scale
$\mu_{\scriptscriptstyle{R}}$ stemming from the quark field-strength renormalization
has cancelled. This is not surprising, since eq. (\ref{eq:UVDGLAPJET_JIMWLK})
contains all contributions ($\mathrm{UV}$, $\mathrm{IS}$, $\mathrm{FS}$
and their conjugates) from the one-loop result for $\mathcal{Z}$
(\ref{eq:Z}), which is independent from $\mu_{\scriptscriptstyle{R}}$. 

\subsection{Virtual contributions}

We will show in the following section (see eq. (\ref{eq:SE_V_no_rap}))
that, although they are individually divergent in the limit of vanishing
gluon plus-momentum, the sums of the UV-subtracted amplitudes $\tilde{\mathcal{M}}_{\mathrm{SE1,sub}}^{0,\lambda}+\tilde{\mathcal{M}}_{\mathrm{V1,sub}}^{0,\lambda}$
and $\tilde{\mathcal{M}}_{\mathrm{SE4,sub}}^{0,\lambda}+\tilde{\mathcal{M}}_{\mathrm{V4,sub}}^{0,\lambda}$
do not contribute any rapidity logarithms to the cross section, nor
is there any dependence on the factorization scale $k_{f}^{+}$. Moreover,
the high-energy logarithms stemming from the field-strength renormalization
diagrams (\ref{eq:Z}) were already taken into account in the previous
subsection. It is, then, very easy to see that the only virtual diagrams
left with a high-energy logarithm are the subtracted diagrams $\tilde{\mathcal{M}}_{\mathrm{SE2,sub}}^{0,\lambda}$
(\ref{eq:MSE2L_sub}) , (\ref{eq:MSE2T_sub}) and $\tilde{\mathcal{M}}_{\mathrm{SE3,sub}}^{0,\lambda}$
(\ref{eq:MSE3L_sub}), (\ref{eq:MSE3T_sub}). Indeed, using that in
the limit $k^{+}\to0$ the Dirac structure (\ref{eq:DiracSE}) becomes:
\begin{equation}
\begin{aligned}\lim_{k^{+}\to0}\mathrm{Dirac}_{\mathrm{SE}}^{\bar{\eta}\eta^{\prime}}(p_{0}^{+}) & =-4\Big(\frac{p_{0}^{+}}{k^{+}}\Big)^{2}\delta^{\bar{\eta}\eta^{\prime}}\;,\end{aligned}
\label{eq:JIMWLK_DiracSE-1}
\end{equation}
we obtain
\begin{equation}
\begin{aligned}\lim_{k^{+}\to0}\tilde{\mathcal{M}}_{\mathrm{SE2,sub}}^{0} & =4\alpha_{s}\frac{1}{M}\frac{p_{0}^{+}\mathbf{P}_{\perp}^{2}-p_{1}^{+}M^{2}}{p_{0}^{+}\mathbf{P}_{\perp}^{2}+p_{1}^{+}M^{2}}\int_{k_{\mathrm{min}}^{+}}^{p_{0}^{+}}\frac{\mathrm{d}k^{+}}{k^{+}}\\
 & \times\int_{\mathbf{x}}e^{-i\mathbf{k}_{\perp}\cdot\mathbf{x}}\Bigg[\int_{\mathbf{z}}A^{i}(\mathbf{x}-\mathbf{z})A^{i}(\mathbf{x}-\mathbf{z})\big(t^{c}U_{\mathbf{x}}U_{\mathbf{z}}^{\dagger}t^{c}U_{\mathbf{z}}-C_{F}\big)\\
 & -\mathcal{A}_{0}(\Delta_{\mathrm{UV}})C_{F}\big(U_{\mathbf{x}}-1\big)\Bigg]\;,
\end{aligned}
\end{equation}
and:
\begin{equation}
\begin{aligned}\lim_{k^{+}\to0}\tilde{\mathcal{M}}_{\mathrm{SE2,sub}}^{\lambda} & =-4\alpha_{s}\frac{q^{+}\mathbf{P}_{\perp}^{\bar{\lambda}}}{p_{0}^{+}\mathbf{P}_{\perp}^{2}+p_{1}^{+}M^{2}}\mathrm{Dirac}_{\mathrm{LO}}^{\lambda\bar{\lambda}}\int_{k_{\mathrm{min}}^{+}}^{p_{0}^{+}}\frac{\mathrm{d}k^{+}}{k^{+}}\\
 & \times\int_{\mathbf{x}}e^{-i\mathbf{k}_{\perp}\cdot\mathbf{x}}\Bigg[\int_{\mathbf{z}}A^{i}(\mathbf{x}-\mathbf{z})A^{i}(\mathbf{x}-\mathbf{z})\big(t^{c}U_{\mathbf{x}}U_{\mathbf{z}}^{\dagger}t^{c}U_{\mathbf{z}}-C_{F}\big)\\
 & -\mathcal{A}_{0}(\Delta_{\mathrm{UV}})C_{F}\big(U_{\mathbf{x}}-1\big)\Bigg]\;.
\end{aligned}
\end{equation}
as well as:
\begin{equation}
\begin{aligned}\lim_{k^{+}\to0}\tilde{\mathcal{M}}_{\mathrm{SE3,sub}}^{0} & =-4\alpha_{s}\frac{p_{0}^{+}\mathbf{q}^{2}-p_{1}^{+}M^{2}}{p_{0}^{+}\mathbf{q}^{2}+p_{1}^{+}M^{2}}\frac{1}{M}\int_{k_{\mathrm{min}}^{+}}^{p_{1}^{+}}\frac{\mathrm{d}k^{+}}{k^{+}}\\
 & \times\int_{\mathbf{x}}e^{-i\mathbf{k}_{\perp}\cdot\mathbf{x}}\Bigg[\int_{\mathbf{z}}A^{i}(\mathbf{x}-\mathbf{z})A^{i}(\mathbf{x}-\mathbf{z})\big(t^{c}U_{\mathbf{x}}U_{\mathbf{z}}^{\dagger}t^{c}U_{\mathbf{z}}-C_{F}\big)\\
 & -\mathcal{A}_{0}(\Delta_{\mathrm{UV}})C_{F}\big(U_{\mathbf{x}}-1\big)\Bigg]\;,
\end{aligned}
\end{equation}
and:
\begin{equation}
\begin{aligned}\lim_{k^{+}\to0}\tilde{\mathcal{M}}_{\mathrm{SE3,sub}}^{\lambda} & =-4\alpha_{s}\frac{q^{+}\mathbf{q}^{\bar{\lambda}}}{p_{0}^{+}\mathbf{q}^{2}+p_{1}^{+}M^{2}}\mathrm{Dirac}_{\mathrm{LO}}^{\lambda\bar{\lambda}}\int_{k_{\mathrm{min}}^{+}}^{p_{1}^{+}}\frac{\mathrm{d}k^{+}}{k^{+}}\\
 & \times\int_{\mathbf{x}}e^{-i\mathbf{k}_{\perp}\cdot\mathbf{x}}\Bigg[\int_{\mathbf{z}}A^{i}(\mathbf{x}-\mathbf{z})A^{i}(\mathbf{x}-\mathbf{z})\big(t^{c}U_{\mathbf{x}}U_{\mathbf{z}}^{\dagger}t^{c}U_{\mathbf{z}}-C_{F}\big)\\
 & -\mathcal{A}_{0}(\Delta_{\mathrm{UV}})C_{F}\big(U_{\mathbf{x}}-1\big)\Bigg]\;.
\end{aligned}
\end{equation}
Cutting off the logarithmic integral over $k^{+}$ at the rapidity
factorization scale $k_{f}^{+}$, the above amplitudes contribute as follows to the cross section:
\begin{equation}
\begin{aligned}\lim_{k^{+}\to0}\mathrm{d}\sigma_{\mathrm{SE2+3}} & =\frac{\mathrm{d}\sigma_{\mathrm{LO}}}{s_{\mathbf{x}\mathbf{x}^{\prime}}+1}\times\ln\frac{k_{f}^{+}}{k_{\mathrm{min}}^{+}} \\
&\times4\alpha_{s}\Bigg[\frac{1}{2}\int_{\mathbf{z}}\Big(A^{i}(\mathbf{x}-\mathbf{z})A^{i}(\mathbf{x}-\mathbf{z})+A^{i}(\mathbf{x}^{\prime}-\mathbf{z})A^{i}(\mathbf{x}^{\prime}-\mathbf{z})\Big)\\
 & \times\Big(N_{c}s_{\mathbf{z}\mathbf{x}^{\prime}}s_{\mathbf{x}\mathbf{z}}-\frac{1}{N_{c}}s_{\mathbf{x}\mathbf{x}^{\prime}}+2C_{F}\Big)-\mathcal{A}_{0}(\Delta_{\mathrm{UV}})2C_{F}(s_{\mathbf{x}\mathbf{x}^{\prime}}+1)\Bigg]\;.
\end{aligned}
\label{eq:virtual_JIMWLK}
\end{equation}

\subsection{\label{subsec:real_JIMWLK}Real contributions}

The real contributions to JIMWLK that were not yet taken into account
in subsection \ref{subsec:jimwlk_leftover}, are the interferences
between the initial-state radiation amplitudes $\mathcal{M}_{\mathrm{IS1}}$
and $\mathcal{M}_{\mathrm{IS3}}$ (\ref{eq:MIS1L}), (\ref{eq:MIS1}),
(\ref{eq:MIS3L}), and (\ref{eq:MIS3}) on the one hand, and the final-state
amplitudes $\mathcal{M}_{\mathrm{FS1}}$ and $\mathcal{M}_{\mathrm{FS3}}$
(\ref{eq:MFS1L_coord}), (\ref{eq:MFS1T_coord}), (\ref{eq:MFS3L_coord})
and (\ref{eq:MFS3T_coord}) on the other.

Indeed, one obtains:
\begin{equation}
\begin{aligned}\lim_{k^{+}\to0}\tilde{\mathcal{M}}_{\mathrm{IS1}+3}^{0\eta} & =-\frac{2}{M}\Big(\frac{p_{0}^{+}\mathbf{P}_{\perp}^{2}-p_{1}^{+}M^{2}}{p_{0}^{+}\mathbf{P}_{\perp}^{2}+p_{1}^{+}M^{2}}-\frac{p_{0}^{+}\mathbf{q}^{2}-p_{1}^{+}M^{2}}{p_{0}^{+}\mathbf{q}{}^{2}+p_{1}^{+}M^{2}}\Big)\\
 & \times\int_{\mathbf{x},\mathbf{z}}iA^{\eta}(\mathbf{x}-\mathbf{z})e^{-i\mathbf{k}_{\perp}\cdot\mathbf{x}}e^{-i\mathbf{p}_{3}\cdot\mathbf{z}}\big(U_{\mathbf{x}}U_{\mathbf{z}}^{\dagger}t^{c}U_{\mathbf{z}}-t^{c}\big)\;,
\end{aligned}
\label{eq:JIMWLK_IS_L}
\end{equation}
and
\begin{equation}
\begin{aligned}\lim_{k^{+}\to0}\tilde{\mathcal{M}}_{\mathrm{FS1+3}}^{0\eta} & =\frac{2}{M}\Big(\frac{p_{0}^{+}\mathbf{P}_{\perp}^{2}-p_{1}^{+}M^{2}}{p_{0}^{+}\mathbf{P}_{\perp}^{2}+p_{1}^{+}M^{2}}-\frac{p_{0}^{+}\mathbf{q}^{2}-p_{1}^{+}M^{2}}{p_{0}^{+}\mathbf{q}^{2}+p_{1}^{+}M^{2}}\Big)\\
 & \times\int_{\mathbf{x},\mathbf{z}}e^{-i\mathbf{p}_{3}\cdot\mathbf{z}}e^{-i\mathbf{k}_{\perp}\cdot\mathbf{x}}iA^{\eta}(\mathbf{x}-\mathbf{z})t^{c}\big(U_{\mathbf{x}}-1\big)\;.
\end{aligned}
\label{eq:JIMWLK_FS_L}
\end{equation}
Similarly, when the photon is transversally polarized:
\begin{equation}
\begin{aligned}\lim_{k^{+}\to0}\tilde{\mathcal{M}}_{\mathrm{IS1+3}}^{\lambda\eta} & =2\mathrm{Dirac}^{\lambda\bar{\lambda}}\big(1+2\frac{p_{1}^{+}}{q^{+}}\big)\Big(\frac{q^{+}\mathbf{P}_{\perp}^{\bar{\lambda}}}{p_{0}^{+}\mathbf{P}_{\perp}^{2}+p_{1}^{+}M^{2}}+\frac{q^{+}\mathbf{q}^{\bar{\lambda}}}{p_{0}^{+}\mathbf{q}^{2}+p_{1}^{+}M^{2}}\Big)\\
 & \times\int_{\mathbf{x},\mathbf{z}}iA^{\eta}(\mathbf{x}-\mathbf{z})e^{-i\mathbf{k}_{\perp}\cdot\mathbf{x}}e^{-i\mathbf{p}_{3}\cdot\mathbf{z}}\big(U_{\mathbf{x}}U_{\mathbf{z}}^{\dagger}t^{c}U_{\mathbf{z}}-t^{c}\big)\;,
\end{aligned}
\label{eq:JIMWLK_IS_T}
\end{equation}
\begin{equation}
\begin{aligned}\lim_{k^{+}\to0}\tilde{\mathcal{M}}_{\mathrm{FS1+3}}^{\lambda\eta} & =-2\mathrm{Dirac}^{\lambda\bar{\lambda}}\big(1+\frac{2p_{1}^{+}}{q^{+}}\big)\Big(\frac{q^{+}\mathbf{P}_{\perp}^{\bar{\lambda}}}{p_{0}^{+}\mathbf{P}_{\perp}^{2}+p_{1}^{+}M^{2}}+\frac{q^{+}\mathbf{q}^{\bar{\lambda}}}{p_{0}^{+}\mathbf{q}^{2}+p_{1}^{+}M^{2}}\Big)\\
 & \times\int_{\mathbf{x},\mathbf{z}}e^{-i\mathbf{p}_{3}\cdot\mathbf{z}}e^{-i\mathbf{k}_{\perp}\cdot\mathbf{x}}iA^{\eta}(\mathbf{x}-\mathbf{z})t^{c}\big(U_{\mathbf{x}}-1\big)\;.
\end{aligned}
\label{eq:JIMWLK_FS_T}
\end{equation}
On the level of the cross section, the interference term due to the
above amplitudes is, independently of the photon polarization:
\begin{equation}
\begin{aligned}\lim_{k^{+}\to0}\mathrm{d}\sigma_{\mathrm{IS-FS}} & =\frac{\mathrm{d}\sigma_{\mathrm{LO}}}{s_{\mathbf{x}\mathbf{x}^{\prime}}+1}\times\ln\frac{k_{f}^{+}}{k_{\mathrm{min}}^{+}}\times4\alpha_{s}\int_{\mathbf{z}}A^{i}(\mathbf{x}-\mathbf{z})A^{i}(\mathbf{x}^{\prime}-\mathbf{z})\\
 & \times\big(\frac{1}{N_{c}}s_{\mathbf{x}\mathbf{x}^{\prime}}-N_{c}s_{\mathbf{z}\mathbf{x}^{\prime}}s_{\mathbf{x}\mathbf{z}}-2C_{F}\big)\;.
\end{aligned}
\label{eq:real_JIMWLK}
\end{equation}

\subsection{JIMWLK}

We can now, finally, add the intermediate results (\ref{eq:UVDGLAPJET_JIMWLK}),
(\ref{eq:virtual_JIMWLK}), and (\ref{eq:real_JIMWLK}):
\begin{equation}
\begin{aligned} & \lim_{k^{+}\to0}\Big(\mathrm{d}\sigma_{\mathrm{UV}}+\mathrm{d}\sigma_{\mathrm{IS}}+\mathrm{d}\sigma_{\mathrm{jet}}+\mathrm{d}\sigma_{\mathrm{SE2+3}}+\mathrm{d}\sigma_{\mathrm{IS-FS}}\Big)\\
 & =\frac{\mathrm{d}\sigma_{\mathrm{LO}}}{s_{\mathbf{x}\mathbf{x}^{\prime}}+1}\times\ln\frac{k_{f}^{+}}{k_{\mathrm{min}}^{+}}\times4\alpha_{s}\Bigg[\frac{1}{2}\int_{\mathbf{z}}\Big(A^{i}(\mathbf{x}-\mathbf{z})A^{i}(\mathbf{x}-\mathbf{z})+A^{i}(\mathbf{x}^{\prime}-\mathbf{z})A^{i}(\mathbf{x}^{\prime}-\mathbf{z})\Big)\\
 & \times\Big(N_{c}s_{\mathbf{z}\mathbf{x}^{\prime}}s_{\mathbf{x}\mathbf{z}}-\frac{1}{N_{c}}s_{\mathbf{x}\mathbf{x}^{\prime}}+2C_{F}\Big)\\
 & +\int_{\mathbf{z}}A^{i}(\mathbf{x}-\mathbf{z})A^{i}(\mathbf{x}^{\prime}-\mathbf{z})\big(\frac{1}{N_{c}}s_{\mathbf{x}\mathbf{x}^{\prime}}-N_{c}s_{\mathbf{z}\mathbf{x}^{\prime}}s_{\mathbf{x}\mathbf{z}}-2C_{F}\big)\\
 & +\Big(\frac{1}{4\pi}\ln\frac{c_{0}^{2}}{\Delta_{\mathrm{UV}}(\mathbf{x}-\mathbf{x}^{\prime})^{2}}-\mathcal{A}_{0}(\Delta_{\mathrm{UV}})\Big)2C_{F}(s_{\mathbf{x}\mathbf{x}^{\prime}}+1)\Bigg]\;.
\end{aligned}
\end{equation}
The terms in the last line of the above expression can combined into:
\begin{equation}
\begin{aligned}\frac{1}{4\pi}\!\ln\frac{c_{0}^{2}}{\Delta_{\mathrm{UV}}(\mathbf{x}\!-\!\mathbf{x}^{\prime})^{2}}\!-\!\mathcal{A}_{0}(\Delta_{\mathrm{UV}}) & \!=\!-\frac{1}{4\pi}\Big(\frac{1}{\epsilon_{\mathrm{UV}}}\!-\!\gamma_{\scriptscriptstyle{E}}\!+\!\ln\frac{4\pi\mu^{2}}{\Delta_{\mathrm{UV}}}\!-\!\ln\frac{c_{0}^{2}}{\Delta_{\mathrm{UV}}(\mathbf{x}\!-\!\mathbf{x}^{\prime})^{2}}\Big)\;,\\
 & =-\frac{1}{4\pi}\Big(\frac{1}{\epsilon_{\mathrm{UV}}}+\gamma_{\scriptscriptstyle{E}}+\ln\pi\mu^{2}(\mathbf{x}-\mathbf{x}^{\prime})^{2}\Big)\;,
\end{aligned}
\end{equation}
where we used the expansion (\ref{eq:A0}) of $\mathcal{A}_{0}$.
But in this result one recognizes:
\begin{equation}
\begin{aligned} &-\frac{1}{4\pi}\Big(\frac{1}{\epsilon_{\mathrm{UV}}}+\gamma_{\scriptscriptstyle{E}}+\ln\pi\mu^{2}(\mathbf{x}-\mathbf{x}^{\prime})^{2}\Big) \\
& =\int_{\boldsymbol{\ell}}\frac{e^{-i\boldsymbol{\ell}\cdot(\mathbf{x}-\mathbf{x}^{\prime})}}{\boldsymbol{\ell}^{2}} =\int_{\mathbf{z}}A^{i}(\mathbf{x}-\mathbf{z})A^{i}(\mathbf{x}^{\prime}-\mathbf{z})\;,
\end{aligned}
\label{eq:AAtrick}
\end{equation}
such that 
\begin{equation}
\begin{aligned} & \lim_{k^{+}\to0}\Big(\mathrm{d}\sigma_{\mathrm{UV}}+\mathrm{d}\sigma_{\mathrm{IS}}+\mathrm{d}\sigma_{\mathrm{jet}}+\mathrm{d}\sigma_{\mathrm{SE2+3}}+\mathrm{d}\sigma_{\mathrm{IS-FS}}\Big)\\
 & =\frac{\mathrm{d}\sigma_{\mathrm{LO}}}{s_{\mathbf{x}\mathbf{x}^{\prime}}+1}\times\ln\frac{k_{f}^{+}}{k_{\mathrm{min}}^{+}}\times4\alpha_{s}\Bigg[\frac{1}{2}\int_{\mathbf{z}}\Big(A^{i}(\mathbf{x}-\mathbf{z})A^{i}(\mathbf{x}-\mathbf{z})+A^{i}(\mathbf{x}^{\prime}-\mathbf{z})A^{i}(\mathbf{x}^{\prime}-\mathbf{z})\Big)\\
 & \times\Big(N_{c}s_{\mathbf{z}\mathbf{x}^{\prime}}s_{\mathbf{x}\mathbf{z}}-\frac{1}{N_{c}}s_{\mathbf{x}\mathbf{x}^{\prime}}+2C_{F}\Big) +\int_{\mathbf{z}}A^{i}(\mathbf{x}-\mathbf{z})A^{i}(\mathbf{x}^{\prime}-\mathbf{z})N_{c}\big(s_{\mathbf{x}\mathbf{x}^{\prime}}-s_{\mathbf{z}\mathbf{x}^{\prime}}s_{\mathbf{x}\mathbf{z}}\big)\Bigg]\;.
\end{aligned}
\label{eq:almostJIMWLK}
\end{equation}
Setting $\mathbf{x}=\mathbf{x}^{\prime}$ in (\ref{eq:AAtrick}),
we obtain a scaleless integral which in dimensional regularization
is, of course, equal to zero:
\begin{equation}
\begin{aligned}\int_{\mathbf{z}}A^{i}(\mathbf{x}-\mathbf{z})A^{i}(\mathbf{x}-\mathbf{z}) & =\int_{\boldsymbol{\ell}}\frac{1}{\boldsymbol{\ell}^{2}}=0\;.\end{aligned}
\end{equation}
We are, therefore, allowed to subtract the following, vanishing, contribution
from (\ref{eq:almostJIMWLK}):
\begin{equation}
\begin{aligned}\int_{\mathbf{z}}\Big(A^{i}(\mathbf{x}-\mathbf{z})A^{i}(\mathbf{x}-\mathbf{z})+A^{i}(\mathbf{x}^{\prime}-\mathbf{z})A^{i}(\mathbf{x}^{\prime}-\mathbf{z})\Big)C_{F}(s_{\mathbf{x}\mathbf{x}^{\prime}}+1) & =0\;,\end{aligned}
\end{equation}
which gives:
\begin{equation}
\begin{aligned} & \lim_{k^{+}\to0}\Big(\mathrm{d}\sigma_{\mathrm{UV}}+\mathrm{d}\sigma_{\mathrm{IS}}+\mathrm{d}\sigma_{\mathrm{jet}}+\mathrm{d}\sigma_{\mathrm{SE2+3}}+\mathrm{d}\sigma_{\mathrm{IS-FS}}\Big)\\
 & =\frac{\mathrm{d}\sigma_{\mathrm{LO}}}{s_{\mathbf{x}\mathbf{x}^{\prime}}+1}\times\ln\frac{k_{f}^{+}}{k_{\mathrm{min}}^{+}}(s_{\mathbf{x}\mathbf{x}^{\prime}}-s_{\mathbf{z}\mathbf{x}^{\prime}}s_{\mathbf{x}\mathbf{z}})\\
 & \times\!4\alpha_{s}N_{c}\!\int_{\mathbf{z}}\Big(\!A^{i}(\mathbf{x}\!-\!\mathbf{z})A^{i}(\mathbf{x}^{\prime}\!-\!\mathbf{z})\!-\!\frac{1}{2}A^{i}(\mathbf{x}\!-\!\mathbf{z})A^{i}(\mathbf{x}\!-\!\mathbf{z})\!-\!\frac{1}{2}A^{i}(\mathbf{x}^{\prime}\!-\!\mathbf{z})A^{i}(\mathbf{x}^{\prime}\!-\!\mathbf{z})\Big)\;.
\end{aligned}
\end{equation}
The final step is to recognize the integration over Weizsäcker-Williams
fields, in the last line, as the BK kernel:
\begin{equation}
\begin{aligned} & \int_{\mathbf{z}}\Big(A^{i}(\mathbf{x}-\mathbf{z})A^{i}(\mathbf{x}^{\prime}-\mathbf{z})-\frac{1}{2}A^{i}(\mathbf{x}-\mathbf{z})A^{i}(\mathbf{x}-\mathbf{z})-\frac{1}{2}A^{i}(\mathbf{x}^{\prime}-\mathbf{z})A^{i}(\mathbf{x}^{\prime}-\mathbf{z})\Big)\\
 & =-\frac{1}{8\pi^{2}}\int_{\mathbf{z}}\frac{(\mathbf{x}-\mathbf{x}^{\prime})^{2}}{(\mathbf{x}-\mathbf{z})^{2}(\mathbf{x}^{\prime}-\mathbf{z})^{2}}\;,
\end{aligned}
\label{eq:BKkernel}
\end{equation}
such that we recover the JIMWLK evolution of a dipole (\ref{eq:dipoleJIMWLK}):
\begin{equation}
\begin{aligned} & \lim_{k^{+}\to0}\Big(\mathrm{d}\sigma_{\mathrm{UV}}+\mathrm{d}\sigma_{\mathrm{IS}}+\mathrm{d}\sigma_{\mathrm{jet}}+\mathrm{d}\sigma_{\mathrm{SE2+3}}+\mathrm{d}\sigma_{\mathrm{IS-FS}}\Big)\\
 & =\frac{\mathrm{d}\sigma_{\mathrm{LO}}}{s_{\mathbf{x}\mathbf{x}^{\prime}}+1}\times\ln\frac{k_{f}^{+}}{k_{\mathrm{min}}^{+}}\times-\frac{\alpha_{s}N_{c}}{2\pi^{2}}\int_{\mathbf{z}}\frac{(\mathbf{x}-\mathbf{x}^{\prime})^{2}}{(\mathbf{x}-\mathbf{z})^{2}(\mathbf{x}^{\prime}-\mathbf{z})^{2}}(s_{\mathbf{x}\mathbf{x}^{\prime}}-s_{\mathbf{z}\mathbf{x}^{\prime}}s_{\mathbf{x}\mathbf{z}})\\
 & =\ln\frac{k_{f}^{+}}{k_{\mathrm{min}}^{+}}\times\hat{H}_{\mathrm{JIMWLK}}\mathrm{d}\sigma_{\mathrm{LO}}\;.
\end{aligned}
\label{eq:JIMWLK}
\end{equation}
Hence, we have proven the validity of eq. (\ref{eq:JIMWLKsubtraction}),
namely that all the rapidity-divergent NLO contributions combine into the JIMWLK Hamiltonian acting on the LO cross section. Therefore,
they can be subtracted and absorbed into the JIMWLK evolution of the
latter, according to eq. (\ref{eq:LO+NLO_finite}).

\paragraph{Some subtleties}

But what about the other real contributions to the high-energy resummation?
First, the squared amplitudes $|\mathcal{M}_{\mathrm{IS1}}|^{2}$
and $|\mathcal{M}_{\mathrm{FS3}}|^{2}$ cause collinear initial- and
final state divergences when integrating over the gluon momentum.
Therefore, we have separately analyzed them in sections \ref{sec:DGLAP}
and \ref{sec:jet}, respectively, and showed how these collinear poles
cancel. The leftovers of these procedures still contain rapidity divergences
that are taken into account via eq. (\ref{eq:UVDGLAPJET_JIMWLK}).

More subtle, however, are the cases of squared amplitudes $|\mathcal{M}_{\mathrm{IS3}}|^{2}$
and $|\mathcal{M}_{\mathrm{FS1}}|^{2}$, as well as the interference
terms $2\mathrm{Re}\mathcal{M}_{\mathrm{IS1}}^{\dagger}\mathcal{M}_{\mathrm{IS3}}$
and $2\mathrm{Re}\mathcal{M}_{\mathrm{FS1}}^{\dagger}\mathcal{M}_{\mathrm{FS3}}$.
They do not contribute to DGLAP or to the jet function and hence are
not taken into account in subsection \ref{subsec:jimwlk_leftover}
At first sight they seem to be missing from (\ref{eq:JIMWLK}). To
solve this puzzle, we should revisit sections \ref{sec:DGLAP} and
\ref{sec:jet}, where we were not able to evaluate the transverse
integrations in amplitudes $\mathcal{M}_{\mathrm{IS2}}$ and $\mathcal{M}_{\mathrm{FS2}}$
exactly and, instead, devised an approximation to extract their collinear
behavior. For definiteness, let us look again at amplitude $\tilde{\mathcal{M}}_{\mathrm{IS2}}^{0}$
(the other cases are similar):
\begin{equation}
\begin{aligned}\tilde{\mathcal{M}}_{\mathrm{IS2}}^{0\eta} & =\frac{1}{M}\mathrm{Dirac}^{\eta\bar{\eta}}\big(1+\frac{2p_{0}^{+}}{p_{3}^{+}}\big)\frac{p_{3}^{+}}{p_{0\scriptscriptstyle{R}}^{+}}\\
 & \times\int_{\boldsymbol{\ell}}\frac{\boldsymbol{\ell}^{\bar{\eta}}}{\boldsymbol{\ell}^{2}}\frac{p_{3}^{+}\big(q^{+}\boldsymbol{\ell}+p_{0}^{+}\mathbf{q}\big)^{2}-p_{1}^{+}p_{3}^{+}p_{0}^{+}M^{2}}{p_{3}^{+}\big(q^{+}\boldsymbol{\ell}+p_{0}^{+}\mathbf{q}\big)^{2}+p_{1}^{+}q^{+}p_{0R}^{+}\boldsymbol{\ell}^{2}+p_{1}^{+}p_{3}^{+}p_{0}^{+}M^{2}}\\
 & \times\int_{\mathbf{x},\mathbf{z}}e^{-i\boldsymbol{\ell}\cdot(\mathbf{\mathbf{x}-\mathbf{z}})}e^{-i\mathbf{k}_{\perp}\cdot\mathbf{x}}e^{-i\mathbf{p}_{3}\cdot\mathbf{z}}\big(U_{\mathbf{x}}U_{\mathbf{z}}^{\dagger}t^{c}U_{\mathbf{z}}-t^{c}\big)\;.
\end{aligned}
\end{equation}
Clearly, this diagram will not lead to rapidity divergences, since
it disappears in the limit $p_{3}^{+}\to0$. Introducing the transformation
$\boldsymbol{\ell}=p_{3}^{+}/q^{+}\tilde{\boldsymbol{\ell}}$ and
only then taking the limit $p_{3}^{+}\to0$ allows us to study the
kinematics where all components of the gluon momentum tend to zero
simultaneously (the \textquoteleft genuine soft' limit $\vec{p}_{3}\to0$,
see e.g. \cite{Taels:2022tza}):
\begin{equation}
\begin{aligned}\lim_{\vec{p}_{3}\to0}\tilde{\mathcal{M}}_{\mathrm{IS2}}^{0\eta} & =\lim_{p_{3}^{+}\to0}\frac{1}{M}\mathrm{Dirac}^{\eta\bar{\eta}}\big(1+\frac{2p_{0}^{+}}{p_{3}^{+}}\big)\frac{p_{3}^{+}}{p_{0\scriptscriptstyle{R}}^{+}}\\
 & \times\int_{\tilde{\boldsymbol{\ell}}}\frac{\tilde{\boldsymbol{\ell}}^{\bar{\eta}}}{\tilde{\boldsymbol{\ell}}^{2}}\frac{p_{3}^{+}}{q^{+}}\frac{\big(p_{3}^{+}\tilde{\boldsymbol{\ell}}+p_{0}^{+}\mathbf{q}\big)^{2}-p_{1}^{+}p_{0}^{+}M^{2}}{\big(p_{3}^{+}\tilde{\boldsymbol{\ell}}+p_{0}^{+}\mathbf{q}\big)^{2}+p_{1}^{+}p_{3}^{+}p_{0R}^{+}\tilde{\boldsymbol{\ell}}^{2}/q^{+}+p_{1}^{+}p_{0}^{+}M^{2}}\\
 & \times\int_{\mathbf{x},\mathbf{z}}e^{-i\frac{p_{3}^{+}}{q^{+}}\tilde{\boldsymbol{\ell}}\cdot(\mathbf{\mathbf{x}-\mathbf{z}})}e^{-i\mathbf{k}_{\perp}\cdot\mathbf{x}}e^{-i\mathbf{p}_{3}\cdot\mathbf{z}}\big(U_{\mathbf{x}}U_{\mathbf{z}}^{\dagger}t^{c}U_{\mathbf{z}}-t^{c}\big) =0\;.
\end{aligned}
\end{equation}
It follows that $\tilde{\mathcal{M}}_{\mathrm{IS2}}^{0}$ cannot generate
logarithmic divergences of the form $\ln p^{+}/k_{\mathrm{min}}^{+}$:
neither rapidity ones, nor soft ones when the gluon transverse momentum
tends to zero simultaneously with its plus-momentum component. However,
in the approximation (\ref{eq:IS_coll_approx}), we have set $\boldsymbol{\ell}\to0$
to extract the collinear behavior:
\begin{equation}
\begin{aligned}\tilde{\mathcal{M}}_{\mathrm{IS2}}^{0\eta} & \Big|_{\mathrm{coll.}}=\frac{1}{M}\frac{p_{0}^{+}\mathbf{q}^{2}-p_{1}^{+}M^{2}}{p_{0}^{+}\mathbf{q}^{2}+p_{1}^{+}M^{2}}\mathrm{Dirac}^{\eta\bar{\eta}}\big(1+\frac{2p_{0}^{+}}{p_{3}^{+}}\big)\frac{p_{3}^{+}}{p_{0\scriptscriptstyle{R}}^{+}}\\
 & \times\int_{\mathbf{x},\mathbf{z}}\int_{\boldsymbol{\ell}}\frac{\boldsymbol{\ell}^{\bar{\eta}}}{\boldsymbol{\ell}^{2}}e^{-i\boldsymbol{\ell}\cdot(\mathbf{\mathbf{x}-\mathbf{z}})}e^{-i\mathbf{k}_{\perp}\cdot\mathbf{x}}e^{-i\mathbf{p}_{3}\cdot\mathbf{z}}\big(U_{\mathbf{x}}U_{\mathbf{z}}^{\dagger}t^{c}U_{\mathbf{z}}-t^{c}\big)\;.
\end{aligned}
\end{equation}
But the above expression \emph{does }lead to a rapidity divergence,
in particular exactly the same as the one from amplitude $\mathcal{M}_{\mathrm{IS3}}^{0\eta}$:
\begin{equation}
\begin{aligned}\lim_{p_{3}^{+}\to0}\mathcal{M}_{\mathrm{IS2}}^{0\eta}\Big|_{\mathrm{coll.}} & =\lim_{p_{3}^{+}\to0}\mathcal{M}_{\mathrm{IS3}}^{0\eta}\;.\end{aligned}
\end{equation}
Thus, the reason that, in section \ref{eq:real_JIMWLK}, we only need
to consider the interference terms $\mathcal{M}_{\mathrm{IS1}+3}^{\dagger}\mathcal{M}_{\mathrm{FS1}+3}^{\dagger}$,
is because the rapidity divergences from $|\mathcal{M}_{\mathrm{IS1+3}}|^{2}$
and $|\mathcal{M}_{\mathrm{FS1+3}}|^{2}$ are accounted for in eq.
\ref{eq:UVDGLAPJET_JIMWLK}. The rapidity-divergent contributions
from $|\mathcal{M}_{\mathrm{IS1}}|^{2}$ and $|\mathcal{M}_{\mathrm{FS3}}|^{2}$
are \textquoteleft physical', while the others: $|\mathcal{M}_{\mathrm{IS3}}|^{2}$,
$|\mathcal{M}_{\mathrm{FS1}}|^{2}$, $\mathcal{M}_{\mathrm{IS1}}^{\dagger}\mathcal{M}_{\mathrm{IS3}}$,
and $\mathcal{M}_{\mathrm{FS1}}^{\dagger}\mathcal{M}_{\mathrm{FS3}}$
are \emph{mimicked} by the artificially large logarithmic terms $|\mathcal{M}_{\mathrm{IS2}}|^{2}\Big|_{\mathrm{coll.}}$,
$|\mathcal{M}_{\mathrm{FS2}}|^{2}\Big|_{\mathrm{coll.}}$, $\mathcal{M}_{\mathrm{IS1}}^{\dagger}\mathcal{M}_{\mathrm{IS2}}\Big|_{\mathrm{coll.}}$,
and $\mathcal{M}_{\mathrm{FS1}}^{\dagger}\mathcal{M}_{\mathrm{FS2}}\Big|_{\mathrm{coll.}}$. 

\section{\label{sec:NLO}Next-to-leading order cross section}

We are now ready to present the full NLO cross section for our process.
It is given by the following sum of separately finite contributions:
\begin{equation}
\begin{aligned}\mathrm{d}\sigma_{\mathrm{LO+NLO}}^{pA\to\gamma^{*}+\mathrm{jet}+X} & =\mathrm{d}\sigma_{\mathrm{LO+DGLAP+JIMWLK}}+\mathrm{d}\sigma_{\mathrm{jet}}+\mathrm{d}\sigma_{\mathrm{IS}}+\mathrm{d}\sigma_{\mathrm{virtual}}+\mathrm{d}\sigma_{\mathrm{real}}\;.\end{aligned}
\label{eq:totalcrosssection}
\end{equation}
The first part of the cross section (\ref{eq:totalcrosssection})
is the leading-order one, eq. (\ref{eq:crossLO}), where the quark
PDF is evolved at least one step with DGLAP (\ref{eq:sigma_LO+DGLAP}),
and where the target average of the dipole is evolved at least one
step with JIMWLK (\ref{eq:LO_JIMWLK}). For all the other terms, it
is understood that the subtraction of high-energy leading logarithms
has been performed according to (\ref{eq:LO+NLO_finite}) in the previous
section.

The term $\mathrm{d}\sigma_{\mathrm{jet}}$ (\ref{eq:jet_final}),
after performing the rapidity subtraction, reads:
\begin{equation}
\begin{aligned}\mathrm{d}\sigma_{\mathrm{jet}} & =\mathrm{d}\sigma_{\mathrm{LO}}\times\frac{\alpha_{s}C_{F}}{\pi}\Big(\frac{3}{4}\ln\big(\frac{4\pi e^{-\gamma_{\scriptscriptstyle{E}}}\mu_{\scriptscriptstyle{R}}^{2}}{\mathbf{p}_{j}^{2}R^{2}}\big)+\frac{13}{4}-\frac{\pi^{2}}{3}\Big)\;.\end{aligned}
\end{equation}
Similarly, the contribution to $\mathrm{d}\sigma_{\mathrm{IS}}$ is
obtained after subtracting the large high-energy logarithm from (\ref{eq:sigma_IS}):
\begin{equation}
\begin{aligned}\mathrm{d}\sigma_{\mathrm{IS}} & =\frac{1}{2(p_{0}^{+})^{2}}\int\mathrm{PS}(\vec{p}_{1},\vec{q})\frac{1}{2}\big\langle\big|\mathcal{M}_{\mathrm{LO}}\big|^{2}\big\rangle\\
 & \times\alpha_{s}\Big(\int_{x_{p}}^{1}\mathrm{d}\xi P_{qq}^{(0)}(\xi)\frac{x_{p}}{\xi}f_{q}^{(0)}\big(\frac{x_{p}}{\xi}\big)\!-\!\frac{3}{2}C_{F}\mathrm{ln}\frac{\mu^{2}}{4\pi e^{-\gamma_{\scriptscriptstyle{E}}}\mu_{\scriptscriptstyle{R}}^{2}}x_{p}f_{q}^{(0)}(x_{p})\!\Big)\!\ln\frac{c_{0}^{2}}{\mu^{2}(\mathbf{x}\!-\!\mathbf{x}^{\prime})^{2}}\;.
\end{aligned}
\end{equation}
The term $\mathrm{d}\sigma_{\mathrm{virtual}}$, the finite leftovers
of the virtual diagrams are collected after cancelling the UV divergences,
and after absorbing the collinear divergences inside the jet definition
and the DGLAP evolution. Moreover, the rapidity subtraction procedure
has been performed explicitly, hence a residual dependence on the
rapidity factorization scale $k_{f}^{+}$ is left. The result reads,
for the longitudinally polarized virtual photon:
\begin{equation}
\begin{aligned}\mathrm{d}\sigma_{\mathrm{virtual}}^{\mathrm{L}} & =x_{p}f_{q}^{(0)}(x_{p})\frac{2\pi}{2(p_{0}^{+})^{2}}\mathrm{PS}(\vec{p}_{1},\vec{q})\frac{1}{2}\\
 & \times\mathrm{Tr}\Big\langle\mathcal{M}_{\mathrm{LO}}^{0\dagger}\Big[\mathcal{M}_{\mathrm{SE1,sub}}^{0}\Big|_{\mathrm{finite}}+\mathcal{M}_{\mathrm{SE4,sub}}^{0}\Big|_{\mathrm{finite}}+\mathcal{M}_{\mathrm{SE2,sub}}^{0}+\mathcal{M}_{\mathrm{SE3,sub}}^{0}\Big)\\
 & +\Big(\mathcal{M}_{\mathrm{SE1,sub}}^{0}+\mathcal{M}_{\mathrm{V1,sub}}^{0}\Big)\Big|_{1/k^{+}}+\mathcal{M}_{\mathrm{V1,sub}}^{0}\Big|_{\mathrm{finite}}+\mathcal{M}_{\mathrm{V2,sub}}^{0}\Big|_{\mathrm{finite}}\\
 & +\mathcal{M}_{\mathrm{V3,sub}}^{0}+\mathcal{M}_{\mathrm{V4,sub}}^{0}+\mathcal{M}_{\mathrm{A1,sub}}^{0}\Big|_{\mathrm{finite}}+\mathcal{M}_{\mathrm{A2,sub}}^{0}\Big|_{\mathrm{finite}}\\
 & +\Big(\mathcal{M}_{\mathrm{V1,sub}}^{0}+\mathcal{M}_{\mathrm{A1,sub}}^{0}\Big)\Big|_{\mathrm{spurious}}+\Big(\mathcal{M}_{\mathrm{V2,sub}}^{0}+\mathcal{M}_{\mathrm{A2,sub}}^{0}\Big)\Big|_{\mathrm{spurious}}\\
 & +\mathcal{M}_{\mathrm{A3,sub}}^{0}+\mathcal{M}_{\mathrm{A4,sub}}^{0}+\mathcal{M}_{\mathrm{Q1,sub}}^{0}+\mathcal{M}_{\mathrm{Q2,sub}}^{0}+\mathcal{M}_{\mathrm{Q3,sub}}^{0}+\mathcal{M}_{\mathrm{Q4,sub}}^{0}\\
 & +\mathcal{M}_{\mathrm{I1,sub}}^{0}+\mathcal{M}_{\mathrm{I2,sub}}^{0}+\mathcal{M}_{\mathrm{I3,sub}}^{0}+\mathcal{M}_{\mathrm{I4,sub}}^{0}\Big]+\mathrm{c.c.}\Big\rangle\;.
\end{aligned}
\label{eq:dsigma_L_virtual}
\end{equation}
The explicit expressions for the different contributions are presented
in subsection \ref{subsec:virtual_L_cross}. In the transversally
polarized case, the \textquoteleft virtual' cross section is given
by:
\begin{equation}
\begin{aligned}\mathrm{d}\sigma_{\mathrm{virtual}}^{\mathrm{T}} & =x_{p}f_{q}^{(0)}(x_{p})\frac{2\pi}{2(p_{0}^{+})^{2}}\mathrm{PS}(\vec{p}_{1},\vec{q})\frac{1}{2}\\
 & \times\mathrm{Tr}\Big\langle\mathcal{M}_{\mathrm{LO}}^{\lambda\dagger}\Big[\mathcal{M}_{\mathrm{SE1,sub}}^{\lambda}\Big|_{\mathrm{finite}}+\mathcal{M}_{\mathrm{SE4,sub}}^{\lambda}\Big|_{\mathrm{finite}}+\mathcal{M}_{\mathrm{SE2,sub}}^{\lambda}+\mathcal{M}_{\mathrm{SE3,sub}}^{\lambda}\\
 & +\Big(\mathcal{M}_{\mathrm{SE1,sub}}^{\lambda}+\mathcal{M}_{\mathrm{V1,sub}}^{\lambda}\Big)\Big|_{1/k^{+}}\\
 & +\mathcal{M}_{\mathrm{V1,sub}}^{\lambda}\Big|_{\mathrm{finite}}+\mathcal{M}_{\mathrm{V2}}^{\lambda}+\mathcal{M}_{\mathrm{V3}}^{\lambda}+\mathcal{M}_{\mathrm{V4,sub}}^{\lambda}\Big|_{\mathrm{finite}}\\
 & +\mathcal{M}_{\mathrm{A1,sub}}^{\lambda}+\mathcal{M}_{\mathrm{A2}}^{\lambda}+\mathcal{M}_{\mathrm{A3}}^{\lambda}+\mathcal{M}_{\mathrm{A4,sub}}^{\lambda}\\
 & +\mathcal{M}_{\mathrm{Q2}}^{\lambda}+\mathcal{M}_{\mathrm{I2}}^{\lambda}+\mathcal{M}_{\mathrm{I3}}^{\lambda}\Big]+\mathrm{c.c.}\Big\rangle\;,
\end{aligned}
\label{eq:dsigma_T_virtual}
\end{equation}
where the expressions for the various terms in are listed in subsection
\ref{subsec:virtual_T_cross}.

The last term, $\mathrm{d}\sigma_{\mathrm{real}}$, collects those contributions from real radiative corrections that were not yet absorbed into $\mathrm{d}\sigma_{\mathrm{jet}}$ or $\mathrm{d}\sigma_{\mathrm{IS}}$, and is given by:
\begin{equation}
\begin{aligned} &\mathrm{d}\sigma_{\mathrm{real}}  =x_{p}f_{q}^{(0)}(x_{p},\mu^{2})\frac{2\pi}{2(p_{1}^{+}+q^{+}+p_{3}^{+})^{2}}\mathrm{PS}(\vec{p}_{1},\vec{q})\int\frac{\mathrm{d}p_{3}^{+}}{4\pi p_{3}^{+}}\frac{1}{2}\\
 & \!\times\!\int_{\mathbf{p}_{3}}\mathrm{Tr}\Big\langle\Big|\mathcal{M}_{\mathrm{IS2,finite}}^{\eta}\Big|^{2}+2\mathrm{Re}\big(\mathcal{M}_{\mathrm{IS1}}^{\eta\dagger}\mathcal{M}_{\mathrm{IS2,finite}}^{\eta}\big)+\Big|\mathcal{M}_{\mathrm{IS3+4}}^{\eta}\Big|^{2}+2\mathrm{Re}\big(\mathcal{M}_{\mathrm{IS1}+2}^{\eta\dagger}\mathcal{M}_{\mathrm{IS3+4}}^{\eta}\big)\\
 & \!+\!\big|\mathcal{M}_{\mathrm{FS2,finite}}^{\eta}\big|^{2}+2\mathrm{Re}\big(\mathcal{M}_{\mathrm{FS2,finite}}^{\eta\dagger}\mathcal{M}_{\mathrm{FS3}}^{\eta}\big)+\big|\mathcal{M}_{\mathrm{FS1}+4}^{\eta}\big|^{2}+2\mathrm{Re}\big(\mathcal{M}_{\mathrm{FS2+3}}^{\eta\dagger}\mathcal{M}_{\mathrm{FS1}+4}^{\eta}\big)\\
 & \!+\!2\mathrm{Re}\big(\mathcal{M}_{\mathrm{IS1}+2}^{\eta\dagger}\mathcal{M}_{\mathrm{FS1}+4}^{\eta}+\mathcal{M}_{\mathrm{IS1}+2}^{\eta\dagger}\mathcal{M}_{\mathrm{FS2}+3}^{\eta}+\mathcal{M}_{\mathrm{IS3}+4}^{\eta\dagger}\mathcal{M}_{\mathrm{FS1}+4}^{\eta}+\mathcal{M}_{\mathrm{IS3}+4}^{\eta\dagger}\mathcal{M}_{\mathrm{FS2}+3}^{\eta}\big)\Big\rangle
\end{aligned}
\label{eq:dsigma_finite_real}
\end{equation}
and have the same structure irregardless of the photon polarization,
which is why the labels $0$ or $\lambda$ are omitted. The explicit
expressions of the terms above are listed in subsections (\ref{subsec:real_fin_L})
and (\ref{subsec:real_fin_T}) for the longitudinal resp. transversally
polarized case. Note that, from plus-momentum conservation, we have
that $x_{p}=(p_{1}^{+}+q^{+}+p_{3}^{+})/p_{p}^{+}$. 

In the following subsections, we present the results for the different terms in eqs.~\eqref{eq:dsigma_L_virtual}, \eqref{eq:dsigma_T_virtual}, and \eqref{eq:dsigma_finite_real}. It is interesting to remark that, despite the complexity of this calculation, most of the result only depends on two simple sets of color operators, namely:
\begin{equation}
\big(s_{\mathbf{x}\mathbf{x}^{\prime}}+1\big)\qquad\mathrm{and}\qquad\big(\frac{N_{c}^{2}}{2}s_{\mathbf{z}\mathbf{x}^{\prime}}s_{\mathbf{x}\mathbf{z}}-\frac{1}{2}s_{\mathbf{x}\mathbf{x}^{\prime}}+C_{F}N_{c}\big)\;.
\label{eq:coloroperators}
\end{equation}
The exceptions to this rule are eqs.~\eqref{eq:A3L_cross}, \eqref{eq:Q2_sub_cross}, \eqref{eq:A3_T_cross}, and \eqref{eq:Q2_T_cross}, due to the virtual amplitudes $\mathcal{M}_{\mathrm{A3}}$ and $\mathcal{M}_{\mathrm{Q2}}$, which also include a quadrupole:
\begin{equation}
\big(\frac{N_{c}^{2}}{2}s_{\mathbf{x}_{1}\mathbf{x}_{2}}s_{\mathbf{x}_{3}\mathbf{x}^{\prime}}-\frac{1}{2}Q_{\mathbf{x}_{1}\mathbf{x}^{\prime}\mathbf{x}_{3}\mathbf{x}_{2}}+C_{F}N_{c}\big)\;.
\label{eq:coloroperators}
\end{equation}

\subsection{Virtual contributions}

\subsubsection{\label{subsec:virtual_L_cross}Longitudinal polarization}

\paragraph{Self-energy corrections}

Diagrams $\tilde{\mathcal{M}}_{\mathrm{SE1,sub}}^{0}$ (\ref{eq:MSE1Lsub_def})
and $\tilde{\mathcal{M}}_{\mathrm{SE4,sub}}^{0}$ (\ref{eq:MSE4_sub})
can be split up in a finite part, and a part that is divergent for
$k^{+}\to0$:
\begin{equation}
\begin{aligned}\tilde{\mathcal{M}}_{\mathrm{SE1,sub}}^{0} & =\tilde{\mathcal{M}}_{\mathrm{LO1}}^{0}\frac{\alpha_{s}C_{F}}{\pi}\Bigg[\int_{0}^{p_{0}^{+}}\mathrm{d}k^{+}\frac{k^{+}-2p_{0}^{+}}{2(p_{0}^{+})^{2}}+\int_{k_{\mathrm{min}}^{+}}^{p_{0}^{+}}\frac{\mathrm{d}k^{+}}{k^{+}}\Bigg]\ln\frac{\Delta_{\mathrm{P}}}{\Delta_{\mathrm{UV}}}\;,\\
 & =\tilde{\mathcal{M}}_{\mathrm{SE1,sub}}^{0}\Big|_{\mathrm{finite}}+\tilde{\mathcal{M}}_{\mathrm{SE1,sub}}^{0}\Big|_{1/k^{+}}\;,\\
\tilde{\mathcal{M}}_{\mathrm{SE4,sub}}^{0} & =\tilde{\mathcal{M}}_{\mathrm{LO2}}^{0}\frac{\alpha_{s}C_{F}}{\pi}\Bigg[\int_{0}^{p_{1}^{+}}\mathrm{d}k^{+}\frac{k^{+}-2p_{1}^{+}}{2(p_{1}^{+})^{2}}+\int_{k_{\mathrm{min}}^{+}}^{p_{1}^{+}}\frac{\mathrm{d}k^{+}}{k^{+}}\Bigg]\ln\frac{\Delta_{\mathrm{q}}}{\Delta_{\mathrm{UV}}}\;,\\
 & =\tilde{\mathcal{M}}_{\mathrm{SE4,sub}}^{0}\Big|_{\mathrm{finite}}+\tilde{\mathcal{M}}_{\mathrm{SE4,sub}}^{0}\Big|_{1/k^{+}}\;.
\end{aligned}
\label{eq:SE14_finite_k}
\end{equation}
Making use of the of the following identities (see \cite{Gradshteyn:1943cpj}),
valid for $a\geq c>0$:
\begin{equation}
\begin{aligned}\int_{0}^{c}\mathrm{d}x\,\ln\big(x(a-x)\big) & =-2c+a\ln a-(a-c)\ln(a-c)+c\ln c\;,\\
\int_{0}^{c}\mathrm{d}x\,x\ln\big(x(a-x)\big) & =\frac{1}{2}\Big(-c(a+c)+a^{2}\ln a+c^{2}\ln c+(c^{2}-a^{2})\ln(a-c)\Big)\;,
\end{aligned}
\label{eq:log_identities}
\end{equation}
(with $\mathrm{Li}_{2}$ the dilogarithm) the integrals over the gluon
plus-momentum can be evaluated, yielding:
\begin{equation}
\begin{aligned}\tilde{\mathcal{M}}_{\mathrm{SE1,sub}}^{0}\Big|_{\mathrm{finite}} & =\tilde{\mathcal{M}}_{\mathrm{LO1}}^{0}\frac{\alpha_{s}C_{F}}{\pi}\!\times\!-\frac{3}{4}\Big(-\!i\pi\!+\!2\ln\frac{p_{0}^{+}}{p_{1}^{+}}\!-\!2\!+\!\ln\frac{p_{1}^{+}\big(p_{0}^{+}\mathbf{P}_{\perp}^{2}\!+\!p_{1}^{+}M^{2}\big)}{p_{0}^{+}q^{+}\Delta_{\mathrm{UV}}}\Big)\;,\\
\tilde{\mathcal{M}}_{\mathrm{SE4,sub}}^{0}\Big|_{\mathrm{finite}} & =\tilde{\mathcal{M}}_{\mathrm{LO2}}^{0,\lambda}\frac{\alpha_{s}C_{F}}{\pi}\times-\frac{3}{4}\Big(\ln\frac{p_{0}^{+}\mathbf{q}^{2}+p_{1}^{+}M^{2}}{q^{+}\Delta_{\mathrm{UV}}}-2\Big)\;.
\end{aligned}
\end{equation}
Hence, we obtain the following virtual contributions to the cross
section:
\begin{equation}
\begin{aligned}\mathrm{Tr}\,\mathcal{M}_{\mathrm{LO}}^{0\dagger}\mathcal{M}_{\mathrm{SE1,sub}}^{0}\Big|_{\mathrm{finite}} & =\mathrm{Tr}\,\big(\mathcal{M}_{\mathrm{LO}}^{0\dagger}\mathcal{M}_{\mathrm{LO1}}^{0}\big)\frac{\alpha_{s}C_{F}}{\pi}\\&\times-\frac{3}{4}\Big(-i\pi+2\ln\frac{p_{0}^{+}}{p_{1}^{+}}-2+\ln\frac{p_{1}^{+}\big(p_{0}^{+}\mathbf{P}_{\perp}^{2}+p_{1}^{+}M^{2}\big)}{p_{0}^{+}q^{+}\Delta_{\mathrm{UV}}}\Big)\;,\end{aligned}
\label{eq:SE1L_fin_cross}
\end{equation}
and:
\begin{equation}
\begin{aligned}\mathrm{Tr}\,\mathcal{M}_{\mathrm{LO}}^{0\dagger}\mathcal{M}_{\mathrm{SE4,sub}}^{0}\Big|_{\mathrm{finite}} & =\mathrm{Tr}\,\big(\mathcal{M}_{\mathrm{LO}}^{0\dagger}\mathcal{M}_{\mathrm{LO2}}^{0}\big)\frac{\alpha_{s}C_{F}}{\pi}\times-\frac{3}{4}\Big(\ln\frac{p_{0}^{+}\mathbf{q}^{2}+p_{1}^{+}M^{2}}{q^{+}\Delta_{\mathrm{UV}}}-2\Big)\;.\end{aligned}
\label{eq:SE4L_fin_cross}
\end{equation}

We will show below that the contributions $\tilde{\mathcal{M}}_{\mathrm{SE1,sub}}^{0}\Big|_{1/k^{+}}$
and $\tilde{\mathcal{M}}_{\mathrm{SE4,sub}}^{0}\Big|_{1/k^{+}}$ nicely
combine with similar contributions from the vertex corrections. First,
let us write down the contributions from the amplitude $\tilde{\mathcal{M}}_{\mathrm{SE2,sub}}^{0}$
(\ref{eq:MSE2L_sub}), which reads after subtracting the high-energy
logarithms and multiplying with the LO amplitude:
\begin{equation}
\begin{aligned}&\mathrm{Tr}\,\mathcal{M}_{\mathrm{LO}}^{0\dagger}\mathcal{M}_{\mathrm{SE2,sub}}^{0} \\& =\frac{g_{\mathrm{em}}^{2}\alpha_{s}}{M^{2}}8p_{1}^{+}p_{0}^{+}\Big(\frac{p_{0}^{+}\mathbf{P}_{\perp}^{2}-p_{1}^{+}M^{2}}{p_{0}^{+}\mathbf{P}_{\perp}^{2}+p_{1}^{+}M^{2}}-\frac{p_{0}^{+}\mathbf{q}^{2}-p_{1}^{+}M^{2}}{p_{0}^{+}\mathbf{q}^{2}+p_{1}^{+}M^{2}}\Big)\frac{p_{0}^{+}\mathbf{P}_{\perp}^{2}-p_{1}^{+}M^{2}}{p_{0}^{+}\mathbf{P}_{\perp}^{2}+p_{1}^{+}M^{2}}\\
 & \times\int_{\mathbf{x},\mathbf{x}^{\prime}}e^{-i\mathbf{k}_{\perp}\cdot(\mathbf{x}-\mathbf{x}^{\prime})}\Bigg\{\int_{k_{\mathrm{min}}^{+}}^{p_{0}^{+}}\frac{\mathrm{d}k^{+}}{k^{+}}\Big(\frac{k^{+}}{p_{0}^{+}}\Big)^{2}\Big(\big(1-\frac{2p_{0}^{+}}{k^{+}}\big)^{2}+1\Big)\\
 & \times\Bigg[\int_{\mathbf{z}}A^{i}(\mathbf{x}-\mathbf{z})A^{i}\big(\mathbf{x}-\mathbf{z},\Delta_{\mathrm{P}}\big)e^{i\frac{k^{+}}{p_{0}^{+}}\mathbf{k}_{\perp}\cdot(\mathbf{x}-\mathbf{z})}\big(\frac{N_{c}^{2}}{2}s_{\mathbf{z}\mathbf{x}^{\prime}}s_{\mathbf{x}\mathbf{z}}-\frac{1}{2}s_{\mathbf{x}\mathbf{x}^{\prime}}+C_{F}N_{c}\big)\\
 & -\mathcal{A}_{0}(\Delta_{\mathrm{UV}})C_{F}N_{c}\big(s_{\mathbf{x}\mathbf{x}^{\prime}}+1\big)\Bigg]\\
 & -4\int_{k_{\mathrm{min}}^{+}}^{k_{f}^{+}}\frac{\mathrm{d}k^{+}}{k^{+}}\Bigg[\int_{\mathbf{z}}A^{i}(\mathbf{x}-\mathbf{z})A^{i}(\mathbf{x}-\mathbf{z})\big(\frac{N_{c}^{2}}{2}s_{\mathbf{z}\mathbf{x}^{\prime}}s_{\mathbf{x}\mathbf{z}}-\frac{1}{2}s_{\mathbf{x}\mathbf{x}^{\prime}}+C_{F}N_{c}\big)\\
 & -\mathcal{A}_{0}(\Delta_{\mathrm{UV}})C_{F}N_{c}\big(s_{\mathbf{x}\mathbf{x}^{\prime}}+1\big)\Bigg]\Bigg\}\;.
\end{aligned}
\label{eq:SE2L_cross}
\end{equation}
Similarly, $\tilde{\mathcal{M}}_{\mathrm{SE3,sub}}^{0}$ (\ref{eq:MSE3L_sub})
leads to the contribution:
\begin{equation}
\begin{aligned}&\mathrm{Tr}\,\mathcal{M}_{\mathrm{LO}}^{0\dagger}\mathcal{M}_{\mathrm{SE3,sub}}^{0} \\& =\frac{g_{\mathrm{em}}^{2}\alpha_{s}}{M^{2}}8p_{1}^{+}p_{0}^{+}\Big(\frac{p_{0}^{+}\mathbf{P}_{\perp}^{2}-p_{1}^{+}M^{2}}{p_{0}^{+}\mathbf{P}_{\perp}^{2}+p_{1}^{+}M^{2}}-\frac{p_{0}^{+}\mathbf{q}^{2}-p_{1}^{+}M^{2}}{p_{0}^{+}\mathbf{q}^{2}+p_{1}^{+}M^{2}}\Big)\Big(-\frac{p_{0}^{+}\mathbf{q}^{2}-p_{1}^{+}M^{2}}{p_{0}^{+}\mathbf{q}^{2}+p_{1}^{+}M^{2}}\Big)\\
 & \times\int_{\mathbf{x},\mathbf{x}^{\prime}}e^{-i\mathbf{k}_{\perp}\cdot(\mathbf{x}-\mathbf{x}^{\prime})}\Bigg\{\int_{k_{\mathrm{min}}^{+}}^{p_{1}^{+}}\frac{\mathrm{d}k^{+}}{k^{+}}\Big(\frac{k^{+}}{p_{1}^{+}}\Big)^{2}\Big(\big(1-\frac{2p_{1}^{+}}{k^{+}}\big)^{2}+1\Big)\\
 & \times\Bigg[\int_{\mathbf{z}}A^{i}(\mathbf{x}-\mathbf{z})A^{i}(\mathbf{x}-\mathbf{z},\Delta_{\mathrm{q}})e^{i\frac{k^{+}}{p_{1}^{+}}\mathbf{k}_{\perp}\cdot(\mathbf{x}-\mathbf{z})}\big(\frac{N_{c}^{2}}{2}s_{\mathbf{z}\mathbf{x}^{\prime}}s_{\mathbf{x}\mathbf{z}}-\frac{1}{2}s_{\mathbf{x}\mathbf{x}^{\prime}}+C_{F}N_{c}\big)\\
 & -\mathcal{A}_{0}(\Delta_{\mathrm{UV}})C_{F}N_{c}\big(s_{\mathbf{x}\mathbf{x}^{\prime}}+1\big)\Bigg]\\
 & -4\int_{k_{\mathrm{min}}^{+}}^{k_{f}^{+}}\frac{\mathrm{d}k^{+}}{k^{+}}\Bigg[\int_{\mathbf{z}}A^{i}(\mathbf{x}-\mathbf{z})A^{i}(\mathbf{x}-\mathbf{z})\big(\frac{N_{c}^{2}}{2}s_{\mathbf{z}\mathbf{x}^{\prime}}s_{\mathbf{x}\mathbf{z}}-\frac{1}{2}s_{\mathbf{x}\mathbf{x}^{\prime}}+C_{F}N_{c}\big)\\
 & -\mathcal{A}_{0}(\Delta_{\mathrm{UV}})C_{F}N_{c}\big(s_{\mathbf{x}\mathbf{x}^{\prime}}+1\big)\Bigg]\Bigg\}\;.
\end{aligned}
\label{eq:SE3_L_cross}
\end{equation}

\paragraph{Vertex corrections}

The parts from $\tilde{\mathcal{M}}_{\mathrm{V1,sub}}^{0}$ (\ref{eq:MV1L_sub})
and $\tilde{\mathcal{M}}_{\mathrm{V4,sub}}^{0}$ (\ref{eq:MV4L_sub})
that contain rapidity divergences are:
\begin{equation}
\begin{aligned}{\mathcal{M}}_{\mathrm{V1,sub}}^{0}\Big|_{1/k^{+}} & ={\mathcal{M}}_{\mathrm{LO1}}^{0}\frac{\alpha_{s}C_{F}}{\pi}\int_{k_{\mathrm{min}}^{+}}^{p_{1}^{+}}\frac{\mathrm{d}k^{+}}{k^{+}}\ln\frac{\Delta_{\mathrm{UV}}}{\Delta_{\mathrm{P}}}\;,\\
{\mathcal{M}}_{\mathrm{V4,sub}}^{0}\Big|_{1/k^{+}} & ={\mathcal{M}}_{\mathrm{LO2}}^{0}\frac{\alpha_{s}C_{F}}{\pi}\int_{k_{\mathrm{min}}^{+}}^{p_{1}^{+}}\frac{\mathrm{d}k^{+}}{k^{+}}\ln\frac{\Delta_{\mathrm{UV}}}{\Delta_{\mathrm{q}}}\;.
\end{aligned}
\label{eq:V14_k}
\end{equation}
When combining them with their counterparts from diagrams $\mathrm{SE1}$
and $\mathrm{SE4}$ (\ref{eq:SE14_finite_k}), and multiplying with
the LO amplitude, we obtain:
\begin{equation}
\begin{aligned}\mathrm{Tr}\mathcal{M}_{\mathrm{LO}}^{0\dagger}\Big(\mathcal{M}_{\mathrm{SE4,sub}}^{0}+\mathcal{M}_{\mathrm{V4,sub}}^{0}\Big)\Big|_{1/k^{+}} & =0\;,\\
\mathrm{Tr}\mathcal{M}_{\mathrm{LO}}^{0\dagger}\Big(\mathcal{M}_{\mathrm{SE1,sub}}^{0}+\mathcal{M}_{\mathrm{V1,sub}}^{0}\Big)\Big|_{1/k^{+}} & =\mathrm{Tr}\big(\mathcal{M}_{\mathrm{LO}}^{0\dagger}\mathcal{M}_{\mathrm{LO1}}^{0}\big)\frac{\alpha_{s}C_{F}}{\pi}\int_{p_{1}^{+}}^{p_{0}^{+}}\frac{\mathrm{d}k^{+}}{k^{+}}\ln\frac{\Delta_{\mathrm{P}}}{\Delta_{\mathrm{UV}}}\;.
\end{aligned}
\label{eq:SE_V_no_rap}
\end{equation}
The plus-momentum integral in the latter expression can be evaluated
explicitly, and yields:
\begin{equation}
\begin{aligned} & \mathrm{Tr}\mathcal{M}_{\mathrm{LO}}^{0\dagger}\Big(\mathcal{M}_{\mathrm{SE1,sub}}^{0}+\mathcal{M}_{\mathrm{V1,sub}}^{0}\Big)\Big|_{1/k^{+}}\\
 & =\mathrm{Tr}\big(\mathcal{M}_{\mathrm{LO}}^{0\dagger}\mathcal{M}_{\mathrm{LO1}}^{0}\big)\frac{\alpha_{s}C_{F}}{\pi}\Bigg[\ln\frac{p_{0}^{+}}{p_{1}^{+}}\Big(-i\pi+\ln\frac{p_{1}^{+}\big(p_{0}^{+}\mathbf{P}_{\perp}^{2}+p_{1}^{+}M^{2}\big)}{p_{0}^{+}q^{+}\Delta_{\mathrm{UV}}}\Big)\\
 & +\frac{1}{2}\ln^{2}\frac{p_{0}^{+}}{p_{1}^{+}}+\ln\frac{p_{0}^{+}}{p_{1}^{+}}\ln\frac{q^{+}}{p_{1}^{+}}-\mathrm{Li}_{2}\frac{q^{+}}{p_{0}^{+}}\Bigg]\;,
\end{aligned}
\label{eq:SE1+V1_k}
\end{equation}
where we used the identity (see \cite{Gradshteyn:1943cpj}):
\begin{equation}
\begin{aligned}\int_{b}^{c}\frac{\mathrm{d}x}{x}\ln\big(x(a-x)\big) & =\frac{1}{2}\ln^{2}c-\frac{1}{2}\ln^{2}b+\ln a\ln\frac{c}{b}-\mathrm{Li}_{2}\frac{c}{a}+\mathrm{Li}_{2}\frac{b}{a}\;.\end{aligned}
\label{eq:log_identity_k}
\end{equation}
Two parts are left of $\tilde{\mathcal{M}}_{\mathrm{V1,sub}}^{0}$
after subtracting the rapidity-divergent part (\ref{eq:V14_k}), namely:
\begin{equation}
\begin{aligned} & \mathrm{Tr}\mathcal{M}_{\mathrm{LO}}^{0\dagger}\mathcal{M}_{\mathrm{V1,sub}}^{0}\Big|_{\mathrm{finite}} \\ & =\mathrm{Tr}\big(\mathcal{M}_{\mathrm{LO}}^{0\dagger}\mathcal{M}_{\mathrm{LO1}}^{0}\big)\frac{\alpha_{s}C_{F}}{\pi}\Bigg\{\frac{2p_{0}^{+}+p_{1}^{+}}{4p_{0}^{+}}\Big(-i\pi+\ln\frac{p_{1}^{+}\big(p_{0}^{+}\mathbf{P}_{\perp}^{2}+p_{1}^{+}M^{2}\big)}{p_{0}^{+}q^{+}\Delta_{\mathrm{UV}}}\Big)\\
 & -\frac{3}{4}\frac{p_{0}^{+}+p_{1}^{+}}{p_{0}^{+}}+\frac{p_{0}^{+}+2p_{1}^{+}}{4p_{1}^{+}}\ln\frac{p_{0}^{+}}{p_{1}^{+}}-\frac{q^{+}(p_{0}^{+}+p_{1}^{+})}{4p_{1}^{+}p_{0}^{+}}\ln\frac{q^{+}}{p_{1}^{+}}\\
 & -\frac{M^{2}\mathbf{P}_{\perp}^{2}}{p_{0}^{+}\mathbf{P}_{\perp}^{2}-p_{1}^{+}M^{2}}\int_{0}^{p_{1}^{+}}\frac{\mathrm{d}k^{+}}{k^{+}}\frac{(k^{+})^{4}}{p_{0}^{+}(p_{1}^{+})^{2}}\Big(\big(2\frac{p_{1}^{+}}{k^{+}}-1\big)\big(2\frac{p_{0}^{+}}{k^{+}}-1\big)+1\Big)\\
 & \times2\pi\mathcal{B}_{1}\big(0,\Delta_{\mathrm{P}},\frac{k^{+}}{p_{1}^{+}}\mathbf{P}_{\perp}\big)\Bigg]\Bigg\}\;,
\end{aligned}
\label{eq:V1L_finite_cross}
\end{equation}
and:
\begin{equation}
\begin{aligned}  &\mathrm{Tr}\mathcal{M}_{\mathrm{LO}}^{0\dagger}\mathcal{M}_{\mathrm{V1,sub}}^{0}\Big|_{\mathrm{spurious}}
  =\mathrm{Tr}\big(\mathcal{M}_{\mathrm{LO}}^{0\dagger}\mathcal{M}_{\mathrm{LO1}}^{0}\big)\frac{\alpha_{s}C_{F}}{4\pi}\frac{p_{0}^{+}\mathbf{P}_{\perp}^{2}+p_{1}^{+}M^{2}}{p_{0}^{+}\mathbf{P}_{\perp}^{2}-p_{1}^{+}M^{2}}\\
 & \times\int_{0}^{p_{1}^{+}}\frac{\mathrm{d}k^{+}}{k^{+}}\frac{(k^{+})^{3}q^{+}}{(p_{0}^{+})^{2}p_{1}^{+}(p_{1}^{+}-k^{+})}\Big(\big(2\frac{p_{1}^{+}}{k^{+}}-1\big)\big(2\frac{p_{0}^{+}}{k^{+}}-1\big)+1\Big)\ln\frac{\Delta_{\mathrm{UV}}}{\Delta_{\mathrm{P}}}\;.
\end{aligned}
\label{eq:V1L_spurious}
\end{equation}
This last contribution still contains a divergence, but it is an unphysical
once. We will show in the next paragraph that it cancels with a similar
spurious pole in $\tilde{\mathcal{M}}_{\mathrm{A1,sub}}^{0}$. Also
the next vertex correction $\mathcal{M}_{\mathrm{V2,sub}}^{0}$ (\ref{eq:MV2L_sub})
yields both a finite and an unphysical divergent contribution to
the amplitude, and also here, it will cancel with a similar contribution
from $\mathcal{M}_{\mathrm{A2,sub}}^{0}$:
\begin{equation}
\begin{aligned}\mathrm{Tr}\mathcal{M}_{\mathrm{LO}}^{0\dagger}\mathcal{M}_{\mathrm{V2,sub}}^{0}\Big|_{\mathrm{finite}} & =\frac{g_{\mathrm{em}}^{2}\alpha_{s}}{M^{2}}8p_{1}^{+}p_{0}^{+}\Bigg(\frac{p_{0}^{+}\mathbf{P}_{\perp}^{2}-p_{1}^{+}M^{2}}{p_{0}^{+}\mathbf{P}_{\perp}^{2}+p_{1}^{+}M^{2}}-\frac{p_{0}^{+}\mathbf{q}^{2}-p_{1}^{+}M^{2}}{p_{0}^{+}\mathbf{q}^{2}+p_{1}^{+}M^{2}}\Bigg)\\
 & \times\int_{0}^{p_{1}^{+}}\frac{\mathrm{d}k^{+}}{k^{+}}\frac{(k^{+})^{3}q^{+}}{(p_{0}^{+})^{2}p_{1}^{+}(p_{1}^{+}-k^{+})}\mathrm{Dirac}_{\mathrm{V}}^{\bar{\eta}0\eta^{\prime}}\\
 & \times\Bigg[\int_{\mathbf{x},\mathbf{z}}iA^{\eta^{\prime}}(\mathbf{x}-\mathbf{z})\int_{\boldsymbol{\ell}}\frac{\boldsymbol{\ell}^{\bar{\eta}}}{\boldsymbol{\ell}^{2}}\frac{e^{i\big(\boldsymbol{\ell}+\frac{k^{+}}{p_{1}^{+}}\mathbf{P}_{\perp}\big)\cdot(\mathbf{x}-\mathbf{z})}}{\big(\boldsymbol{\ell}+\frac{k^{+}}{p_{1}^{+}}\mathbf{P}_{\perp}\big)^{2}+\Delta_{\mathrm{P}}}\\
 & \times\Big(\big(\boldsymbol{\ell}-\frac{p_{0}^{+}(p_{1}^{+}-k^{+})}{q^{+}p_{1}^{+}}\mathbf{P}_{\perp}\big)^{2}-\frac{(p_{1}^{+}-k^{+})(p_{0}^{+}-k^{+})}{(q^{+})^{2}}M^{2}\Big)\\
 & \times e^{-i\mathbf{k}_{\perp}\cdot\Big(\frac{p_{0}^{+}-k^{+}}{p_{0}^{+}}\mathbf{x}+\frac{k^{+}}{p_{0}^{+}}\mathbf{z}\Big)}\Big(\frac{N_{c}^{2}}{2}s_{\mathbf{z}\mathbf{x}^{\prime}}s_{\mathbf{x}\mathbf{z}}-\frac{1}{2}s_{\mathbf{x}\mathbf{x}^{\prime}}+C_{F}N_{c}\Big)\Bigg]\;,
\end{aligned}
\label{eq:V2L_finite_cross}
\end{equation}
and:
\begin{equation}
\begin{aligned}\mathrm{Tr}\mathcal{M}_{\mathrm{LO}}^{0\dagger}\mathcal{M}_{\mathrm{V2,sub}}^{0}\Big|_{\mathrm{spurious}} & =\frac{g_{\mathrm{em}}^{2}\alpha_{s}}{M^{2}}8p_{1}^{+}p_{0}^{+}\Bigg(\frac{p_{0}^{+}\mathbf{P}_{\perp}^{2}-p_{1}^{+}M^{2}}{p_{0}^{+}\mathbf{P}_{\perp}^{2}+p_{1}^{+}M^{2}}-\frac{p_{0}^{+}\mathbf{q}^{2}-p_{1}^{+}M^{2}}{p_{0}^{+}\mathbf{q}^{2}+p_{1}^{+}M^{2}}\Bigg)\\
 & \times\int_{0}^{p_{1}^{+}}\frac{\mathrm{d}k^{+}}{k^{+}}\frac{(k^{+})^{3}q^{+}}{(p_{0}^{+})^{2}p_{1}^{+}(p_{1}^{+}-k^{+})}\Big(\big(2\frac{p_{1}^{+}}{k^{+}}-1\big)\big(2\frac{p_{0}^{+}}{k^{+}}-1\big)+1\Big)\\
 & \times\mathcal{A}_{0}(\Delta_{\mathrm{UV}})\int_{\mathbf{x},\mathbf{x}^{\prime}}e^{-i\mathbf{k}_{\perp}\cdot(\mathbf{x}-\mathbf{x}^{\prime})}C_{F}N_{c}\big(s_{\mathbf{x}\mathbf{x}^{\prime}}+1\big)\;.
\end{aligned}
\label{eq:V2L_spurious}
\end{equation}
The last two vertex-correction amplitudes $\mathcal{M}_{\mathrm{V3,sub}}^{0}$
(\ref{eq:MV3L_sub}) and $\mathcal{M}_{\mathrm{V4,sub}}^{0}$ (\ref{eq:MV4L_sub})
lead to the following contributions to the cross section:
\begin{equation}
\begin{aligned}\mathrm{Tr}\mathcal{M}_{\mathrm{LO}}^{0\dagger}\mathcal{M}_{\mathrm{V3,sub}}^{0} & =-g_{\mathrm{em}}^{2}\alpha_{s}8p_{1}^{+}p_{0}^{+}\Bigg(\frac{p_{0}^{+}\mathbf{P}_{\perp}^{2}-p_{1}^{+}M^{2}}{p_{0}^{+}\mathbf{P}_{\perp}^{2}+p_{1}^{+}M^{2}}-\frac{p_{0}^{+}\mathbf{q}^{2}-p_{1}^{+}M^{2}}{p_{0}^{+}\mathbf{q}^{2}+p_{1}^{+}M^{2}}\Bigg)\frac{1}{M^{2}}\\
 & \times\int_{\mathbf{x},\mathbf{x}^{\prime}}e^{-i\mathbf{k}_{\perp}\cdot(\mathbf{x}-\mathbf{x}^{\prime})}\int_{0}^{p_{1}^{+}}\frac{\mathrm{d}k^{+}}{k^{+}}\frac{(k^{+})^{3}q^{+}}{(p_{1}^{+})^{2}p_{0}^{+}(p_{0}^{+}-k^{+})}\mathrm{Dirac}_{\mathrm{V}}^{\bar{\eta}0\eta^{\prime}}\\
 & \times\Bigg[\int_{\mathbf{x},\mathbf{z}}iA^{\bar{\eta}}(\mathbf{x}-\mathbf{z})\int_{\boldsymbol{\mathbf{\ell}}}\frac{\boldsymbol{\ell}^{\eta^{\prime}}}{\boldsymbol{\ell}^{2}}\frac{e^{i\big(\boldsymbol{\ell}+\frac{k^{+}}{p_{1}^{+}}\mathbf{q}\big)\cdot(\mathbf{z}-\mathbf{x})}}{\big(\boldsymbol{\ell}+\frac{k^{+}}{p_{1}^{+}}\mathbf{q}\big)^{2}+\Delta_{\mathrm{q}}}\\
 & \times\Big(\big(\boldsymbol{\ell}+\frac{p_{0}^{+}-k^{+}}{q^{+}}\mathbf{q}\big)^{2}-\frac{(p_{1}^{+}-k^{+})(p_{0}^{+}-k^{+})}{(q^{+})^{2}}M^{2}\Big)\\
 & \times e^{i\frac{k^{+}}{p_{1}^{+}}\mathbf{k}_{\perp}\cdot(\mathbf{x}-\mathbf{z})}\Big(\frac{N_{c}^{2}}{2}s_{\mathbf{z}\mathbf{x}^{\prime}}s_{\mathbf{x}\mathbf{z}}-\frac{1}{2}s_{\mathbf{x}\mathbf{x}^{\prime}}+C_{F}N_{c}\Big)\\
 & +\frac{\delta^{\bar{\eta}\eta^{\prime}}}{2}\mathcal{A}_{0}(\Delta_{\mathrm{UV}})C_{F}N_{c}\big(s_{\mathbf{x}\mathbf{x}^{\prime}}+1\big)\Bigg]\;,
\end{aligned}
\label{eq:V3L_cross}
\end{equation}
and:
\begin{equation}
\begin{aligned} & \mathrm{Tr}\mathcal{M}_{\mathrm{LO}}^{0\dagger}\mathcal{M}_{\mathrm{V4,sub}}^{0} \\ & =\mathrm{Tr}\big(\mathcal{M}_{\mathrm{LO}}^{0\dagger}\mathcal{M}_{\mathrm{LO2}}^{0}\big)\frac{\alpha_{s}C_{F}}{\pi}\Bigg\{-\frac{2p_{0}^{+}+p_{1}^{+}}{2p_{0}^{+}}\Big(1+\frac{1}{2}\ln\frac{q^{+}\Delta_{\mathrm{UV}}}{p_{0}^{+}\mathbf{q}^{2}+p_{1}^{+}M^{2}}\Big)\\
 & \times \frac{-1}{p_{0}^{+}\mathbf{q}^{2}-p_{1}^{+}M^{2}}\Bigg[-\frac{q^{+}\big(p_{1}^{+}+2p_{0}^{+}\ln p_{0}^{+}/q^{+}\big)}{4p_{1}^{+}p_{0}^{+}}(p_{0}^{+}\mathbf{q}^{2}+p_{1}^{+}M^{2})\ln\frac{q^{+}\Delta_{\mathrm{UV}}}{(p_{0}^{+}\mathbf{q}^{2}+p_{1}^{+}M^{2})}\\
 & +\frac{q^{+}}{6p_{0}^{+}}\Big(-3+\frac{p_{0}^{+}}{p_{1}^{+}}\pi^{2}+\frac{3}{2}\frac{p_{0}^{+}}{p_{1}^{+}}\ln\frac{q^{+}}{p_{0}^{+}}\big(4i\pi-3\ln\frac{q^{+}}{p_{1}^{+}}-\ln\frac{p_{0}^{+}}{p_{1}^{+}}\big)-6\frac{p_{0}^{+}}{p_{1}^{+}}\mathrm{Li}_{2}\frac{p_{0}^{+}}{q^{+}}\Big)\Bigg]\\
 & -\int_{0}^{p_{1}^{+}}\frac{\mathrm{d}k^{+}}{k^{+}}\frac{4(k^{+})^{2}\big((k^{+})^{2}+2p_{1}^{+}p_{0}^{+}-k^{+}(p_{0}^{+}+p_{1}^{+})\big)}{p_{1}^{+}(p_{0}^{+})^{2}}\mathbf{q}^{2}M^{2}\pi\mathcal{B}_{1}\big(0,\Delta_{\mathrm{q}},\frac{k^{+}}{p_{1}^{+}}\mathbf{q}\big)\Bigg\}\;.
\end{aligned}
\label{eq:V4L_cross}
\end{equation}

\paragraph{Antiquark vertex corrections}

As already announced in the previous paragraph, $\tilde{\mathcal{M}}_{\mathrm{A1,sub}}^{0}$
(\ref{eq:MA1L_sub}) can be split in both a finite and a divergent
contribution to the cross section:
\begin{equation}
\begin{aligned} & \mathrm{Tr}\mathcal{M}_{\mathrm{LO}}^{0\dagger}\mathcal{M}_{\mathrm{A1,sub}}^{0}\Big|_{\mathrm{finite}} \\ 
& =\mathrm{Tr}\big(\mathcal{M}_{\mathrm{LO}}^{0}\mathcal{M}_{\mathrm{LO1}}^{0}\big)\frac{\alpha_{s}C_{F}}{\pi}\int_{p_{1}^{+}}^{p_{0}^{+}}\frac{\mathrm{d}k^{+}}{k^{+}}\frac{k^{+}p_{1}^{+}(p_{0}^{+}-k^{+})}{p_{0}^{+}q^{+}}\Big(\big(2\frac{p_{1}^{+}}{k^{+}}-1\big)\big(2\frac{p_{0}^{+}}{k^{+}}-1\big)+1\Big)\\
 & \times\frac{-1}{p_{0}^{+}\mathbf{P}_{\perp}^{2}-p_{1}^{+}M^{2}}\Bigg\{\frac{M^{2}p_{1}^{+}}{2}\Bigg[\Big(\frac{(7p_{0}^{+}+p_{1}^{+})}{p_{0}^{+}}-\frac{2(p_{0}^{+}+3p_{1}^{+})}{q^{+}}\ln\frac{p_{0}^{+}}{p_{1}^{+}}\Big)\Big(-i\mathrm{\pi}+\ln\frac{\Delta_{\mathrm{UV}}}{M^{2}}\Big)\\
 & +6+2\frac{p_{1}^{+}}{p_{0}^{+}}+4\frac{p_{0}^{+}+p_{1}^{+}}{q^{+}}\ln\frac{p_{0}^{+}}{p_{1}^{+}}\\
 & +2\frac{p_{0}^{+}+3p_{1}^{+}}{q^{+}}\big(-\frac{\pi^{2}}{6}+\ln^{2}\frac{p_{0}^{+}}{p_{1}^{+}}+3\ln\frac{p_{0}^{+}}{p_{1}^{+}}\ln\frac{q^{+}}{p_{1}^{+}}+\mathrm{Li}_{2}\frac{q^{+}}{p_{1}^{+}}+\mathrm{Li}_{2}\frac{p_{1}^{+}}{p_{0}^{+}}\big)\Big)\Bigg]\\
 & -2M^{2}\Delta_{\mathrm{P}}\pi\mathcal{B}_{0}\big(\Delta_{\mathrm{P}},\hat{M}^{2},\frac{p_{0}^{+}-k^{+}}{q^{+}}\mathbf{P}_{\perp}\big)\\
 & +2\frac{k^{+}}{p_{1}^{+}}\frac{p_{0}^{+}-k^{+}}{q^{+}}M^{2}\mathbf{P}_{\perp}^{2}\pi\mathcal{B}_{1}\big(\Delta_{\mathrm{P}},\hat{M}^{2},\frac{p_{0}^{+}-k^{+}}{q^{+}}\mathbf{P}_{\perp}\big)\Bigg\}\;.
\end{aligned}
\label{eq:A1L_cross-1}
\end{equation}
and:
\begin{equation}
\begin{aligned} & \mathrm{Tr}\mathcal{M}_{\mathrm{LO}}^{0\dagger}\mathcal{M}_{\mathrm{A1,sub}}^{0}\Big|_{\mathrm{spurious}}  =\mathrm{Tr}\big(\mathcal{M}_{\mathrm{LO}}^{0\dagger}\mathcal{M}_{\mathrm{LO1}}^{0}\big)\frac{\alpha_{s}C_{F}}{\pi}\frac{p_{0}^{+}\mathbf{P}_{\perp}^{2}+p_{1}^{+}M^{2}}{p_{0}^{+}\mathbf{P}_{\perp}^{2}-p_{1}^{+}M^{2}}\\
 & \times\int_{p_{1}^{+}}^{p_{0}^{+}}\frac{\mathrm{d}k^{+}}{k^{+}}\frac{(k^{+})^{2}(p_{0}^{+}-k^{+})}{(p_{0}^{+})^{2}(p_{1}^{+}-k^{+})}\Big(\big(2\frac{p_{1}^{+}}{k^{+}}-1\big)\big(2\frac{p_{0}^{+}}{k^{+}}-1\big)+1\Big)\frac{1}{4}\ln\frac{\Delta_{\mathrm{UV}}}{\Delta_{\mathrm{P}}}\;.
\end{aligned}
\label{eq:A1L_spurious_cross}
\end{equation}
Combining eqs. (\ref{eq:A1L_spurious_cross}) and (\ref{eq:V1L_spurious}),
we see that the pathological integrals can be combined and evaluated
with the help of the Sokhotski--Plemelj theorem: 
\begin{equation}
\begin{aligned} & \Bigg[\int_{0}^{p_{1}^{+}}\frac{\mathrm{d}k^{+}}{k^{+}}\frac{(k^{+})^{3}q^{+}}{(p_{0}^{+})^{2}p_{1}^{+}(p_{1}^{+}\!-\!k^{+})}\!+\!\int_{p_{1}^{+}}^{p_{0}^{+}}\frac{\mathrm{d}k^{+}}{k^{+}}\frac{(k^{+})^{2}(p_{0}^{+}\!-\!k^{+})}{(p_{0}^{+})^{2}(p_{1}^{+}\!-\!k^{+})}\Bigg]\!\Big(\!\big(2\frac{p_{1}^{+}}{k^{+}}\!-\!1\big)\!\big(2\frac{p_{0}^{+}}{k^{+}}\!-\!1\big)\!+\!1\!\Big)\\
 & =\frac{q^{+}(2p_{0}^{+}-p_{1}^{+})}{(p_{0}^{+})^{2}}-\frac{q^{+}(3p_{0}^{+}-p_{1}^{+})}{(p_{0}^{+})^{2}}+4\ln\frac{p_{0}^{+}}{p_{1}^{+}}-\frac{2q^{+}}{p_{0}^{+}}\int_{0}^{p_{0}^{+}}\frac{\mathrm{d}k^{+}}{k^{+}-p_{1}^{+}+i0^{+}}\\
 & =4\ln\frac{p_{0}^{+}}{p_{1}^{+}}-\frac{2q^{+}}{p_{0}^{+}}\Big(\frac{1}{2}-i\pi+\ln\frac{q^{+}}{p_{1}^{+}}\Big)\;,
\end{aligned}
\label{eq:Plemelj_cross}
\end{equation}
and:
\begin{equation}
\begin{aligned} & \Bigg[\int_{0}^{p_{1}^{+}}\frac{\mathrm{d}k^{+}}{k^{+}}\frac{(k^{+})^{3}q^{+}}{(p_{0}^{+})^{2}p_{1}^{+}(p_{1}^{+}-k^{+})}+\int_{p_{1}^{+}}^{p_{0}^{+}}\frac{\mathrm{d}k^{+}}{k^{+}}\frac{(k^{+})^{2}(p_{0}^{+}-k^{+})}{(p_{0}^{+})^{2}(p_{1}^{+}-k^{+})}\Bigg] \\
& \times \Big(\big(2\frac{p_{1}^{+}}{k^{+}}-1\big)\big(2\frac{p_{0}^{+}}{k^{+}}-1\big)+1\Big)\ln\frac{k^{+}(p_{0}^{+}-k^{+})}{(p_{1}^{+})^{2}}\\
 & =\frac{3q^{+}}{p_{0}^{+}}+\frac{p_{0}^{+}-4p_{1}^{+}}{p_{1}^{+}}\ln\frac{p_{0}^{+}}{p_{1}^{+}}+\frac{(p_{1}^{+})^{2}-(p_{0}^{+})^{2}}{p_{0}^{+}p_{1}^{+}}\ln\frac{q^{+}}{p_{1}^{+}}+2\ln^{2}\frac{p_{0}^{+}}{p_{1}^{+}}+4\ln\frac{p_{0}^{+}}{p_{1}^{+}}\ln\frac{q^{+}}{p_{1}^{+}}\\
 & -4\mathrm{Li}_{2}\frac{q^{+}}{p_{0}^{+}}-\frac{2q^{+}}{p_{0}^{+}}\Big(\ln^{2}\frac{q^{+}}{p_{1}^{+}}+\mathrm{Li}_{2}\big(-\frac{p_{1}^{+}}{q^{+}}\big)-\mathrm{Li}_{2}\big(-\frac{q^{+}}{p_{1}^{+}}\big)+i\pi\ln\frac{q^{+}}{p_{1}^{+}}\Big)\;.
\end{aligned}
\label{eq:Plemelj_log_cross}
\end{equation}
We end up with the well-behaved result:
\begin{equation}
\begin{aligned} & \mathrm{Tr}\mathcal{M}_{\mathrm{LO}}^{0\dagger}\Big(\mathcal{M}_{\mathrm{V1,sub}}^{0}+\mathcal{M}_{\mathrm{A1,sub}}^{0}\Big)\Big|_{\mathrm{spurious}}\\
 & =\mathrm{Tr}\big(\mathcal{M}_{\mathrm{LO}}^{0\dagger}\mathcal{M}_{\mathrm{LO1}}^{0}\big)\frac{\alpha_{s}C_{F}}{4\pi}\frac{p_{0}^{+}\mathbf{P}_{\perp}^{2}+p_{1}^{+}M^{2}}{p_{0}^{+}\mathbf{P}_{\perp}^{2}-p_{1}^{+}M^{2}}\\
 & \times\Bigg[\Big(4\ln\frac{p_{0}^{+}}{p_{1}^{+}}-\frac{2q^{+}}{p_{0}^{+}}\big(\frac{1}{2}+i\pi+\ln\frac{q^{+}}{p_{1}^{+}}\big)\Big)\Big(i\pi+\ln\frac{p_{0}^{+}q^{+}\Delta_{\mathrm{UV}}}{p_{1}^{+}\big(p_{0}^{+}\mathbf{P}_{\perp}^{2}+p_{1}^{+}M^{2}\big)}\Big)\\
 & -\frac{3q^{+}}{p_{0}^{+}}-\frac{p_{0}^{+}-4p_{1}^{+}}{p_{1}^{+}}\ln\frac{p_{0}^{+}}{p_{1}^{+}}-\frac{(p_{1}^{+})^{2}-(p_{0}^{+})^{2}}{p_{0}^{+}p_{1}^{+}}\ln\frac{q^{+}}{p_{1}^{+}}-2\ln^{2}\frac{p_{0}^{+}}{p_{1}^{+}}-4\ln\frac{p_{0}^{+}}{p_{1}^{+}}\ln\frac{q^{+}}{p_{1}^{+}}\\
 &+4\mathrm{Li}_{2}\frac{q^{+}}{p_{0}^{+}} +\frac{2q^{+}}{p_{0}^{+}}\Big(\ln^{2}\frac{q^{+}}{p_{1}^{+}}+\mathrm{Li}_{2}\big(-\frac{p_{1}^{+}}{q^{+}}\big)-\mathrm{Li}_{2}\big(-\frac{q^{+}}{p_{1}^{+}}\big)+i\pi\ln\frac{q^{+}}{p_{1}^{+}}\Big)\Bigg]\;.
\end{aligned}
\label{eq:V1+A1_spurious}
\end{equation}
Similarly, for amplitude $\mathcal{M}_{\mathrm{A2,sub}}^{0}$ (\ref{eq:MA2L_sub}),
we find the finite part:
\begin{equation}
\begin{aligned}& \mathrm{Tr}\mathcal{M}_{\mathrm{LO}}^{0\dagger}\mathcal{M}_{\mathrm{A2,sub}}^{0}\Big|_{\mathrm{finite}} \\ & =g_{\mathrm{em}}^{2}\alpha_{s}8p_{1}^{+}p_{0}^{+}\Bigg(\frac{p_{0}^{+}\mathbf{P}_{\perp}^{2}-p_{1}^{+}M^{2}}{p_{0}^{+}\mathbf{P}_{\perp}^{2}+p_{1}^{+}M^{2}}-\frac{p_{0}^{+}\mathbf{q}^{2}-p_{1}^{+}M^{2}}{p_{0}^{+}\mathbf{q}^{2}+p_{1}^{+}M^{2}}\Bigg)\frac{1}{M^{2}}\\
 & \times-\int_{p_{1}^{+}}^{p_{0}^{+}}\frac{\mathrm{d}k^{+}}{k^{+}}\frac{(k^{+})^{2}(p_{0}^{+}-k^{+})}{(p_{0}^{+})^{2}(p_{1}^{+}-k^{+})}\mathrm{Dirac}_{\mathrm{V}}^{\bar{\eta}0\eta^{\prime}}\int_{\mathbf{x},\mathbf{x}^{\prime}}e^{-i\mathbf{k}_{\perp}\cdot(\mathbf{x}-\mathbf{x}^{\prime})}\\
 & \times\int_{\mathbf{x},\mathbf{z}}iA^{\eta^{\prime}}(\mathbf{x}-\mathbf{z})\int_{\boldsymbol{\ell}}e^{-i\boldsymbol{\ell}\cdot(\mathbf{x}-\mathbf{z})}\frac{\boldsymbol{\ell}^{\bar{\eta}}+\frac{k^{+}}{p_{1}^{+}}\mathbf{P}_{\perp}^{\bar{\eta}}}{\boldsymbol{\ell}^{2}+\Delta_{\mathrm{P}}}\frac{\Big(\boldsymbol{\ell}+\frac{p_{0}^{+}-k^{+}}{q^{+}}\mathbf{P}_{\perp}\Big)^{2}-\hat{M}^{2}}{\Big(\boldsymbol{\ell}+\frac{p_{0}^{+}-k^{+}}{q^{+}}\mathbf{P}_{\perp}\Big)^{2}+\hat{M}^{2}}\\
 & \times e^{i\frac{k^{+}}{p_{0}^{+}}\mathbf{k}_{\perp}\cdot(\mathbf{x}-\mathbf{z})}\big(\frac{N_{c}^{2}}{2}s_{\mathbf{x}\mathbf{z}}s_{\mathbf{z}\mathbf{x}^{\prime}}-\frac{1}{2}s_{\mathbf{x}\mathbf{x}^{\prime}}+C_{F}N_{c}\big)\;,
\end{aligned}
\label{eq:A2L_finite_cross}
\end{equation}
while the term with the spurious singularity is:
\begin{equation}
\begin{aligned}\mathrm{Tr}\mathcal{M}_{\mathrm{LO}}^{0\dagger}\mathcal{M}_{\mathrm{A2,sub}}^{0}\Big|_{\mathrm{spurious}} & =g_{\mathrm{em}}^{2}\alpha_{s}8p_{1}^{+}p_{0}^{+}\Bigg(\frac{p_{0}^{+}\mathbf{P}_{\perp}^{2}-p_{1}^{+}M^{2}}{p_{0}^{+}\mathbf{P}_{\perp}^{2}+p_{1}^{+}M^{2}}-\frac{p_{0}^{+}\mathbf{q}^{2}-p_{1}^{+}M^{2}}{p_{0}^{+}\mathbf{q}^{2}+p_{1}^{+}M^{2}}\Bigg)\frac{1}{M^{2}}\\
 & \times\int_{p_{1}^{+}}^{p_{0}^{+}}\frac{\mathrm{d}k^{+}}{k^{+}}\frac{(k^{+})^{2}(p_{0}^{+}-k^{+})}{(p_{0}^{+})^{2}(p_{1}^{+}-k^{+})}\Big(\big(2\frac{p_{1}^{+}}{k^{+}}-1\big)\big(2\frac{p_{0}^{+}}{k^{+}}-1\big)+1\Big)\\
 & \times\mathcal{A}_{0}(\Delta_{\mathrm{UV}})\int_{\mathbf{x},\mathbf{x}^{\prime}}e^{-i\mathbf{k}_{\perp}\cdot(\mathbf{x}-\mathbf{x}^{\prime})}C_{F}N_{c}\big(s_{\mathbf{x}\mathbf{x}^{\prime}}+1\big)\;.
\end{aligned}
\label{eq:A2L_spurious_cross}
\end{equation}
Combining the above with (\ref{eq:V2L_spurious}) and applying (\ref{eq:Plemelj_cross}):
\begin{equation}
\begin{aligned} & \mathrm{Tr}\mathcal{M}_{\mathrm{LO}}^{0\dagger}\Big(\mathcal{M}_{\mathrm{V2,sub}}^{0}+\mathcal{M}_{\mathrm{A2,sub}}^{0}\Big)\Big|_{\mathrm{spurious}} \\ & =\frac{g_{\mathrm{em}}^{2}\alpha_{s}}{M^{2}}8p_{1}^{+}p_{0}^{+}\Bigg(\frac{p_{0}^{+}\mathbf{P}_{\perp}^{2}-p_{1}^{+}M^{2}}{p_{0}^{+}\mathbf{P}_{\perp}^{2}+p_{1}^{+}M^{2}}-\frac{p_{0}^{+}\mathbf{q}^{2}-p_{1}^{+}M^{2}}{p_{0}^{+}\mathbf{q}^{2}+p_{1}^{+}M^{2}}\Bigg)\\
 & \times\Big(4\ln\frac{p_{0}^{+}}{p_{1}^{+}}-\frac{2q^{+}}{p_{0}^{+}}\big(\frac{1}{2}-i\pi+\ln\frac{q^{+}}{p_{1}^{+}}\big)\Big)\mathcal{A}_{0}(\Delta_{\mathrm{UV}})\int_{\mathbf{x},\mathbf{x}^{\prime}}e^{-i\mathbf{k}_{\perp}\cdot(\mathbf{x}-\mathbf{x}^{\prime})}C_{F}N_{c}\big(s_{\mathbf{x}\mathbf{x}^{\prime}}+1\big)\;.
\end{aligned}
\label{eq:V2L_A2L_spurious}
\end{equation}
The next contribution is due to amplitude $\mathcal{M}_{\mathrm{A3,sub}}^{0}$
(\ref{eq:MA3L_sub}):
\begin{equation}
\begin{aligned} & \mathrm{Tr}\mathcal{M}_{\mathrm{LO}}^{0\dagger}\mathcal{M}_{\mathrm{A3,sub}}^{0}   =\frac{g_{\mathrm{em}}^{2}\alpha_{s}}{M^{2}}8p_{1}^{+}p_{0}^{+}\Bigg(\frac{p_{0}^{+}\mathbf{P}_{\perp}^{2}-p_{1}^{+}M^{2}}{p_{0}^{+}\mathbf{P}_{\perp}^{2}+p_{1}^{+}M^{2}}-\frac{p_{0}^{+}\mathbf{q}^{2}-p_{1}^{+}M^{2}}{p_{0}^{+}\mathbf{q}^{2}+p_{1}^{+}M^{2}}\Bigg)\\
 & \times\Bigg\{\mathrm{Dirac}_{\mathrm{V}}^{\bar{\eta}0\eta^{\prime}}\int_{p_{1}^{+}}^{p_{0}^{+}}\frac{\mathrm{d}k^{+}}{k^{+}}\frac{k^{+}(p_{0}^{+}-k^{+})}{q^{+}p_{0}^{+}}\int_{\mathbf{x}_{1},\mathbf{x}_{2},\mathbf{x}_{3}}2\hat{M}^{2}\mathcal{K}\big(\mathbf{x}_{1}-\mathbf{x}_{2},\hat{M}^{2}\big)\\
 & \times\int_{\boldsymbol{\ell},\boldsymbol{\ell}_{2}}e^{-i\boldsymbol{\ell}\cdot\mathbf{x}_{12}}e^{-i\boldsymbol{\ell}_{2}\cdot\mathbf{x}_{23}}\frac{\boldsymbol{\ell}^{\eta^{\prime}}}{\mathbf{\boldsymbol{\ell}}^{2}}\frac{\boldsymbol{\ell}^{\bar{\eta}}-\frac{k^{+}}{p_{1}^{+}}\boldsymbol{\ell}_{2}^{\bar{\eta}}}{\Big(\boldsymbol{\ell}-\frac{p_{0}^{+}-k^{+}}{q^{+}}\boldsymbol{\ell}_{2}\Big)^{2}-\frac{p_{0}^{+}(p_{0}^{+}-k^{+})(p_{1}^{+}-k^{+})}{p_{1}^{+}(q^{+})^{2}}\boldsymbol{\ell}_{2}^{2}}\\
 & \times e^{-i\mathbf{p}_{1}\cdot\mathbf{x}_{3}}e^{-i\mathbf{q}\cdot\Big(\frac{p_{0}^{+}-k^{+}}{q^{+}}\mathbf{x}_{1}-\frac{p_{1}^{+}-k^{+}}{q^{+}}\mathbf{x}_{2}\Big)}\big(\frac{N_{c}^{2}}{2}s_{\mathbf{x}_{3}\mathbf{x}^{\prime}}s_{\mathbf{x}_{1}\mathbf{x}_{2}}-\frac{1}{2}Q_{\mathbf{x}_{1}\mathbf{x}^{\prime}\mathbf{x}_{3}\mathbf{x}_{2}}t^{c}+C_{F}N_{c}\big)\\
 & -\Big(\frac{7p_{0}^{+}+p_{1}^{+}}{p_{0}^{+}}+2\frac{p_{0}^{+}+3p_{1}^{+}}{q^{+}}\ln\frac{p_{1}^{+}}{p_{0}^{+}}\Big)\mathcal{A}_{0}(\Delta_{\mathrm{UV}})\int_{\mathbf{x},\mathbf{x}^{\prime}}e^{-i\mathbf{k}_{\perp}\cdot(\mathbf{x}-\mathbf{x}^{\prime})}C_{F}N_{c}\big(s_{\mathbf{x}\mathbf{x}^{\prime}}+1\big)\Bigg\}\;.
\end{aligned}
\label{eq:A3L_cross}
\end{equation}
Finally:
\begin{equation}
\begin{aligned} & \mathrm{Tr}\mathcal{M}_{\mathrm{LO}}^{0\dagger}\mathcal{M}_{\mathrm{A4,sub}}^{0}  =\mathrm{Tr}\big(\mathcal{M}_{\mathrm{LO}}^{0\dagger}\mathcal{M}_{\mathrm{LO2}}^{0}\big)\frac{\alpha_{s}C_{F}}{\pi}\\
 & \times\Bigg[\Big(\frac{7}{4}+\frac{p_{1}^{+}}{4p_{0}^{+}}-\frac{2(p_{0}^{+}+3p_{1}^{+})}{4q^{+}}\ln\frac{p_{0}^{+}}{p_{1}^{+}}\Big)\ln\frac{(q^{+})^{2}\Delta_{\mathrm{UV}}}{p_{0}^{+}p_{1}^{+}\mathbf{q}^{2}}+\frac{3p_{0}^{+}+p_{1}^{+}}{2p_{0}^{+}}-\frac{7p_{0}^{+}+p_{1}^{+}}{2p_{0}^{+}}\ln\frac{q^{+}}{p_{1}^{+}}\\
 & -\frac{p_{0}^{+}+p_{1}^{+}+(p_{0}^{+}+3p_{1}^{+})\ln\frac{q^{+}}{p_{1}^{+}}}{q^{+}}\ln\frac{p_{0}^{+}}{p_{1}^{+}}-\frac{p_{0}^{+}+3p_{1}^{+}}{2q^{+}}\Big(\mathrm{Li}_{2}\Big(-\frac{q^{+}}{p_{1}^{+}}\Big)-\mathrm{Li}_{2}\frac{q^{+}}{p_{0}^{+}}\Big)\\
 & +\int_{p_{1}^{+}}^{p_{0}^{+}}\frac{\mathrm{d}k^{+}}{k^{+}}\frac{(k^{+})^{2}(p_{0}^{+}\!-\!k^{+})^{2}}{p_{1}^{+}p_{0}^{+}(q^{+})^{2}}\Big(\big(2\frac{p_{1}^{+}}{k^{+}}\!-\!1\big)\big(2\frac{p_{0}^{+}}{k^{+}}\!-\!1\big)\!+\!1\Big)\mathbf{q}^{2}\pi\mathcal{B}_{1}\big(0,\hat{Q}^{2},\frac{p_{0}^{+}\!-\!k^{+}}{q^{+}}\mathbf{q}\big)\Bigg]\;.
\end{aligned}
\label{eq:A4L_cross}
\end{equation}

\paragraph{Instantaneous four-fermion interaction}

We obtain the following cross-section contributions, which are all finite:
\begin{equation}
\begin{aligned} & \mathrm{Tr}\,\mathcal{M}_{\mathrm{LO}}^{0\dagger}\mathcal{M}_{\mathrm{Q1,sub}}^{0}  =\mathrm{Tr}\big(\mathcal{M}_{\mathrm{LO}}^{0\dagger}\mathcal{M}_{\mathrm{LO1}}^{0}\big)\\
& \times\frac{\alpha_{s}C_{F}}{\pi}\frac{M^{2}}{p_{0}^{+}\mathbf{P}_{\perp}^{2}-p_{1}^{+}M^{2}}\frac{2p_{1}^{+}}{q^{+}}\Bigg[\Big(2q^{+}-(p_{0}^{+}+p_{1}^{+})\ln\frac{p_{0}^{+}}{p_{1}^{+}}\Big)\Big(i\pi+\ln\frac{\Delta_{\mathrm{UV}}}{M^{2}}\Big)\\
 & +2q^{+}+(p_{0}^{+}+p_{1}^{+})\Big(\ln\frac{p_{0}^{+}}{p_{1}^{+}}+\mathrm{Li}_{2}\Big(-\frac{q^{+}}{p_{1}^{+}}\Big)-\mathrm{Li}_{2}\Big(\frac{q^{+}}{p_{0}^{+}}\Big)\Big)\Bigg]\;,
\end{aligned}
\label{eq:Q1_sub_cross}
\end{equation}
and:
\begin{equation}
\begin{aligned} & \mathrm{Tr}\,\mathcal{M}_{\mathrm{LO}}^{0\dagger}\mathcal{M}_{\mathrm{Q2,sub}}^{0} \\
& =g_{\mathrm{em}}^{2}\Big(\frac{p_{0}^{+}\mathbf{P}_{\perp}^{2}-p_{1}^{+}M^{2}}{p_{0}^{+}\mathbf{P}_{\perp}^{2}+p_{1}^{+}M^{2}}-\frac{p_{0}^{+}\mathbf{q}^{2}-p_{1}^{+}M^{2}}{p_{0}^{+}\mathbf{q}^{2}+p_{1}^{+}M^{2}}\Big)8p_{1}^{+}p_{0}^{+}\\
 & \times\frac{4\alpha_{s}}{M^{2}}\Bigg[\int_{0}^{q^{+}}\mathrm{d}\ell_{1}^{+}\frac{1}{(p_{0}^{+}-\ell_{1}^{+})^{2}}\frac{p_{1}^{+}\ell_{1}^{+}}{p_{1}^{+}+\ell_{1}^{+}}\\
 & \times\int_{\mathbf{x}_{1},\mathbf{x}_{2},\mathbf{x}_{3}}\int_{\boldsymbol{\ell}}e^{-i\boldsymbol{\ell}\cdot\mathbf{x}_{12}}\frac{\boldsymbol{\ell}^{2}+\frac{\ell_{1}^{+}(q^{+}-\ell_{1}^{+})}{(q^{+})^{2}}M^{2}}{\boldsymbol{\ell}^{2}-\frac{\ell_{1}^{+}(q^{+}-\ell_{1}^{+})}{(q^{+})^{2}}M^{2}}\int_{\boldsymbol{\ell}_{1},\boldsymbol{\ell}_{4}}\frac{e^{i\boldsymbol{\ell}_{1}\cdot\mathbf{x}_{13}}e^{i\boldsymbol{\ell}_{4}\cdot\mathbf{x}_{23}}}{\Big(\boldsymbol{\ell}_{1}+\frac{\ell_{1}^{+}}{\ell_{1}^{+}+p_{1}^{+}}\boldsymbol{\ell}_{4}\Big)^{2}+\frac{p_{1}^{+}\ell_{1}^{+}p_{0}^{+}\boldsymbol{\ell}_{4}^{2}}{(q^{+}-\ell_{1}^{+})(p_{1}^{+}+\ell_{1}^{+})^{2}}}\\
 & \times e^{-i\mathbf{q}\cdot\Big(\frac{q^{+}-\ell_{1}^{+}}{q^{+}}\mathbf{x}_{2}+\frac{\ell_{1}^{+}}{q^{+}}\mathbf{x}_{1}\Big)}e^{-i\mathbf{p}_{1}\cdot\mathbf{x}_{3}}\int_{\mathbf{x}^{\prime}}e^{i\mathbf{k}_{\perp}\cdot\mathbf{x}^{\prime}}\big(\frac{N_{c}^{2}}{2}s_{\mathbf{x}_{1}\mathbf{x}_{2}}s_{\mathbf{x}_{3}\mathbf{x}^{\prime}}-\frac{1}{2}Q_{\mathbf{x}_{1}\mathbf{x}^{\prime}\mathbf{x}_{3}\mathbf{x}_{2}}+C_{F}N_{c}\big)\\
 & +\Big(2-\frac{(p_{0}^{+}+p_{1}^{+})}{q^{+}}\ln\frac{p_{0}^{+}}{p_{1}^{+}}\Big)\mathcal{A}_{0}(\Delta_{\mathrm{UV}})\int_{\mathbf{x},\mathbf{x}^{\prime}}e^{-i\mathbf{k}_{\perp}\cdot(\mathbf{x}-\mathbf{x}^{\prime})}C_{F}N_{c}\big(s_{\mathbf{x}\mathbf{x}^{\prime}}+1\big)\Bigg]\;,
\end{aligned}
\label{eq:Q2_sub_cross}
\end{equation}
and finally:
\begin{equation}
\begin{aligned} & \mathrm{Tr}\,\mathcal{M}_{\mathrm{LO}}^{0\dagger}\mathcal{M}_{\mathrm{Q3,sub}}^{0}  =\mathrm{Tr}\big(\mathcal{M}_{\mathrm{LO}}^{0\dagger}\mathcal{M}_{\mathrm{LO2}}^{0}\big)\\
&\times\frac{\alpha_{s}C_{F}}{\pi}\Bigg[\Big(-2+\frac{p_{0}^{+}+p_{1}^{+}}{q^{+}}\ln\frac{p_{0}^{+}}{p_{1}^{+}}\Big)\ln\frac{p_{1}^{+}\Delta_{\mathrm{UV}}}{p_{0}^{+}\mathbf{q}^{2}}\\
 & -2+(2\frac{p_{0}^{+}}{q^{+}}-1)\Big(\ln\Big(\frac{p_{1}^{+}}{p_{0}^{+}}\Big)\Big(1+i\pi+\ln\frac{p_{0}^{+}p_{1}^{+}}{(q^{+})^{2}}\Big)-\mathrm{Li}_{2}\big(\frac{p_{1}^{+}}{p_{0}^{+}}\big)+\mathrm{Li}_{2}\big(\frac{p_{0}^{+}}{p_{1}^{+}}\big)\Big)\Bigg]\;.
\end{aligned}
\label{eq:Q3_sub_cross}
\end{equation}
To obtain the expression in the last line of the above result, we
made use of the identity:
\begin{equation}
\begin{aligned} & -\int_{0}^{1}\mathrm{d}x\,\frac{x(1-x)}{(c-x)^{2}}\ln\big(x(1-x)-i0^{+}\big)\\
 & =-2+(2c-1)\Big[\ln\Big(\frac{c-1}{c}\Big)\Big(1+i\pi+\ln c(c-1)\Big)-\mathrm{Li}_{2}\big(\frac{c-1}{c}\big)+\mathrm{Li}_{2}\big(\frac{c}{c-1}\big)\Big]
\end{aligned}
\end{equation}
for $c>1$.

\paragraph{Instantaneous $gq\gamma q$ interaction}

Here, we list the four contributions to the cross section due to the
amplitudes with an instantaneous $gq\gamma q$ interaction. They are
all finite. The first one is:
\begin{equation}
\begin{aligned} & \mathrm{Tr}\mathcal{M}_{\mathrm{LO}}^{0\dagger}\mathcal{M}_{\mathrm{I1,sub}}^{0} \\
& =\frac{g_{\mathrm{em}}^{2}N_{c}}{M^{2}}8p_{1}^{+}p_{0}^{+}\Big(\frac{p_{0}^{+}\mathbf{P}_{\perp}^{2}-p_{1}^{+}M^{2}}{p_{0}^{+}\mathbf{P}_{\perp}^{2}+p_{1}^{+}M^{2}}-\frac{p_{0}^{+}\mathbf{q}^{2}-p_{1}^{+}M^{2}}{p_{0}^{+}\mathbf{q}^{2}+p_{1}^{+}M^{2}}\Big)\int_{\mathbf{x},\mathbf{x}^{\prime}}e^{-i\mathbf{k}_{\perp}\cdot(\mathbf{x}-\mathbf{x}^{\prime})}\big(s_{\mathbf{x}\mathbf{x}^{\prime}}+1\big)\\
 & \!\times\!\frac{\alpha_{s}C_{F}}{\pi}\frac{q^{+}}{2p_{0}^{+}}\Big[\big(\!\ln\!\frac{q^{+}}{p_{1}^{+}}\!-\!i\pi\big)\Big(\!i\pi\!+\!\ln\!\frac{p_{0}^{+}\Delta_{\mathrm{UV}}}{p_{0}^{+}\mathbf{P}_{\perp}^{2}\!+\!p_{1}^{+}M^{2}}\Big)\!-\!\mathrm{Li}_{2}\Big(\!-\frac{p_{1}^{+}}{q^{+}}\Big)\!+\!\mathrm{Li}_{2}\Big(\!-\frac{q^{+}}{p_{1}^{+}}\Big)\Big]\;,
\end{aligned}
\label{eq:I1L_cross}
\end{equation}
and the second:
\begin{equation}
\begin{aligned} & \mathrm{Tr}\mathcal{M}_{\mathrm{LO}}^{0\dagger}\mathcal{M}_{\mathrm{I2,sub}}^{0} \\
& =\frac{g_{\mathrm{em}}^{2}}{M^{2}}8p_{1}^{+}p_{0}^{+}\Big(\frac{p_{0}^{+}\mathbf{P}_{\perp}^{2}-p_{1}^{+}M^{2}}{p_{0}^{+}\mathbf{P}_{\perp}^{2}+p_{1}^{+}M^{2}}-\frac{p_{0}^{+}\mathbf{q}^{2}-p_{1}^{+}M^{2}}{p_{0}^{+}\mathbf{q}^{2}+p_{1}^{+}M^{2}}\Big)\int_{\mathbf{x},\mathbf{x}^{\prime}}e^{-i\mathbf{k}_{\perp}\cdot(\mathbf{x}-\mathbf{x}^{\prime})}\\
 & \times\alpha_{s}\int_{0}^{p_{0}^{+}}\frac{\mathrm{d}k^{+}}{k^{+}}\frac{k^{+}}{p_{0}^{+}}\Bigg[\frac{k^{+}}{p_{0}^{+}}\frac{p_{0}^{+}-k^{+}}{p_{1}^{+}-k^{+}}\int_{\mathbf{z}}iA^{\eta^{\prime}}(\mathbf{x}-\mathbf{z})\\
 & \times\!\Bigg(\!\frac{q^{+}iA^{\eta^{\prime}}(\mathbf{x}\!-\!\mathbf{z},\Delta_{\mathrm{P}})}{p_{0}^{+}\!-\!k^{+}}\frac{2p_{0}^{+}}{k^{+}}\!-\!\frac{2p_{1}^{+}\!-\!k^{+}}{p_{1}^{+}}\mathbf{P}_{\perp}^{i}\mathcal{K}\big(\mathbf{x}\!-\!\mathbf{z},\Delta_{\mathrm{P}}\big)\!\Big(\!\frac{2p_{0}^{+}}{k^{+}}\!-\!1\!+\!\frac{k^{+}}{2p_{1}^{+}\!-\!k^{+}}\!\Big)\!\Bigg)\\
 & \times \!e^{i\frac{k^{+}}{p_{0}^{+}}\mathbf{k}_{\perp}\cdot(\mathbf{x}\!-\!\mathbf{z})}\big(\frac{N_{c}^{2}}{2}s_{\mathbf{x}\mathbf{z}}s_{\mathbf{z}\mathbf{x}^{\prime}}\!-\!\frac{1}{2}s_{\mathbf{x}\mathbf{x}^{\prime}}\!+\!C_{F}N_{c}\big) \!+\!\frac{2q^{+}}{k^{+}\!-\!p_{1}^{+}}\mathcal{A}_{0}(\Delta_{\mathrm{UV}})C_{F}N_{c}\big(s_{\mathbf{x}\mathbf{x}^{\prime}}\!+\!1\big)\!\Bigg]\;.
\end{aligned}
\label{eq:I2_cross}
\end{equation}
Moreover, we have:
\begin{equation}
\begin{aligned} & \mathrm{Tr}\mathcal{M}_{\mathrm{LO}}^{0\dagger}\mathcal{M}_{\mathrm{I3,sub}}^{0} \\
& =\frac{g_{\mathrm{em}}^{2}}{M^{2}}8p_{1}^{+}p_{0}^{+}\Big(\frac{p_{0}^{+}\mathbf{P}_{\perp}^{2}-p_{1}^{+}M^{2}}{p_{0}^{+}\mathbf{P}_{\perp}^{2}+p_{1}^{+}M^{2}}-\frac{p_{0}^{+}\mathbf{q}^{2}-p_{1}^{+}M^{2}}{p_{0}^{+}\mathbf{q}^{2}+p_{1}^{+}M^{2}}\Big)\int_{\mathbf{x},\mathbf{x}^{\prime}}e^{-i\mathbf{k}_{\perp}\cdot(\mathbf{x}-\mathbf{x}^{\prime})}\\
 & \times\alpha_{s}\int_{0}^{p_{1}^{+}}\frac{\mathrm{d}k^{+}}{p_{0}^{+}-k^{+}}\Bigg[\int_{\mathbf{z}}\frac{k^{+}(p_{1}^{+}-k^{+})}{(p_{1}^{+})^{2}}iA^{i}(\mathbf{x}-\mathbf{z})\\
 & \times\Bigg(-\frac{2p_{1}^{+}}{k^{+}}\frac{q^{+}iA^{i}(\mathbf{x}-\mathbf{z},\Delta_{\mathrm{q}})}{p_{1}^{+}-k^{+}}+\frac{k^{+}}{p_{1}^{+}}\mathbf{q}^{i}\mathcal{K}\big(\mathbf{x}-\mathbf{z},\Delta_{\mathrm{P}}\big)\Big(\frac{2p_{0}^{+}-k^{+}}{k^{+}}\big(1-\frac{2p_{1}^{+}}{k^{+}}\big)-1\Big)\Bigg)\\
 & \times e^{i\frac{k^{+}}{p_{1}^{+}}\mathbf{k}_{\perp}\cdot(\mathbf{x}-\mathbf{z})}\big(\frac{N_{c}^{2}}{2}s_{\mathbf{x}\mathbf{z}}s_{\mathbf{z}\mathbf{x}^{\prime}}-\frac{1}{2}s_{\mathbf{x}\mathbf{x}^{\prime}}+C_{F}N_{c}\big)-\frac{2q^{+}}{p_{1}^{+}}\mathcal{A}_{0}(\Delta_{\mathrm{UV}})C_{F}N_{c}\big(s_{\mathbf{x}\mathbf{x}^{\prime}}+1\big)\Bigg]\;,
\end{aligned}
\label{eq:I3L_cross}
\end{equation}
and finally:
\begin{equation}
\begin{aligned}\mathrm{Tr}\mathcal{M}_{\mathrm{LO}}^{0\dagger}\mathcal{M}_{\mathrm{I4,sub}}^{0} & =\frac{g_{\mathrm{em}}^{2}N_{c}}{M^{2}}8p_{1}^{+}p_{0}^{+}\Big(\frac{p_{0}^{+}\mathbf{P}_{\perp}^{2}-p_{1}^{+}M^{2}}{p_{0}^{+}\mathbf{P}_{\perp}^{2}+p_{1}^{+}M^{2}}-\frac{p_{0}^{+}\mathbf{q}^{2}-p_{1}^{+}M^{2}}{p_{0}^{+}\mathbf{q}^{2}+p_{1}^{+}M^{2}}\Big)\\
 & \times\frac{\alpha_{s}C_{F}}{\pi}\frac{q^{+}}{2p_{1}^{+}}\Big(\ln\frac{q^{+}}{p_{0}^{+}}\ln\frac{q^{+}\Delta_{\mathrm{UV}}}{p_{0}^{+}\mathbf{q}^{2}+p_{1}^{+}M^{2}}-\mathrm{Li}_{2}\big(\frac{p_{1}^{+}}{p_{0}^{+}}\big)+\mathrm{Li}_{2}\big(-\frac{p_{1}^{+}}{q^{+}}\big)\Big)\\
 & \times\int_{\mathbf{x},\mathbf{x}^{\prime}}e^{-i\mathbf{k}_{\perp}\cdot(\mathbf{x}-\mathbf{x}^{\prime})}\big(s_{\mathbf{x}\mathbf{x}^{\prime}}+1\big)\;,
\end{aligned}
\label{eq:I4L_cross}
\end{equation}

\subsubsection{\label{subsec:virtual_T_cross}Transverse polarization}

\paragraph{Self-energy corrections}

The contributions $\mathrm{Tr}\mathcal{M}_{\mathrm{LO}}^{\lambda\dagger}\mathcal{M}_{\mathrm{SE1,sub}}^{\lambda}\Big|_{\mathrm{finite}}$
and $\mathrm{Tr}\mathcal{M}_{\mathrm{LO}}^{\lambda\dagger}\mathcal{M}_{\mathrm{SE4,sub}}^{\lambda}\Big|_{\mathrm{finite}}$
are the same as their counterparts (\ref{eq:SE1L_fin_cross}) and
(\ref{eq:SE4L_fin_cross}) after changing the polarization of the
LO amplitudes which they are proportional to. A similar principle
holds for amplitudes and $\mathcal{M}_{\mathrm{SE2,sub}}$ (\ref{eq:MSE2T_sub})
and $\mathcal{M}_{\mathrm{SE3,sub}}$ (\ref{eq:MSE3T_sub}), but here
the LO amplitudes cannot be completely factorized out. We, therefore,
present the expressions explicitly:
\begin{equation}
\begin{aligned} & \mathrm{Tr}\mathcal{M}_{\mathrm{LO}}^{\lambda\dagger}\mathcal{M}_{\mathrm{SE2,sub}}^{\lambda} \\
& =g_{\mathrm{em}}^{2}\alpha_{s}8p_{1}^{+}p_{0}^{+}\Big(\frac{q^{+}\mathbf{P}_{\perp}^{\bar{\lambda}}}{p_{0}^{+}\mathbf{P}_{\perp}^{2}+p_{1}^{+}M^{2}}+\frac{q^{+}\mathbf{q}^{\bar{\lambda}}}{p_{0}^{+}\mathbf{q}^{2}+p_{1}^{+}M^{2}}\Big)\frac{q^{+}\mathbf{P}_{\perp}^{\bar{\lambda}}}{p_{0}^{+}\mathbf{P}_{\perp}^{2}+p_{1}^{+}M^{2}}\\
 & \times\Big(\big(1+2\frac{p_{1}^{+}}{q^{+}}\big)^{2}+1\Big)\int_{\mathbf{x},\mathbf{x}^{\prime}}e^{-i\mathbf{k}_{\perp}\cdot(\mathbf{x}-\mathbf{x}^{\prime})}\Bigg\{\int_{0}^{p_{0}^{+}}\frac{\mathrm{d}k^{+}}{k^{+}}\Big(\frac{k^{+}}{p_{0}^{+}}\Big)^{2}\Big(\big(1-\frac{2p_{0}^{+}}{k^{+}}\big)^{2}+1\Big)\\
 & \times\Bigg[\int_{\mathbf{z}}e^{i\frac{k^{+}}{p_{0}^{+}}\mathbf{k}_{\perp}\cdot(\mathbf{x}-\mathbf{z})}A^{i}(\mathbf{x}-\mathbf{z})A^{i}\big(\mathbf{x}-\mathbf{z},\Delta_{\mathrm{P}}\big)\big(\frac{N_{c}^{2}}{2}s_{\mathbf{z}\mathbf{x}^{\prime}}s_{\mathbf{x}\mathbf{z}}-\frac{1}{2}s_{\mathbf{x}\mathbf{x}^{\prime}}+C_{F}N_{c}\big)\\
 & -\mathcal{A}_{0}(\Delta_{\mathrm{UV}})C_{F}N_{c}\big(s_{\mathbf{x}\mathbf{x}^{\prime}}+1\big)\Bigg]\\
 & -4\int_{k_{\mathrm{min}}^{+}}^{k_{f}^{+}}\frac{\mathrm{d}k^{+}}{k^{+}}\Bigg[\int_{\mathbf{z}}A^{i}(\mathbf{x}-\mathbf{z})A^{i}(\mathbf{x}-\mathbf{z})\big(\frac{N_{c}^{2}}{2}s_{\mathbf{z}\mathbf{x}^{\prime}}s_{\mathbf{x}\mathbf{z}}-\frac{1}{2}s_{\mathbf{x}\mathbf{x}^{\prime}}+C_{F}N_{c}\big)\\
 & -\mathcal{A}_{0}(\Delta_{\mathrm{UV}})C_{F}N_{c}\big(s_{\mathbf{x}\mathbf{x}^{\prime}}+1\big)\Bigg]\Bigg\}\;,
\end{aligned}
\label{eq:SE2_T_cross}
\end{equation}
and:
\begin{equation}
\begin{aligned} & \mathrm{Tr}\mathcal{M}_{\mathrm{LO}}^{\lambda\dagger}\mathcal{M}_{\mathrm{SE3,sub}}^{\lambda} \\ 
& =g_{\mathrm{em}}^{2}\alpha_{s}8p_{1}^{+}p_{0}^{+}\Big(\frac{q^{+}\mathbf{P}_{\perp}^{\bar{\lambda}}}{p_{0}^{+}\mathbf{P}_{\perp}^{2}+p_{1}^{+}M^{2}}+\frac{q^{+}\mathbf{q}^{\bar{\lambda}}}{p_{0}^{+}\mathbf{q}^{2}+p_{1}^{+}M^{2}}\Big)\frac{q^{+}\mathbf{q}^{\bar{\lambda}}}{p_{0}^{+}\mathbf{q}^{2}+p_{1}^{+}M^{2}}\\
 & \times\Big(\big(1+2\frac{p_{1}^{+}}{q^{+}}\big)^{2}+1\Big)\int_{\mathbf{x},\mathbf{x}^{\prime}}e^{-i\mathbf{k}_{\perp}\cdot(\mathbf{x}-\mathbf{x}^{\prime})}\Bigg\{\int_{0}^{p_{1}^{+}}\frac{\mathrm{d}k^{+}}{k^{+}}\Big(\frac{k^{+}}{p_{1}^{+}}\Big)^{2}\Big(\big(1-\frac{2p_{1}^{+}}{k^{+}}\big)^{2}+1\Big)\\
 & \times\Bigg[\int_{\mathbf{z}}e^{i\frac{k^{+}}{p_{1}^{+}}\mathbf{k}_{\perp}\cdot(\mathbf{x}-\mathbf{z})}A^{i}(\mathbf{x}-\mathbf{z})A^{i}(\mathbf{x}-\mathbf{z},\Delta_{\mathrm{q}})\big(\frac{N_{c}^{2}}{2}s_{\mathbf{z}\mathbf{x}^{\prime}}s_{\mathbf{x}\mathbf{z}}-\frac{1}{2}s_{\mathbf{x}\mathbf{x}^{\prime}}+C_{F}N_{c}\big)\\
 & -\mathcal{A}_{0}(\Delta_{\mathrm{UV}})C_{F}N_{c}\big(s_{\mathbf{x}\mathbf{x}^{\prime}}+1\big)\Bigg]\\
 & -4\int_{k_{\mathrm{min}}^{+}}^{k_{f}^{+}}\frac{\mathrm{d}k^{+}}{k^{+}}\Bigg[\int_{\mathbf{z}}A^{i}(\mathbf{x}-\mathbf{z})A^{i}(\mathbf{x}-\mathbf{z})\big(\frac{N_{c}^{2}}{2}s_{\mathbf{z}\mathbf{x}^{\prime}}s_{\mathbf{x}\mathbf{z}}-\frac{1}{2}s_{\mathbf{x}\mathbf{x}^{\prime}}+C_{F}N_{c}\big)\\
 & -\mathcal{A}_{0}(\Delta_{\mathrm{UV}})C_{F}N_{c}\big(s_{\mathbf{x}\mathbf{x}^{\prime}}+1\big)\Bigg]\Bigg\}\;.
\end{aligned}
\label{eq:SE3_T_cross}
\end{equation}

\paragraph{Vertex corrections }

In an attempt to make the expressions below a bit more compact, we
introduce the following notation for the spinor trace over the Dirac
structures, evaluated with the help of identity~\eqref{eq:DiracTraceBossV}
\begin{equation}
\begin{aligned} & \mathrm{Tr}\Big(\bar{u}_{G}(p_{0}^{+})\gamma^{+}\mathrm{Dirac}^{\lambda\lambda^{\prime}}\big(1+\frac{2p_{1}^{+}}{q^{+}}\big)^{\dagger}u_{G}(p_{1}^{+})\bar{u}_{G}(p_{1}^{+})\gamma^{+}\mathrm{Dirac}_{\mathrm{V}}^{\bar{\eta}\bar{\lambda}\eta^{\prime}}u_{G}(p_{0}^{+})\Big)\\
 & =-8p_{0}^{+}p_{1}^{+}\Big[\mathcal{E}_{\mathrm{V}}\delta^{\bar{\eta}\eta^{\prime}}\delta^{\bar{\lambda}\lambda^{\prime}}+\mathcal{O}_{\mathrm{V}}\epsilon^{\bar{\eta}\eta^{\prime}}\epsilon^{\bar{\lambda}\lambda^{\prime}}\Big]\;,
\end{aligned}
\label{eq:trace_Dirac_V}
\end{equation}
with:
\begin{equation}
\begin{aligned}\mathcal{E}_{\mathrm{V}} & \equiv\Big(\big(1+\frac{2p_{1}^{+}}{q^{+}}\big)\big(2\frac{k^{+}-p_{1}^{+}}{q^{+}}-1\big)-1\Big)\Big(\big(2\frac{p_{1}^{+}}{k^{+}}-1\big)\big(1-2\frac{p_{0}^{+}}{k^{+}}\big)-1\Big)\;,\\
\mathcal{\mathcal{O}}_{\mathrm{V}} & \equiv4\frac{(p_{0}^{+}+p_{1}^{+}-k^{+})^{2}}{k^{+}q^{+}}\;.
\end{aligned}
\end{equation}
We can show that the UV-subtracted vertex corrections $\mathcal{M}_{\mathrm{V1,sub}}^{\lambda}$
(\ref{eq:MV1_final}) and $\mathcal{M}_{\mathrm{V4,sub}}^{\lambda}$
(\ref{eq:MV4T_sub}) contribute as follows to the cross section:
\begin{equation}
\begin{aligned} & \mathrm{Tr}\,\mathcal{M}_{\mathrm{LO}}^{\lambda\dagger}\mathcal{M}_{\mathrm{V1,sub}}^{\lambda} \\
& =\mathrm{Tr}\,\mathcal{M}_{\mathrm{LO}}^{\lambda\dagger}\mathcal{M}_{\mathrm{LO1}}^{\lambda}\frac{\alpha_{s}C_{F}}{\pi}\frac{1}{2}\frac{(q^{+})^{2}}{(p_{0}^{+})^{2}+(p_{1}^{+})^{2}}\\
 & \times\Bigg\{2\frac{(p_{0}^{+})^{2}+(p_{1}^{+})^{2}}{(q^{+})^{2}}\int_{k_{\mathrm{min}}^{+}}^{p_{1}^{+}}\frac{\mathrm{d}k^{+}}{k^{+}}\ln\frac{\Delta_{\mathrm{UV}}}{\Delta_{\mathrm{P}}} +2\int_{0}^{p_{1}^{+}}\mathrm{d}k^{+}\frac{k^{+}-2p_{1}^{+}-q^{+}}{(q^{+})^{2}}\ln\frac{\Delta_{\mathrm{UV}}}{\Delta_{\mathrm{P}}}\\
 & +\int_{0}^{p_{1}^{+}}\mathrm{d}k^{+}\Big(\frac{k^{+}}{p_{1}^{+}(p_{0}^{+}-k^{+})}\Big(\frac{k^{+}}{p_{1}^{+}}\Big)^{2}\mathbf{P}_{\perp}^{2}\\
 & +4\frac{(p_{1}^{+})^{2}+(p_{0}^{+}-k^{+})^{2}}{p_{1}^{+}q^{+}(p_{0}^{+}-k^{+})}\Big(\Delta_{\mathrm{P}}+\Big(\frac{k^{+}}{p_{1}^{+}}\Big)^{2}\mathbf{P}_{\perp}^{2}\Big)\Big)\pi\mathcal{B}_{1}\big(0,\Delta_{\mathrm{P}},\frac{k^{+}}{p_{1}^{+}}\mathbf{P}_{\perp}\big)\Bigg\}\;,
\end{aligned}
\label{eq:V1_sub_T_cross_new}
\end{equation}
and:
\begin{equation}
\begin{aligned} &\mathrm{Tr}\,\mathcal{M}_{\mathrm{LO}}^{\lambda\dagger}\mathcal{M}_{\mathrm{V4,sub}}^{\lambda}  =\mathrm{Tr}\,\mathcal{M}_{\mathrm{LO}}^{\lambda\dagger}\mathcal{M}_{\mathrm{LO2}}^{\lambda}\frac{\alpha_{s}C_{F}}{\pi}\frac{1}{2}\frac{(q^{+})^{2}}{(p_{0}^{+})^{2}+(p_{1}^{+})^{2}}\\
 & \times\Bigg\{2\frac{(p_{0}^{+})^{2}+(p_{1}^{+})^{2}}{(q^{+})^{2}}\int_{k_{\mathrm{min}}^{+}}^{p_{1}^{+}}\frac{\mathrm{d}k^{+}}{k^{+}}\ln\frac{\Delta_{\mathrm{UV}}}{\Delta_{\mathrm{q}}} +2\int_{0}^{p_{1}^{+}}\mathrm{d}k^{+}\frac{k^{+}-p_{0}^{+}-p_{1}^{+}}{(q^{+})^{2}}\ln\frac{\Delta_{\mathrm{UV}}}{\Delta_{\mathrm{q}}}\\
 & +4\int_{0}^{p_{1}^{+}}\mathrm{d}k^{+}\Bigg[\frac{(p_{0}^{+}+p_{1}^{+}-k^{+})^{2}}{p_{0}^{+}q^{+}(p_{1}^{+}-k^{+})}\Big(\frac{k^{+}}{p_{1}^{+}}\Big)^{2}\mathbf{q}^{2}\\
 & +\frac{(p_{0}^{+})^{2}+(p_{1}^{+}-k^{+})^{2}}{q^{+}p_{0}^{+}(p_{1}^{+}-k^{+})}\big(\Delta_{\mathrm{q}}+\Big(\frac{k^{+}}{p_{1}^{+}}\Big)^{2}\mathbf{q}^{2}\big)\Bigg]\pi\mathcal{B}_{1}\big(0,\Delta_{\mathrm{q}},\frac{k^{+}}{p_{1}^{+}}\mathbf{q}\big)\Bigg\}\;.
\end{aligned}
\label{eq:V4_sub_T_cross_total_new}
\end{equation}
The second lines of the above two expressions contain the rapidity-divergent
parts. These terms nicely combine with the similar parts of $\mathcal{M}_{\mathrm{SE1,sub}}^{\lambda}\Big|_{1/k^{+}}$
into completely finite expressions that are independent of the rapidity
factorization scale. Indeed, we have:
\begin{equation}
\begin{aligned}\mathrm{Tr}\,\mathcal{M}_{\mathrm{LO}}^{\lambda\dagger}\mathcal{M}_{\mathrm{V1,sub}}^{\lambda}\Big|_{1/k^{+}} & =\mathrm{Tr}\,\mathcal{M}_{\mathrm{LO}}^{\lambda\dagger}\mathcal{M}_{\mathrm{LO1}}^{\lambda}\frac{\alpha_{s}C_{F}}{\pi}\int_{k_{\mathrm{min}}^{+}}^{p_{1}^{+}}\frac{\mathrm{d}k^{+}}{k^{+}}\ln\frac{\Delta_{\mathrm{UV}}}{\Delta_{\mathrm{P}}}\;,\end{aligned}
\end{equation}
but also:
\begin{equation}
\begin{aligned}\mathrm{Tr}\,\mathcal{M}_{\mathrm{LO}}^{\lambda\dagger}\mathcal{M}_{\mathrm{SE1,sub}}^{\lambda}\Big|_{1/k^{+}} & =\mathrm{Tr}\,\mathcal{M}_{\mathrm{LO}}^{\lambda\dagger}\mathcal{M}_{\mathrm{LO1}}^{\lambda}\frac{\alpha_{s}C_{F}}{\pi}\int_{k_{\mathrm{min}}^{+}}^{p_{0}^{+}}\frac{\mathrm{d}k^{+}}{k^{+}}\ln\frac{\Delta_{\mathrm{P}}}{\Delta_{\mathrm{UV}}}\;.\end{aligned}
\end{equation}
Their sum is:
\begin{equation}
\begin{aligned} & \mathrm{Tr}\,\mathcal{M}_{\mathrm{LO}}^{\lambda\dagger}\Big(\mathcal{M}_{\mathrm{SE1,sub}}^{\lambda}+\mathcal{M}_{\mathrm{V1,sub}}^{\lambda}\Big)\Big|_{1/k^{+}}\\
 & =\mathrm{Tr}\,\mathcal{M}_{\mathrm{LO}}^{\lambda\dagger}\mathcal{M}_{\mathrm{LO1}}^{\lambda}\frac{\alpha_{s}C_{F}}{\pi}\int_{p_{1}^{+}}^{p_{0}^{+}}\frac{\mathrm{d}k^{+}}{k^{+}}\ln\frac{\Delta_{\mathrm{P}}}{\Delta_{\mathrm{UV}}}\;,\\
 & =\mathrm{Tr}\,\mathcal{M}_{\mathrm{LO}}^{\lambda\dagger}\mathcal{M}_{\mathrm{LO1}}^{\lambda}\frac{\alpha_{s}C_{F}}{\pi}\\
 & \times\Big[\ln\frac{p_{0}^{+}}{p_{1}^{+}}\Big(i\pi+\ln\frac{p_{0}^{+}q^{+}\Delta_{\mathrm{P}}}{p_{1}^{+}\big(p_{0}^{+}\mathbf{P}_{\perp}^{2}+p_{1}^{+}M^{2}\big)}\Big)-\ln\frac{p_{0}^{+}}{p_{1}^{+}}\ln\frac{q^{+}}{p_{1}^{+}}-\frac{1}{2}\ln^{2}\frac{p_{0}^{+}}{p_{1}^{+}}+\mathrm{Li}_{2}\frac{q^{+}}{p_{0}^{+}}\Big]\;,
\end{aligned}
\label{eq:SE1+V1_k_T_cross}
\end{equation}
in which all dependence on the rapidity cutoff or the rapidity factorization
scale has disappeared. Likewise:
\begin{equation}
\begin{aligned}\mathrm{Tr}\,\mathcal{M}_{\mathrm{LO}}^{\lambda\dagger}\mathcal{M}_{\mathrm{V4,sub}}^{\lambda}\Big|_{1/k^{+}} & =\mathrm{Tr}\,\mathcal{M}_{\mathrm{LO}}^{\lambda\dagger}\mathcal{M}_{\mathrm{LO2}}^{\lambda}\frac{\alpha_{s}C_{F}}{\pi}\int_{k_{\mathrm{min}}^{+}}^{p_{1}^{+}}\frac{\mathrm{d}k^{+}}{k^{+}}\ln\frac{\Delta_{\mathrm{UV}}}{\Delta_{\mathrm{q}}}\;,\end{aligned}
\end{equation}
cancels with:
\begin{equation}
\begin{aligned}\mathrm{Tr}\,\mathcal{M}_{\mathrm{LO}}^{\lambda\dagger}\mathcal{M}_{\mathrm{SE2,sub}}^{\lambda}\Big|_{1/k^{+}} & =\mathrm{Tr}\,\mathcal{M}_{\mathrm{LO}}^{\lambda\dagger}\mathcal{M}_{\mathrm{LO2}}^{\lambda}\frac{\alpha_{s}C_{F}}{\pi}\int_{k_{\mathrm{min}}^{+}}^{p_{1}^{+}}\frac{\mathrm{d}k^{+}}{k^{+}}\ln\frac{\Delta_{\mathrm{q}}}{\Delta_{\mathrm{UV}}}\;,\\
 & =-\mathrm{Tr}\,\mathcal{M}_{\mathrm{LO}}^{\lambda\dagger}\mathcal{M}_{\mathrm{V4,sub}}^{\lambda}\Big|_{1/k^{+}}\;.
\end{aligned}
\end{equation}
The completely finite leftovers of (\ref{eq:V1_sub_T_cross_new})
and (\ref{eq:V4_sub_T_cross_total_new}) are:
\begin{equation}
\begin{aligned} & \mathrm{Tr}\,\mathcal{M}_{\mathrm{LO}}^{\lambda\dagger}\mathcal{M}_{\mathrm{V1,sub}}^{\lambda}\Big|_{\mathrm{finite}} \\
 & =\mathrm{Tr}\,\mathcal{M}_{\mathrm{LO}}^{\lambda\dagger}\mathcal{M}_{\mathrm{LO1}}^{\lambda}\frac{\alpha_{s}C_{F}}{\pi}\frac{1}{2}\frac{(q^{+})^{2}}{(p_{0}^{+})^{2}+(p_{1}^{+})^{2}}\\
 & \times\Bigg\{-\frac{p_{1}^{+}(2p_{0}^{+}+p_{1}^{+})}{(q^{+})^{2}}\Big(i\pi+\ln\frac{p_{0}^{+}q^{+}\Delta_{\mathrm{P}}}{p_{1}^{+}\big(p_{0}^{+}\mathbf{P}_{\perp}^{2}+p_{1}^{+}M^{2}\big)}\Big)\\
 & -\frac{3p_{1}^{+}(p_{0}^{+}+p_{1}^{+})+\big((p_{0}^{+})^{2}-(p_{1}^{+})^{2}\big)\ln\frac{q^{+}}{p_{1}^{+}}-p_{0}^{+}(p_{0}^{+}+2p_{1}^{+})\ln\frac{p_{0}^{+}}{p_{1}^{+}}}{(q^{+})^{2}}\\
 & +\int_{0}^{p_{1}^{+}}\mathrm{d}k^{+}\Big(\frac{k^{+}}{p_{1}^{+}(p_{0}^{+}-k^{+})}\Big(\frac{k^{+}}{p_{1}^{+}}\Big)^{2}\mathbf{P}_{\perp}^{2}\\
 & +4\frac{(p_{1}^{+})^{2}+(p_{0}^{+}-k^{+})^{2}}{p_{1}^{+}q^{+}(p_{0}^{+}-k^{+})}\Big(\Delta_{\mathrm{P}}+\Big(\frac{k^{+}}{p_{1}^{+}}\Big)^{2}\mathbf{P}_{\perp}^{2}\Big)\Big)\pi\mathcal{B}_{1}\big(0,\Delta_{\mathrm{P}},\frac{k^{+}}{p_{1}^{+}}\mathbf{P}_{\perp}\big)\Bigg\}\;,
\end{aligned}
\label{eq:V1_sub_T_cross_finite}
\end{equation}
and:
\begin{equation}
\begin{aligned} & \mathrm{Tr}\,\mathcal{M}_{\mathrm{LO}}^{\lambda\dagger}\mathcal{M}_{\mathrm{V4,sub}}^{\lambda}\Big|_{\mathrm{finite}} \\ & =\mathrm{Tr}\,\mathcal{M}_{\mathrm{LO}}^{\lambda\dagger}\mathcal{M}_{\mathrm{LO2}}^{\lambda}\frac{\alpha_{s}C_{F}}{\pi}\frac{1}{2}\frac{(q^{+})^{2}}{(p_{0}^{+})^{2}+(p_{1}^{+})^{2}}\\
 & \times\Bigg\{-\frac{p_{1}^{+}(2p_{0}^{+}+p_{1}^{+})}{(q^{+})^{2}}\ln\frac{q^{+}\Delta_{\mathrm{UV}}}{p_{0}^{+}\mathbf{q}^{2}+p_{1}^{+}M^{2}}-\frac{2p_{1}^{+}(2p_{0}^{+}+p_{1}^{+})}{(q^{+})^{2}}\\
 & +4\int_{0}^{p_{1}^{+}}\mathrm{d}k^{+}\Bigg[\frac{(p_{0}^{+}+p_{1}^{+}-k^{+})^{2}}{p_{0}^{+}q^{+}(p_{1}^{+}-k^{+})}\Big(\frac{k^{+}}{p_{1}^{+}}\Big)^{2}\mathbf{q}^{2}\\
 & +\frac{(p_{0}^{+})^{2}+(p_{1}^{+}-k^{+})^{2}}{q^{+}p_{0}^{+}(p_{1}^{+}-k^{+})}\big(\Delta_{\mathrm{q}}+\Big(\frac{k^{+}}{p_{1}^{+}}\Big)^{2}\mathbf{q}^{2}\big)\Bigg]\pi\mathcal{B}_{1}\big(0,\Delta_{\mathrm{q}},\frac{k^{+}}{p_{1}^{+}}\mathbf{q}\big)\Bigg\}\;.
\end{aligned}
\label{eq:V4_sub_T_cross_finite}
\end{equation}
The other contributions to the cross section stemming from vertex
correction diagrams never exhibited divergences to begin with:
\begin{equation}
\begin{aligned} & \mathrm{Tr}\,\mathcal{M}_{\mathrm{LO}}^{\lambda\dagger}\mathcal{M}_{\mathrm{V2}}^{\lambda} \\
& =g_{\mathrm{em}}^{2}\alpha_{s}8p_{0}^{+}p_{1}^{+}\Big(\frac{q^{+}\mathbf{P}_{\perp}^{\lambda^{\prime}}}{p_{0}^{+}\mathbf{P}_{\perp}^{2}+p_{1}^{+}M^{2}}+\frac{q^{+}\mathbf{q}^{\lambda^{\prime}}}{p_{0}^{+}\mathbf{q}^{2}+p_{1}^{+}M^{2}}\Big)\int_{0}^{p_{1}^{+}}\frac{\mathrm{d}k^{+}}{k^{+}}\frac{q^{+}(k^{+})^{3}}{p_{1}^{+}(p_{0}^{+})^{2}(p_{1}^{+}-k^{+})}\\
 & \times\Bigg[\Big(\big(1+\frac{2p_{1}^{+}}{q^{+}}\big)\big(2\frac{k^{+}-p_{1}^{+}}{q^{+}}-1\big)-1\Big)\Big(\big(2\frac{p_{1}^{+}}{k^{+}}-1\big)\big(1-2\frac{p_{0}^{+}}{k^{+}}\big)-1\Big)\delta^{\bar{\eta}\eta^{\prime}}\delta^{\bar{\lambda}\lambda^{\prime}}\\
 & +4\frac{(p_{0}^{+}+p_{1}^{+}-k^{+})^{2}}{k^{+}q^{+}}\epsilon^{\bar{\eta}\eta^{\prime}}\epsilon^{\bar{\lambda}\lambda^{\prime}}\Bigg]\\
 & \times\int_{\mathbf{x},\mathbf{x}^{\prime},\mathbf{z}}iA^{\eta^{\prime}}(\mathbf{x}-\mathbf{z})\int_{\boldsymbol{\ell}}e^{i\boldsymbol{\ell}\cdot(\mathbf{x}-\mathbf{z})}\frac{\boldsymbol{\ell}^{\bar{\eta}}-\frac{k^{+}}{p_{1}^{+}}\mathbf{P}_{\perp}^{\bar{\eta}}}{\big(\boldsymbol{\ell}-\frac{k^{+}}{p_{1}^{+}}\mathbf{P}_{\perp}\big)^{2}}\frac{\boldsymbol{\ell}^{\bar{\lambda}}-\frac{p_{0}^{+}-k^{+}}{q^{+}}\mathbf{P}_{\perp}^{\bar{\lambda}}}{\boldsymbol{\ell}^{2}+\Delta_{\mathrm{P}}}\\
 & \times e^{i\frac{k^{+}}{p_{0}^{+}}\mathbf{k}_{\perp}\cdot(\mathbf{x}-\mathbf{z})}e^{-i\mathbf{k}_{\perp}\cdot(\mathbf{x}-\mathbf{z})}\big(\frac{N_{c}^{2}}{2}s_{\mathbf{x}\mathbf{z}}s_{\mathbf{z}\mathbf{x}^{\prime}}-\frac{1}{2}s_{\mathbf{x}\mathbf{x}^{\prime}}+C_{F}N_{c}\big)\;,
\end{aligned}
\label{eq:V2_T_cross}
\end{equation}
and similarly:
\begin{equation}
\begin{aligned} & \mathrm{Tr}\,\mathcal{M}_{\mathrm{LO}}^{\lambda\dagger}\mathcal{M}_{\mathrm{V3}}^{\lambda} \\
& =g_{\mathrm{ew}}^{2}\alpha_{s}8p_{0}^{+}p_{1}^{+}\Big(\frac{q^{+}\mathbf{P}_{\perp}^{\lambda^{\prime}}}{p_{0}^{+}\mathbf{P}_{\perp}^{2}+p_{1}^{+}M^{2}}+\frac{q^{+}\mathbf{q}^{\lambda^{\prime}}}{p_{0}^{+}\mathbf{q}^{2}+p_{1}^{+}M^{2}}\Big)\int_{0}^{p_{1}^{+}}\frac{\mathrm{d}k^{+}}{k^{+}}\frac{q^{+}(k^{+})^{3}}{p_{0}^{+}(p_{1}^{+})^{2}(p_{0}^{+}-k^{+})}\\
 & \times\Bigg[\Big(\big(1+\frac{2p_{1}^{+}}{q^{+}}\big)\big(2\frac{k^{+}-p_{1}^{+}}{q^{+}}-1\big)-1\Big)\Big(\big(2\frac{p_{1}^{+}}{k^{+}}-1\big)\big(1-2\frac{p_{0}^{+}}{k^{+}}\big)-1\Big)\delta^{\bar{\eta}\eta^{\prime}}\delta^{\bar{\lambda}\lambda^{\prime}}\\
 & +4\frac{(p_{0}^{+}+p_{1}^{+}-k^{+})^{2}}{k^{+}q^{+}}\epsilon^{\bar{\eta}\eta^{\prime}}\epsilon^{\bar{\lambda}\lambda^{\prime}}\Bigg]\\
 & \times\int_{\mathbf{x},\mathbf{x}^{\prime},\mathbf{z}}iA^{\bar{\eta}}(\mathbf{z}-\mathbf{x})\int_{\boldsymbol{\ell}}e^{-i\boldsymbol{\ell}\cdot(\mathbf{x}-\mathbf{z})}\frac{\boldsymbol{\ell}^{\eta^{\prime}}}{\boldsymbol{\ell}^{2}}\frac{\boldsymbol{\ell}^{\bar{\lambda}}+\frac{p_{0}^{+}-k^{+}}{q^{+}}\mathbf{q}^{\bar{\lambda}}}{\big(\boldsymbol{\ell}+\frac{k^{+}}{p_{1}^{+}}\mathbf{q}\big)^{2}+\Delta_{\mathrm{q}}}\\
 & \times e^{i\frac{k^{+}}{p_{1}^{+}}\mathbf{p}_{1}\cdot(\mathbf{x}-\mathbf{z})}e^{-i\mathbf{k}_{\perp}\cdot(\mathbf{x}-\mathbf{x}^{\prime})}\big(\frac{N_{c}^{2}}{2}s_{\mathbf{x}\mathbf{z}}s_{\mathbf{z}\mathbf{x}^{\prime}}-\frac{1}{2}s_{\mathbf{x}\mathbf{x}^{\prime}}+C_{F}N_{c}\big)\;.
\end{aligned}
\label{eq:V3_T_cross}
\end{equation}

\paragraph{Antiquark vertex corrections }

The contribution to the cross section coming from the UV-subtracted
diagram $\mathrm{A1,sub}$ is given by the following, quite complicated,
formula:
\begin{equation}
\begin{aligned} & \mathrm{Tr}\mathcal{M}_{\mathrm{LO}}^{\lambda\dagger}\mathcal{M}_{\mathrm{A1,sub}}^{\lambda} \\
& =\mathrm{Tr}\big(\mathcal{M}_{\mathrm{LO}}^{\lambda\dagger}\mathcal{M}_{\mathrm{LO1}}^{\lambda}\big)\times\frac{1}{2}\frac{(q^{+})^{2}}{(p_{1}^{+})^{2}+(p_{0}^{+})^{2}}\frac{\alpha_{s}C_{F}}{\pi}\\
 & \times\Bigg\{\frac{-q^{+}(2p_{0}^{+}+q^{+})+2\big((p_{1}^{+})^{2}+(p_{0}^{+})^{2}\big)\ln(p_{0}^{+}/p_{1}^{+})}{2(q^{+})^{2}}\Big(-i\pi+\ln\frac{\Delta_{\mathrm{UV}}}{M^{2}}\Big)\\
 & +\frac{(p_{1}^{+})^{2}+(p_{0}^{+})^{2}}{(q^{+})^{2}}\Big(-\frac{\pi^{2}}{6}+\ln^{2}\frac{p_{0}^{+}}{p_{1}^{+}}+3\ln\frac{p_{0}^{+}}{p_{1}^{+}}\ln\frac{q^{+}}{p_{1}^{+}}+\mathrm{Li}_{2}\Big(-\frac{q^{+}}{p_{1}^{+}}\Big)+\mathrm{Li}_{2}\Big(\frac{p_{1}^{+}}{p_{0}^{+}}\Big)\Big)\\
 & -\frac{2p_{0}^{+}+q^{+}}{q^{+}}\Big(-1+2\ln\frac{q^{+}}{p_{1}^{+}}\Big)\\
 & +\int_{p_{1}^{+}}^{p_{0}^{+}}\frac{\mathrm{d}k^{+}}{k^{+}}\frac{k^{+}p_{1}^{+}}{p_{0}^{+}(p_{1}^{+}-k^{+})}\Big(-\frac{p_{0}^{+}-k^{+}}{q^{+}}\mathcal{E}_{\mathrm{V}}+\frac{k^{+}}{p_{1}^{+}}\mathcal{O}_{\mathrm{V}}\Big)\Delta_{\mathrm{P}}\pi\mathcal{B}_{0}\big(\Delta_{\mathrm{P}},\hat{M}^{2},\frac{p_{0}^{+}-k^{+}}{q^{+}}\mathbf{P}_{\perp}\big)\\
 & +\int_{p_{1}^{+}}^{p_{0}^{+}}\frac{\mathrm{d}k^{+}}{k^{+}}\frac{k^{+}p_{1}^{+}}{p_{0}^{+}(p_{1}^{+}-k^{+})}\frac{k^{+}(p_{0}^{+}-k^{+})}{2p_{1}^{+}(q^{+})^{2}}\Bigg(\frac{2p_{1}^{+}(p_{0}^{+}-k^{+})-k^{+}q^{+}}{p_{0}^{+}p_{1}^{+}}\Big(p_{0}^{+}\mathbf{P}_{\perp}^{2}+p_{1}^{+}M^{2}\Big)\\
 & +\Big((p_{0}^{+}-k^{+}\big)\mathbf{P}_{\perp}^{2}-(p_{1}^{+}-k^{+})M^{2}\Big)\Bigg)\mathcal{E}_{\mathrm{V}}\pi\mathcal{B}_{1}\big(\Delta_{\mathrm{P}},\hat{M}^{2},\frac{p_{0}^{+}-k^{+}}{q^{+}}\mathbf{P}_{\perp}\big)\\
 & \!+\!\!\int_{p_{1}^{+}}^{p_{0}^{+}}\!\frac{\mathrm{d}k^{+}}{k^{+}}\!\frac{(k^{+})^2}{p_{0}^{+}(2p_{1}^{+}\!-\!k^{+})}\!\Big(\!\Delta_{\mathrm{P}}\!-\!\hat{M}^{2}\!-\!\Big(\!\frac{p_{0}^{+}\!-\!k^{+}}{q^{+}}\mathbf{P}_{\perp}\!\Big)^{\!2}\Big)\mathcal{O}_{\mathrm{V}}\pi\mathcal{B}_{1}\!\big(\!\Delta_{\mathrm{P}},\hat{M}^{2},\frac{p_{0}^{+}\!-\!k^{+}}{q^{+}}\mathbf{P}_{\perp}\!\big)\!\Bigg\}\;.
\end{aligned}
\label{eq:A1_T_cross_new}
\end{equation}
The contribution due to the, finite, diagram $\mathrm{A2}$ reads:
\begin{equation}
\begin{aligned}\mathrm{Tr}\mathcal{M}_{\mathrm{LO}}^{\lambda\dagger}\mathcal{M}_{\mathrm{A2}}^{\lambda} & =g_{\mathrm{em}}^{2}\alpha_{s}8p_{1}^{+}p_{0}^{+}\Big(\frac{q^{+}\mathbf{P}_{\perp}^{\lambda^{\prime}}}{p_{0}^{+}\mathbf{P}_{\perp}^{2}+p_{1}^{+}M^{2}}+\frac{q^{+}\mathbf{q}^{\lambda^{\prime}}}{p_{0}^{+}\mathbf{q}^{2}+p_{1}^{+}M^{2}}\Big)\\
 & \times\int_{p_{1}^{+}}^{p_{0}^{+}}\frac{\mathrm{d}k^{+}}{k^{+}}\frac{(k^{+})^{2}(p_{0}^{+}-k^{+})}{(p_{0}^{+})^{2}(p_{1}^{+}-k^{+})}\Bigg[\mathcal{E}_{\mathrm{V}}\delta^{\bar{\eta}\eta^{\prime}}\delta^{\bar{\lambda}\lambda^{\prime}}+\mathcal{O}_{\mathrm{V}}\epsilon^{\bar{\eta}\eta^{\prime}}\epsilon^{\bar{\lambda}\lambda^{\prime}}\Bigg]\\
 & \times\int_{\mathbf{x},\mathbf{x}^{\prime},\mathbf{z}}iA^{\eta^{\prime}}(\mathbf{x}-\mathbf{z})\int_{\boldsymbol{\ell}}e^{-i\boldsymbol{\ell}\cdot(\mathbf{x}-\mathbf{z})}\frac{\boldsymbol{\ell}^{\bar{\eta}}+\frac{k^{+}}{p_{1}^{+}}\mathbf{P}_{\perp}^{\bar{\eta}}}{\boldsymbol{\ell}^{2}+\Delta_{\mathrm{P}}}\frac{\Big(\boldsymbol{\ell}^{\bar{\lambda}}+\frac{p_{0}^{+}-k^{+}}{q^{+}}\mathbf{P}_{\perp}^{\bar{\lambda}}\Big)}{\Big(\boldsymbol{\ell}+\frac{p_{0}^{+}-k^{+}}{q^{+}}\mathbf{P}_{\perp}\Big)^{2}+\hat{M}^{2}}\\
 & \times\int_{\mathbf{x}^{\prime}}e^{i\mathbf{k}_{\perp}\cdot(\mathbf{x}-\mathbf{x}^{\prime})}e^{i\frac{k^{+}}{p_{0}^{+}}\mathbf{k}_{\perp}\cdot(\mathbf{x}-\mathbf{z})}\big(\frac{N_{c}^{2}}{2}s_{\mathbf{x}\mathbf{z}}s_{\mathbf{z}\mathbf{x}^{\prime}}-\frac{1}{2}s_{\mathbf{x}\mathbf{x}^{\prime}}+C_{F}N_{c}\big)\;.
\end{aligned}
\label{eq:A2_T_cross}
\end{equation}
Amplitude $\mathcal{M}_{\mathrm{A3}}^{\lambda}$, which did not exhibit
any divergences neither, yields the following term:
\begin{equation}
\begin{aligned} & \mathrm{Tr}\mathcal{M}_{\mathrm{LO}}^{\lambda\dagger}\mathcal{M}_{\mathrm{A3}}^{\lambda}\\
 & =g_{\mathrm{ew}}^{2}\alpha_{s}8p_{0}^{+}p_{1}^{+}\Big(\frac{q^{+}\mathbf{P}_{\perp}^{\lambda^{\prime}}}{p_{0}^{+}\mathbf{P}_{\perp}^{2}+p_{1}^{+}M^{2}}+\frac{q^{+}\mathbf{q}^{\lambda^{\prime}}}{p_{0}^{+}\mathbf{q}^{2}+p_{1}^{+}M^{2}}\Big)\\
 & \times\int_{p_{1}^{+}}^{p_{0}^{+}}\frac{\mathrm{d}k^{+}}{k^{+}}\frac{k^{+}(p_{0}^{+}-k^{+})}{q^{+}p_{0}^{+}}\Bigg[\mathcal{E}_{\mathrm{V}}\delta^{\bar{\eta}\eta^{\prime}}\delta^{\bar{\lambda}\lambda^{\prime}}+\mathcal{O}_{\mathrm{V}}\epsilon^{\bar{\eta}\eta^{\prime}}\epsilon^{\bar{\lambda}\lambda^{\prime}}\Bigg]\\
 & \times\int_{\mathbf{x}_{1},\mathbf{x}_{2},\mathbf{x}_{3},\mathbf{x}^{\prime}}iA^{\bar{\lambda}}\big(\mathbf{x}_{1}-\mathbf{x}_{2},\hat{M}^{2}\big)\\
 & \times\int_{\mathbf{k},\boldsymbol{\ell}}e^{-i\mathbf{k}\cdot\mathbf{x}_{12}}e^{-i\boldsymbol{\ell}\cdot\mathbf{x}_{23}}\frac{\mathbf{k}^{\eta^{\prime}}}{\mathbf{k}^{2}}\frac{\Big(\mathbf{k}^{\bar{\eta}}-\frac{k^{+}}{p_{1}^{+}}\boldsymbol{\ell}^{\bar{\eta}}\Big)}{\Big(\mathbf{k}-\frac{p_{0}^{+}-k^{+}}{q^{+}}\boldsymbol{\ell}\Big)^{2}-\frac{p_{0}^{+}(p_{0}^{+}-k^{+})(p_{1}^{+}-k^{+})}{p_{1}^{+}(q^{+})^{2}}\boldsymbol{\ell}^{2}}\\
 & \times e^{i\mathbf{k}_{\perp}\cdot\mathbf{x}^{\prime}}e^{-i\mathbf{p}_{1}\cdot\mathbf{x}_{3}-i\mathbf{q}\cdot\Big(\frac{p_{0}^{+}-k^{+}}{q^{+}}\mathbf{x}_{1}-\frac{p_{1}^{+}-k^{+}}{q^{+}}\mathbf{x}_{2}\Big)}\big(\frac{N_{c}^{2}}{2}s_{\mathbf{x}_{1}\mathbf{x}_{2}}s_{\mathbf{x}_{3}\mathbf{x}^{\prime}}-\frac{1}{2}Q_{\mathbf{x}_{1}\mathbf{x}^{\prime}\mathbf{x}_{3}\mathbf{x}_{2}}+C_{F}N_{c}\big)\;.
\end{aligned}
\label{eq:A3_T_cross}
\end{equation}
Finally, the contribution due to the UV-subtracted amplitude $\mathcal{M}_{\mathrm{A4,sub}}^{\lambda}$ is given by:
\begin{equation}
\begin{aligned} & \mathrm{Tr}\mathcal{M}_{\mathrm{LO}}^{\lambda\dagger}\mathcal{M}_{\mathrm{A4,sub}}^{\lambda} \\
 & =\mathrm{Tr}\big(\mathcal{M}_{\mathrm{LO}}^{\lambda\dagger}\mathcal{M}_{\mathrm{LO2}}^{\lambda}\big)\times\frac{1}{2}\frac{(q^{+})^{2}}{(p_{1}^{+})^{2}+(p_{0}^{+})^{2}}\frac{\alpha_{s}C_{F}}{\pi}\\
 & \times\Bigg\{\frac{-q^{+}(2p_{0}^{+}+q^{+})+2\big((p_{1}^{+})^{2}+(p_{0}^{+})^{2}\big)\ln(p_{0}^{+}/p_{1}^{+})}{2(q^{+})^{2}}\ln\frac{p_{1}^{+}\Delta_{\mathrm{UV}}}{p_{0}^{+}\mathbf{q}^{2}}\\
 & +\frac{(p_{1}^{+})^{2}+(p_{0}^{+})^{2}}{(q^{+})^{2}}\Big(-\frac{\pi^{2}}{6}+\ln^{2}\frac{p_{0}^{+}}{p_{1}^{+}}+3\ln\frac{p_{0}^{+}}{p_{1}^{+}}\ln\frac{q^{+}}{p_{1}^{+}}+\mathrm{Li}_{2}\Big(-\frac{q^{+}}{p_{1}^{+}}\Big)+\mathrm{Li}_{2}\Big(\frac{p_{1}^{+}}{p_{0}^{+}}\Big)\Big)\\
 & -\frac{2p_{0}^{+}+q^{+}}{q^{+}}\Big(-1+2\ln\frac{q^{+}}{p_{1}^{+}}\Big)\\
 & -\int_{p_{1}^{+}}^{p_{0}^{+}}\frac{\mathrm{d}k^{+}}{k^{+}}\frac{(k^{+})^{2}}{2p_{0}^{+}(p_{1}^{+}-k^{+})}\Bigg[\Big(\hat{Q}^{2}-\Big(\frac{p_{0}^{+}-k^{+}}{q^{+}}\mathbf{q}\Big)^{2}\Big)\mathcal{E}_{V}+\frac{k^{+}(p_{0}^{+}-k^{+})}{p_{1}^{+}q^{+}}\mathbf{q}^{2}\mathcal{O}_{\mathrm{V}}\Bigg]\\
 & \times\pi\mathcal{B}_{1}\big(0,\hat{Q}^{2},\frac{p_{0}^{+}-k^{+}}{q^{+}}\mathbf{q}\big)\Bigg\}\;.
\end{aligned}
\label{eq:A4_T_cross}
\end{equation}

\paragraph{Instantaneous interactions}

We conclude this overview of virtual NLO contributions to the cross
section with those coming from instantaneous interactions. For a transversely
polarized photon, only a couple of them are nonzero; for the four-fermion
vertex we have:
\begin{equation}
\begin{aligned}\mathrm{Tr}\mathcal{M}_{\mathrm{LO}}^{\lambda\dagger}\mathcal{M}_{\mathrm{Q2}}^{\lambda} & =g_{\mathrm{ew}}^{2}\alpha_{s}8p_{1}^{+}p_{0}^{+}\Big(\frac{q^{+}\mathbf{P}_{\perp}^{i}}{p_{0}^{+}\mathbf{P}_{\perp}^{2}+p_{1}^{+}M^{2}}+\frac{q^{+}\mathbf{q}^{i}}{p_{0}^{+}\mathbf{q}^{2}+p_{1}^{+}M^{2}}\Big)\\
 & \times\int_{0}^{q^{+}}\mathrm{d}k^{+}\frac{4}{(p_{0}^{+}-k^{+})^{2}}\frac{p_{1}^{+}k^{+}}{p_{1}^{+}+k^{+}}\Big(\big(1+\frac{2p_{1}^{+}}{q^{+}}\big)\Big(\frac{2k^{+}}{q^{+}}-1\Big)+1\Big)\\
 & \times\int_{\mathbf{x}_{1},\mathbf{x}_{2},\mathbf{x}_{3},\mathbf{x}^{\prime}}iA^{i}\big(\mathbf{x}_{12},\tilde{M}^{2}\big)\\
 & \times\int_{\boldsymbol{\ell}_{1},\boldsymbol{\ell}_{2}}\frac{e^{i\boldsymbol{\ell}_{1}\cdot\mathbf{x}_{13}}e^{i\boldsymbol{\ell}_{2}\cdot\mathbf{x}_{23}}}{\Big(\boldsymbol{\ell}_{1}+\frac{k^{+}}{k^{+}+p_{1}^{+}}\boldsymbol{\ell}_{2}\Big)^{2}+\frac{p_{1}^{+}k^{+}p_{0}^{+}\boldsymbol{\ell}_{2}^{2}}{(q^{+}-k^{+})(p_{1}^{+}+k^{+})^{2}}}\\
 & \times\int e^{i\mathbf{k}_{\perp}\cdot\mathbf{x}^{\prime}}e^{-i\mathbf{q}\cdot\Big(\frac{q^{+}-k^{+}}{q^{+}}\mathbf{x}_{2}+\frac{k^{+}}{q^{+}}\mathbf{x}_{1}\Big)}e^{-i\mathbf{p}_{1}\cdot\mathbf{x}_{3}}\\
 & \times\big(\frac{N_{c}^{2}}{2}s_{\mathbf{x}_{1}\mathbf{x}_{2}}s_{\mathbf{x}_{3}\mathbf{x}^{\prime}}-\frac{1}{2}Q_{\mathbf{x}_{1}\mathbf{x}^{\prime}\mathbf{x}_{3}\mathbf{x}_{2}}+C_{F}N_{c}\big)\;.
\end{aligned}
\label{eq:Q2_T_cross}
\end{equation}
The non-vanishing contributions due to the instantaneous $gq\gamma q$ vertex read:
\begin{equation}
\begin{aligned} & \mathrm{Tr}\mathcal{M}_{\mathrm{LO}}^{\lambda\dagger}\mathcal{M}_{\mathrm{I2}}^{\lambda} \\ & =-g_{\mathrm{ew}}^{2}\alpha_{s}8p_{1}^{+}p_{0}^{+}\Big(\frac{q^{+}\mathbf{P}_{\perp}^{i}}{p_{0}^{+}\mathbf{P}_{\perp}^{2}+p_{1}^{+}M^{2}}+\frac{q^{+}\mathbf{q}^{i}}{p_{0}^{+}\mathbf{q}^{2}+p_{1}^{+}M^{2}}\Big)\\
 & \times4\int_{0}^{p_{0}^{+}}\frac{\mathrm{d}k^{+}}{k^{+}}\Big(\frac{k^{+}}{p_{0}^{+}}\Big)^{2}(p_{0}^{+}-k^{+})\frac{p_{0}^{+}\big((p_{1}^{+})^{2}+(p_{0}^{+})^{2}\big)+p_{1}^{+}k^{+}(k^{+}-p_{0}^{+}-p_{1}^{+})}{k^{+}q^{+}p_{0}^{+}(p_{1}^{+}-k^{+})}\\
 &\! \times\!\int_{\mathbf{x},\mathbf{x}^{\prime},\mathbf{z}}iA^{i}(\mathbf{x}\!-\!\mathbf{z})\mathcal{K}\big(\mathbf{x}\!-\!\mathbf{z},\Delta_{\mathrm{P}}\big)e^{i\mathbf{k}_{\perp}\cdot(\mathbf{x}\!-\!\mathbf{x}^{\prime})}e^{i\frac{k^{+}}{p_{0}^{+}}\mathbf{k}_{\perp}\cdot(\mathbf{x}\!-\!\mathbf{z})}\big(\frac{N_{c}^{2}}{2}s_{\mathbf{x}\mathbf{z}}s_{\mathbf{z}\mathbf{x}^{\prime}}\!-\!\frac{1}{2}s_{\mathbf{x}\mathbf{x}^{\prime}}\!+\!C_{F}N_{c}\big)\;,
\end{aligned}
\label{eq:I2_T_cross}
\end{equation}
and:
\begin{equation}
\begin{aligned} & \mathrm{Tr}\mathcal{M}_{\mathrm{LO}}^{\lambda\dagger}\mathcal{M}_{\mathrm{I3}}^{\lambda} \\
& =-g_{\mathrm{ew}}^{2}\alpha_{s}8p_{1}^{+}p_{0}^{+}\Big(\frac{q^{+}\mathbf{P}_{\perp}^{i}}{p_{0}^{+}\mathbf{P}_{\perp}^{2}+p_{1}^{+}M^{2}}+\frac{q^{+}\mathbf{q}^{i}}{p_{0}^{+}\mathbf{q}^{2}+p_{1}^{+}M^{2}}\Big)\\
 & \times\int_{0}^{p_{1}^{+}}\mathrm{d}k^{+}\frac{2(k^{+}-p_{1}^{+})\Big(q^{+}(k^{+})^{2}-k^{+}(p_{0}^{+}+p_{1}^{+})^{2}+2p_{1}^{+}\big((p_{0}^{+})^{2}+(p_{1}^{+})^{2}\big)\Big)}{(p_{1}^{+})^{3}q^{+}(p_{0}^{+}-k^{+})}\\
 & \!\times\!\int_{\mathbf{x},\mathbf{x}^{\prime},\mathbf{z}}iA^{i}(\mathbf{x}\!-\!\mathbf{z})\mathcal{K}\big(\mathbf{x}\!-\!\mathbf{z},\Delta_{\mathrm{P}}\big)e^{i\mathbf{k}_{\perp}\cdot(\mathbf{x}\!-\!\mathbf{x}^{\prime})}e^{i\frac{k^{+}}{p_{1}^{+}}\mathbf{k}_{\perp}\cdot(\mathbf{x}\!-\!\mathbf{z})}\big(\frac{N_{c}^{2}}{2}s_{\mathbf{x}\mathbf{z}}s_{\mathbf{z}\mathbf{x}^{\prime}}\!-\!\frac{1}{2}s_{\mathbf{x}\mathbf{x}^{\prime}}\!+\!C_{F}N_{c}\big)\;.
\end{aligned}
\label{eq:I3_T_cross}
\end{equation}

\subsection{Real NLO corrections}

\subsubsection{\label{subsec:real_fin_L}Longitudinal polarization}

\paragraph{Initial-state radiation}

During the analysis of initial-state radiation in section \ref{sec:DGLAP},
we have extracted the part of amplitude $\mathcal{M}_{\mathrm{IS2}}^{0\eta}$
that, upon squaring or multiplying with amplitude $\mathcal{M}_{\mathrm{IS1}}^{0\eta}$,
would lead to collinear divergences. The collinear-safe leftovers
yield the following finite contribution to the cross section:
\begin{equation}
\begin{aligned} & \int\mathrm{PS}(\vec{p}_{3})\mathrm{Tr}\Big(2\mathrm{Re}\big(\mathcal{M}_{\mathrm{IS1}}^{0\eta\dagger}\mathcal{M}_{\mathrm{IS2,finite}}^{0\eta}\big)+\Big|\mathcal{M}_{\mathrm{IS2,finite}}^{0\eta}\Big|^{2}\Big)\\
 & =\frac{g_{\mathrm{em}}^{2}\alpha_{s}N_{c}C_{F}}{M^{2}}8p_{1}^{+}p_{0}^{+}\int_{\mathbf{x},\mathbf{x}^{\prime}}\int_{\boldsymbol{\ell}}\frac{1}{\boldsymbol{\ell}^{2}}e^{-i\boldsymbol{\ell}\cdot(\mathbf{\mathbf{x}}-\mathbf{x}^{\prime})}e^{-i\mathbf{k}_{\perp}\cdot(\mathbf{x}-\mathbf{x}^{\prime})}\big(s_{\mathbf{x}^{\prime}\mathbf{x}}+1\big)\\
 & \times\int_{0}^{+\infty}\frac{\mathrm{d}p_{3}^{+}}{p_{3}^{+}}\Big(\frac{p_{3}^{+}}{p_{0\scriptscriptstyle{R}}^{+}}\Big)^{2}\frac{p_{0\scriptscriptstyle{R}}^{+}}{p_{0}^{+}}\Big(\big(1+\frac{2p_{0}^{+}}{p_{3}^{+}}\big)^{2}+1\Big)\\
 & \times\Bigg[-2\frac{p_{0}^{+}\mathbf{P}_{\perp}^{2}-p_{1}^{+}M^{2}}{p_{0}^{+}\mathbf{P}_{\perp}^{2}+p_{1}^{+}M^{2}}\\
 & \times\Big(\frac{p_{3}^{+}\big(q^{+}\boldsymbol{\ell}+p_{0}^{+}\mathbf{q}\big)^{2}-p_{0}^{+}p_{1}^{+}p_{3}^{+}M^{2}}{p_{3}^{+}\big(q^{+}\boldsymbol{\ell}+p_{0}^{+}\mathbf{q}\big)^{2}+p_{1}^{+}q^{+}p_{0R}^{+}\boldsymbol{\ell}^{2}+p_{0}^{+}p_{1}^{+}p_{3}^{+}M^{2}}-\frac{p_{0}^{+}\mathbf{q}^{2}-p_{1}^{+}M^{2}}{p_{0}^{+}\mathbf{q}^{2}+p_{1}^{+}M^{2}}\Big)\\
 & +\Big(\frac{p_{3}^{+}\big(q^{+}\boldsymbol{\ell}+p_{0}^{+}\mathbf{q}\big)^{2}-p_{0}^{+}p_{1}^{+}p_{3}^{+}M^{2}}{p_{3}^{+}\big(q^{+}\boldsymbol{\ell}+p_{0}^{+}\mathbf{q}\big)^{2}+p_{1}^{+}q^{+}p_{0R}^{+}\boldsymbol{\ell}^{2}+p_{0}^{+}p_{1}^{+}p_{3}^{+}M^{2}}-\frac{p_{0}^{+}\mathbf{q}^{2}-p_{1}^{+}M^{2}}{p_{0}^{+}\mathbf{q}^{2}+p_{1}^{+}M^{2}}\Big)^{2}\Bigg]\;.
\end{aligned}
\label{eq:DGLAP_finite}
\end{equation}
Diagrams $\mathcal{M}_{\mathrm{IS3}}^{0\eta}$ and $\mathcal{M}_{\mathrm{IS4}}^{0\eta}$
never lead to collinear singularities. Their sum, multiplied with
its complex conjugate, traced and integrated over the gluon transverse
momentum, reads:
\begin{equation}
\begin{aligned} & \int\mathrm{PS}(\vec{p}_{3})\mathrm{Tr}\Big|\mathcal{M}_{\mathrm{IS3+4}}^{0\eta}\Big|^{2} \\
& =\frac{g_{\mathrm{em}}^{2}\alpha_{s}C_{F}N_{c}}{M^{2}}8p_{1}^{+}p_{0}^{+}\int_{\mathbf{x},\mathbf{x}^{\prime}}e^{-i\mathbf{k}_{\perp}\cdot(\mathbf{x}-\mathbf{x}^{\prime})}\big(s_{\mathbf{x}\mathbf{x}^{\prime}}+1\big)\\
 & \times\Bigg\{\int_{0}^{+\infty}\frac{\mathrm{d}p_{3}^{+}}{p_{3}^{+}}\Big(\frac{p_{3}^{+}}{p_{1}^{+}+p_{3}^{+}}\Big)^{2}\frac{p_{0\scriptscriptstyle{R}}^{+}}{p_{0}^{+}}e^{i\frac{p_{3}^{+}}{p+p_{3}^{+}}\mathbf{q}\cdot(\mathbf{x}-\mathbf{x}^{\prime})}\\
 & \times\Bigg[\Bigg(\Big(\frac{p_{0\scriptscriptstyle{R}}^{+}\mathbf{q}^{2}-(p_{1}^{+}+p_{3}^{+})M^{2}}{p_{0\scriptscriptstyle{R}}^{+}\mathbf{q}^{2}+(p_{1}^{+}+p_{3}^{+})M^{2}}\Big)^{2}\Big(\big(1+\frac{2p_{1}^{+}}{p_{3}^{+}}\big)^{2}+1\Big)\\
 & -\frac{4q^{+}(p_{1}^{+}+p_{3}^{+})}{p_{3}^{+}p_{0}^{+}}\frac{p_{0\scriptscriptstyle{R}}^{+}\mathbf{q}^{2}-(p_{1}^{+}+p_{3}^{+})M^{2}}{p_{0\scriptscriptstyle{R}}^{+}\mathbf{q}^{2}+(p_{1}^{+}+p_{3}^{+})M^{2}}+2\Big(\frac{q^{+}}{p_{0}^{+}}\Big)^{2}\Bigg)\mathcal{K}\big(\mathbf{x}-\mathbf{x}^{\prime},\Delta_{\mathrm{IS}}\big)\\
 & +2\frac{(p_{1}^{+})^{2}\big((p_{0}^{+})^{2}+(p_{0}^{+}+p_{3}^{+})^{2}\big)}{(p_{0}^{+})^{2}(p_{1}^{+}+p_{3}^{+})^{2}}\mathbf{q}^{2}\int_{\boldsymbol{\ell}}\frac{e^{i\boldsymbol{\ell}\cdot(\mathbf{x}-\mathbf{x}^{\prime})}}{(\boldsymbol{\ell}^{2}+\Delta_{\mathrm{IS}})^{2}}\\
 & +\frac{4p_{1}^{+}(p_{0}^{+}+p_{3}^{+})}{p_{0}^{+}(p_{1}^{+}+p_{3}^{+})}\int_{\boldsymbol{\ell}}\frac{\mathbf{q}\cdot\boldsymbol{\ell}}{(\boldsymbol{\ell}^{2}+\Delta_{\mathrm{IS}})^{2}}e^{-i\boldsymbol{\ell}\cdot(\mathbf{x}-\mathbf{x}^{\prime})}\\
 & \times\Big(\frac{p_{0\scriptscriptstyle{R}}^{+}\mathbf{q}^{2}-(p_{1}^{+}+p_{3}^{+})M^{2}}{p_{0\scriptscriptstyle{R}}^{+}\mathbf{q}^{2}+(p_{1}^{+}+p_{3}^{+})M^{2}}\frac{2(p_{1}^{+})^{2}+(2p_{1}^{+}+p_{3}^{+})(q^{+}+p_{3}^{+})}{p_{3}^{+}p_{0\scriptscriptstyle{R}}^{+}}-\frac{q^{+}}{p_{0}^{+}}\Big)\Bigg]\\
 & -4\int_{k_{\mathrm{min}}^{+}}^{k_{f}^{+}}\frac{\mathrm{d}p_{3}^{+}}{p_{3}^{+}}\Big(\frac{p_{0}^{+}\mathbf{q}^{2}-p_{1}^{+}M^{2}}{p_{0}^{+}\mathbf{q}^{2}+p_{1}^{+}M^{2}}\Big)^{2}\int_{\boldsymbol{\ell}}\frac{e^{i\boldsymbol{\ell}\cdot(\mathbf{\mathbf{x}-\mathbf{x}^{\prime}})}}{\boldsymbol{\ell}^{2}}\Bigg\}\;.
\end{aligned}
\label{eq:IS3+4L_cross}
\end{equation}
Note that the integrals that appear in the above formula have analytical
solutions in terms of Macdonald functions, which are easily obtained
using the Schwinger trick together with formula (\ref{eq:Bessel}):
\begin{equation}
\begin{aligned}\mathcal{K}\big(\mathbf{x},\Delta\big) & =\frac{1}{2\pi}K_{0}\big(\sqrt{\mathbf{x}^{2}\Delta}\big)\;,\\
\int_{\boldsymbol{\ell}}\frac{e^{i\boldsymbol{\ell}\cdot\mathbf{x}}}{(\boldsymbol{\ell}^{2}+\Delta)^{2}} & =\frac{1}{4\pi}\sqrt{\frac{\mathbf{x}^{2}}{\Delta}}K{}_{1}\big(\sqrt{\mathbf{x}^{2}\Delta}\big)\;,\\
\int_{\boldsymbol{\ell}}\frac{\boldsymbol{\ell}^{i}e^{i\boldsymbol{\ell}\cdot\mathbf{x}}}{(\boldsymbol{\ell}^{2}\!+\!\Delta)^{2}} & \!=\!\frac{1}{4\pi}\frac{\mathbf{x}^{i}}{\sqrt{\mathbf{x}^{2}\Delta}}\!\Big(\!K{}_{1}\big(\sqrt{\mathbf{x}^{2}\Delta}\big)\!-\!\frac{1}{2}\sqrt{\mathbf{x}^{2}\Delta}K{}_{0}\big(\sqrt{\mathbf{x}^{2}\Delta}\big)\!-\!\frac{1}{2}\sqrt{\mathbf{x}^{2}\Delta}K_{2}\big(\sqrt{\mathbf{x}^{2}\Delta}\big)\!\Big)\,.
\end{aligned}
\end{equation}
Moreover, we have a contribution from the following interference term:
\begin{equation}
\begin{aligned} & \int\mathrm{PS}(\vec{p}_{3})\mathrm{Tr}\,2\mathrm{Re}\big(\mathcal{M}_{\mathrm{IS1}+2}^{0\eta\dagger}\mathcal{M}_{\mathrm{IS3+4}}^{0\eta}\big)\\
 & =\frac{g_{\mathrm{em}}^{2}\alpha_{s}C_{F}N_{c}}{M^{2}}8p_{1}^{+}p_{0}^{+}2\mathrm{Re}\int_{\mathbf{x},\mathbf{x}^{\prime}}e^{-i\mathbf{k}_{\perp}\cdot(\mathbf{x}-\mathbf{x}^{\prime})}\big(s_{\mathbf{x}\mathbf{x}^{\prime}}+1\big)\\
 & \times\Bigg\{\int_{0}^{+\infty}\frac{\mathrm{d}p_{3}^{+}}{p_{3}^{+}}\frac{p_{3}^{+}}{p_{0}^{+}}\frac{p_{3}^{+}}{p_{1}^{+}+p_{3}^{+}}\int_{\mathbf{z}}e^{i\frac{p_{3}^{+}}{p_{1}^{+}+p_{3}^{+}}\mathbf{q}\cdot(\mathbf{x}-\mathbf{z})}\int_{\boldsymbol{\ell}_{2}}\frac{1}{\boldsymbol{\ell}_{2}^{2}+\Delta_{\mathrm{IS}}}e^{-i\boldsymbol{\ell}_{2}\cdot(\mathbf{\mathbf{x}-\mathbf{z}})}\int_{\boldsymbol{\ell}}\frac{\boldsymbol{\ell}^{j}}{\boldsymbol{\ell}^{2}}e^{i\boldsymbol{\ell}\cdot(\mathbf{x}^{\prime}-\mathbf{\mathbf{z}})}\\
 & \times\Bigg[-\frac{p_{0}^{+}\mathbf{P}_{\perp}^{2}-p_{1}^{+}M^{2}}{p_{0}^{+}\mathbf{P}_{\perp}^{2}+p_{1}^{+}M^{2}}+\frac{p_{3}^{+}\big(q^{+}\boldsymbol{\ell}+p_{0}^{+}\mathbf{q}\big)^{2}-p_{3}^{+}p_{1}^{+}p_{0}^{+}M^{2}}{p_{3}^{+}\big(q^{+}\boldsymbol{\ell}+p_{0}^{+}\mathbf{q}\big)^{2}+p_{1}^{+}q^{+}p_{0R}^{+}\boldsymbol{\ell}^{2}+p_{1}^{+}p_{3}^{+}p_{0}^{+}M^{2}}\Bigg]\\
 & \times\Bigg[\Bigg(\frac{p_{0\scriptscriptstyle{R}}^{+}\mathbf{q}^{2}-(p_{1}^{+}+p_{3}^{+})M^{2}}{p_{0\scriptscriptstyle{R}}^{+}\mathbf{q}^{2}+(p_{1}^{+}+p_{3}^{+})M^{2}}\Big(\big(1+\frac{2p_{0}^{+}}{p_{3}^{+}}\big)\big(1+\frac{2p_{1}^{+}}{p_{3}^{+}}\big)+1\Big)-\frac{q^{+}}{p_{0}^{+}}\frac{2(p_{0}^{+}+p_{3}^{+})}{p_{3}^{+}}\Bigg)\boldsymbol{\ell}_{2}^{j}\\
 & -\frac{2p_{1}^{+}\big((p_{0}^{+})^{2}+(p_{3}^{+}+p_{0}^{+})^{2}\big)}{p_{3}^{+}p_{0}^{+}(p_{1}^{+}+p_{3}^{+})}\mathbf{q}^{i}\Bigg]\\
 & +4\int_{k_{\mathrm{min}}^{+}}^{k_{f}^{+}}\frac{\mathrm{d}p_{3}^{+}}{p_{3}^{+}}\frac{p_{0}^{+}\mathbf{P}_{\perp}^{2}-p_{1}^{+}M^{2}}{p_{0}^{+}\mathbf{P}_{\perp}^{2}+p_{1}^{+}M^{2}}\frac{p_{0}^{+}\mathbf{q}^{2}-p_{1}^{+}M^{2}}{p_{0}^{+}\mathbf{q}^{2}+p_{1}^{+}M^{2}}\int_{\boldsymbol{\ell}}\frac{e^{-i\boldsymbol{\ell}\cdot(\mathbf{x}-\mathbf{x}^{\prime})}}{\boldsymbol{\ell}^{2}}\Bigg\}\;.
\end{aligned}
\label{eq:IS1+2-IS3+4L_cross}
\end{equation}

\paragraph{Final-state radiation}

In the analysis of collinear divergences in final-state radiation
\ref{sec:jet}, we have noted that the amplitudes $\mathcal{M}_{\mathrm{FS2}}^{0\eta}$
and $\mathcal{M}_{\mathrm{FS3}}^{0\eta}$ exhibit collinear-like behavior
even when the gluon and the quark are not grouped inside the jet.
Even though the jet function cuts off the collinear pole in this \textquoteleft outside-jet'
configuration, there are large collinear logarithms that combine with
the \textquoteleft inside-jet' contributions. We have extracted the
part of $\mathcal{M}_{\mathrm{FS2}}^{0\eta}$ responsible for these
large logs outside the jet, and its leftover, as well as its interference
with $\mathcal{M}_{\mathrm{FS3}}^{0\eta}$, is completely finite.
This results in the following, finite, contribution to the cross section:
\begin{equation}
\begin{aligned} & \int\mathrm{PS}(\vec{p}_{3})\mathrm{Tr}\Big(\big|\mathcal{M}_{\mathrm{FS2,finite}}^{0\eta}\big|^{2}+2\mathrm{Re}\big(\mathcal{M}_{\mathrm{FS2,finite}}^{0\eta\dagger}\mathcal{M}_{\mathrm{FS3}}^{0\eta}\big)\Big)\\
 & =\frac{g_{\mathrm{em}}^{2}g_{s}^{2}C_{F}N_{c}}{M^{2}}8p_{1}^{+}p_{0}^{+}\int_{\mathbf{x},\mathbf{x}^{\prime}}e^{-i\mathbf{k}_{\perp}\cdot(\mathbf{x}-\mathbf{x}^{\prime})}\big(s_{\mathbf{x}\mathbf{x}^{\prime}}+1\big)\\
 & \times\int_{0}^{+\infty}\frac{\mathrm{d}p_{3}^{+}}{p_{3}^{+}}\Big(\frac{p_{3}^{+}}{p_{1}^{+}}\Big)^{2}\frac{p_{0\scriptscriptstyle{R}}^{+}}{p_{0}^{+}}\Big(\big(1+\frac{2p_{1}^{+}}{p_{3}^{+}}\big)^{2}+1\Big)\int_{\mathbf{p}_{3}}\frac{e^{-i\mathbf{p}_{3}\cdot(\mathbf{x}-\mathbf{x}^{\prime})}}{\big(\mathbf{p}_{3}-\frac{p_{3}^{+}}{p_{1}^{+}}\mathbf{p}_{1}\big)^{2}}\\
 & \times\Bigg[\Big(\frac{p_{0\scriptscriptstyle{R}}^{+}p_{3}^{+}}{q^{+}p_{0}^{+}}\frac{\frac{\big((p_{1}^{+}+p_{3}^{+})\mathbf{q}-q^{+}(\mathbf{p}_{1}+\mathbf{p}_{3})\big)^{2}}{p_{0\scriptscriptstyle{R}}^{+}(p_{1}^{+}+p_{3}^{+})}-M^{2}}{\big(\mathbf{p}_{3}-\frac{p_{3}^{+}}{p_{0}^{+}}\mathbf{k}_{\perp}\big)^{2}+\Delta_{\mathrm{FS}}}-\frac{p_{1}^{+}\frac{p_{1}^{+}+p_{3}^{+}}{p_{0\scriptscriptstyle{R}}^{+}}\Big(\frac{p_{0}^{+}}{p_{1}^{+}}\Big)^{2}\mathbf{P}_{\perp}^{2}-p_{1}^{+}M^{2}}{p_{1}^{+}\frac{p_{1}^{+}+p_{3}^{+}}{p_{0\scriptscriptstyle{R}}^{+}}\Big(\frac{p_{0}^{+}}{p_{1}^{+}}\Big)^{2}\mathbf{P}_{\perp}^{2}+p_{1}^{+}M^{2}}\Big)^{2}\\
 & -2\frac{p_{0\scriptscriptstyle{R}}^{+}\mathbf{q}^{2}-(p_{1}^{+}+p_{3}^{+})M^{2}}{p_{0\scriptscriptstyle{R}}^{+}\mathbf{q}^{2}+(p_{1}^{+}+p_{3}^{+})M^{2}}\\
 & \times \Big(\frac{p_{0\scriptscriptstyle{R}}^{+}p_{3}^{+}}{q^{+}p_{0}^{+}}\frac{\frac{\big((p_{1}^{+}+p_{3}^{+})\mathbf{q}-q^{+}(\mathbf{p}_{1}+\mathbf{p}_{3})\big)^{2}}{p_{0\scriptscriptstyle{R}}^{+}(p_{1}^{+}+p_{3}^{+})}-M^{2}}{\big(\mathbf{p}_{3}-\frac{p_{3}^{+}}{p_{0}^{+}}\mathbf{k}_{\perp}\big)^{2}+\Delta_{\mathrm{FS}}}-\frac{p_{1}^{+}\frac{p_{1}^{+}+p_{3}^{+}}{p_{0\scriptscriptstyle{R}}^{+}}\Big(\frac{p_{0}^{+}}{p_{1}^{+}}\Big)^{2}\mathbf{P}_{\perp}^{2}-p_{1}^{+}M^{2}}{p_{1}^{+}\frac{p_{1}^{+}+p_{3}^{+}}{p_{0\scriptscriptstyle{R}}^{+}}\Big(\frac{p_{0}^{+}}{p_{1}^{+}}\Big)^{2}\mathbf{P}_{\perp}^{2}+p_{1}^{+}M^{2}}\Big)\Bigg]\;.
\end{aligned}
\label{eq:FS2+3L_finite_cross}
\end{equation}
Moreover, we have the following finite contributions to the cross
section stemming from real final-state radiation:
\begin{equation}
\begin{aligned} & \int\mathrm{PS}(\vec{p}_{3})\mathrm{Tr}\big|\mathcal{M}_{\mathrm{FS1}+4}^{0\eta}\big|^{2}\\
 & =\frac{g_{\mathrm{em}}^{2}\alpha_{s}N_{c}C_{F}}{M^{2}}8p_{1}^{+}p_{0}^{+}\int_{\mathbf{x},\mathbf{x}^{\prime}}e^{-i\mathbf{k}_{\perp}\cdot(\mathbf{x}-\mathbf{x}^{\prime})}\big(s_{\mathbf{x}\mathbf{x}^{\prime}}+1\big)\\
 & \times\Bigg\{\int_{0}^{+\infty}\frac{\mathrm{d}p_{3}^{+}}{p_{3}^{+}}\Big(\frac{p_{3}^{+}}{p_{0}^{+}}\Big)^{2}\frac{p_{0\scriptscriptstyle{R}}^{+}}{p_{0}^{+}}e^{-i\frac{p_{3}^{+}}{p_{0}^{+}}\mathbf{k}_{\perp}\cdot\mathbf{x}}\Bigg[\Bigg(\Big(\frac{p_{0}^{+}\mathbf{P}_{\perp}^{2}-p_{1}^{+}M^{2}}{p_{0}^{+}\mathbf{P}_{\perp}^{2}+p_{1}^{+}M^{2}}\Big)^{2}\Big(\big(1+\frac{2p_{0}^{+}}{p_{3}^{+}}\big)^{2}+1\Big)\\
 & +2\Big(\frac{q^{+}}{p_{1}^{+}+p_{3}^{+}}\Big)^{2}+\frac{4q^{+}p_{0}^{+}}{p_{3}^{+}(p_{1}^{+}+p_{3}^{+})}\frac{p_{0}^{+}\mathbf{P}_{\perp}^{2}-p_{1}^{+}M^{2}}{p_{0}^{+}\mathbf{P}_{\perp}^{2}+p_{1}^{+}M^{2}}\Bigg)\mathcal{K}(\mathbf{x}-\mathbf{x}^{\prime},\Delta_{\mathrm{FS}})\\
 & +\frac{2(p_{0\scriptscriptstyle{R}}^{+})^{2}\big((p_{1}^{+})^{2}+(p_{1}^{+}+p_{3}^{+})^{2}\big)}{(p_{1}^{+})^{2}(p_{1}^{+}+p_{3}^{+})^{2}}\mathbf{P}_{\perp}^{2}\int_{\boldsymbol{\ell}}\frac{e^{i\boldsymbol{\ell}\cdot(\mathbf{x}-\mathbf{x}^{\prime})}}{(\boldsymbol{\ell}^{2}+\Delta_{\mathrm{FS}})^{2}}\\
 & +\frac{4p_{0\scriptscriptstyle{R}}^{+}}{p_{1}^{+}+p_{3}^{+}}\Big(\frac{p_{0}^{+}\mathbf{P}_{\perp}^{2}-p_{1}^{+}M^{2}}{p_{0}^{+}\mathbf{P}_{\perp}^{2}+p_{1}^{+}M^{2}}\frac{2(p_{1}^{+})^{2}+(2p_{1}^{+}+p_{3}^{+})(p_{3}^{+}+q^{+})}{p_{1}^{+}p_{3}^{+}}-\frac{q^{+}}{p_{1}^{+}+p_{3}^{+}}\Big)\\
 & \times\int_{\boldsymbol{\ell}}\frac{\mathbf{P}_{\perp}\cdot\boldsymbol{\ell}}{(\boldsymbol{\ell}^{2}+\Delta_{\mathrm{FS}})^{2}}e^{-i\boldsymbol{\ell}\cdot(\mathbf{x}-\mathbf{x}^{\prime})}\Bigg]\\
 & -4\int_{k_{\mathrm{min}}^{+}}^{k_{f}^{+}}\frac{\mathrm{d}p_{3}^{+}}{p_{3}^{+}}\Big(\frac{p_{0}^{+}\mathbf{P}_{\perp}^{2}-p_{1}^{+}M^{2}}{p_{0}^{+}\mathbf{P}_{\perp}^{2}+p_{1}^{+}M^{2}}\Big)^{2}\int_{\boldsymbol{\ell}}\frac{e^{i\boldsymbol{\ell}\cdot(\mathbf{x}-\mathbf{x}^{\prime})}}{\boldsymbol{\ell}^{2}}\Bigg\}\;,
\end{aligned}
\label{eq:FS1+4L_cross}
\end{equation}
and:
\begin{equation}
\begin{aligned} & \int\mathrm{PS}(\vec{p}_{3})\mathrm{Tr}\,2\mathrm{Re}\big(\mathcal{M}_{\mathrm{FS2+3}}^{0\eta\dagger}\mathcal{M}_{\mathrm{FS1}+4}^{0\eta}\big) \\
& =\frac{g_{\mathrm{em}}^{2}\alpha_{s}N_{c}C_{F}}{M^{2}}8p_{1}^{+}p_{0}^{+}2\mathrm{Re}\int_{\mathbf{x},\mathbf{x}^{\prime}}e^{-i\mathbf{k}_{\perp}\cdot(\mathbf{x}-\mathbf{x}^{\prime})}\big(s_{\mathbf{x}\mathbf{x}^{\prime}}+1\big)\\
 & \times\Bigg\{\int_{0}^{+\infty}\frac{\mathrm{d}p_{3}^{+}}{p_{3}^{+}}\int_{\mathbf{z}}\frac{(p_{3}^{+})^{2}}{p_{1}^{+}p_{0}^{+}}\frac{p_{0\scriptscriptstyle{R}}^{+}}{p_{0}^{+}}e^{-i\frac{p_{3}^{+}}{p_{0}^{+}}\mathbf{k}_{\perp}\cdot(\mathbf{x}-\mathbf{z})}\int_{\mathbf{p}_{3}}e^{i\mathbf{p}_{3}\cdot(\mathbf{x}^{\prime}-\mathbf{z})}\\
 & \!\times\!\frac{\mathbf{p}_{3}^{i}\!-\!\frac{p_{3}^{+}}{p_{1}^{+}}\mathbf{p}_{1}^{i}}{\big(\mathbf{p}_{3}\!-\!\frac{p_{3}^{+}}{p_{1}^{+}}\mathbf{p}_{1}\big)^{2}}\!\Bigg[\!\frac{p_{3}^{+}\big(\mathbf{p}_{3}\!+\!\mathbf{p}_{1}\!-\!\frac{p_{1}^{+}\!+p_{3}^{+}}{q^{+}}\mathbf{q}\big)^{2}\!-\!p_{3}^{+}\frac{p_{0\scriptscriptstyle{R}}^{+}(p_{1}^{+}\!+p_{3}^{+})}{(q^{+})^{2}}M^{2}}{\frac{p_{1}^{+}p_{0\scriptscriptstyle{R}}^{+}}{q^{+}}\big(\mathbf{p}_{3}\!-\!\frac{p_{3}^{+}}{p_{1}^{+}}\mathbf{p}_{1}\big)^{2}\!+\!p_{3}^{+}\big(\mathbf{p}_{3}\!+\!\mathbf{p}_{1}\!-\!\frac{p_{1}^{+}\!+p_{3}^{+}}{q^{+}}\mathbf{q}\big)^{2}\!+\!p_{3}^{+}\frac{p_{0\scriptscriptstyle{R}}^{+}(p_{1}^{+}\!+p_{3}^{+})}{(q^{+})^{2}}M^{2}}\\
 & -\frac{p_{0\scriptscriptstyle{R}}^{+}\mathbf{q}^{2}-(p_{1}^{+}+p_{3}^{+})M^{2}}{p_{0\scriptscriptstyle{R}}^{+}\mathbf{q}^{2}+(p_{1}^{+}+p_{3}^{+})M^{2}}\Bigg]\\
 & \times\Bigg[\Bigg(\frac{p_{0}^{+}\mathbf{P}_{\perp}^{2}-p_{1}^{+}M^{2}}{p_{0}^{+}\mathbf{P}_{\perp}^{2}+p_{1}^{+}M^{2}}\Big(\big(1+\frac{2p_{1}^{+}}{p_{3}^{+}}\big)\big(1+\frac{2p_{0}^{+}}{p_{3}^{+}}\big)+1\Big)-\frac{q^{+}}{p_{1}^{+}+p_{3}^{+}}\frac{2p_{1}^{+}}{p_{3}^{+}}\Bigg)iA^{i}(\mathbf{x}-\mathbf{z},\Delta_{\mathrm{FS}})\\
 & -p_{0\scriptscriptstyle{R}}^{+}\frac{2p_{1}^{+}+p_{3}^{+}}{p_{1}^{+}(p_{1}^{+}+p_{3}^{+})}\mathcal{K}(\mathbf{x}-\mathbf{z},\Delta_{\mathrm{FS}})\mathbf{P}_{\perp}^{i}\Big(1+\frac{2p_{1}^{+}}{p_{3}^{+}}+\frac{p_{3}^{+}}{2p_{1}^{+}+p_{3}^{+}}\Big)\Bigg]\\
 & +4\int_{k_{\mathrm{min}}^{+}}^{k_{f}^{+}}\frac{\mathrm{d}p_{3}^{+}}{p_{3}^{+}}\frac{p_{0}^{+}\mathbf{q}^{2}-p_{1}^{+}M^{2}}{p_{0}^{+}\mathbf{q}^{2}+p_{1}^{+}M^{2}}\frac{p_{0}^{+}\mathbf{P}_{\perp}^{2}-p_{1}^{+}M^{2}}{p_{0}^{+}\mathbf{P}_{\perp}^{2}+p_{1}^{+}M^{2}}\int_{\boldsymbol{\ell}}\frac{e^{i\boldsymbol{\ell}\cdot(\mathbf{x}-\mathbf{x}^{\prime})}}{\boldsymbol{\ell}^{2}}\Bigg\}\;.
\end{aligned}
\label{eq:FS2+3-FS1+4L_cross}
\end{equation}

\paragraph{Initial-final state radiation interference}

Finally, in our calculation, the interference between initial- and
final state radiation never leads to collinear divergences, although
it constitutes an important contribution to the high-energy resummation.
We have:
\begin{equation}
\begin{aligned} & \int\mathrm{PS}(\vec{p}_{3})\mathrm{Tr}\,2\mathrm{Re}\big(\mathcal{M}_{\mathrm{IS1}+2}^{0\eta\dagger}\mathcal{M}_{\mathrm{FS1}+4}^{0\eta}\big)\\
 & =\frac{g_{\mathrm{em}}^{2}\alpha_{s}}{M^{2}}8p_{1}^{+}p_{0}^{+}2\mathrm{Re}\int_{\mathbf{x},\mathbf{x}^{\prime},\mathbf{z}}e^{-i\mathbf{k}_{\perp}\cdot(\mathbf{x}-\mathbf{x}^{\prime})}\big(\frac{N_{c}^{2}}{2}s_{\mathbf{z}\mathbf{x}^{\prime}}s_{\mathbf{x}\mathbf{z}}-\frac{1}{2}s_{\mathbf{x}\mathbf{x}^{\prime}}+C_{F}N_{c}\big)\\
 & \times\Bigg\{\int_{0}^{+\infty}\frac{\mathrm{d}p_{3}^{+}}{p_{3}^{+}}\Big(\frac{p_{3}^{+}}{p_{0}^{+}}\Big)^{2}\int_{\mathbf{p}_{3}}\frac{e^{-i\mathbf{p}_{3}\cdot(\mathbf{x}-\mathbf{z})}}{\big(\mathbf{p}_{3}-\frac{p_{3}^{+}}{p_{0}^{+}}\mathbf{k}_{\perp}\big)^{2}+\Delta_{\mathrm{FS}}}\int_{\boldsymbol{\ell}}\frac{\boldsymbol{\ell}^{j}}{\boldsymbol{\ell}^{2}}e^{i\boldsymbol{\ell}\cdot(\mathbf{\mathbf{x}^{\prime}-\mathbf{z}})}\\
 & \times\Bigg[-\frac{p_{0}^{+}\mathbf{P}_{\perp}^{2}-p_{1}^{+}M^{2}}{p_{0}^{+}\mathbf{P}_{\perp}^{2}+p_{1}^{+}M^{2}}+\frac{p_{3}^{+}\big(q^{+}\boldsymbol{\ell}+p_{0}^{+}\mathbf{q}\big)^{2}-p_{1}^{+}p_{0}^{+}p_{3}^{+}M^{2}}{p_{3}^{+}\big(q^{+}\boldsymbol{\ell}+p_{0}^{+}\mathbf{q}\big)^{2}+p_{1}^{+}q^{+}p_{0R}^{+}\boldsymbol{\ell}^{2}+p_{1}^{+}p_{3}^{+}p_{0}^{+}M^{2}}\Bigg]\\
 & \times\Bigg[\Bigg(\frac{p_{0}^{+}\mathbf{P}_{\perp}^{2}-p_{1}^{+}M^{2}}{p_{0}^{+}\mathbf{P}_{\perp}^{2}+p_{1}^{+}M^{2}}\Big(\big(1+\frac{2p_{0}^{+}}{p_{3}^{+}}\big)^{2}+1\Big)-\frac{q^{+}}{p_{1}^{+}+p_{3}^{+}}\frac{2p_{0}^{+}}{p_{3}^{+}}\Bigg)\big(\mathbf{p}_{3}-\frac{p_{3}^{+}}{p_{0}^{+}}\mathbf{k}_{\perp}\big)^{j}\\
 & -p_{0\scriptscriptstyle{R}}^{+}\frac{2p_{1}^{+}+p_{3}^{+}}{p_{1}^{+}(p_{1}^{+}+p_{3}^{+})}\mathbf{P}_{\perp}^{j}\Big(1+\frac{2p_{0}^{+}}{p_{3}^{+}}+\frac{p_{3}^{+}}{2p_{1}^{+}+p_{3}^{+}}\Big)\Bigg]\\
 & +4\int_{k_{\mathrm{min}}^{+}}^{k_{f}^{+}}\frac{\mathrm{d}p_{3}^{+}}{p_{3}^{+}}\Big(\frac{p_{0}^{+}\mathbf{P}_{\perp}^{2}-p_{1}^{+}M^{2}}{p_{0}^{+}\mathbf{P}_{\perp}^{2}+p_{1}^{+}M^{2}}\Big)^{2}A^{i}(\mathbf{x}-\mathbf{z})A^{i}(\mathbf{x}^{\prime}-\mathbf{z})\Bigg\}\;.
\end{aligned}
\label{eq:IS1+2-FS1+4L_cross}
\end{equation}
Moreover:
\begin{equation}
\begin{aligned} & \int\mathrm{PS}(\vec{p}_{3})\mathrm{Tr}\,2\mathrm{Re}\big(\mathcal{M}_{\mathrm{IS1}+2}^{0\eta\dagger}\mathcal{M}_{\mathrm{FS2+3}}^{0\eta}\big)\\
 & =\frac{g_{\mathrm{em}}^{2}\alpha_{s}}{M^{2}}8p_{1}^{+}p_{0}^{+}2\mathrm{Re}\int_{\mathbf{x},\mathbf{x}^{\prime},\mathbf{z}}e^{i\mathbf{k}_{\perp}\cdot(\mathbf{x}-\mathbf{x}^{\prime})}\big(\frac{N_{c}^{2}}{2}s_{\mathbf{z}\mathbf{x}^{\prime}}s_{\mathbf{x}\mathbf{z}}-\frac{1}{2}s_{\mathbf{x}\mathbf{x}^{\prime}}+C_{F}N_{c}\big)\\
 & \times\Bigg\{\int_{0}^{+\infty}\frac{\mathrm{d}p_{3}^{+}}{p_{3}^{+}}\frac{p_{3}^{+}}{p_{0}^{+}}\frac{p_{3}^{+}}{p_{1}^{+}}\Big(\big(1+\frac{2p_{0}^{+}}{p_{3}^{+}}\big)\big(1+\frac{2p_{1}^{+}}{p_{3}^{+}}\big)+1\Big)\\
 & \times\int_{\mathbf{p}_{3}}e^{-i\mathbf{p}_{3}\cdot(\mathbf{x}-\mathbf{z})}\frac{\mathbf{p}_{3}^{i}-\frac{p_{3}^{+}}{p_{1}^{+}}\mathbf{p}_{1}^{i}}{\big(\mathbf{p}_{3}-\frac{p_{3}^{+}}{p_{1}^{+}}\mathbf{p}_{1}\big)^{2}}\int_{\boldsymbol{\ell}}e^{i\boldsymbol{\ell}\cdot(\mathbf{x}^{\prime}-\mathbf{z})}\frac{\boldsymbol{\ell}^{i}}{\boldsymbol{\ell}^{2}}\\
 & \times\Bigg[-\frac{p_{0}^{+}\mathbf{P}_{\perp}^{2}-p_{1}^{+}M^{2}}{p_{0}^{+}\mathbf{P}_{\perp}^{2}+p_{1}^{+}M^{2}}+\frac{p_{3}^{+}\big(q^{+}\boldsymbol{\ell}+p_{0}^{+}\mathbf{q}\big)^{2}-p_{1}^{+}p_{3}^{+}p_{0}^{+}M^{2}}{p_{1}^{+}q^{+}p_{0R}^{+}\boldsymbol{\ell}^{2}+p_{3}^{+}\big(q^{+}\boldsymbol{\ell}+p_{0}^{+}\mathbf{q}\big)^{2}+p_{1}^{+}p_{3}^{+}p_{0}^{+}M^{2}}\Bigg]\\
 & \times\Bigg[\frac{p_{3}^{+}\big(\mathbf{p}_{3}+\mathbf{p}_{1}-\frac{p_{1}^{+}+p_{3}^{+}}{q^{+}}\mathbf{q}\big)^{2}-p_{3}^{+}\frac{p_{0\scriptscriptstyle{R}}^{+}(p_{1}^{+}+p_{3}^{+})}{(q^{+})^{2}}M^{2}}{\frac{p_{1}^{+}p_{0\scriptscriptstyle{R}}^{+}}{q^{+}}\big(\mathbf{p}_{3}-\frac{p_{3}^{+}}{p_{1}^{+}}\mathbf{p}_{1}\big)^{2}+p_{3}^{+}\big(\mathbf{p}_{3}+\mathbf{p}_{1}-\frac{p_{1}^{+}+p_{3}^{+}}{q^{+}}\mathbf{q}\big)^{2}+p_{3}^{+}\frac{p_{0\scriptscriptstyle{R}}^{+}(p_{1}^{+}+p_{3}^{+})}{(q^{+})^{2}}M^{2}}\\
 & -\frac{p_{0\scriptscriptstyle{R}}^{+}\mathbf{q}^{2}-(p_{1}^{+}+p_{3}^{+})M^{2}}{p_{0\scriptscriptstyle{R}}^{+}\mathbf{q}^{2}+(p_{1}^{+}+p_{3}^{+})M^{2}}\Bigg]\\
 & -4\int_{k_{\mathrm{min}}^{+}}^{k_{f}^{+}}\frac{\mathrm{d}p_{3}^{+}}{p_{3}^{+}}\frac{p_{0}^{+}\mathbf{P}_{\perp}^{2}-p_{1}^{+}M^{2}}{p_{0}^{+}\mathbf{P}_{\perp}^{2}+p_{1}^{+}M^{2}}\frac{p_{0}^{+}\mathbf{q}^{2}-p_{1}^{+}M^{2}}{p_{0}^{+}\mathbf{q}^{2}+p_{1}^{+}M^{2}}A^{i}(\mathbf{x}-\mathbf{z})A^{i}(\mathbf{x}^{\prime}-\mathbf{z})\Bigg\}\;.
\end{aligned}
\label{eq:IS1+2-FS2+3L_cross}
\end{equation}
Furthermore:
\begin{equation}
\begin{aligned} & \int\mathrm{PS}(\vec{p}_{3})\mathrm{Tr}\,2\mathrm{Re}\big(\mathcal{M}_{\mathrm{IS3+4}}^{0\eta\dagger}\mathcal{M}_{\mathrm{FS1}+4}^{0\eta}\big)\\
 & =\frac{g_{\mathrm{em}}^{2}\alpha_{s}}{M^{2}}8p_{1}^{+}p_{0}^{+}2\mathrm{Re}\int_{\mathbf{x},\mathbf{x}^{\prime},\mathbf{z}}e^{-i\mathbf{k}_{\perp}\cdot(\mathbf{x}-\mathbf{x}^{\prime})}\big(\frac{N_{c}^{2}}{2}s_{\mathbf{z}\mathbf{x}^{\prime}}s_{\mathbf{x}\mathbf{z}}-\frac{1}{2}s_{\mathbf{x}\mathbf{x}^{\prime}}+C_{F}N_{c}\big)\\
 & \Bigg\{\int_{0}^{+\infty}\frac{\mathrm{d}p_{3}^{+}}{p_{3}^{+}}\frac{p_{0\scriptscriptstyle{R}}^{+}}{p_{1}^{+}+p_{3}^{+}}\Big(\frac{p_{3}^{+}}{p_{0}^{+}}\Big)^{2}e^{-i\frac{p_{3}^{+}}{p_{1}^{+}+p_{3}^{+}}\mathbf{q}\cdot(\mathbf{x}^{\prime}-\mathbf{z})}e^{-i\frac{p_{3}^{+}}{p_{0}^{+}}\mathbf{k}_{\perp}\cdot(\mathbf{x}-\mathbf{z})}\\
 & \times\Bigg[\Bigg(\frac{p_{0\scriptscriptstyle{R}}^{+}\mathbf{q}^{2}-(p_{1}^{+}+p_{3}^{+})M^{2}}{p_{0\scriptscriptstyle{R}}^{+}\mathbf{q}^{2}+(p_{1}^{+}+p_{3}^{+})M^{2}}\frac{p_{0}^{+}\mathbf{P}_{\perp}^{2}-p_{1}^{+}M^{2}}{p_{0}^{+}\mathbf{P}_{\perp}^{2}+p_{1}^{+}M^{2}}\Big(\big(1+\frac{2p_{1}^{+}}{p_{3}^{+}}\big)\big(1+\frac{2p_{0}^{+}}{p_{3}^{+}}\big)+1\Big)\\
 & -2\frac{p_{0}^{+}+p_{3}^{+}}{p_{3}^{+}}\frac{q^{+}}{p_{0}^{+}}\frac{p_{0}^{+}\mathbf{P}_{\perp}^{2}-p_{1}^{+}M^{2}}{p_{0}^{+}\mathbf{P}_{\perp}^{2}+p_{1}^{+}M^{2}}-\frac{2p_{1}^{+}}{p_{3}^{+}}\frac{q^{+}}{p_{1}^{+}+p_{3}^{+}}\frac{p_{0\scriptscriptstyle{R}}^{+}\mathbf{q}^{2}-(p_{1}^{+}+p_{3}^{+})M^{2}}{p_{0\scriptscriptstyle{R}}^{+}\mathbf{q}^{2}+(p_{1}^{+}+p_{3}^{+})M^{2}}\Bigg)\\
 & \times A^{i}(\mathbf{x}^{\prime}-\mathbf{z},\Delta_{\mathrm{IS}})A^{i}(\mathbf{x}-\mathbf{z},\Delta_{\mathrm{FS}})\\
 & +2\frac{p_{0\scriptscriptstyle{R}}^{+}}{p_{1}^{+}}\Bigg(\frac{p_{0\scriptscriptstyle{R}}^{+}\mathbf{q}^{2}-(p_{1}^{+}+p_{3}^{+})M^{2}}{p_{0\scriptscriptstyle{R}}^{+}\mathbf{q}^{2}+(p_{1}^{+}+p_{3}^{+})M^{2}}\frac{(p_{1}^{+})^{2}+(p_{1}^{+}+p_{3}^{+})^{2}}{p_{3}^{+}(p_{1}^{+}+p_{3}^{+})}-\frac{q^{+}}{p_{0}^{+}}\Bigg)\\
 & \times\mathbf{P}_{\perp}^{i}iA^{i}(\mathbf{x}^{\prime}-\mathbf{z},\Delta_{\mathrm{IS}})\mathcal{K}(\mathbf{x}-\mathbf{z},\Delta_{\mathrm{FS}})\\
 & -\frac{2p_{1}^{+}}{p_{1}^{+}+p_{3}^{+}}\Bigg(\frac{p_{0}^{+}\mathbf{P}_{\perp}^{2}-p_{1}^{+}M^{2}}{p_{0}^{+}\mathbf{P}_{\perp}^{2}+p_{1}^{+}M^{2}}\frac{2(p_{0}^{+})^{2}+p_{3}^{+}(p_{3}^{+}+2p_{0}^{+})}{p_{3}^{+}p_{0}^{+}}-\frac{q^{+}}{p_{1}^{+}+p_{3}^{+}}\Bigg)\\
 & \times\mathbf{q}^{i}iA^{i}(\mathbf{x}-\mathbf{z},\Delta_{\mathrm{FS}})\mathcal{K}\big(\mathbf{x}^{\prime}-\mathbf{z},\Delta_{\mathrm{IS}})\\
 & +\frac{2p_{0\scriptscriptstyle{R}}^{+}\big(2p_{1}^{+}p_{0\scriptscriptstyle{R}}^{+}+p_{3}^{+}q^{+}\big)}{p_{0}^{+}(p_{1}^{+}+p_{3}^{+})^{2}}\mathbf{q}\cdot\mathbf{P}_{\perp}\times\mathcal{K}\big(\mathbf{x}^{\prime}-\mathbf{z},\Delta_{\mathrm{IS}})\mathcal{K}(\mathbf{x}-\mathbf{z},\Delta_{\mathrm{FS}})\Bigg]\\
 & -4\int_{k_{\mathrm{min}}^{+}}^{k_{f}^{+}}\frac{\mathrm{d}p_{3}^{+}}{p_{3}^{+}}\frac{p_{0}^{+}\mathbf{q}^{2}-p_{1}^{+}M^{2}}{p_{0}^{+}\mathbf{q}^{2}+p_{1}^{+}M^{2}}\frac{p_{0}^{+}\mathbf{P}_{\perp}^{2}-p_{1}^{+}M^{2}}{p_{0}^{+}\mathbf{P}_{\perp}^{2}+p_{1}^{+}M^{2}}A^{i}(\mathbf{x}^{\prime}-\mathbf{z})A^{i}(\mathbf{x}-\mathbf{z})\Bigg\}\;.
\end{aligned}
\label{eq:IS34FS14L_cross}
\end{equation}
The final contribution to the finite real NLO corrections is:
\begin{equation}
\begin{aligned} & \int\mathrm{PS}(\vec{p}_{3})\mathrm{Tr}\,2\mathrm{Re}\big(\mathcal{M}_{\mathrm{IS3+4}}^{0\eta\dagger}\mathcal{M}_{\mathrm{FS2+3}}^{0\eta}\big)\\
 & =\frac{g_{\mathrm{em}}^{2}\alpha_{s}}{M^{2}}8p_{1}^{+}p_{0}^{+}2\mathrm{Re}\int_{\mathbf{x},\mathbf{x}^{\prime},\mathbf{z}}e^{-i\mathbf{k}_{\perp}(\mathbf{x}-\mathbf{x}^{\prime})}\big(\frac{N_{c}^{2}}{2}s_{\mathbf{z}\mathbf{x}^{\prime}}s_{\mathbf{x}\mathbf{z}}-\frac{1}{2}s_{\mathbf{x}\mathbf{x}^{\prime}}+C_{F}N_{c}\big)\\
 & \times\Bigg\{\int_{0}^{+\infty}\frac{\mathrm{d}p_{3}^{+}}{p_{3}^{+}}\frac{p_{3}^{+}}{p_{1}^{+}+p_{3}^{+}}\frac{p_{3}^{+}}{p_{1}^{+}}\frac{p_{0\scriptscriptstyle{R}}^{+}}{p_{0}^{+}}\,e^{-i\frac{p_{3}^{+}}{p_{1}^{+}+p_{3}^{+}}\mathbf{q}\cdot(\mathbf{x}^{\prime}-\mathbf{z})}\int_{\mathbf{p}_{3}}\frac{\mathbf{p}_{3}^{j}-\frac{p_{3}^{+}}{p_{1}^{+}}\mathbf{p}_{1}^{j}}{\big(\mathbf{p}_{3}-\frac{p_{3}^{+}}{p_{1}^{+}}\mathbf{p}_{1}\big)^{2}}e^{-i\mathbf{p}_{3}\cdot(\mathbf{x}-\mathbf{z})}\\
 & \!\times\!\Bigg[\!\Bigg(\!-\frac{p_{0\scriptscriptstyle{R}}^{+}\mathbf{q}^{2}\!-\!(p_{1}^{+}\!+\!p_{3}^{+})M^{2}}{p_{0\scriptscriptstyle{R}}^{+}\mathbf{q}^{2}\!+\!(p_{1}^{+}\!+\!p_{3}^{+})M^{2}}\!\Big(\big(1\!+\!\frac{2p_{1}^{+}}{p_{3}^{+}}\big)^{2}\!+\!1\Big)\!+\!2\frac{q^{+}}{p_{0}^{+}}\frac{p_{1}^{+}\!+\!p_{3}^{+}}{p_{3}^{+}}\big)\!\Bigg)\!iA^{\bar{\eta}}(\mathbf{x}^{\prime}\!-\!\mathbf{z},\Delta_{\mathrm{IS}})\\
 & -p_{1}^{+}\frac{2p_{0}^{+}+p_{3}^{+}}{p_{0}^{+}(p_{1}^{+}+p_{3}^{+})}\mathbf{q}^{\bar{\eta}}\mathcal{K}\big(\mathbf{x}^{\prime}-\mathbf{z},\Delta_{\mathrm{IS}})\big(1+\frac{2p_{1}^{+}}{p_{3}^{+}}+\frac{p_{3}^{+}}{2p_{0}^{+}+p_{3}^{+}}\big)\Bigg]\\
 & \times\Bigg[\frac{p_{3}^{+}\big(\mathbf{p}_{3}+\mathbf{p}_{1}-\frac{p_{1}^{+}+p_{3}^{+}}{q^{+}}\mathbf{q}\big)^{2}-p_{3}^{+}\frac{p_{0\scriptscriptstyle{R}}^{+}(p_{1}^{+}+p_{3}^{+})}{(q^{+})^{2}}M^{2}}{\frac{p_{1}^{+}p_{0\scriptscriptstyle{R}}^{+}}{q^{+}}\big(\mathbf{p}_{3}-\frac{p_{3}^{+}}{p_{1}^{+}}\mathbf{p}_{1}\big)^{2}+p_{3}^{+}\big(\mathbf{p}_{3}+\mathbf{p}_{1}-\frac{p_{1}^{+}+p_{3}^{+}}{q^{+}}\mathbf{q}\big)^{2}+p_{3}^{+}\frac{p_{0\scriptscriptstyle{R}}^{+}(p_{1}^{+}+p_{3}^{+})}{(q^{+})^{2}}M^{2}}\\
 &-\frac{p_{0\scriptscriptstyle{R}}^{+}\mathbf{q}^{2}-(p_{1}^{+}+p_{3}^{+})M^{2}}{p_{0\scriptscriptstyle{R}}^{+}\mathbf{q}^{2}+(p_{1}^{+}+p_{3}^{+})M^{2}}\Bigg] \!+\!4\!\int_{k_{\mathrm{min}}^{+}}^{k_{f}^{+}}\frac{\mathrm{d}p_{3}^{+}}{p_{3}^{+}}\Bigg(\frac{p_{0}^{+}\mathbf{q}^{2}\!-\!p_{1}^{+}M^{2}}{p_{0}^{+}\mathbf{q}^{2}\!+\!p_{1}^{+}M^{2}}\!\Bigg)^{2}A^{i}(\mathbf{x}^{\prime}\!-\!\mathbf{z})A^{i}(\mathbf{x}\!-\!\mathbf{z})\Bigg\}\;.
\end{aligned}
\label{eq:IS3+4-FS2+3L_cross}
\end{equation}

\subsubsection{\label{subsec:real_fin_T}Transverse polarization}

\paragraph{Initial-state radiation}

Firstly, we have the finite leftovers of those initial-state diagrams
that constituted the real part of the DGLAP equations:
\begin{equation}
\begin{aligned} & \int\mathrm{PS}(\vec{p}_{3})\mathrm{Tr}\Big(2\mathrm{Re}\big(\mathcal{M}_{\mathrm{IS1}}^{\lambda\eta\dagger}\mathcal{M}_{\mathrm{IS2,finite}}^{\lambda\eta}\big)+\Big|\mathcal{M}_{\mathrm{IS2,finite}}^{\lambda\eta}\Big|^{2}\Big)\\
 & =g_{\mathrm{em}}^{2}\alpha_{s}N_{c}C_{F}8p_{0}^{+}p_{1}^{+}\Big(\big(1+\frac{2p_{1}^{+}}{q^{+}}\big)^{2}+1\Big)\int_{\mathbf{x},\mathbf{x}^{\prime}}e^{-i\mathbf{k}_{\perp}\cdot(\mathbf{x}-\mathbf{x}^{\prime})}\big(s_{\mathbf{x}^{\prime}\mathbf{x}}+1\big)\int_{\boldsymbol{\ell}}\frac{e^{-i\boldsymbol{\ell}\cdot(\mathbf{\mathbf{x}}-\mathbf{x}^{\prime})}}{\boldsymbol{\ell}^{2}}\\
 & \times\int_{0}^{+\infty}\frac{\mathrm{d}p_{3}^{+}}{p_{3}^{+}}\frac{(p_{3}^{+})^{2}}{p_{0\scriptscriptstyle{R}}^{+}p_{0}^{+}}\Big(\big(1+\frac{2p_{0}^{+}}{p_{3}^{+}}\big)^{2}+1\Big)\\
 & \times\Bigg[\Big(\frac{(q^{+})^{2}p_{3}^{+}\big(\boldsymbol{\ell}+\frac{p_{0}^{+}}{q^{+}}\mathbf{q}\big)}{p_{3}^{+}\big(q^{+}\boldsymbol{\ell}+p_{0}^{+}\mathbf{q}\big)^{2}+p_{1}^{+}q^{+}p_{0R}^{+}\boldsymbol{\ell}^{2}+p_{1}^{+}p_{3}^{+}p_{0}^{+}M^{2}}-\frac{q^{+}\mathbf{q}}{p_{0}^{+}\mathbf{q}^{2}+p_{1}^{+}M^{2}}\Big)\\
 & +2\frac{q^{+}\mathbf{P}_{\perp}^{\bar{\lambda}}}{p_{0}^{+}\mathbf{P}_{\perp}^{2}+p_{1}^{+}M^{2}}\\
 & \times\Big(\frac{(q^{+})^{2}p_{3}^{+}\big(\boldsymbol{\ell}^{\bar{\lambda}}+\frac{p_{0}^{+}}{q^{+}}\mathbf{q}^{\bar{\lambda}}\big)}{p_{3}^{+}\big(q^{+}\boldsymbol{\ell}+p_{0}^{+}\mathbf{q}\big)^{2}+p_{1}^{+}q^{+}p_{0R}^{+}\boldsymbol{\ell}^{2}+p_{1}^{+}p_{3}^{+}p_{0}^{+}M^{2}}-\frac{q^{+}\mathbf{q}^{\bar{\lambda}}}{p_{0}^{+}\mathbf{q}^{2}+p_{1}^{+}M^{2}}\Big)\Bigg]
\end{aligned}
\label{eq:DGLAP_finite_T}
\end{equation}
where we used eq.~\eqref{eq:DiracTraceBossIS}. Similarly:
\begin{equation}
\begin{aligned} & \int\mathrm{PS}(\vec{p}_{3})\mathrm{Tr}\big|\tilde{\mathcal{M}}_{\mathrm{IS3+4}}^{\lambda\eta}\big|^{2}\\
 & \!=\!\int\frac{\mathrm{d}p_{3}^{+}}{p_{3}^{+}}g_{\mathrm{em}}^{2}\alpha_{s}\Big(\frac{p_{3}^{+}}{p_{1}^{+}+p_{3}^{+}}\Big)^{2}8p_{1}^{+}p_{0\scriptscriptstyle{R}}^{+}\int_{\mathbf{x},\mathbf{z}}e^{i\frac{p_{3}^{+}}{p_{1}^{+}+p_{3}^{+}}\mathbf{q}\cdot(\mathbf{x}-\mathbf{x}^{\prime})}e^{-i\mathbf{k}_{\perp}\cdot(\mathbf{x}-\mathbf{x}^{\prime})}C_{F}N_{c}\big(s_{\mathbf{x}\mathbf{x}^{\prime}}+1\big)\\
 &\!\times \!\Bigg[\!\Big(\big(1\!+\!\frac{2p_{1}^{+}}{p_{3}^{+}}\big)^{2}\!+\!1\Big)\!\Big(\big(1\!+\!2\frac{p_{1}^{+}\!+\!p_{3}^{+}}{q^{+}}\big)^{2}\!+\!1\Big) \\
 & \!\times\!\frac{(q^{+})^{2}\mathbf{q}^{2}}{\big(p_{0\scriptscriptstyle{R}}^{+}\mathbf{q}^{2}\!+\!(p_{1}^{+}\!+\!p_{3}^{+})M^{2}\big)^{2}}A^{i}(\mathbf{x}\!-\!\mathbf{z},\Delta_{\mathrm{IS}})A^{i}(\mathbf{x}^{\prime}\!-\!\mathbf{z},\Delta_{\mathrm{IS}})\\
 & \!+\!4\Big(\frac{p_{1}^{+}}{p_{0}^{+}(p_{1}^{+}+p_{3}^{+})}\Big)^{2}\Big((p_{0}^{+})^{2}+(p_{1}^{+}+p_{3}^{+})^{2}\Big)\mathcal{K}\big(\mathbf{x}-\mathbf{z},\Delta_{\mathrm{IS}}\big)\mathcal{K}\big(\mathbf{x}^{\prime}-\mathbf{z},\Delta_{\mathrm{IS}}\big)\\
 & \!+\!\frac{p_{1}^{+}q^{+}\mathbf{q}^{i}}{p_{0\scriptscriptstyle{R}}^{+}\mathbf{q}^{2}\!+\!(p_{1}^{+}\!+\!p_{3}^{+})M^{2}}\\
 &\!\times\!\Big(\!iA^{i}(\mathbf{x}^{\prime}\!-\!\mathbf{z},\Delta_{\mathrm{IS}})\mathcal{K}\big(\mathbf{x}\!-\!\mathbf{z},\Delta_{\mathrm{IS}}\big)\!-\!iA^{i}(\mathbf{x}\!-\!\mathbf{z},\Delta_{\mathrm{IS}})\mathcal{K}\big(\mathbf{x}^{\prime}\!-\!\mathbf{z},\Delta_{\mathrm{IS}}\big)\!\Big)\!\\
 &\! \times\!\Bigg(\Big(\frac{1}{p_{1}^{+}+p_{3}^{+}}+\frac{1}{p_{0}^{+}}\Big)\Big(\big(1+\frac{2p_{1}^{+}}{p_{3}^{+}}\big)\big(1+2\frac{p_{1}^{+}+p_{3}^{+}}{q^{+}}\big)-1\Big)\\
 & \!+\!\Big(\frac{1}{p_{1}^{+}+p_{3}^{+}}-\frac{1}{p_{0}^{+}}\Big)\Big(\frac{2p_{1}^{+}}{p_{3}^{+}}-2\frac{p_{1}^{+}+p_{3}^{+}}{q^{+}}\Big)\Bigg)\Bigg]\;.
\end{aligned}
\label{eq:IS3+4_T_cross}
\end{equation}
Furthermore, we have, adapting identity~\eqref{eq:DiracTraceBossV}:
\begin{equation}
\begin{aligned} & \int\mathrm{PS}(\vec{p}_{3})\mathrm{Tr}\,2\mathrm{Re}\big(\mathcal{M}_{\mathrm{IS1+2}}^{\lambda\eta\dagger}\mathcal{M}_{\mathrm{IS3+4}}^{\lambda\eta}\big)\\
 & =g_{\mathrm{em}}^{2}\alpha_{s}8p_{1}^{+}p_{0}^{+}2\mathrm{Re}\int_{\mathbf{x},\mathbf{x}^{\prime},\mathbf{z}}e^{-i\mathbf{k}_{\perp}\cdot(\mathbf{x}-\mathbf{x}^{\prime})}C_{F}N_{c}\big(s_{\mathbf{x}\mathbf{x}^{\prime}}+1\big)\int_{\boldsymbol{\ell}}e^{i\boldsymbol{\ell}\cdot(\mathbf{x}^{\prime}-\mathbf{z})}\frac{\boldsymbol{\ell}^{\eta^{\prime}}}{\boldsymbol{\ell}^{2}}\\
 & \times\Bigg\{\int_{0}^{+\infty}\frac{\mathrm{d}p_{3}^{+}}{p_{3}^{+}}\frac{p_{3}^{+}}{p_{1}^{+}+p_{3}^{+}}\frac{p_{3}^{+}}{p_{0}^{+}}e^{i\frac{p_{3}^{+}}{p_{1}^{+}+p_{3}^{+}}\mathbf{q}\cdot(\mathbf{x}-\mathbf{z})}\\
 & \times\Bigg[\frac{q^{+}\mathbf{P}_{\perp}^{\lambda^{\prime}}}{p_{0}^{+}\mathbf{P}_{\perp}^{2}+p_{1}^{+}M^{2}}+\frac{(q^{+})^{2}p_{3}^{+}\big(\boldsymbol{\ell}^{\lambda^{\prime}}+\frac{p_{0}^{+}}{q^{+}}\mathbf{q}^{\lambda^{\prime}}\big)}{p_{3}^{+}\big(q^{+}\boldsymbol{\ell}+p_{0}^{+}\mathbf{q}\big)^{2}+p_{1}^{+}q^{+}p_{0R}^{+}\boldsymbol{\ell}^{2}+p_{1}^{+}p_{3}^{+}p_{0}^{+}M^{2}}\Bigg]\\
 & \times\Bigg[\frac{q^{+}\mathbf{q}^{\bar{\lambda}}}{p_{0\scriptscriptstyle{R}}^{+}\mathbf{q}^{2}+(p_{1}^{+}+p_{3}^{+})M^{2}}\Big(\big(1+\frac{2p_{1}^{+}}{q^{+}}\big)\big(1+2\frac{p_{1}^{+}+p_{3}^{+}}{q^{+}}\big)+1\Big)\\
 & \times\Big(\big(1+\frac{2p_{0}^{+}}{p_{3}^{+}}\big)\big(1+\frac{2p_{1}^{+}}{p_{3}^{+}}\big)+1\Big)\delta^{\bar{\lambda}\lambda^{\prime}}\delta^{\bar{\eta}\eta^{\prime}}iA^{\bar{\eta}}(\mathbf{x}-\mathbf{z},\Delta_{\mathrm{IS}})\\
 & -\frac{\mathbf{q}^{\bar{\lambda}}}{p_{0\scriptscriptstyle{R}}^{+}\mathbf{q}^{2}+(p_{1}^{+}+p_{3}^{+})M^{2}}\frac{4(p_{0}^{+}+p_{1}^{+}+p_{3}^{+})^{2}}{p_{3}^{+}}\epsilon^{\bar{\lambda}\lambda^{\prime}}\epsilon^{\bar{\eta}\eta^{\prime}}iA^{\bar{\eta}}(\mathbf{x}-\mathbf{z},\Delta_{\mathrm{IS}})\\
 & -\Bigg(\Big(\frac{1}{p_{1}^{+}+p_{3}^{+}}+\frac{1}{p_{0}^{+}}\Big)\Big(\big(1+\frac{2p_{1}^{+}}{q^{+}}\big)\big(1+\frac{2p_{0}^{+}}{p_{3}^{+}}\big)-1\Big)\\
 & +\Big(\frac{1}{p_{1}^{+}+p_{3}^{+}}-\frac{1}{p_{0}^{+}}\Big)\big(\frac{2p_{0}^{+}}{p_{3}^{+}}-\frac{2p_{1}^{+}}{q^{+}}\big)\Bigg)p_{1}^{+}\delta^{\lambda^{\prime}\eta^{\prime}}\mathcal{K}\big(\mathbf{x}-\mathbf{z},\Delta_{\mathrm{IS}}\big)\Bigg]\Bigg\}\\
 & -4\int_{0}^{+\infty}\frac{\mathrm{d}p_{3}^{+}}{p_{3}^{+}}\Big(\big(1+\frac{2p_{1}^{+}}{q^{+}}\big)^{2}+1\Big)iA^{\eta^{\prime}}(\mathbf{x}-\mathbf{z})\\
 & \times\frac{q^{+}\mathbf{P}_{\perp}^{\lambda^{\prime}}}{p_{0}^{+}\mathbf{P}_{\perp}^{2}+p_{1}^{+}M^{2}}\frac{q^{+}\mathbf{q}^{\lambda^{\prime}}}{p_{0\scriptscriptstyle{R}}^{+}\mathbf{q}^{2}+(p_{1}^{+}+p_{3}^{+})M^{2}}\Bigg\}\;.
\end{aligned}
\label{eq:IS1+2-IS3+4_T_cross}
\end{equation}

\paragraph{Final-state radiation}

Using identity~\eqref{eq:DiracTraceBossIS}, what is left after the analysis of collinear final-state divergences can be cast in the following form:
\begin{equation}
\begin{aligned} & \int\mathrm{PS}(\vec{p}_{3})\mathrm{Tr}\Big(\big|\mathcal{M}_{\mathrm{FS2,finite}}^{0\eta}\big|^{2}+2\mathrm{Re}\big(\mathcal{M}_{\mathrm{FS2,finite}}^{0\eta\dagger}\mathcal{M}_{\mathrm{FS3}}^{0\eta}\big)\Big)\\
 & =\int_{0}^{\infty}\frac{\mathrm{d}p_{3}^{+}}{p_{3}^{+}}g_{\mathrm{em}}^{2}\alpha_{s}8p_{1}^{+}p_{0\scriptscriptstyle{R}}^{+}\Big(\frac{p_{3}^{+}}{p_{1}^{+}}\Big)^{2}\Big(\big(1+2\frac{p_{1}^{+}+p_{3}^{+}}{q^{+}}\big)^{2}+1\Big)\Big(\big(1+\frac{2p_{1}^{+}}{p_{3}^{+}}\big)^{2}+1\Big)\\
 & \times\int_{\mathbf{x},\mathbf{x}^{\prime}}e^{i\mathbf{k}_{\perp}\cdot(\mathbf{x}-\mathbf{x}^{\prime})}C_{F}N_{c}\big(s_{\mathbf{x}\mathbf{x}^{\prime}}+1\big)\int_{\boldsymbol{\ell}}\frac{e^{i\big(\boldsymbol{\ell}+\frac{p_{3}^{+}}{p_{1}^{+}}\mathbf{p}_{1}\big)\cdot(\mathbf{x}-\mathbf{x}^{\prime})}}{\boldsymbol{\ell}^{2}}\\
 & \times\Bigg[\Bigg(\frac{p_{3}^{+}}{p_{0}^{+}}\frac{q^{+}}{p_{1}^{+}+p_{3}^{+}}\frac{\boldsymbol{\ell}+\frac{p_{0}^{+}}{p_{1}^{+}}\frac{p_{1}^{+}+p_{3}^{+}}{q^{+}}\mathbf{P}_{\perp}}{\big(\boldsymbol{\ell}+\frac{p_{3}^{+}}{p_{1}^{+}}\mathbf{P}_{\perp}\big)^{2}+\Delta_{\mathrm{FS}}}-\frac{q^{+}\mathbf{P}_{\perp}}{(p_{1}^{+}+p_{3}^{+})\frac{p_{0}^{+}}{p_{1}^{+}}\mathbf{P}_{\perp}^{2}+\frac{p_{0\scriptscriptstyle{R}}^{+}p_{1}^{+}}{p_{0}^{+}}M^{2}}\Bigg)^{2}\\
 & +2\frac{q^{+}\mathbf{q}^{\bar{\lambda}}}{p_{0\scriptscriptstyle{R}}^{+}\mathbf{q}^{2}+(p_{1}^{+}+p_{3}^{+})M^{2}}\Bigg(\frac{p_{3}^{+}}{p_{0}^{+}}\frac{q^{+}}{p_{1}^{+}+p_{3}^{+}}\frac{\boldsymbol{\ell}^{\bar{\lambda}}+\frac{p_{0}^{+}}{p_{1}^{+}}\frac{p_{1}^{+}+p_{3}^{+}}{q^{+}}\mathbf{P}_{\perp}^{\bar{\lambda}}}{\big(\boldsymbol{\ell}+\frac{p_{3}^{+}}{p_{1}^{+}}\mathbf{P}_{\perp}\big)^{2}+\Delta_{\mathrm{FS}}}\\
 & -\frac{q^{+}\mathbf{P}_{\perp}^{\bar{\lambda}}}{(p_{1}^{+}+p_{3}^{+})\frac{p_{0}^{+}}{p_{1}^{+}}\mathbf{P}_{\perp}^{2}+\frac{p_{0\scriptscriptstyle{R}}^{+}p_{1}^{+}}{p_{0}^{+}}M^{2}}\Bigg)\Bigg]\;.
\end{aligned}
\label{eq:FS2+3T_finite_cross}
\end{equation}
Likewise, one obtains for the amplitudes that never contributed to the jet function:
\begin{equation}
\begin{aligned} & \int\mathrm{PS}(\vec{p}_{3})\big|\mathcal{M}_{\mathrm{FS1+4}}^{\lambda\eta\dagger}\big|^{2} \\
& =g_{\mathrm{em}}^{2}\alpha_{s}8p_{1}^{+}p_{0}^{+}\int_{\mathbf{x},\mathbf{x}^{\prime}}e^{i\mathbf{k}_{\perp}\cdot(\mathbf{x}-\mathbf{x}^{\prime})}C_{F}N_{c}\big(s_{\mathbf{x}\mathbf{x}^{\prime}}+1\big)\\
 & \times\int_{0}^{+\infty}\frac{\mathrm{d}p_{3}^{+}}{p_{3}^{+}}\Bigg\{\Big(\frac{p_{3}^{+}}{p_{0}^{+}}\Big)^{2}\frac{p_{0\scriptscriptstyle{R}}^{+}}{p_{0}^{+}}e^{i\frac{p_{3}^{+}}{p_{0}^{+}}\mathbf{k}_{\perp}\cdot(\mathbf{x}-\mathbf{x}^{\prime})}\\
 & \times\Bigg[\Big(\frac{q^{+}\mathbf{P}_{\perp}}{p_{0}^{+}\mathbf{P}_{\perp}^{2}+p_{1}^{+}M^{2}}\Big)^{2}\Big(\big(1+\frac{2p_{0}^{+}}{p_{3}^{+}}\big)^{2}+1\Big)\Big(\big(1+\frac{2p_{1}^{+}}{q^{+}}\big)^{2}+1\Big)\int_{\boldsymbol{\ell}}\frac{\boldsymbol{\ell}^{2}e^{-i\boldsymbol{\ell}\cdot(\mathbf{x}-\mathbf{x}^{\prime})}}{(\boldsymbol{\ell}^{2}+\Delta_{\mathrm{FS}})^{2}}\\
 & +4\frac{(p_{0\scriptscriptstyle{R}}^{+})^{2}\big((p_{1}^{+}+p_{3}^{+})^{2}+(p_{0}^{+})^{2}\big)}{(p_{1}^{+}+p_{3}^{+})^{2}(p_{0}^{+})^{2}}\int_{\boldsymbol{\ell}}\frac{e^{-i\boldsymbol{\ell}\cdot(\mathbf{x}-\mathbf{x}^{\prime})}}{(\boldsymbol{\ell}^{2}+\Delta_{\mathrm{FS}})^{2}}\\
 & -2\frac{p_{0\scriptscriptstyle{R}}^{+}q^{+}\mathbf{P}_{\perp}^{i}}{p_{0}^{+}\mathbf{P}_{\perp}^{2}+p_{1}^{+}M^{2}}\int_{\boldsymbol{\ell}}\frac{\boldsymbol{\ell}^{i}e^{-i\boldsymbol{\ell}\cdot(\mathbf{x}-\mathbf{x}^{\prime})}}{(\boldsymbol{\ell}^{2}+\Delta_{\mathrm{FS}})^{2}}\\
 & \times\Bigg(\Big(\frac{1}{p_{1}^{+}+p_{3}^{+}}+\frac{1}{p_{0}^{+}}\Big)\Big(\big(1+\frac{2p_{0}^{+}}{p_{3}^{+}}\big)\big(1+\frac{2p_{1}^{+}}{q^{+}}\big)-1\Big)\\
 & +\Big(\frac{2p_{0}^{+}}{p_{3}^{+}}-\frac{2p_{1}^{+}}{q^{+}}\Big)\Big(\frac{1}{p_{1}^{+}+p_{3}^{+}}-\frac{1}{p_{0}^{+}}\Big)\Bigg)\Bigg]\\
 & -4\int_{k_{\mathrm{min}}^{+}}^{k_{f}^{+}}\frac{\mathrm{d}p_{3}^{+}}{p_{3}^{+}}\Big(\big(1+\frac{2p_{1}^{+}}{q^{+}}\big)^{2}+1\Big)\Big(\frac{q^{+}\mathbf{P}_{\perp}}{p_{0}^{+}\mathbf{P}_{\perp}^{2}+p_{1}^{+}M^{2}}\Big)^{2}\int_{\boldsymbol{\ell}}\frac{e^{-i\boldsymbol{\ell}\cdot(\mathbf{x}-\mathbf{x}^{\prime})}}{\boldsymbol{\ell}^{2}}\Bigg\}\;.
\end{aligned}
\label{eq:MFS1+4_T_cross}
\end{equation}
Next, we find the following result for the interference terms, using the trace identity~\eqref{eq:DiracTraceBossV}:
\begin{equation}
\begin{aligned} & \int\mathrm{PS}(\vec{p}_{3})\mathrm{Tr}\,2\mathrm{Re}\big(\mathcal{M}_{\mathrm{FS2+3}}^{\lambda\eta\dagger}\mathcal{M}_{\mathrm{FS1+4}}^{\lambda\eta}\big)\\
 & =-g_{\mathrm{em}}^{2}\alpha_{s}8p_{1}^{+}p_{0}^{+}\int_{\mathbf{x},\mathbf{x}^{\prime}}e^{-i\mathbf{k}_{\perp}\cdot(\mathbf{x}-\mathbf{x}^{\prime})}C_{F}N_{c}\big(s_{\mathbf{x}\mathbf{x}^{\prime}}+1\big)\\
 & \times\Bigg\{\int_{0}^{+\infty}\frac{\mathrm{d}p_{3}^{+}}{p_{3}^{+}}\frac{p_{3}^{+}}{p_{1}^{+}}\frac{p_{3}^{+}}{p_{0}^{+}}\frac{p_{0\scriptscriptstyle{R}}^{+}}{p_{0}^{+}}e^{-i\frac{p_{3}^{+}}{p_{0}^{+}}\mathbf{k}_{\perp}\cdot(\mathbf{x}-\mathbf{z})}e^{i\frac{p_{3}^{+}}{p_{1}^{+}}\mathbf{p}_{1}(\mathbf{z}-\mathbf{x}^{\prime})}\int_{\mathbf{z}}\int_{\boldsymbol{\ell}}\frac{\boldsymbol{\ell}^{\eta^{\prime}}e^{-i\boldsymbol{\ell}\cdot(\mathbf{z}-\mathbf{x}^{\prime})}}{\boldsymbol{\ell}^{2}}\\
 & \times\Bigg[\frac{p_{3}^{+}}{p_{0}^{+}}\frac{q^{+}}{p_{1}^{+}+p_{3}^{+}}\frac{\boldsymbol{\ell}^{\lambda^{\prime}}+\frac{p_{0}^{+}}{p_{1}^{+}}\frac{p_{1}^{+}+p_{3}^{+}}{q^{+}}\mathbf{P}_{\perp}^{\lambda^{\prime}}}{\big(\boldsymbol{\ell}+\frac{p_{3}^{+}}{p_{1}^{+}}\mathbf{P}_{\perp}\big)^{2}+\Delta_{\mathrm{FS}}}+\frac{q^{+}\mathbf{q}^{\lambda^{\prime}}}{p_{0\scriptscriptstyle{R}}^{+}\mathbf{q}^{2}+(p_{1}^{+}+p_{3}^{+})M^{2}}\Bigg]\\
 & \times\Bigg[-\frac{q^{+}\mathbf{P}_{\perp}^{\bar{\lambda}}}{p_{0}^{+}\mathbf{P}_{\perp}^{2}+p_{1}^{+}M^{2}}iA^{\bar{\eta}}(\mathbf{x}-\mathbf{z},\Delta_{\mathrm{FS}})\Bigg(\Big(\big(1+2\frac{p_{1}^{+}+p_{3}^{+}}{q^{+}}\big)\big(1+\frac{2p_{1}^{+}}{q^{+}}\big)+1\Big)\\
 &\times\Big(\big(1+\frac{2p_{1}^{+}}{p_{3}^{+}}\big)\big(1+\frac{2p_{0}^{+}}{p_{3}^{+}}\big)+1\Big)\delta^{\bar{\eta}\eta^{\prime}}\delta^{\bar{\lambda}\lambda^{\prime}}\\
 & -4\frac{(p_{0}^{+}+p_{1}^{+}+p_{3}^{+})^{2}}{p_{3}^{+}q^{+}}\epsilon^{\bar{\eta}\eta^{\prime}}\epsilon^{\bar{\lambda}\lambda^{\prime}}\Bigg)\\
 & +\delta^{\eta^{\prime}\lambda^{\prime}}p_{0\scriptscriptstyle{R}}^{+}\mathcal{K}(\mathbf{x}-\mathbf{z},\Delta_{\mathrm{FS}})\Bigg(\Big(\frac{1}{p_{1}^{+}+p_{3}^{+}}+\frac{1}{p_{0}^{+}}\Big)\Big(\big(1+2\frac{p_{1}^{+}+p_{3}^{+}}{q^{+}}\big)\big(1+\frac{2p_{1}^{+}}{p_{3}^{+}}\big)-1\Big)\\
 & +2\big(\frac{p_{1}^{+}}{p_{3}^{+}}-\frac{p_{1}^{+}+p_{3}^{+}}{q^{+}}\big)\Big(\frac{1}{p_{1}^{+}+p_{3}^{+}}-\frac{1}{p_{0}^{+}}\Big)\Bigg)\Bigg]\\
 & +4\int_{0}^{+\infty}\frac{\mathrm{d}p_{3}^{+}}{p_{3}^{+}}\Big(\big(1+2\frac{p_{1}^{+}}{q^{+}}\big)^{2}+1\Big)\frac{q^{+}\mathbf{q}^{\bar{\lambda}}}{p_{0}^{+}\mathbf{q}^{2}+p_{1}^{+}M^{2}}\frac{q^{+}\mathbf{P}_{\perp}^{\bar{\lambda}}}{p_{0}^{+}\mathbf{P}_{\perp}^{2}+p_{1}^{+}M^{2}}\int_{\boldsymbol{\ell}}\frac{e^{-i\boldsymbol{\ell}\cdot(\mathbf{x}-\mathbf{x}^{\prime})}}{\boldsymbol{\ell}^{2}}\Bigg\}\;.
\end{aligned}
\label{eq:FS2+3-FS1+4_T_cross}
\end{equation}

\paragraph{Initial-final state radiation interference}

The last real NLO contributions are due to the interference between
initial- and final state radiation in the transversely polarized case:
\begin{equation}
\begin{aligned} & \int\mathrm{PS}(\vec{p}_{3})\mathrm{Tr}\,2\mathrm{Re}\big(\mathcal{M}_{\mathrm{IS1+2}}^{\lambda\eta\dagger}\mathcal{M}_{\mathrm{FS1+4}}^{\lambda\eta}\big)\\
 & =g_{\mathrm{em}}^{2}\alpha_{s}8p_{1}^{+}p_{0}^{+}2\mathrm{Re}\int_{\mathbf{x},\mathbf{x}^{\prime},\mathbf{z}}e^{-i\mathbf{k}_{\perp}\cdot(\mathbf{x}-\mathbf{x}^{\prime})}\big(\frac{N_{c}^{2}}{2}s_{\mathbf{z}\mathbf{x}^{\prime}}s_{\mathbf{x}\mathbf{z}}-\frac{1}{2}s_{\mathbf{x}\mathbf{x}^{\prime}}+C_{F}N_{c}\big)\\
 & \times\Bigg\{\int_{0}^{+\infty}\frac{\mathrm{d}p_{3}^{+}}{p_{3}^{+}}\Big(\frac{p_{3}^{+}}{p_{0}^{+}}\Big)^{2}e^{-i\frac{p_{3}^{+}}{p_{0}^{+}}\mathbf{k}_{\perp}\cdot(\mathbf{x}-\mathbf{z})}\int_{\boldsymbol{\ell}}e^{i\boldsymbol{\ell}\cdot(\mathbf{x}^{\prime}-\mathbf{z})}\frac{\boldsymbol{\ell}^{\eta^{\prime}}}{\boldsymbol{\ell}^{2}}\\
 & \times\Bigg[\frac{q^{+}\mathbf{P}_{\perp}^{\lambda^{\prime}}}{p_{0}^{+}\mathbf{P}_{\perp}^{2}+p_{1}^{+}M^{2}}+\frac{(q^{+})^{2}p_{3}^{+}\big(\boldsymbol{\ell}^{\lambda^{\prime}}+\frac{p_{0}^{+}}{q^{+}}\mathbf{q}^{\lambda^{\prime}}\big)}{p_{3}^{+}\big(q^{+}\boldsymbol{\ell}+p_{0}^{+}\mathbf{q}\big)^{2}+p_{1}^{+}q^{+}p_{0R}^{+}\boldsymbol{\ell}^{2}+p_{1}^{+}p_{3}^{+}p_{0}^{+}M^{2}}\Bigg]\\
 & \times\Bigg[-\frac{q^{+}\mathbf{P}_{\perp}^{\bar{\lambda}}}{p_{0}^{+}\mathbf{P}_{\perp}^{2}+p_{1}^{+}M^{2}}iA^{\bar{\eta}}(\mathbf{x}-\mathbf{z},\Delta_{\mathrm{FS}})\\
 & \times\Bigg(\Big(\big(1+\frac{2p_{1}^{+}}{q^{+}}\big)^{2}+1\Big)\Big(\big(1+\frac{2p_{0}^{+}}{p_{3}^{+}}\big)^{2}+1\Big)\delta^{\bar{\eta}\eta^{\prime}}\delta^{\bar{\lambda}\lambda^{\prime}}-4\big(1+\frac{2p_{1}^{+}}{q^{+}}\big)\big(1+\frac{2p_{0}^{+}}{p_{3}^{+}}\big)\epsilon^{\bar{\eta}\eta^{\prime}}\epsilon^{\bar{\lambda}\lambda^{\prime}}\Bigg)\\
 & +\delta^{\eta^{\prime}\lambda^{\prime}}p_{0\scriptscriptstyle{R}}^{+}\mathcal{K}(\mathbf{x}-\mathbf{z},\Delta_{\mathrm{FS}})\Bigg(\Big(\frac{1}{p_{1}^{+}+p_{3}^{+}}+\frac{1}{p_{0}^{+}}\Big)\Big(\big(1+\frac{2p_{1}^{+}}{q^{+}}\big)\big(1+\frac{2p_{0}^{+}}{p_{3}^{+}}\big)-1\Big)\\
 & +\Big(\frac{2p_{0}^{+}}{p_{3}^{+}}-\frac{2p_{1}^{+}}{q^{+}}\Big)\Big(\frac{1}{p_{1}^{+}+p_{3}^{+}}-\frac{1}{p_{0}^{+}}\Big)\Bigg)\Bigg]\\
 & +4\int_{k_{\mathrm{min}}^{+}}^{k_{f}^{+}}\frac{\mathrm{d}p_{3}^{+}}{p_{3}^{+}}\Big(\big(1+\frac{2p_{1}^{+}}{q^{+}}\big)^{2}+1\Big)\Big(\frac{q^{+}\mathbf{P}_{\perp}}{p_{0}^{+}\mathbf{P}_{\perp}^{2}+p_{1}^{+}M^{2}}\Big)^{2}A^{i}(\mathbf{x}^{\prime}-\mathbf{z})A^{i}(\mathbf{x}-\mathbf{z})\Bigg\}\;,
\end{aligned}
\label{eq:IS1+2-FS1+4_T_cross}
\end{equation}
where, once again, we relied on \eqref{eq:DiracTraceBossIS}.
Moreover, using \eqref{eq:DiracTraceBossV}: 
\begin{equation}
\begin{aligned} & \int\mathrm{PS}(\vec{p}_{3})\mathrm{Tr}\,2\mathrm{Re}\big(\mathcal{M}_{\mathrm{IS1+2}}^{\lambda\eta\dagger}\mathcal{M}_{\mathrm{FS2+3}}^{\lambda\eta}\big)\\
 & =-g_{\mathrm{em}}^{2}\alpha_{s}8p_{1}^{+}p_{0}^{+}2\mathrm{Re}\int_{\mathbf{x},\mathbf{x}^{\prime},\mathbf{z}}e^{i\mathbf{k}_{\perp}\cdot(\mathbf{x}-\mathbf{x}^{\prime})}\big(\frac{N_{c}^{2}}{2}s_{\mathbf{z}\mathbf{x}^{\prime}}s_{\mathbf{x}\mathbf{z}}-\frac{1}{2}s_{\mathbf{x}\mathbf{x}^{\prime}}+C_{F}N_{c}\big)\\
 & \times\Bigg\{\int_{0}^{+\infty}\frac{\mathrm{d}p_{3}^{+}}{p_{3}^{+}}\frac{p_{3}^{+}}{p_{1}^{+}}\frac{p_{3}^{+}}{p_{0}^{+}}e^{-i\frac{p_{3}^{+}}{p_{1}^{+}}\mathbf{p}_{1}\cdot(\mathbf{x}-\mathbf{z})}\int_{\boldsymbol{\ell}}e^{i\boldsymbol{\ell}\cdot(\mathbf{x}^{\prime}-\mathbf{z})}\frac{\boldsymbol{\ell}^{\eta^{\prime}}}{\boldsymbol{\ell}^{2}}\int_{\boldsymbol{\ell}_{2}}\frac{\boldsymbol{\ell}_{2}^{\bar{\eta}}}{\boldsymbol{\ell}_{2}^{2}}e^{-i\boldsymbol{\ell}_{2}\cdot(\mathbf{x}-\mathbf{z})}\\
 & \times\Bigg[\Big(\big(1+\frac{2p_{1}^{+}}{q^{+}}\big)\big(1+2\frac{p_{1}^{+}+p_{3}^{+}}{q^{+}}\big)+1\Big)\Big(\big(1+\frac{2p_{0}^{+}}{p_{3}^{+}}\big)\big(1+\frac{2p_{1}^{+}}{p_{3}^{+}}\big)+1\Big)\delta^{\bar{\lambda}\lambda^{\prime}}\delta^{\bar{\eta}\eta^{\prime}}\\
 &-4\frac{(p_{0\scriptscriptstyle{R}}^{+}+p_{1}^{+})^{2}}{p_{3}^{+}q^{+}}\epsilon^{\bar{\lambda}\lambda^{\prime}}\epsilon^{\bar{\eta}\eta^{\prime}}\Bigg]\\
 & \times\Bigg[\frac{q^{+}\mathbf{P}_{\perp}^{\lambda^{\prime}}}{p_{0}^{+}\mathbf{P}_{\perp}^{2}+p_{1}^{+}M^{2}}+\frac{(q^{+})^{2}p_{3}^{+}\big(\boldsymbol{\ell}^{\lambda^{\prime}}+\frac{p_{0}^{+}}{q^{+}}\mathbf{q}^{\lambda^{\prime}}\big)}{p_{3}^{+}\big(q^{+}\boldsymbol{\ell}+p_{0}^{+}\mathbf{q}\big)^{2}+p_{1}^{+}q^{+}p_{0R}^{+}\boldsymbol{\ell}^{2}+p_{1}^{+}p_{3}^{+}p_{0}^{+}M^{2}}\Bigg]\\
 & \times\Bigg[\frac{p_{3}^{+}}{p_{0}^{+}}\frac{q^{+}}{p_{1}^{+}+p_{3}^{+}}\frac{\boldsymbol{\ell}_{2}^{\bar{\lambda}}+\frac{p_{0}^{+}}{p_{1}^{+}}\frac{p_{1}^{+}+p_{3}^{+}}{q^{+}}\mathbf{P}_{\perp}^{\bar{\lambda}}}{\big(\boldsymbol{\ell}_{2}+\frac{p_{3}^{+}}{p_{1}^{+}}\mathbf{P}_{\perp}\big)^{2}+\Delta_{\mathrm{FS}}}+\frac{q^{+}\mathbf{q}^{\bar{\lambda}}}{p_{0\scriptscriptstyle{R}}^{+}\mathbf{q}^{2}+(p_{1}^{+}+p_{3}^{+})M^{2}}\Bigg]\\
 & \!-\!4\int_{k_{\mathrm{min}}^{+}}^{k_{f}^{+}}\!\frac{\mathrm{d}p_{3}^{+}}{p_{3}^{+}}\!\Big(\big(1\!+\!\frac{2p_{1}^{+}}{q^{+}}\big)^{2}\!+\!1\!\Big)\!A^{i}(\mathbf{x}^{\prime}\!-\!\mathbf{z})A^{i}(\mathbf{x}\!-\!\mathbf{z})\frac{q^{+}\mathbf{P}_{\perp}^{\bar{\lambda}}}{p_{0}^{+}\mathbf{P}_{\perp}^{2}\!+\!p_{1}^{+}M^{2}}\frac{q^{+}\mathbf{q}^{\bar{\lambda}}}{p_{0}^{+}\mathbf{q}^{2}\!+\!p_{1}^{+}M^{2}}\!\Bigg\}\;,
\end{aligned}
\label{eq:IS1+2-FS2+3_T_cross}
\end{equation}
and:
\begin{equation}
\begin{aligned} & \int\mathrm{PS}(\vec{p}_{3})\mathrm{Tr}\,2\mathrm{Re}\big(\mathcal{M}_{\mathrm{IS3+4}}^{\lambda\eta\dagger}\mathcal{M}_{\mathrm{FS1+4}}^{\lambda\eta}\big)\\
 & =g_{\mathrm{em}}^{2}\alpha_{s}8p_{1}^{+}p_{0\scriptscriptstyle{R}}^{+}2\mathrm{Re}\int_{\mathbf{x},\mathbf{x}^{\prime},\mathbf{z}}e^{i\mathbf{k}_{\perp}\cdot(\mathbf{x}-\mathbf{x}^{\prime})}\big(\frac{N_{c}^{2}}{2}s_{\mathbf{z}\mathbf{x}^{\prime}}s_{\mathbf{x}\mathbf{z}}-\frac{1}{2}s_{\mathbf{x}\mathbf{x}^{\prime}}+C_{F}N_{c}\big)\\
 & \times\Bigg\{\int_{0}^{+\infty}\frac{\mathrm{d}p_{3}^{+}}{p_{3}^{+}}\frac{p_{3}^{+}}{p_{1}^{+}+p_{3}^{+}}\frac{p_{3}^{+}}{p_{0}^{+}}e^{-i\frac{p_{3}^{+}}{p_{1}^{+}+p_{3}^{+}}\mathbf{q}\cdot(\mathbf{x}^{\prime}-\mathbf{z})}e^{-i\frac{p_{3}^{+}}{p_{0}^{+}}\mathbf{k}_{\perp}\cdot(\mathbf{x}-\mathbf{z})}\\
 & \times\Bigg[\frac{q^{+}\mathbf{q}^{\lambda^{\prime}}}{p_{0\scriptscriptstyle{R}}^{+}\mathbf{q}^{2}+(p_{1}^{+}+p_{3}^{+})M^{2}}\frac{q^{+}\mathbf{P}_{\perp}^{\bar{\lambda}}}{p_{0}^{+}\mathbf{P}_{\perp}^{2}+p_{1}^{+}M^{2}}iA^{\eta^{\prime}}(\mathbf{x}^{\prime}-\mathbf{z},\Delta_{\mathrm{IS}})iA^{\bar{\eta}}(\mathbf{x}-\mathbf{z},\Delta_{\mathrm{FS}})\\
 & \times\Bigg(\Big(\big(1+2\frac{p_{1}^{+}+p_{3}^{+}}{q^{+}}\big)\big(1+\frac{2p_{1}^{+}}{q^{+}}\big)+1\Big)\Big(\big(1+\frac{2p_{1}^{+}}{p_{3}^{+}}\big)\big(1+\frac{2p_{0}^{+}}{p_{3}^{+}}\big)+1\Big)\delta^{\bar{\eta}\eta^{\prime}}\delta^{\bar{\lambda}\lambda^{\prime}}\\
 &-4\frac{(p_{0\scriptscriptstyle{R}}^{+}+p_{1}^{+})}{p_{3}^{+}q^{+}}\epsilon^{\bar{\eta}\eta^{\prime}}\epsilon^{\bar{\lambda}\lambda^{\prime}}\Bigg)\\
 & -4p_{1}^{+}(p_{0\scriptscriptstyle{R}}^{+})^{3}\frac{(p_{1}^{+}+p_{3}^{+})^{2}+(p_{0}^{+})^{2}}{(p_{1}^{+}+p_{3}^{+})^{2}(p_{0}^{+})^{2}}\mathcal{K}\big(\mathbf{x}^{\prime}-\mathbf{z},\Delta_{\mathrm{IS}}\big)\mathcal{K}(\mathbf{x}-\mathbf{z},\Delta_{\mathrm{FS}})\\
 & -p_{0\scriptscriptstyle{R}}^{+}\frac{q^{+}\mathbf{q}^{i}}{p_{0\scriptscriptstyle{R}}^{+}\mathbf{q}^{2}+(p_{1}^{+}+p_{3}^{+})M^{2}}iA^{i}(\mathbf{x}^{\prime}-\mathbf{z},\Delta_{\mathrm{IS}})\mathcal{K}(\mathbf{x}-\mathbf{z},\Delta_{\mathrm{FS}})\\
 & \times\Bigg(\Big(\frac{1}{p_{1}^{+}+p_{3}^{+}}+\frac{1}{p_{0}^{+}}\Big)\Big(\big(1+2\frac{p_{1}^{+}+p_{3}^{+}}{q^{+}}\big)\big(1+\frac{2p_{1}^{+}}{p_{3}^{+}}\big)-1\Big)\\
 &+2\Big(\frac{p_{1}^{+}}{p_{3}^{+}}-\frac{p_{1}^{+}+p_{3}^{+}}{q^{+}}\Big)\Big(\frac{1}{p_{1}^{+}+p_{3}^{+}}-\frac{1}{p_{0}^{+}}\Big)\Bigg)\\
 & +\frac{p_{1}^{+}q^{+}\mathbf{P}_{\perp}^{i}}{p_{0}^{+}\mathbf{P}_{\perp}^{2}+p_{1}^{+}M^{2}}iA^{i}(\mathbf{x}-\mathbf{z},\Delta_{\mathrm{FS}})\mathcal{K}\big(\mathbf{x}^{\prime}-\mathbf{z},\Delta_{\mathrm{IS}})\\
 & \times\Bigg(\Big(\frac{1}{p_{1}^{+}+p_{3}^{+}}+\frac{1}{p_{0}^{+}}\Big)\Big(\big(1+\frac{2p_{1}^{+}}{q^{+}}\big)\big(1+\frac{2p_{0}^{+}}{p_{3}^{+}}\big)-1\Big)\\
 &-2\Big(\frac{p_{1}^{+}}{q^{+}}-\frac{p_{0}^{+}}{p_{3}^{+}}\Big)\Big(\frac{1}{p_{1}^{+}+p_{3}^{+}}-\frac{1}{p_{0}^{+}}\Big)\Bigg)\Bigg]\\
 & \!+\!4\!\int_{k_{\mathrm{min}}^{+}}^{k_{f}^{+}}\!\frac{\mathrm{d}p_{3}^{+}}{p_{3}^{+}}\Big(\big(1\!+\!\frac{2p_{1}^{+}}{q^{+}}\big)^{2}\!+\!1\Big)\frac{q^{+}\mathbf{q}^{j}}{p_{0}^{+}\mathbf{q}^{2}\!+\!p_{1}^{+}M^{2}}\frac{q^{+}\mathbf{P}_{\perp}^{j}}{p_{0}^{+}\mathbf{P}_{\perp}^{2}\!+\!p_{1}^{+}M^{2}}A^{i}(\mathbf{x}^{\prime}\!-\!\mathbf{z})A^{i}(\mathbf{x}\!-\!\mathbf{z})\!\Bigg\}\;.
\end{aligned}
\label{eq:IS3+4-FS1+4_T_cross}
\end{equation}
The following monstrosity, obtained with the help of identity \eqref{eq:DiracTraceBossIS} concludes this section:
\begin{equation}
\begin{aligned} & \int\mathrm{PS}(\vec{p}_{3})\mathrm{Tr}\,2\mathrm{Re}\big(\mathcal{M}_{\mathrm{IS3+4}}^{\lambda\eta\dagger}\mathcal{M}_{\mathrm{FS2+3}}^{\lambda\eta}\big)\\
 & =-g_{\mathrm{em}}^{2}\alpha_{s}8p_{1}^{+}p_{0}^{+}2\mathrm{Re}\int_{\mathbf{x},\mathbf{x}^{\prime},\mathbf{z}}e^{i\mathbf{k}_{\perp}\cdot(\mathbf{x}-\mathbf{x}^{\prime})}\big(\frac{N_{c}^{2}}{2}s_{\mathbf{z}\mathbf{x}^{\prime}}s_{\mathbf{x}\mathbf{z}}-\frac{1}{2}s_{\mathbf{x}\mathbf{x}^{\prime}}+C_{F}N_{c}\big)\\
 & \times\Bigg\{\int_{0}^{+\infty}\frac{\mathrm{d}p_{3}^{+}}{p_{3}^{+}}\frac{p_{3}^{+}}{p_{1}^{+}}\frac{p_{0\scriptscriptstyle{R}}^{+}}{p_{0}^{+}}\frac{p_{3}^{+}}{p_{1}^{+}+p_{3}^{+}}e^{-i\frac{p_{3}^{+}}{p_{1}^{+}+p_{3}^{+}}\mathbf{q}\cdot(\mathbf{x}^{\prime}-\mathbf{z})}e^{-i\frac{p_{3}^{+}}{p_{1}^{+}}\mathbf{p}_{1}\cdot(\mathbf{x}-\mathbf{z})}\int_{\boldsymbol{\ell}_{2}}\frac{\boldsymbol{\ell}_{2}^{\bar{\eta}}}{\boldsymbol{\ell}_{2}^{\bar{\lambda}}}e^{-i\boldsymbol{\ell}_{2}\cdot(\mathbf{x}-\mathbf{z})}\\
 & \times\Bigg[\frac{p_{3}^{+}}{p_{0}^{+}}\frac{q^{+}}{p_{1}^{+}+p_{3}^{+}}\frac{\boldsymbol{\ell}_{2}^{\bar{\lambda}}+\frac{p_{0}^{+}}{p_{1}^{+}}\frac{p_{1}^{+}+p_{3}^{+}}{q^{+}}\mathbf{P}_{\perp}^{\bar{\lambda}}}{\big(\boldsymbol{\ell}_{2}+\frac{p_{3}^{+}}{p_{1}^{+}}\mathbf{P}_{\perp}\big)^{2}+\Delta_{\mathrm{FS}}}+\frac{q^{+}\mathbf{q}^{\bar{\lambda}}}{p_{0\scriptscriptstyle{R}}^{+}\mathbf{q}^{2}+(p_{1}^{+}+p_{3}^{+})M^{2}}\Bigg]\\
 & \times\Bigg[-\frac{q^{+}\mathbf{q}^{\lambda^{\prime}}}{p_{0\scriptscriptstyle{R}}^{+}\mathbf{q}^{2}+(p_{1}^{+}+p_{3}^{+})M^{2}}iA^{\eta^{\prime}}(\mathbf{x}^{\prime}-\mathbf{z},\Delta_{\mathrm{IS}})\\
 & \times\Bigg(\Big(\big(1+2\frac{p_{1}^{+}+p_{3}^{+}}{q^{+}}\big)^{2}+1\Big)\Big(\big(1+\frac{2p_{1}^{+}}{p_{3}^{+}}\big)^{2}+1\Big)\delta^{\bar{\lambda}\lambda^{\prime}}\delta^{\bar{\eta}\eta^{\prime}}\\
&-4\big(1+2\frac{p_{1}^{+}+p_{3}^{+}}{q^{+}}\big)\big(1+\frac{2p_{1}^{+}}{p_{3}^{+}}\big)\epsilon^{\bar{\lambda}\lambda^{\prime}}\epsilon^{\bar{\eta}\eta^{\prime}}\Bigg)\\
 & -p_{1}^{+}\delta^{\bar{\eta}\bar{\lambda}}\mathcal{K}\big(\mathbf{x}^{\prime}-\mathbf{z},\Delta_{\mathrm{IS}}\big)\Bigg(\Big(\frac{1}{p_{1}^{+}+p_{3}^{+}}+\frac{1}{p_{0}^{+}}\Big)\Big(\big(1+\frac{2p_{1}^{+}}{p_{3}^{+}}\big)\big(1+2\frac{p_{1}^{+}+p_{3}^{+}}{q^{+}}\big)-1\Big)\\
 & -2\Big(\frac{1}{p_{1}^{+}+p_{3}^{+}}-\frac{1}{p_{0}^{+}}\Big)\Big(\frac{p_{1}^{+}+p_{3}^{+}}{q^{+}}-\frac{p_{1}^{+}}{p_{3}^{+}}\Big)\Bigg)\Bigg]\\
 & -4\int_{k_{\mathrm{min}}^{+}}^{k_{f}^{+}}\frac{\mathrm{d}p_{3}^{+}}{p_{3}^{+}}\Big(\big(1+2\frac{p_{1}^{+}}{q^{+}}\big)^{2}+1\Big)\Big(\frac{q^{+}\mathbf{q}}{p_{0}^{+}\mathbf{q}^{2}+p_{1}^{+}M^{2}}\Big)^{2}A^{i}(\mathbf{x}^{\prime}-\mathbf{z})A^{i}(\mathbf{x}-\mathbf{z})\Bigg\}\;.
\end{aligned}
\label{eq:IS3+4-FS2+3_T_cross}
\end{equation}

\section{\label{sec:conclusions}Conclusions}
We have presented the first next-to-leading order calculation of the cross section for the $p+A\to\gamma^{*}+\mathrm{jet}+X$ process within the hybrid dilute-dense formalism. Replacing the electromagnetic coupling constant, the result can be carried over to the production of a Z-boson. The Drell-Yan + jet cross section is obtained trivially by multiplying with the $\gamma^{*}/Z\to\ell^{+}+\ell^{-}$ branching ratio, unless angular correlations between both leptons are measured. In that case, the leptons might be sensitive to the interference between the production of a longitudinally or transversally polarized virtual boson, which is not calculated explicitly in this work but can in principle be computed using our intermediate results presented here.

\begin{figure}[t]
\begin{centering}
\includegraphics[scale=0.3]{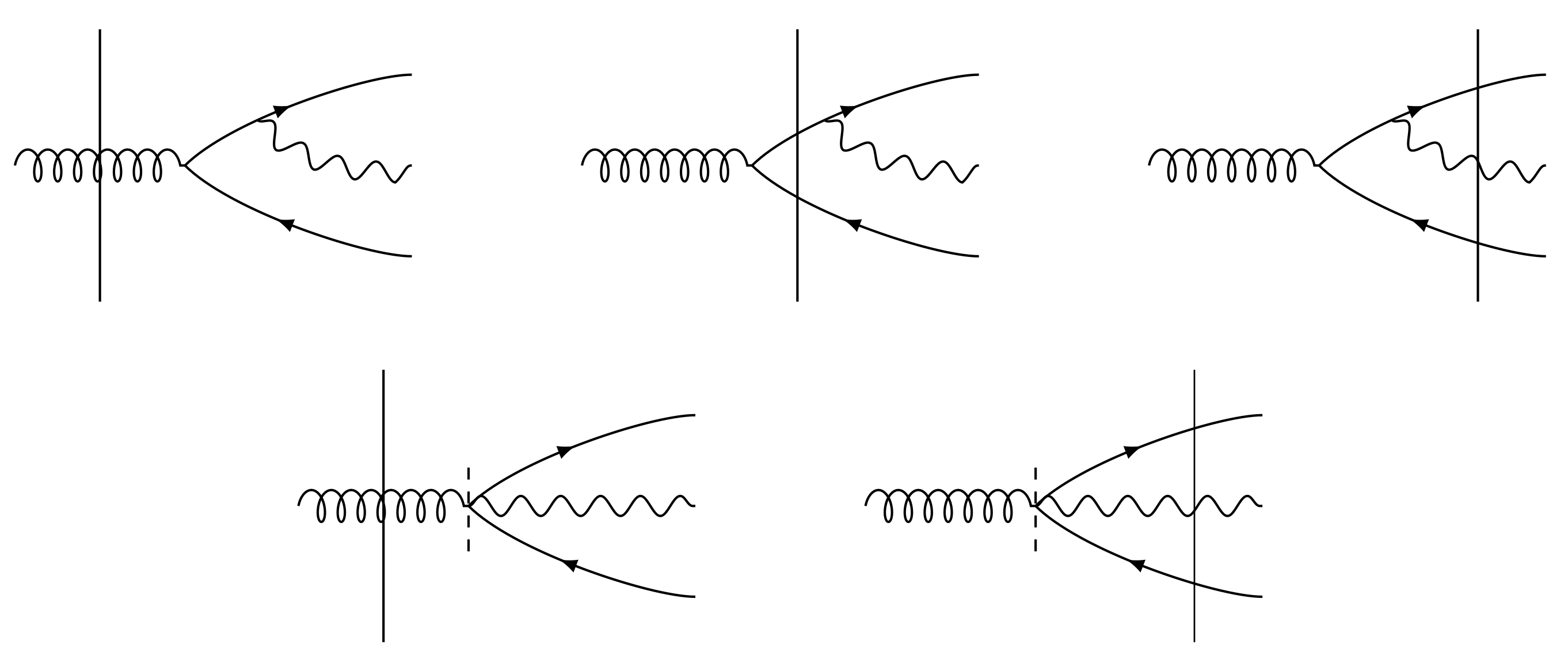}
\par\end{centering}
\centering{}\caption{\label{fig:gluonchannel}The representative Feynman diagrams for the
gluon channel, i.e. the partonic process $g+A\to\gamma^{*}+q+\bar{q}$.
Only the second and third diagram (and its $q\leftrightarrow\bar{q}$ counterparts)
contain a singularity, namely when the antiquark (quark) is integrated
out. These contributions can then be viewed as the LO quark channel
amplitudes, where the incoming quark is generated from the $g\to q\bar{q}$
splitting of the gluon PDF. Hence, these collinear divergences should
be absorbed into the DGLAP evolution of the LO cross section.}
\end{figure}

For phenomenological applications, the gluon channel $g+A\to\gamma^{*}+q+\bar{q}$ should be included. The representative Feynman diagrams are depicted in figure~\ref{fig:gluonchannel}. Assuming the photon has nonvanishing plus-momentum and virtuality, the amplitudes where the incoming gluon interacts with the shockwave are completely finite, as are the amplitudes with instantaneous $g\to\gamma^* q \bar{q}$ splittings. However, the remaining amplitudes, corresponding to the diagrams where the quark-antiquark pair  scatters off the shockwave before or after emitting the photon, are not. Indeed, integrating over the momentum of the antiquark (quark) will yield a singularity when this antiquark (quark) is collinear to its parent gluon. These contributions are proportional to the leading-order amplitudes for the quark (antiquark) channel, where the incoming quark (antiquark) comes from the $g\to q \bar{q}$ splitting in the DGLAP evolution of the PDF. Indeed, these terms were not taken into account in our analysis in section~\ref{sec:DGLAP}. We leave the gluon channel for future work. 

Our result encompasses the NLO cross section for prompt photon plus jet hadroproduction as an important by-product, easily obtained by taking the real photon limit. This process has been studied quite extensively in the literature at leading order~\cite{Jalilian-Marian:2012wwi,Rezaeian:2012wa,Benic:2017znu,Goncalves:2020tvh,Benic:2022ixp,Ganguli:2023joy} and is planned to me measured with the proposed forward calorimeter (FoCal) of the ALICE experiment~\cite{ALICE:2020mso,ALICE:2023fov}. The next-to-leading cross section we provide will greatly increase the potential of this study.

One of the major motivations behind this work is the prospect of studying the region where the virtual photon and the jet are back-to-back in the transverse plane~\cite{Boer:2017xpy}, hence ${\mathbf{P}^2_\perp \gg \mathbf{k}^2_\perp}$. It was observed in ref.~\cite{Dominguez:2010xd} that in this regime the leading-order cross section factorizes into a convolution of a hard part with the quark PDF and a (dipole-type) gluon TMD PDF~\cite{Mulders:2000sh}. Proving such a factorization at next-to-leading order would be much more complex, complicated amongst others by the emergence of large Sudakov logarithms in the ratio ${\mathbf{P}^2_\perp/\mathbf{k}^2_\perp}$, which should be absorbed into the Collins-Soper-Sterman~\cite{Collins:2011zzd} evolution of the gluon TMD. However, should the above nontrivial scenario indeed take place, it would be a further confirmation of the CGC and TMD factorization being compatible at one loop. First steps in this direction were taken in~\cite{Mueller:2012uf,Mueller:2013wwa}. In our recent work~\cite{Taels:2022tza} we considered the case of dijet photoproduction, and provided the first proof that large Sudakov double logarithms can be extracted in a way consistent with high-energy resummation provided the latter is kinematically improved. In a series of papers~\cite{Caucal:2022ulg,Caucal:2023nci,Caucal:2023fsf}, complete next-to-leading order TMD factorization was then established for back-to-back dijet production in deep-inelastic scattering in the CGC. We should remark that the study of combined low-$x$ and Sudakov resummation is a very active field, undertaken within several different formalisms, see e.g.,~\cite{Hautmann:2008vd,Deak:2009xt,Sun:2011iw,Deak:2011ga,Dooling:2014kia,Balitsky:2015qba,Marzani:2015oyb,Hentschinski:2016wya,Bury:2017jxo,Xiao:2017yya,Hentschinski:2017ayz,Deak:2018obv,Zhou:2018lfq,Stasto:2018rci,Blanco:2019qbm,Marquet:2019ltn,vanHameren:2019ysa,Zheng:2019zul,vanHameren:2020rqt,Hentschinski:2020tbi,Balitsky:2020jzt,Nefedov:2020ugj,Hentschinski:2021lsh,Nefedov:2021vvy,Boer:2022njw,Al-Mashad:2022zbq,Shi:2022hee,Balitsky:2022vnb,vanHameren:2023oiq,Balitsky:2023hmh,Shi:2023ejp}.

Finally, another interesting future direction is the study of the inclusive Drell-Yan process, obtained from our result by integrating also the outgoing quark out. This integration, albeit nontrivial, should not lead to new divergences. Experimental data on dilepton production at low invariant mass is currently being analyzed by the ATLAS collaboration~\cite{Giuli}. Moreover, the intermediate results of our calculation, one could then extend to next-to-leading order the higher-twist analysis of the Drell-Yan structure functions performed in~\cite{Motyka:2014lya,Motyka:2016lta,Brzeminski:2016lwh}. 

\section*{Acknowledgements}
First, I am very grateful to Guillaume Beuf for sharing some of his profound insight in LCPT and higher-order calculations with me. Moreover, I would like to thank Valerio Bertone, Dani\"el Boer, Renaud Boussarie, Francesco Giuli, Cyrille Marquet, and Pierre Van Mechelen for interesting discussions. My work is supported by a postdoctoral fellowship fundamental research of the Research Foundation Flanders (FWO) no. 1233422N. 
\appendix
\section{Transverse integral identities}

A standard tool for evaluating the different transverse integrals
in our calculation is the Schwinger trick:
\begin{equation}
\begin{aligned}\frac{\Gamma(\alpha)}{A^{\alpha}} & =\int_{0}^{\infty}\mathrm{d}t\,t^{\alpha-1}e^{-At}\;,\end{aligned}
\label{eq:Schwinger}
\end{equation}
using which we find, for instance, the following expressions for the
Weizsäcker-Williams field (taking the derivative with respect
to $\mathbf{x}$): 
\begin{equation}
\begin{aligned} iA^{i}(\mathbf{x})\equiv\int_{\mathbf{k}}e^{-i\mathbf{k}\cdot\mathbf{x}}\frac{\mathbf{k}^{i}}{\mathbf{k}^{2}}  
  =\frac{-i}{2\pi^{\frac{D}{2}-1}}\frac{\mathbf{x}^{i}}{(\mathbf{x}^{2})^{\frac{D}{2}-1}}\Gamma(\frac{D}{2}-1)\overset{D\to4}{=}\frac{-i}{2\pi}\frac{\mathbf{x}^{i}}{\mathbf{x}^{2}}\;,
\end{aligned}
\label{eq:WWfield}
\end{equation}
and its generalization:
\begin{equation}
\begin{aligned}iA^{i}(\mathbf{x},\Delta) &\equiv \int_{\mathbf{k}}e^{-i\mathbf{k}\cdot\mathbf{x}}\frac{\mathbf{k}^{i}}{\mathbf{k}^{2}+\Delta} \;,\\ & =-i\frac{\mu^{4-D}}{(2\pi)^{\frac{D-2}{2}}}(\sqrt{\Delta})^{\frac{D-2}{2}}\frac{\mathbf{x}^{i}}{|\mathbf{x}|^{\frac{D-2}{2}}}\mathrm{K}_{\frac{D-2}{2}}\big(\sqrt{\mathbf{x}^{2}\Delta}\big)\;,\\
 & \overset{D\to4}{=}\frac{-i}{2\pi}\frac{\mathbf{x}^{i}}{|\mathbf{x}|}\sqrt{\Delta}\mathrm{K}_{1}\big(\sqrt{\mathbf{x}^{2}\Delta}\big)\;.
\end{aligned}
\label{eq:WWfieldmass}
\end{equation}
Note that the latter satisfies the following identity, for a constant
$\chi$:
\begin{equation}
\begin{aligned}iA^{i}(\chi\mathbf{x},\Delta) & =\chi^{3-D}iA^{i}(\mathbf{x},\chi^{2}\Delta)\;.\end{aligned}
\label{eq:WWlinear}
\end{equation}
A second very useful identity is the following:
\begin{equation}
\int_{0}^{\infty}\mathrm{d}s\,s^{\nu-1}e^{-B/s}e^{-Cs}=2\Big(\frac{B}{C}\Big)^{\frac{\nu}{2}}\mathrm{K}_{-\nu}\big(2\sqrt{BC}\big)\;,\label{eq:Bessel}
\end{equation}
which we can use to solve, e.g.:
\begin{equation}
\begin{aligned}\mathcal{K}(\mathbf{x},\Delta)  \equiv\int_{\boldsymbol{\ell}}\frac{e^{i\boldsymbol{\ell}\cdot\mathbf{x}}}{\boldsymbol{\ell}^{2}+\Delta} =\frac{\mu^{4-D}}{(2\pi)^{\frac{D-2}{2}}}\Big(\frac{\mathbf{x}^{2}}{\Delta}\Big)^{\frac{4-D}{4}}K_{\frac{D-4}{2}}\big(\sqrt{\mathbf{x}^{2}\Delta}\big)\;.
\end{aligned}
\label{eq:integralK}
\end{equation}
The above integral is finite when $\Delta\neq0$, since $\Delta$
acts as an infrared regulator, while the phase cuts off ultraviolet
divergences. We can, therefore, set $D=4$ and take the derivative
with respect to $\mathbf{x}$ to obtain:
\begin{equation}
\begin{aligned}\int_{\boldsymbol{\ell}}e^{i\boldsymbol{\ell}\cdot\mathbf{x}}\frac{\boldsymbol{\ell}^{2}-\Delta}{\boldsymbol{\ell}^{2}+\Delta} & =-2\Delta\mathcal{K}(\mathbf{x},\Delta)\;.\end{aligned}
\end{equation}

The following UV-divergent loop integral is omnipresent in our calculation,
and is very straightforward to solve (see e.g. \cite{Peskin:1995ev}) using
dimensional regularization with $D=4-2\epsilon_{\mathrm{UV}}$:
\begin{equation}
\begin{aligned}\mathcal{A}_{0}(\Delta) & \equiv\int_{\boldsymbol{\ell}}\frac{1}{\boldsymbol{\ell}^{2}+\Delta} =\frac{1}{4\pi}\Gamma\big(\frac{4-D}{2}\big)\Big(\frac{\Delta}{4\pi\mu^{2}}\Big)^{\frac{D-4}{2}}\;,\\
 & =\frac{1}{4\pi}\Big(\frac{1}{\epsilon_{\mathrm{UV}}}-\gamma_{E}+\ln\frac{4\pi\mu^{2}}{\Delta}\Big)+\mathcal{O}(\epsilon_{\mathrm{UV}})\;.
\end{aligned}
\label{eq:A0}
\end{equation}
Likewise:
\begin{equation}
\begin{aligned}\int_{\boldsymbol{\ell}}\frac{\boldsymbol{\ell}^{2}}{\boldsymbol{\ell}^{2}+\Delta} & =-\frac{\mu^{4-D}}{(4\pi)^{\frac{D-2}{2}}}\Gamma\Big(\frac{4-D}{2}\Big)\Delta^{\frac{D-2}{2}}\;,\\
 & =-\frac{\Delta}{4\pi}\Big(\frac{1}{\epsilon_{\mathrm{UV}}}-\gamma_{E}+\ln\frac{4\pi\mu^{2}}{\Delta}\Big)+\mathcal{O}(\epsilon_{\mathrm{UV}})\;,\\
 & =-\Delta\mathcal{A}_{0}(\Delta)\;.
\end{aligned}
\label{eq:SEloop}
\end{equation}
In the virtual amplitudes, we encounter many transverse loop integrals
that require a bit more of machinery to tackle them. This machinery
is provided by the Passarino-Veltman reduction procedure (ref.~\cite{Passarino:1978jh},
see also \cite{Beuf:2016wdz} for a clear outline), using which we
obtain the following identities:
\begin{equation}
\begin{aligned}\int_{\boldsymbol{\ell}}\frac{1}{\boldsymbol{\ell}^{2}+\Omega}\frac{1}{\big(\mathbf{\boldsymbol{\ell}}\!+\mathbf{k}\big)^{2}+\Delta} & \!=\!\mathcal{B}_{0}(\Omega,\Delta,\mathbf{k})\;,\\
\int_{\boldsymbol{\ell}}\frac{\mathbf{\boldsymbol{\ell}}^{i}}{\boldsymbol{\ell}^{2}+\Omega}\frac{1}{\big(\mathbf{\boldsymbol{\ell}}+\mathbf{k}\big)^{2}+\Delta} 
& \!\equiv\!\mathbf{k}^{i}\mathcal{B}_{1}(\Omega,\Delta,\mathbf{k})\;,\\
 & \!=\!\frac{\mathbf{k}^{i}}{2\mathbf{k}^{2}}\Big(\mathcal{A}_{0}(\Omega)-\mathcal{A}_{0}(\Delta)+(\Omega-\Delta-\mathbf{k}^{2})\mathcal{B}_{0}(\Omega,\Delta,\mathbf{k})\Big)\;,\\
\int_{\boldsymbol{\ell}}\frac{\mathbf{\boldsymbol{\ell}}^{i}\mathbf{\boldsymbol{\ell}}^{j}}{\boldsymbol{\ell}^{2}+\Omega}\frac{1}{\big(\mathbf{\boldsymbol{\ell}}+\mathbf{k}\big)^{2}+\Delta} 
& \!=\!\frac{\mathbf{k}^{i}\mathbf{k}^{j}}{\mathbf{k}^{2}}\frac{1}{D-3}\Big(\frac{D-4}{2}\mathcal{A}_{0}(\Delta)+\Omega\mathcal{B}_{0}(\Omega,\Delta,\mathbf{k})\\
 & \!+\!\frac{D-2}{2}(\Omega-\Delta-\mathbf{k}^{2})\mathcal{B}_{1}(\Omega,\Delta,\mathbf{k})\Big) \!+\!\frac{\delta^{ij}}{D\!-\!3}\Big(\frac{1}{2}\mathcal{A}_{0}(\Delta)\\
 &-\Omega\mathcal{B}_{0}(\Omega,\Delta,\mathbf{k})-\frac{1}{2}(\Omega\!-\!\Delta\!-\!\mathbf{k}^{2})\mathcal{B}_{1}(\Omega,\Delta,\mathbf{k})\Big)\;,\\
\mathbf{k}^{j}\!\int_{\boldsymbol{\ell}}\frac{\mathbf{\boldsymbol{\ell}}^{i}\mathbf{\boldsymbol{\ell}}^{j}}{\boldsymbol{\ell}^{2}+\Omega}\frac{1}{\big(\mathbf{\boldsymbol{\ell}}+\mathbf{k}\big)^{2}+\Delta} 
& \!=\!\frac{\mathbf{k}^{i}}{2}\Big(\mathcal{A}_{0}(\Delta)+(\Omega-\Delta-\mathbf{k}^{2})\mathcal{B}_{1}(\Omega,\Delta,\mathbf{k})\Big)\;,\\
\mathcal{B}_{0}(\Delta,\mathbf{k}) 
& \!\equiv\!\mathcal{B}_{0}(0,\Delta,\mathbf{k})=\mathcal{B}_{0}(\Delta,0,\mathbf{k})\;,\\
\mathcal{B}_{0}(\Delta) 
& \!\equiv\!\mathcal{B}_{0}(0,\Delta,0)=\mathcal{B}_{0}(\Delta,0,0)\;.
\end{aligned}
\label{eq:IntegralIdentitiesB}
\end{equation}
All the above integrals all are reduced to functions of the simple
scalar integrals $\mathcal{A}_{0}(\Delta)$ and $\mathcal{B}_{0}(\Omega,\Delta,\mathbf{k})$.
As already explained above, $\mathcal{A}_{0}(\Delta)$ is divergent
in the ultraviolet. In the limit $\Delta\to0$, a second, infrared
divergence appears, which exactly cancels the UV one, at least within
dimensional regularization. Another way to see this is that $\mathcal{A}_{0}(0)$
is a scaleless integral:
\begin{equation}
\begin{aligned}\mathcal{A}_{0}(0) & =\int_{\boldsymbol{\ell}}\frac{1}{\boldsymbol{\ell}^{2}}=\frac{1}{4\pi}\Big(\frac{1}{\epsilon_{\mathrm{UV}}}-\frac{1}{\epsilon_{\mathrm{IR}}}\Big)=0\;.\end{aligned}
\end{equation}
The above property plays a central role in the analysis of field-strength
renormalization amplitudes, section \ref{subsec:Z}. 

From simple power counting it follows that the integral $\mathcal{B}_{0}(\Omega,\Delta,\mathbf{k})$,
defined in the first line of eq. (\ref{eq:IntegralIdentitiesB}),
is divergent in the infrared when one of its first two arguments tends
to zero. Indeed, introducing Feynman parameters (see e.g. \cite{Peskin:1995ev}):
\begin{equation}
\begin{aligned}\frac{1}{A_{1}^{m_{1}}A_{2}^{m_{2}}...A_{n}^{m_{n}}} & \!=\!\int_{0}^{1}\!\mathrm{d}x_{1}...\mathrm{d}x_{n}\delta(\sum x_{i}\!-\!1)\frac{x_{1}^{m_{1}-1}\!\cdot\! x_{2}^{m_{2}-1}\!\cdot\!...\!\cdot\! x_{n}^{m_{n}-1}}{\big(x_{1}A_{1}\!+\!x_{2}A_{2}\!+\!...\!+\!x_{n}A_{n}\big)^{(m_{1}\!+\!m_{2}\!+\!...\!+\!m_{n})}}\\
&\times\frac{\Gamma(m_{1}+m_{2}+...+m_{n})}{\Gamma(m_{1})\Gamma(m_{2})...\Gamma(m_{n})}\;,\end{aligned}
\label{eq:Feynman}
\end{equation}
and writing $D=4-2\epsilon_{\mathrm{IR}}$, we obtain in the $\overline{\mathrm{MS}}$
scheme (where $\mu^{2}\to\mu^{2}e^{\gamma_{E}}/4\pi$):
\begin{equation}
\begin{aligned}\mathcal{B}_{0}(0,\Delta,\mathbf{k}) & =\mu^{4-D}\Big(\frac{1}{4\pi}\Big)^{\frac{D-2}{2}}\Gamma\big(\frac{6-D}{2}\big)\int_{0}^{1}\mathrm{d}x\frac{1}{x^{\frac{6-D}{2}}\big((1-x)\mathbf{k}^{2}+\Delta\big)^{\frac{6-D}{2}}}\;,\\
 & =\frac{1}{4\pi}\frac{-1}{\mathbf{k}^{2}+\Delta}\Bigg(\frac{1}{\epsilon_{\mathrm{IR}}}-\ln\frac{\Delta+\mathbf{k}^{2}}{\mu^{2}}-\,_{2}F_{1}^{(0,0,1,0)}\big(0,1,1,\frac{\mathbf{k}^{2}}{\mathbf{k}^{2}+\Delta}\big)\\
 & +\,_{2}F_{1}^{(0,1,0,0)}\big(0,1,1,\frac{\mathbf{k}^{2}}{\mathbf{k}^{2}+\Delta}\big)-\,_{2}F_{1}^{(1,0,0,0)}\big(0,1,1,\frac{\mathbf{k}^{2}}{\mathbf{k}^{2}+\Delta}\big)\Bigg)+\mathcal{O}(\epsilon_{\mathrm{IR}})\;,\\
 & =\mathcal{B}_{0}(\Delta,0,\mathbf{k})\;,
\end{aligned}
\label{eq:B0_explicit}
\end{equation}
where $_{2}F_{1}{}^{(0,0,1,0)}$ is the derivative of the third argument
of the hypergeometric function $_{2}F_{1}$. The above calculation
serves to prove that all the other expressions in (\ref{eq:IntegralIdentitiesB})
are finite when either $\Delta$ or $\Omega$ is zero.
For example:
\begin{equation}
\begin{aligned}\lim_{\Omega\to0}\mathcal{B}_{1}(\Omega,\Delta,\mathbf{k}) & =\frac{1}{2\mathbf{k}^{2}}\Big(\mathcal{A}_{0}(0)-\mathcal{A}_{0}(\Delta)+(-\Delta-\mathbf{k}^{2})\mathcal{B}_{0}(0,\Delta,\mathbf{k})\Big)\;,\\
 & =\frac{1}{2\mathbf{k}^{2}}\Big(\mathcal{A}_{0}(0)-\mathcal{A}_{0}(\Delta)+\frac{1}{4\pi}\frac{1}{\epsilon_{\mathrm{IR}}}\Big)+\mathrm{finite}\;,\\
 & =\frac{1}{2\mathbf{k}^{2}}\frac{1}{4\pi}\Big(\frac{1}{\epsilon_{\mathrm{UV}}}-\frac{1}{\epsilon_{\mathrm{IR}}}-\frac{1}{\epsilon_{\mathrm{UV}}}+\frac{1}{\epsilon_{\mathrm{IR}}}\Big)+\mathrm{finite}\;.
\end{aligned}
\label{eq:B1}
\end{equation}
A second useful set of integrals is:
\begin{equation}
\begin{aligned}\int_{\boldsymbol{\ell}}\frac{1}{\boldsymbol{\ell}^{2}}\frac{1}{\mathbf{\boldsymbol{\ell}}^{2}\!+\!\Delta}\frac{1}{\big(\mathbf{\boldsymbol{\ell}}\!+\!\mathbf{k}\big)^{2}+\Pi} &\! =\!\mathcal{C}_{0}(\Delta,\Pi,\mathbf{k})\;,\\
\int_{\boldsymbol{\ell}}\frac{\mathbf{\boldsymbol{\ell}}^{i}}{\boldsymbol{\ell}^{2}}\frac{1}{\mathbf{\boldsymbol{\ell}}^{2}\!+\!\Delta}\frac{1}{\big(\mathbf{\boldsymbol{\ell}}\!+\!\mathbf{k}\big)^{2}\!+\!\Pi} & \!=\mathbf{k}^{i}\mathcal{C}_{1}(\Delta,\Pi,\mathbf{k})\;,\\
 & \!=\!\frac{\mathbf{k}^{i}}{2\mathbf{k}^{2}}\Big(\mathcal{B}_{0}(\Delta)\!-\!\mathcal{B}_{0}(\Delta,\Pi,\mathbf{k})\!-\!(\mathbf{k}^{2}\!+\!\Pi)\mathcal{C}_{0}(\Delta,\Pi,\mathbf{k})\Big)\;,\\
 \mathbf{k}^{j}\!\int_{\boldsymbol{\ell}}\frac{\mathbf{\boldsymbol{\ell}}^{i}\mathbf{\boldsymbol{\ell}}^{j}}{\boldsymbol{\ell}^{2}}\frac{1}{\mathbf{\boldsymbol{\ell}}^{2}\!+\!\Delta}\frac{1}{\big(\mathbf{\boldsymbol{\ell}}\!+\!\mathbf{k}\big)^{2}\!+\!\Pi} & \!=\!-\frac{\mathbf{k}^{i}}{2}\Big(\mathcal{B}_{1}(\Delta,\Pi,\mathbf{k})\!+\!(\mathbf{k}^{2}\!+\!\Pi)\mathcal{C}_{1}(\Delta,\Pi,\mathbf{k})\Big)\;,\\
\int_{\boldsymbol{\ell}}\frac{\mathbf{\boldsymbol{\ell}}^{i}\mathbf{\boldsymbol{\ell}}^{j}}{\boldsymbol{\ell}^{2}}\frac{1}{\mathbf{\boldsymbol{\ell}}^{2}\!+\!\Delta}\frac{1}{\big(\mathbf{\boldsymbol{\ell}}\!+\!\mathbf{k}\big)^{2}\!+\!\Pi} & =\mathcal{C}^{ij}(\Delta,\Pi,\mathbf{k}) =\mathbf{k}^{i}\mathbf{k}^{j}\mathcal{C}_{21}(\Delta,\Pi,\mathbf{k})\!+\!\delta^{ij}\mathcal{C}_{22}(\Delta,\Pi,\mathbf{k})\;,
\end{aligned}
\label{eq:IntegralIdentitiesC}
\end{equation}
with:
\begin{equation}
\begin{aligned}
\mathcal{C}_{22}(\Delta,\Pi,\mathbf{k}) & =\frac{1}{D-3}\Big(\frac{1}{2}\mathcal{B}_{1}(\Delta,\Pi,\mathbf{k})+\frac{1}{2}(\mathbf{k}^{2}+\Pi)\mathcal{C}_{1}(\Delta,\Pi,\mathbf{k})+\mathcal{B}_{0}(\Delta,\Pi,\mathbf{k})\Big)\;,\\
\mathcal{C}_{21}(\Delta,\Pi,\mathbf{k}) & =-\frac{1}{D-3}\frac{1}{\mathbf{k}^{2}}\Big(\mathcal{B}_{0}(\Delta,\Pi,\mathbf{k})\\
&+\frac{D-2}{2}\mathcal{B}_{1}(\Delta,\Pi,\mathbf{k})+\frac{D-2}{2}(\mathbf{k}^{2}+\Pi)\mathcal{C}_{1}(\Delta,\Pi,\mathbf{k})\Big)\;.
\end{aligned}
\label{eq:IntegralIdentitiesCii}
\end{equation}
Explicitly, we find that:
\begin{equation}
\begin{aligned}\mathcal{C}_{0}(\Delta,\mathbf{k}) &\equiv \mathcal{C}_{0}(\Delta,0,\mathbf{k}) =\mathcal{C}_{0}(0,\Delta,\mathbf{k}) \;,\\
& =\frac{\mu^{4-D}}{(4\pi)^{\frac{D-2}{2}}}\Gamma\big(\frac{8-D}{2}\big)\int_{0}^{1}\mathrm{d}y\int_{0}^{1-y}\mathrm{d}z\frac{1}{\big(y(1-y)\mathbf{k}^{2}+z\Delta\big)^{\frac{8-D}{2}}}\;,\\
 & =-\frac{1}{\mathbf{k}^{2}\Delta}\Big(1+\frac{\Delta}{\mathbf{k}^{2}+\Delta}\Big)\frac{1}{4\pi\epsilon_{\mathrm{IR}}}+\mathcal{O}(\epsilon_{\mathrm{IR}}^{0})\;.
\end{aligned}
\label{eq:C0}
\end{equation}

We conclude this section with a list of the divergent (either ultraviolet or infrared)
parts of transverse integrals commonly encountered in the calculation,
in particular in the evaluation of the virtual vertex corrections
in section \ref{subsec:V}:
\begin{equation}
\begin{aligned}\mathcal{A}_{0}(\Delta) & =\int_{\boldsymbol{\ell}}\frac{1}{\mathbf{\boldsymbol{\ell}}^{2}+\Delta}=\frac{1}{4\pi}\frac{1}{\epsilon_{\mathrm{UV}}}+\mathcal{O}(\epsilon^{0})\;,\\
\mathcal{B}_{0}(\Delta,\mathbf{k}) & =\int_{\boldsymbol{\ell}}\frac{1}{\big(\mathbf{\boldsymbol{\ell}}+\mathbf{k}\big)^{2}}\frac{1}{\mathbf{\boldsymbol{\ell}}^{2}+\Delta}=-\frac{1}{\mathbf{k}^{2}+\Delta}\frac{1}{4\pi\epsilon_{\mathrm{IR}}}+\mathcal{O}(\epsilon^{0})\;,\\
\mathcal{C}_{0}(\Delta,\mathbf{k}) & =\int_{\boldsymbol{\ell}}\frac{1}{\boldsymbol{\ell}^{2}}\frac{1}{\big(\mathbf{\boldsymbol{\ell}}+\mathbf{k}\big)^{2}}\frac{1}{\mathbf{\boldsymbol{\ell}}^{2}+\Delta}=\frac{\mathbf{k}^{2}+2\Delta}{\mathbf{k}^{2}\Delta}\mathcal{B}_{0}(\Delta,\mathbf{k})+\mathcal{O}(\epsilon^{0})\;,\\
\mathbf{k}^{i}\mathcal{C}_{1}(\Delta,\mathbf{k}) & =\int_{\boldsymbol{\ell}}\frac{\mathbf{\boldsymbol{\ell}}^{i}}{\boldsymbol{\ell}^{2}}\frac{1}{\big(\mathbf{\boldsymbol{\ell}}+\mathbf{k}\big)^{2}}\frac{1}{\mathbf{\boldsymbol{\ell}}^{2}+\Delta}=-\frac{\mathbf{k}^{i}}{\mathbf{k}^{2}}\mathcal{B}_{0}(\Delta,\mathbf{k})+\mathcal{O}(\epsilon^{0})\;,\\
\mathcal{C}^{ij}(\Delta,\mathbf{k}) & =\int_{\boldsymbol{\ell}}\frac{\mathbf{\boldsymbol{\ell}}^{i}\mathbf{\boldsymbol{\ell}}^{j}}{\boldsymbol{\ell}^{2}}\frac{1}{\big(\mathbf{\boldsymbol{\ell}}+\mathbf{k}\big)^{2}}\frac{1}{\mathbf{\boldsymbol{\ell}}^{2}+\Delta}=\frac{\mathbf{k}^{i}\mathbf{k}^{j}}{\mathbf{k}^{2}}\mathcal{B}_{0}(\Delta,\mathbf{k})+\mathcal{O}(\epsilon^{0})\;.
\end{aligned}
\label{eq:DivInts}
\end{equation}

\section{\label{sec:Dirac}Dirac algebra}
In this section, we remind the reader of some key gamma matrix identities in transverse $D-2$-dimensional Euclidean space. 

First, it follows from the definition $\{\gamma^{\mu},\gamma^{\nu}\}=2g^{\mu\nu}\mathds{1}_{4}$ that:
\begin{equation}
\{\bgamma^{i},\bgamma^{j}\}=-2\delta^{ij}\mathds{1}_{4}\;,\label{eq:gammacommutatorD-2}
\end{equation}
with the trivial corollary:
\begin{equation}
\begin{aligned}\bgamma^{i}\bgamma^{i}\mathds{1}_{4} & =-(D-2)\mathds{1}_{4}\;.\end{aligned}
\end{equation}
Repeated application of \eqref{eq:gammacommutatorD-2} brings us to the following important identity:
\begin{equation}
\begin{aligned}\bgamma^{i}\bgamma^{j}\bgamma^{k}\bgamma^{l} & =\bgamma^{k}\bgamma^{l}\bgamma^{i}\bgamma^{j}+2\delta^{il}\bgamma^{k}\bgamma^{j}-2\delta^{ik}\bgamma^{l}\bgamma^{j}+2\delta^{jl}\bgamma^{i}\bgamma^{k}-2\delta^{jk}\bgamma^{i}\bgamma^{l}\;.\end{aligned}
\end{equation}
The Dirac sigma in $D-2$ dimensions: $\sigma^{ij}=(i/2)[\bgamma^{i},\bgamma^{j}]$, can then be shown to satisfy the following commutation relation:
\begin{equation}
\begin{aligned}\big[\sigma^{ij},\sigma^{kl}\big] & =2i\delta^{il}\sigma^{kj}-2i\delta^{ik}\sigma^{lj}+2i\delta^{jl}\sigma^{ik}-2i\delta^{jk}\sigma^{il}\;,\end{aligned}
\label{eq:sigmacommutator}
\end{equation}
while contracting two Dirac sigmas gives:
\begin{equation}
\begin{aligned}\sigma^{ij}\sigma^{il} & =(D-3)\delta^{jl}\mathds{1}_{4}+i(D-4)\sigma^{jl}\;.\end{aligned}
\label{eq:sigmasigma}
\end{equation}
The following identities will be useful as well:
\begin{equation}
\begin{aligned}\sigma^{\eta\eta^{\prime}}\sigma^{\lambda\bar{\lambda}}\sigma^{\eta\bar{\eta}} & =2i(D-3)\delta^{\lambda\bar{\eta}}\delta^{\eta^{\prime}\bar{\lambda}}-2(D-4)\delta^{\lambda\bar{\eta}}\sigma^{\eta^{\prime}\bar{\lambda}}-2i\sigma^{\lambda\eta^{\prime}}\sigma^{\bar{\eta}\bar{\lambda}}-2i(D-3)\delta^{\bar{\lambda}\bar{\eta}}\delta^{\eta^{\prime}\lambda}\\
 & +2(D-4)\delta^{\bar{\lambda}\bar{\eta}}\sigma^{\eta^{\prime}\lambda}-2i\sigma^{\bar{\lambda}\eta^{\prime}}\sigma^{\lambda\bar{\eta}}+(D-3)\delta^{\eta^{\prime}\bar{\eta}}\sigma^{\lambda\bar{\lambda}}+i(D-4)\sigma^{\eta^{\prime}\bar{\eta}}\sigma^{\lambda\bar{\lambda}}\;,
\end{aligned}
\label{eq:3sigma}
\end{equation}
and:
\begin{equation}
\begin{aligned}\sigma^{\eta\bar{\eta}}\sigma^{\lambda\bar{\lambda}}\sigma^{\eta\bar{\eta}} & =\Big[(D-3)(D-2)-8(D-4)\Big]\sigma^{\lambda\bar{\lambda}}\;.\end{aligned}
\label{eq:3sigmadelta}
\end{equation}
Dirac traces can be often simplified using the fact that $\gamma^{+}$
and $\gamma^{-}$ commute with the transverse gamma matrices and thus also with $\sigma^{ij}$, and then applying the completeness relation
\begin{equation}
\begin{aligned}u_{G}^{s}(q^{+})\bar{u}_{G}^{s}(q^{+})\gamma^{+} & =2q^{+}\mathcal{P}_{G}\;,\end{aligned}
\end{equation}
with $\mathcal{P}_{G}=\gamma^{-}\gamma^{+}/2$ the projector on good spinors.

Using the above definitions, together with the cyclic permutation property
of the spinor trace and the fact that one can reverse the order of
gamma matrices inside a trace:
\begin{equation}
\mathrm{Tr}(\gamma^{\mu}\gamma^{\nu}\gamma^{\rho}...)=\mathrm{Tr}(...\gamma^{\rho}\gamma^{\nu}\gamma^{\mu})\;,
\end{equation}
it is straightforward to establish the identities: 
\begin{equation}
\begin{aligned}\mathrm{Tr}\big(\mathcal{P}_{G}\big) & =2\;,\\
\mathrm{Tr}\big(\mathcal{P}_{G}\sigma^{ij}\big) & =0\;,\\
\mathrm{Tr}\big(\mathcal{P}_{G}\sigma^{ij}\sigma^{kl}\big) & =2(\delta^{ik}\delta^{jl}-\delta^{il}\delta^{jk})\overset{D\to4}{=}2\epsilon^{ij}\epsilon^{kl}\;.
\end{aligned}
\label{eq:traceP}
\end{equation}
Moreover, using the above relations as well as the commutation relation
(\ref{eq:sigmacommutator}), it is straightforward to prove the following
relations: 
\begin{equation}
\begin{aligned} & \mathrm{Tr}\Big\{\mathcal{P}_{G}\big(a\delta^{\lambda\lambda^{\prime}}+i\sigma^{\lambda\lambda^{\prime}}\big)\big(b\delta^{\eta\bar{\eta}}-i\sigma^{\eta\bar{\eta}}\big)\big(c\delta^{\lambda\bar{\lambda}}-i\sigma^{\lambda\bar{\lambda}}\big)\big(d\delta^{\eta\eta^{\prime}}-i\sigma^{\eta\eta^{\prime}}\big)\Big\}\\
 & =2\Big[\big(ac+D-3\big)\big(bd-(D-3)\big)\delta^{\bar{\eta}\eta^{\prime}}\delta^{\bar{\lambda}\lambda^{\prime}}\\
 & +\Big((a+c)(b-d)+(D-4)\big(c-a-b-d\big)+(D-4)^{2}\Big)\epsilon^{\bar{\eta}\eta^{\prime}}\epsilon^{\bar{\lambda}\lambda^{\prime}}\Big]\;,
\end{aligned}
\label{eq:DiracTraceBossV}
\end{equation}
and:
\begin{equation}
\begin{aligned} & \mathrm{Tr}\Big\{\mathcal{P}_{G}\big(a\delta^{\eta\eta^{\prime}}+i\sigma^{\eta\eta^{\prime}}\big)\big(b\delta^{\lambda\lambda^{\prime}}+i\sigma^{\lambda\lambda^{\prime}}\big)\big(c\delta^{\lambda\bar{\lambda}}-i\sigma^{\lambda\bar{\lambda}}\big)\big(d\delta^{\eta\bar{\eta}}-i\sigma^{\eta\bar{\eta}}\big)\Big\}\\
 & \!=\!2\Big[\big(ad\!+\!D\!-\!3\big)\big(bc\!+\!D\!-\!3\big)\delta^{\bar{\eta}\eta^{\prime}}\delta^{\bar{\lambda}\lambda^{\prime}}\!-\!\big(a\!+\!d\!+\!D\!-\!4\big)\big(b\!+\!c\!-\!(D\!-\!4)\big)\epsilon^{\bar{\eta}\eta^{\prime}}\epsilon^{\bar{\lambda}\lambda^{\prime}}\Big]\;.
\end{aligned}
\label{eq:DiracTraceBossIS}
\end{equation}
\section{\label{sec:Example-calculation}Example calculation}

\begin{figure}[t]
\begin{centering}
\includegraphics[scale=0.2]{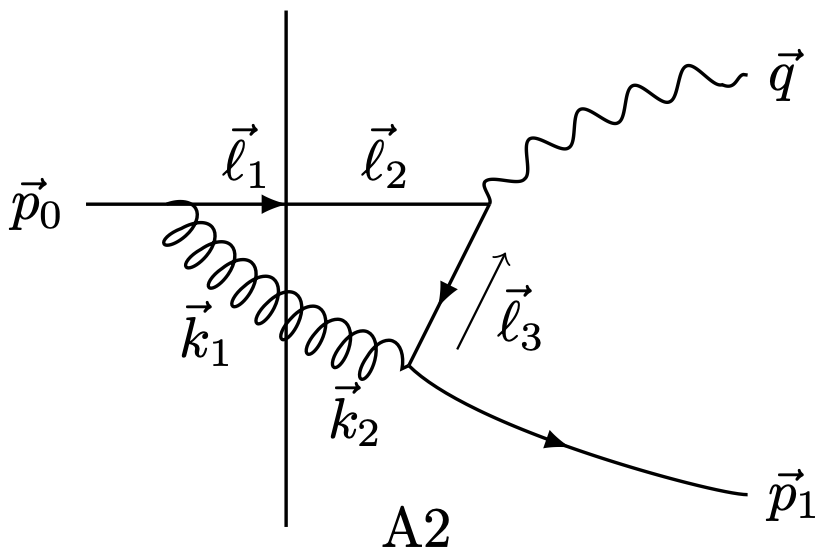}
\par\end{centering}
\caption{\label{fig:A2}Feynman diagram for the production of a virtual photon
and a quark via a gluon-quark-antiquark loop. The virtual gluon and
quark scatter off the shockwave.}
\end{figure}

In this section, we describe in full detail the calculation of the
amplitude corresponding to the diagram A2, depicted in figure \ref{fig:A2}.
We follow the Bjorken-Kogut-Soper formulation of light-cone perturbation
theory~\cite{Bjorken:1970ah}, and approach the CGC in the spirit of the dipole approach
~\cite{Mueller:1989st,Nikolaev:1990ja}. In particular, we draw heavily from the excellent introduction to LCPT applied to
the CGC in the appendix of ref.~\cite{Beuf:2016wdz}.

We remind the reader that the shockwave is treated as an
external static potential. In this background field, we then perturbatively calculate the projectile
dynamics, which take place on a much shorter timescale than those
of the \textquoteleft frozen' target. The leading-order and virtual amplitudes
$\mathcal{M}$ in our calculation are defined as:
\begin{equation}
\begin{aligned}\,_{f}\big\langle\mathbf{q}(\vec{p}_{1})\boldsymbol{\gamma}^{*}(\vec{q})\big|\hat{F}-1\big|\mathbf{q}(\vec{p}_{0})\big\rangle_{i} & =2\pi\delta(p_{0}^{+}-p_{1}^{+}-q^{+})\mathcal{M}\;,\end{aligned}
\label{eq:Mdefapx}
\end{equation}
where $\hat{F}-1$ is the external potential evaluated between the Fock states of the incoming quark
and outgoing quark with a virtual photon or vector boson. 

The Fock states appearing in eq.~\eqref{eq:Mdefapx}
are obtained from the perturbative evolution of the asymptotic eigenstates of the free Hamiltonian $\hat{H}_{0}$ :
\begin{equation}
\begin{aligned}\big|\mathbf{q}(\vec{p}_{0})\big\rangle_{i} & =\hat{\mathcal{U}}(0,-\infty)\big|\mathbf{q}(\vec{p}_{0})\big\rangle\;,\\
\,_{f}\big\langle\mathbf{q}(\vec{p}_{1})\boldsymbol{\gamma}^{*}(\vec{q})\big| & =\big\langle\mathbf{q}(\vec{p}_{1})\boldsymbol{\gamma}^{*}(\vec{q})\big|\hat{\mathcal{U}}(+\infty,0)\;,
\end{aligned}
\end{equation}
where $x^{+}=0$ is the space-time position of the shockwave.
Next, remember from eq.~\eqref{eq:timeevo} that:
\begin{equation}
\begin{aligned}\hat{\mathcal{U}}(0,-\infty) & \equiv\hat{T}\exp\Bigl(-i\int_{-\infty}^{0}\mathrm{d}x^{+}\hat{\mathcal{H}}(x^{+})\Bigr)\;,\end{aligned}
\end{equation}
where the Hamiltonian has the usual time dependence of the interaction picture:
\begin{equation}
\begin{aligned}\hat{\mathcal{H}}(x^{+}) & =e^{i\hat{H}_{0}x^{+}}\hat{V}e^{-i\hat{H}_{0}x^{+}}\;.\end{aligned}
\end{equation}
We now obtain:
\begin{equation}
\begin{aligned}\hat{\mathcal{U}}(0,-\infty)\big|\mathbf{q}(\vec{p}_{0})\big\rangle & \!=\!\big|\mathbf{q}(\vec{p}_{0})\big\rangle\\
&\!-\!i\!\int\mathrm{PS}(\vec{\ell_{1}},\vec{k}_{1})\int_{-\infty}^{0}\mathrm{d}x^{+}\big\langle\mathbf{q}(\vec{\ell}_{1})\mathbf{g}(\vec{k}_{1})\big|e^{i(\ell_{1}^{-}+k^{-}_1)x^{+}}\hat{V}e^{-ip^{-}_0x^{+}}\big|\mathbf{q}(\vec{p}_{0})\big\rangle\\
 &\! \times\!\big|\mathbf{q}(\vec{\ell}_{1})\mathbf{g}(\vec{k}_{1})\big\rangle+\mathcal{O}(g_s^2)\;,\\
 & \!=\!\big|\mathbf{q}(\vec{p}_{0})\big\rangle\!+\!\!\!\int\!\!\mathrm{PS}(\vec{\ell}_{1},\vec{k}_{1})\frac{\big\langle\mathbf{q}(\vec{\ell}_{1})\mathbf{g}(\vec{k}_{1})\big|\hat{V}\big|\mathbf{q}(\vec{p}_{0})\big\rangle}{p_{0}^{-}\!-\!\ell_{1}^{-}\!-\!k^{-}\!+\!i0^{+}}\big|\mathbf{q}(\vec{\ell}_{1})\mathbf{g}(\vec{k}_{1})\big\rangle\!+\!\mathcal{O}(g_s^2)\;.
\end{aligned}
\label{eq:forwardsevo}
\end{equation}
Likewise, being careful with the time-ordering operator $\hat{T}$:
\begin{equation}
\begin{aligned} & \big\langle\mathbf{q}(\vec{p}_{1})\boldsymbol{\gamma}^{*}(\vec{q})\big|\hat{\mathcal{U}}(+\infty,0)=\big\langle\mathbf{q}(\vec{p}_{1})\boldsymbol{\gamma}^{*}(\vec{q})\big|+\int\mathrm{PS}(\vec{\ell}_{2},\vec{\ell}_{3})\big\langle\mathbf{q}(\vec{p}_{1})\mathbf{q}(\vec{\ell}_{2})\bar{\mathbf{q}}(\vec{\ell}_{3})\big|\\
 & \times\big\langle\mathbf{q}(\vec{p}_{1})\boldsymbol{\gamma}^{*}(\vec{q})\big|\exp\Bigl(-i\int_{0}^{+\infty}\mathrm{d}x^{+}\hat{\mathcal{H}}(x^{+})\Bigr)\big|\mathbf{q}(\vec{p}_{1})\mathbf{q}(\vec{\ell}_{2})\bar{\mathbf{q}}(\vec{\ell}_{3})\big\rangle+(...)\;,\\
 & =\big\langle\mathbf{q}(\vec{p}_{1})\boldsymbol{\gamma}^{*}(\vec{q})\big|+\int\mathrm{PS}(\vec{k}_{2},\vec{\ell}_{2},\vec{\ell}_{3})\big\langle\mathbf{q}(\vec{\ell}_{2})\mathbf{g}(\vec{k}_{2})\big|\\
 & \times\big\langle\mathbf{q}(\vec{p}_{1})\boldsymbol{\gamma}^{*}(\vec{q})\big|\exp\Bigl(-i\int_{0}^{+\infty}\mathrm{d}x^{+}\hat{\mathcal{H}}(x^{+})\Bigr)\big|\mathbf{q}(\vec{p}_{1})\mathbf{q}(\vec{\ell}_{2})\bar{\mathbf{q}}(\vec{\ell}_{3})\big\rangle\\
 & \times\big\langle\mathbf{q}(\vec{p}_{1})\mathbf{q}(\vec{\ell}_{2})\bar{\mathbf{q}}(\vec{\ell}_{3})\big|\exp\Bigl(-i\int_{0}^{x^{+}}\mathrm{d}y^{+}\hat{\mathcal{H}}(y^{+})\Bigr)\big|\mathbf{q}(\vec{\ell}_{2})\mathbf{g}(\vec{k}_{2})\big\rangle+(...)\;,\\
 & =\big\langle\mathbf{q}(\vec{p}_{1})\boldsymbol{\gamma}^{*}(\vec{q})\big|+\int\mathrm{PS}(\vec{k}_{2},\vec{\ell}_{2},\vec{\ell}_{3})\big\langle\mathbf{q}(\vec{\ell}_{2})\mathbf{g}(\vec{k}_{2})\big|\\
 & \times\frac{\big\langle\mathbf{q}(\vec{p}_{1})\boldsymbol{\gamma}^{*}(\vec{q})\big|\hat{V}\big|\mathbf{q}(\vec{p}_{1})\mathbf{q}(\vec{\ell}_{2})\bar{\mathbf{q}}(\vec{\ell}_{3})\big\rangle}{q^{-}-\ell_{2}^{-}-\ell_{3}^{-}+i0^{+}}\frac{\big\langle\mathbf{q}(\vec{p}_{1})\mathbf{q}(\vec{\ell}_{2})\bar{\mathbf{q}}(\vec{\ell}_{3})\big|\hat{V}\big|\mathbf{q}(\vec{\ell}_{2})\mathbf{g}(\vec{k}_{2})\big\rangle}{p_{1}^{+}+q^{-}-\ell_{2}^{-}-k_{2}^{-}+i0^{+}}+(...)\;,
\end{aligned}
\label{eq:backwardsevo}
\end{equation}
where (...) stands for all other possible diagrams, generated by inserting
different intermediate Fock states.
Combining \eqref{eq:Mdefapx} with \eqref{eq:forwardsevo} and \eqref{eq:backwardsevo},
we obtain:
\begin{equation}
\begin{aligned} & \,_{f}\big\langle\mathbf{q}(\vec{p}_{1})\boldsymbol{\gamma}^{*}(\vec{q})\big|\hat{F}-1\big|\mathbf{q}(\vec{p}_{0})\big\rangle_{i}=\big\langle\mathbf{q}(\vec{p}_{1})\boldsymbol{\gamma}^{*}(\vec{q})\big|\hat{F}-1\big|\mathbf{q}(\vec{p}_{0})\big\rangle\\
 & +\int\mathrm{PS}(\vec{k}_{1},\vec{k}_{2},\vec{\ell}_{1},\vec{\ell}_{2},\vec{\ell}_{3})\frac{\big\langle\mathbf{q}(\vec{p}_{1})\boldsymbol{\gamma}^{*}(\vec{q})\big|\hat{V}\big|\mathbf{q}(\vec{p}_{1})\mathbf{q}(\vec{\ell}_{2})\bar{\mathbf{q}}(\vec{\ell}_{3})\big\rangle}{q^{-}-\ell_{2}^{-}-\ell_{3}^{-}+i0^{+}}\\
 & \times\frac{\big\langle\mathbf{q}(\vec{p}_{1})\mathbf{q}(\vec{\ell}_{2})\bar{\mathbf{q}}(\vec{\ell}_{3})\big|\hat{V}\big|\mathbf{q}(\vec{\ell}_{2})\mathbf{g}(\vec{k}_{2})\big\rangle}{p_{1}^{+}+q^{-}-\ell_{2}^{-}-k_{2}^{-}+i0^{+}}\frac{\big\langle\mathbf{q}(\vec{\ell}_{1})\mathbf{g}(\vec{k}_{1})\big|\hat{V}\big|\mathbf{q}(\vec{p}_{0})\big\rangle}{p_{0}^{-}-\ell_{1}^{-}-k^{-}+i0^{+}}\\
 & \times\big\langle\mathbf{q}(\vec{\ell}_{2})\mathbf{g}(\vec{k}_{2})\big|\hat{F}-1\big|\mathbf{q}(\vec{\ell}_{1})\mathbf{g}(\vec{k}_{1})\big\rangle+(...)\;.
\end{aligned}
\label{eq:Fockint}
\end{equation}
Neglecting the leading-order part,
there are three different numerators to evaluate. The interaction
terms of the light-cone Hamiltonian for QCD and QED of interest are:
\begin{equation}
\begin{aligned}\hat{V} & =\int\mathrm{d}^{D-3}\vec{x}:g_{\mathrm{em}}\bar{\psi}(\vec{x})\cancel{A}(\vec{x})\psi(\vec{x})+g_{s}t^{c}\bar{\psi}(\vec{x})\cancel{A}^{c}(\vec{x})\psi(\vec{x}):\;.\end{aligned}
\end{equation}
We then obtain
\begin{equation}
\begin{aligned}\big\langle\mathbf{q}(\vec{\ell}_{1})\mathbf{g}(\vec{k}_{1})\big|\hat{V}\big|\mathbf{q}(\vec{p}_{0})\big\rangle & =\big\langle0\big|b_{\vec{\ell}_{1}}\,a_{\vec{k}_{1}}\,\hat{V}\,b_{\vec{p}_{0}}^{\dagger}\big|0\big\rangle\;,\\
 & =g_{s}t^{c}\int\mathrm{d}^{D-3}\vec{x}\big\langle0\big|b_{\vec{\ell}_{1}}a_{\vec{k}_{1}}:\bar{\psi}(\vec{x})\cancel{A}(\vec{x})\psi(\vec{x}):b_{\vec{p}_{0}}^{\dagger}\big|0\big\rangle\;,\\
 & =(2\pi)^{D-1}\delta^{(D-1)}\big(\vec{p}_{0}-\vec{\ell}_{1}-\vec{k}_{1}\big)g_{s}t^{c}\bar{u}(\vec{\ell}_{1})\cancel{\epsilon}^{*}(\vec{k}_{2})u(\vec{p}_{0})\;.
\end{aligned}
\label{eq:Vaction1}
\end{equation}
Likewise, the other numerators give:
\begin{equation}
\begin{aligned}\big\langle\mathbf{q}(\vec{p}_{1})\mathbf{q}(\vec{\ell}_{2})\bar{\mathbf{q}}(\vec{\ell}_{3})\big|\hat{V}\big|\mathbf{q}(\vec{\ell}_{2})\mathbf{g}(\vec{k}_{2})\big\rangle & \!=\!(2\pi)^{D\!-\!1}\delta^{(D\!-\!1)}\big(\vec{p}_{1}\!+\!\vec{\ell}_{3}\!-\!\vec{k}_{2}\big)g_{s}t^{c}\bar{u}(\vec{p}_{1})\cancel{\epsilon}(\vec{k}_{2})v(\vec{\ell}_{3})\;,\\
\big\langle\mathbf{q}(\vec{p}_{1})\boldsymbol{\gamma}^{*}(\vec{q})\big|\hat{V}\big|\mathbf{q}(\vec{p}_{1})\mathbf{q}(\vec{\ell}_{2})\bar{\mathbf{q}}(\vec{\ell}_{3})\big\rangle & \!=\!-(2\pi)^{D\!-\!1}\delta^{(D\!-\!1)}\big(\vec{\ell}_{2}\!+\!\vec{\ell}_{3}\!-\!\vec{q}\big)g_{\mathrm{em}}\bar{v}(\vec{\ell}_{3})\cancel{\epsilon}^{*}(\vec{q})u(\vec{p}_{1})\;.
\end{aligned}
\label{eq:Vaction2}
\end{equation}
The minus sign in the last line comes from the anticommutation of
the fermion field operators. Note that one should be careful to make
sure that $\big\langle\mathbf{q}(\vec{p}_{1})\mathbf{q}(\vec{\ell}_{2})\bar{\mathbf{q}}(\vec{\ell}_{3})\big|$
and $\big|\mathbf{q}(\vec{p}_{1})\mathbf{q}(\vec{\ell}_{2})\bar{\mathbf{q}}(\vec{\ell}_{3})\big\rangle$
are indeed each other's conjugate.\footnote{I am grateful to Guillaume Beuf for pointing me to this possible source
of error, which allowed me to identify and correct a sign mistake.}

We will shortly discuss the evaluation of the Dirac algebra, but let
us first turn to the interaction of the projectile with the shockwave,
encoded in the last line of~\eqref{eq:Fockint}. With the help of
the (anti-)commutation relations~\eqref{eq:anticommutation} we derive the normalization of the Fock states:
\begin{equation}
\begin{aligned}\big\langle\mathbf{q}(\vec{\ell}_{2})\mathbf{g}(\vec{k}_{2})\big|\mathbf{q}(\vec{\ell}_{1})\mathbf{g}(\vec{k}_{1})\big\rangle & \!=\!2k_{1}^{+}(2\pi)^{D\!-\!1}\delta^{(D\!-\!1)}(\vec{k}_{1}\!-\!\vec{k}_{2})2\ell_{1}^{+}(2\pi)^{D\!-\!1}\delta^{(D\!-\!1)}(\vec{\ell}_{1}\!-\!\vec{\ell}_{2})\;.\end{aligned}
\label{eq:trivialinteraction}
\end{equation}
Moreover, the CGC dictates that, in the eikonal approximation, the
shockwave is built from Wilson lines, and, therefore, the action of
the external potential becomes:
\begin{equation}
\begin{aligned}\big\langle\mathbf{q}(\vec{\ell}_{2})\mathbf{g}(\vec{k}_{2})\big|\hat{F}\big|\mathbf{q}(\vec{\ell}_{1})\mathbf{g}(\vec{k}_{1})\big\rangle & =2k_{1}^{+}2\pi\delta(k_{1}^{+}-k_{2}^{+})2\ell_{1}^{+}2\pi\delta(\ell_{1}^{+}-\ell_{2}^{+})\\
 & \times\int_{\mathbf{x},\mathbf{z}}e^{-i\mathbf{x}\cdot(\boldsymbol{\ell}_{2}-\boldsymbol{\ell}_{1})}e^{-i\mathbf{z}\cdot(\mathbf{k}_{2}-\mathbf{k}_{1})}U_{\mathbf{x}}W_{\mathbf{z}}^{c}\;,
\end{aligned}
\label{eq:CGCpotential}
\end{equation}
where $U$ and $W$ are Wilson lines and the fundamental and adjoint
representation, respectively. Filling in the implicit color factors and using the Fierz identity:
\begin{equation}
t^{d}W^{dc}=U^{\dagger}t^{c}U\;,\label{eq:Fierz}
\end{equation}
equations~\eqref{eq:trivialinteraction} and (\ref{eq:CGCpotential}) can be conveniently combined into:
\begin{equation}
\begin{aligned}\big\langle\mathbf{q}(\vec{\ell}_{2})\mathbf{g}(\vec{k}_{2})\big|\hat{F}-1\big|\mathbf{q}(\vec{\ell}_{1})\mathbf{g}(\vec{k}_{1})\big\rangle & =2k_{1}^{+}2\pi\delta(k_{1}^{+}-k_{2}^{+})2\ell_{1}^{+}2\pi\delta(\ell_{1}^{+}-\ell_{2}^{+})\\
 & \times\int_{\mathbf{x},\mathbf{z}}e^{-i\mathbf{x}\cdot(\boldsymbol{\ell}_{2}-\boldsymbol{\ell}_{1})}e^{-i\mathbf{z}\cdot(\mathbf{k}_{2}-\mathbf{k}_{1})}\big(t^{c}U_{\mathbf{x}}U_{\mathbf{z}}^\dagger t^{c}U_{\mathbf{z}}-C_{F}\big)\;,
\end{aligned}
\end{equation}
We are now in a position to extract from definition~\eqref{eq:Mdefapx}
the expression for the amplitude $\mathcal{M}_{\mathrm{A2}}$, by
plugging the above equation together with (\ref{eq:Vaction1}) and
(\ref{eq:Vaction2}) into (\ref{eq:Fockint}), and using the delta
functions to eliminate as many intermediate momenta as possible 
\begin{equation}
\begin{aligned}\mathcal{M}_{\mathrm{A2}} & =-g_{\mathrm{em}}g_{s}^{2}\int_{p_{1}^{+}}^{p_{0}^{+}}\frac{\mathrm{d}k^{+}}{2\pi2k^{+}}\frac{1}{2\ell_{2}^{+}2\ell_{3}^{+}}\int_{\mathbf{k}_{1},\mathbf{k}_{2}}\frac{\bar{u}(\vec{p}_{1})\cancel{\epsilon}(\vec{k}_{2})v(\vec{\ell}_{3})}{q^{-}-\ell_{2}^{-}-\ell_{3}^{-}+i0^{+}}\\
 & \times\frac{\bar{v}(\vec{\ell}_{3})\cancel{\epsilon}^{*}(\vec{q})u(\vec{p}_{1})}{p_{1}^{+}+q^{-}-\ell_{2}^{-}-k_{2}^{-}+i0^{+}}\frac{\bar{u}(\vec{\ell}_{1})\cancel{\epsilon}^{*}(\vec{k}_{2})u(\vec{p}_{0})}{p_{0}^{-}-\ell_{1}^{-}-k^{-}+i0^{+}}\\
 & \times\int_{\mathbf{x},\mathbf{z}}e^{-i\mathbf{x}\cdot(\boldsymbol{\ell}_{2}-\boldsymbol{\ell}_{1})}e^{-i\mathbf{z}\cdot(\mathbf{k}_{2}-\mathbf{k}_{1})}\big(t^{c}U_{\mathbf{x}}U_{\mathbf{z}}^\dagger t^{c}U_{\mathbf{z}}-C_{F}\big)\;,
\end{aligned}
\label{eq:MA2_ex_0}
\end{equation}
where it is understood that
\begin{equation}
\begin{aligned}\vec{\ell}_{3} & =\vec{k}_{2}-\vec{p}_{1}\;,\qquad\vec{\ell}_{2}  =\vec{q}+\vec{p}_{1}-\vec{k}_{2}\;,\\
\vec{\ell}_{1}  &=\vec{p}_{0}-\vec{k}_{1}\;,\qquad k^{+} =k_{1}^{+}=k_{2}^{+}\;.
\end{aligned}
\label{eq:A2momenta}
\end{equation}
Note that the Heaviside theta functions hidden in the phase-space
integrations of~\eqref{eq:Fockint} result in upper and lower limits
for the integration over the gluon plus momentum.

The above expression can be simplified further by exploiting the properties
of the projectors $\mathcal{P}_{G,B}$, such as $u_{G}=\mathcal{P}_{G}u_{G}$,
$\bar{u}_{G}\mathcal{P}_{B}=0$, $\gamma^{-}\mathcal{P}_{G}=\mathcal{P}_{G}\gamma^{+}=0$,
and the fact that they commute with the transverse gamma matrices:
$[\mathcal{P}_{G,B},\boldsymbol{\gamma}^{i}]=0$. A generic spinor product then further simplifies into:
\begin{equation}
\begin{aligned}&\bar{u}(\vec{k}_{1})\cancel{\epsilon}(\vec{k}_{3})u(\vec{k}_{2})  \!=\!\bar{u}_{G}(k_{1}^{+})\Big[\gamma^{+}\epsilon^{-}(\vec{k}_{3})-\mathbf{k}_{1}\cdot\boldsymbol{\gamma}\frac{\gamma^{+}}{2k_{1}^{+}}\boldsymbol{\gamma}\cdot\boldsymbol{\epsilon}(\vec{k}_{3})-\boldsymbol{\gamma}\cdot\boldsymbol{\epsilon}(\vec{k}_{3})\frac{\gamma^{+}}{2k_{2}^{+}}\mathbf{k}_{2}\cdot\boldsymbol{\gamma}\\
 & \!+\!\mathbf{k}_{1}\cdot\boldsymbol{\gamma}\frac{\gamma^{+}}{2k_{1}^{+}}\gamma^{-}\frac{\gamma^{+}}{2k_{2}^{+}}\mathbf{k}_{2}\cdot\boldsymbol{\gamma}\epsilon^{+}(\vec{k}_{3})\Big]u_{G}(k_{2}^{+})\;,\\
 & \!=\!\bar{u}_{G}(k_{1}^{+})\gamma^{+}\Big[\!\epsilon^{-}(\vec{k}_{3})\!+\!\Big(\!\frac{\mathbf{k}_{1}^{i}\boldsymbol{\epsilon}^{j}(\vec{k}_{3})}{2k_{1}^{+}}\!+\!\frac{\boldsymbol{\epsilon}^{i}(\vec{k}_{3})\mathbf{k}_{2}^{j}}{2k_{2}^{+}}\!-\!\epsilon^{+}(\vec{k}_{3})\frac{\mathbf{k}_{1}^{i}\mathbf{k}_{2}^{j}}{2k_{1}^{+}k_{2}^{+}}\!\Big)\mathcal{\bgamma}^{i}\bgamma^{j}\!\Big]u_{G}(k_{2}^{+})\;.
\end{aligned}
\end{equation}
Finally, introducing $\sigma^{ij}=(i/2)[\bgamma^{i},\bgamma^{j}]$ which
allows us to write $\mathcal{\bgamma}^{i}\bgamma^{j}=-\delta^{ij}-i\sigma^{ij}$,
we obtain the expression:
\begin{equation}
\begin{aligned}\bar{u}(\vec{k}_{1})\cancel{\epsilon}(\vec{k}_{3})u(\vec{k}_{2}) & =\bar{u}_{G}(k_{1}^{+})\gamma^{+}\Bigg[\Big(\epsilon^{-}(\vec{k}_{3})-\frac{\mathbf{k}_{1}^{i}\boldsymbol{\epsilon}^{j}(\vec{k}_{3})}{2k_{1}^{+}}-\frac{\boldsymbol{\epsilon}^{i}(\vec{k}_{3})\mathbf{k}_{2}^{j}}{2k_{2}^{+}}+\epsilon^{+}(\vec{k}_{3})\frac{\mathbf{k}_{1}^{i}\mathbf{k}_{2}^{j}}{2k_{1}^{+}k_{2}^{+}}\Big)\delta^{ij}\\
 & -i\sigma^{ij}\Big(\frac{\mathbf{k}_{1}^{i}\boldsymbol{\epsilon}^{j}(\vec{k}_{3})}{2k_{1}^{+}}+\frac{\boldsymbol{\epsilon}^{i}(\vec{k}_{3})\mathbf{k}_{2}^{j}}{2k_{2}^{+}}-\epsilon^{+}(\vec{k}_{3})\frac{\mathbf{k}_{1}^{i}\mathbf{k}_{2}^{j}}{2k_{1}^{+}k_{2}^{+}}\Big)\Bigg]u_{G}(k_{2}^{+})\;.
\end{aligned}
\label{eq:ueu_generic}
\end{equation}
With a suitable choice of polarization vectors, and taking the relation
between the different momenta into account, the above formula gives
rise to remarkably compact expressions. For instance, returning to
amplitude (\ref{eq:MA2_ex_0}), we have the numerator:
\begin{equation}
\bar{u}(\vec{\ell}_{1}=\vec{p}_{0}-\vec{k}_{2})\cancel{\epsilon}_{\eta}^{*}(\vec{k}_{2})u(\vec{p}_{0})\;,\label{eq:spinor_ex}
\end{equation}
where $\eta=1,2$ is the (transverse) polarization of the gluon. Choosing
the polarization vectors for the gluon to be linearly polarized: $\boldsymbol{\epsilon}_{\eta}^{i}=\delta^{i\eta}$,
we have:
\begin{equation}
\begin{aligned}\epsilon_{\eta}^{\mu}(\vec{k}) & =\big(0,\frac{\mathbf{k}\cdot\boldsymbol{\epsilon}_{\eta}}{k^{+}},\boldsymbol{\epsilon}_{\eta}\big)=\big(0,\frac{\mathbf{k}^{\eta}}{k^{+}},\delta^{i\eta}\big)\;,\end{aligned}
\end{equation}
applying (\ref{eq:ueu_generic}) to the spinor product (\ref{eq:spinor_ex})
gives:
\begin{equation}
\begin{aligned} & \bar{u}(\vec{\ell}_{1}\!=\!\vec{p}_{0}-\vec{k}_{1})\cancel{\epsilon}_{\eta}^{*}(\vec{k}_{1})u(\vec{p}_{0})\\
 & \!=\!\bar{u}_{G}(\ell_{1}^{+})\gamma^{+}\Big[\!\Big(\frac{\mathbf{k}_{1}^{\eta^{\prime}}}{k^{+}}\!-\!\frac{(\mathbf{p}_{0}-\mathbf{k}_{1})^{\eta^{\prime}}}{2(p_{0}^{+}\!-\!k^{+})}\!-\!\frac{\mathbf{p}_{0}^{\eta^{\prime}}}{2p_{0}^{+}}\Big)\delta^{\eta\eta^{\prime}}\!-\!i\sigma^{\eta\eta^{\prime}}\Big(\frac{\mathbf{p}_{0}^{\eta^{\prime}}}{2p_{0}^{+}}\!-\!\frac{(\mathbf{p}_{0}\!-\!\mathbf{k}_{1})^{\eta^{\prime}}}{2(p_{0}^{+}-k^{+})}\Big)\!\Big]u_{G}(p_{0}^{+})\;,\\
 & \!=\!\frac{p_{0}^{+}\mathbf{k}_{1}^{\eta^{\prime}}-k^{+}\mathbf{p}_{0}^{\eta^{\prime}}}{2p_{0}^{+}(p_{0}^{+}-k^{+})}\bar{u}_{G}(\ell_{1}^{+})\gamma^{+}\Big[\big(2\frac{p_{0}^{+}}{k^{+}}-1\big)\delta^{\eta\eta^{\prime}}-i\sigma^{\eta\eta^{\prime}}\Big]u_{G}(p_{0}^{+})\;,\\
 & \!=\!\frac{p_{0}^{+}\mathbf{k}_{1}^{\eta^{\prime}}-k^{+}\mathbf{p}_{0}^{\eta^{\prime}}}{2p_{0}^{+}(p_{0}^{+}-k^{+})}\bar{u}_{G}(\ell_{1}^{+})\gamma^{+}\mathrm{Dirac}^{\eta\eta^{\prime}}\big(2\frac{p_{0}^{+}}{k^{+}}-1\big)u_{G}(p_{0}^{+})\;.
\end{aligned}
\label{eq:A2spinor1}
\end{equation}
Likewise:
\begin{equation}
\begin{aligned}\bar{u}(\vec{p}_{1})\cancel{\epsilon}_{\eta}(\vec{k}_{2})v(\vec{\ell}_{3}) & =-\frac{\boldsymbol{\ell}^{\bar{\eta}}-\frac{k^{+}}{p_{1}^{+}}\mathbf{P}_{\perp}^{\bar{\eta}}}{2(p_{1}^{+}-k^{+})}\bar{u}_{G}(p_{1}^{+})\gamma^{+}\mathrm{Dirac}^{\eta\bar{\eta}}\Big(1-\frac{2p_{1}^{+}}{k^{+}}\Big)v_{G}(\ell_{3}^{+})\;,\end{aligned}
\label{eq:A2spinor2}
\end{equation}
where, for later convenience, we introduced the momentum combination:
\begin{equation}
\begin{aligned}\boldsymbol{\ell} & \equiv\mathbf{k}_{2}-\frac{k^{+}}{p_{0}^{+}}\mathbf{k}_{\perp}\;.\end{aligned}
\label{eq:A2ldef}
\end{equation}
The last spinor product in~\eqref{eq:MA2_ex_0} involves the external virtual photon. When the latter is transversely polarized, we obtain a similar expression to the ones with a gluon:
\begin{equation}
\begin{aligned}\bar{v}_{G}(\vec{\ell}_{3})\cancel{\epsilon}_{\lambda}^{*}(\vec{q})u_{G}(\vec{\ell}_{2}) & =\frac{q^{+}\Big(\boldsymbol{\ell}^{\bar{\lambda}}-\frac{p_{0}^{+}-k^{+}}{q^{+}}\mathbf{P}_{\perp}^{\bar{\lambda}}\Big)}{2(p_{1}^{+}-k^{+})(p_{0}^{+}-k^{+})}\\
&\times\bar{v}_{G}(\ell_{3}^{+})\gamma^{+}\mathrm{Dirac}^{\lambda\bar{\lambda}}\Big(1+\frac{2(p_{1}^{+}-k^{+})}{q^{+}}\Big)u_{G}(\ell_{2}^{+})\;.\end{aligned}
\label{eq:A2spinorT}
\end{equation}
However, in the longitudinal case, the spinor product looks quite
different:
\begin{equation}
\begin{aligned}\bar{v}_{G}(\vec{\ell}_{3})\cancel{\epsilon}_{0}^{*}(\vec{q})u_{G}(\vec{\ell}_{2}) & \!=\!\frac{q^{+}}{2(p_{1}^{+}\!-\!k^{+})(p_{0}^{+}\!-\!k^{+})M}\\
&\times\Big[\!\Big(\boldsymbol{\ell}\!-\!\frac{p_{0}^{+}-k^{+}}{q^{+}}\mathbf{P}_{\perp}\!\Big)^{2}\!-\!\hat{M}^{2}\!\Big]\bar{v}_{G}(\ell_{3}^{+})\gamma^{+}u_{G}(\ell_{2}^{+})\;,\end{aligned}
\label{eq:A2spinorL}
\end{equation}
where $\hat{M}^{2}$ as defined in~\eqref{eq:defM}.

Finally, it is convenient to cast the energy denominators in a form dictated
by the momentum structure of the spinor products~\eqref{eq:A2spinor1},
(\ref{eq:A2spinor2}), and (\ref{eq:A2spinorT}), yielding after some
algebra:
\begin{equation}
\begin{aligned}\frac{1}{p_{0}^{-}-k_{1}^{-}-\ell_{1}^{-}} & =\frac{-2k^{+}(p_{0}^{+}-k^{+})}{p_{0}^{+}\mathbf{k}_{1}^{2}}\;,\\
\frac{1}{q^{-}-\ell_{2}^{-}-\ell_{3}^{-}} & =\frac{2(p_{1}^{+}-k^{+})(p_{0}^{+}-k^{+})}{q^{+}}\frac{1}{\Big(\boldsymbol{\ell}-\frac{p_{0}^{+}-k^{+}}{q^{+}}\mathbf{P}_{\perp}\Big)^{2}+\hat{M}^{2}}\;,\\
\frac{1}{q^{-}+p_{1}^{-}-\ell_{2}^{-}-k_{2}^{-}} & =\frac{-2k^{+}(p_{0}^{+}-k^{+})}{p_{0}^{+}}\frac{1}{\boldsymbol{\ell}^{2}+\Delta_{\mathrm{P}}}\;,
\end{aligned}
\label{eq:A2denominators}
\end{equation}
where $\Delta_{\mathrm{P}}$ was defined in equation~\eqref{eq:DeltaP}.

Gathering our expressions (\ref{eq:A2spinor1}), (\ref{eq:A2spinor2}),
(\ref{eq:A2spinorT}), and (\ref{eq:A2spinorL}) for the numerators,
together with the calculation (\ref{eq:A2denominators}) of the denominators,
using (\ref{eq:A2momenta}) and (\ref{eq:A2ldef}) to rewrite the
phases, we finally obtain:
\begin{equation}
\begin{aligned}\mathcal{M}_{\mathrm{A2}}^{0} & =\frac{1}{M}\int_{p_{1}^{+}}^{p_{0}^{+}}\frac{\mathrm{d}k^{+}}{2\pi2k^{+}}\frac{(k^{+})^{2}(p_{0}^{+}-k^{+})}{(p_{0}^{+})^{2}(p_{1}^{+}-k^{+})}\mathrm{Dirac}_{\mathrm{V}}^{\bar{\eta}0\eta^{\prime}}\\
 & \times\int_{\mathbf{x},\mathbf{z}}\int_{\mathbf{k}_{1}}\frac{\mathbf{k}_{1}^{\eta^{\prime}}}{\mathbf{k}_{1}^{2}}e^{-i\mathbf{k}_{1}\cdot(\mathbf{x}-\mathbf{z})}\int_{\boldsymbol{\ell}}e^{-i\boldsymbol{\ell}\cdot(\mathbf{x}-\mathbf{z})}\frac{\boldsymbol{\ell}^{\bar{\eta}}+\frac{k^{+}}{p_{1}^{+}}\mathbf{P}_{\perp}^{\bar{\eta}}}{\boldsymbol{\ell}^{2}+\Delta_{\mathrm{P}}}\frac{\Big(\boldsymbol{\ell}+\frac{p_{0}^{+}-k^{+}}{q^{+}}\mathbf{P}_{\perp}\Big)^{2}-\hat{M}^{2}}{\Big(\boldsymbol{\ell}+\frac{p_{0}^{+}-k^{+}}{q^{+}}\mathbf{P}_{\perp}\Big)^{2}+\hat{M}^{2}}\\
 & \times e^{-i\mathbf{k}_{\perp}\cdot\Big(\frac{p_{0}^{+}-k^{+}}{p_{0}^{+}}\mathbf{x}+\frac{k^{+}}{p_{0}^{+}}\mathbf{z}\Big)}\big(t^{c}U_{\mathbf{x}}U_{\mathbf{z}}^{\dagger}t^{c}U_{\mathbf{z}}-C_{F}\big)\;,
\end{aligned}
\label{eq:MA2L_0-1}
\end{equation}
for the longitudinal polarization, and 
\begin{equation}
\begin{aligned}\mathcal{M}_{\mathrm{A2}}^{\lambda} & =-\int_{p_{1}^{+}}^{p_{0}^{+}}\frac{\mathrm{d}k^{+}}{2\pi2k^{+}}\frac{(k^{+})^{2}(p_{0}^{+}-k^{+})}{(p_{0}^{+})^{2}(p_{1}^{+}-k^{+})}\mathrm{Dirac}_{\mathrm{V}}^{\bar{\eta}\bar{\lambda}\eta^{\prime}}\\
 & \times\int_{\mathbf{x},\mathbf{z}}\int_{\mathbf{k}_{1}}\frac{\mathbf{k}_{1}^{\eta^{\prime}}}{\mathbf{k}_{1}^{2}}e^{-i\mathbf{k}_{1}\cdot(\mathbf{x}-\mathbf{z})}\int_{\boldsymbol{\ell}}e^{-i\boldsymbol{\ell}\cdot(\mathbf{x}-\mathbf{z})}\frac{\boldsymbol{\ell}^{\bar{\eta}}+\frac{k^{+}}{p_{1}^{+}}\mathbf{P}_{\perp}^{\bar{\eta}}}{\boldsymbol{\ell}^{2}+\Delta_{\mathrm{P}}}\frac{\Big(\boldsymbol{\ell}^{\bar{\lambda}}+\frac{p_{0}^{+}-k^{+}}{q^{+}}\mathbf{P}_{\perp}^{\bar{\lambda}}\Big)}{\Big(\boldsymbol{\ell}+\frac{p_{0}^{+}-k^{+}}{q^{+}}\mathbf{P}_{\perp}\Big)^{2}+\hat{M}^{2}}\\
 & \times e^{-i\mathbf{k}_{\perp}\cdot\Big(\frac{p_{0}^{+}-k^{+}}{p_{0}^{+}}\mathbf{x}+\frac{k^{+}}{p_{0}^{+}}\mathbf{z}\Big)}\big(t^{c}U_{\mathbf{x}}U_{\mathbf{z}}^{\dagger}t^{c}U_{\mathbf{z}}-C_{F}\big)\;,
\end{aligned}
\label{eq:MA2_0-1}
\end{equation}
in the transverse case.

\section{\label{sec:MV2_UV}Ultraviolet behavior of amplitude $\mathcal{M}_{\mathrm{V2}}$}

To investigate whether the transverse integral in (\ref{eq:MV2})
exhibits an UV divergence in the $\mathbf{z}\to\mathbf{x}$ limit,
we set $\mathbf{z}=\mathbf{x}$ in the last line, after which $\mathbf{z}$
can be integrated over in the second line. We obtain:
\begin{equation}
\begin{aligned}\lim_{\mathbf{z}\to\mathbf{x}}\tilde{\mathcal{M}}_{\mathrm{V2}}^{\lambda} & =\alpha_{s}C_{F}\int_{k_{\mathrm{min}}^{+}}^{p_{1}^{+}}\frac{\mathrm{d}k^{+}}{k^{+}}\frac{(k^{+})^{3}q^{+}}{p_{1}^{+}(p_{0}^{+})^{2}(p_{1}^{+}-k^{+})}\mathrm{Dirac}_{\mathrm{V}}^{\bar{\eta}\bar{\lambda}\eta^{\prime}}\\
 & \times\int_{\boldsymbol{\ell}}\frac{\boldsymbol{\ell}^{\eta^{\prime}}}{\boldsymbol{\ell}^{2}}\frac{\boldsymbol{\ell}^{\bar{\eta}}-\frac{k^{+}}{p_{1}^{+}}\mathbf{P}_{\perp}^{\bar{\eta}}}{\big(\boldsymbol{\ell}-\frac{k^{+}}{p_{1}^{+}}\mathbf{P}_{\perp}\big)^{2}}\frac{\boldsymbol{\ell}^{\bar{\lambda}}-\frac{p_{0}^{+}-k^{+}}{q^{+}}\mathbf{P}_{\perp}^{\bar{\lambda}}}{\boldsymbol{\ell}^{2}+\Delta_{\mathrm{P}}}\\
 & \times\int_{\mathbf{x}}e^{-i\mathbf{k}_{\perp}\cdot\mathbf{x}}[U_{\mathbf{x}}-1]\;.
\end{aligned}
\label{eq:MV2_limit}
\end{equation}
The integral over $\boldsymbol{\ell}$, which we denote $\mathcal{I}_{\mathrm{V2}}^{\eta^{\prime}\bar{\eta}\lambda}$,
contains three open transverse indices. However, when multiplying
(\ref{eq:MV2L_limit}) with $\mathcal{M}_{\mathrm{LO}}^{\lambda^{\prime}\dagger}$
to obtain the cross section, the trace of the Dirac structures $\mathrm{Dirac}_{\mathrm{LO}}^{\lambda^{\prime}\dagger}\mathrm{Dirac}_{\mathrm{V}}^{\bar{\eta}\bar{\lambda}\eta^{\prime}}$will
yield only the Lorentz structures $\delta^{\bar{\eta}\eta^{\prime}}$
and $\epsilon^{\bar{\eta}\eta^{\prime}}$. We will use this knowledge
and only calculate the projections $\delta^{\bar{\eta}\eta^{\prime}}\mathcal{I}_{\mathrm{V2}}^{\eta^{\prime}\bar{\eta}\lambda}$
and $\epsilon^{\bar{\eta}\eta^{\prime}}\mathcal{I}_{\mathrm{V2}}^{\eta^{\prime}\bar{\eta}\lambda}$,
dramatically simplifying the evaluation. Using the identities (\ref{eq:IntegralIdentitiesB})
and (\ref{eq:IntegralIdentitiesC}), one finds:
\begin{equation}
\begin{aligned}\delta^{\bar{\eta}\eta^{\prime}}\mathcal{I}_{\mathrm{V2}}^{\eta^{\prime}\bar{\eta}\lambda} & =\int_{\boldsymbol{\ell}}\frac{1}{\boldsymbol{\ell}^{2}}\frac{\boldsymbol{\ell}\cdot\big(\boldsymbol{\ell}-\frac{k^{+}}{p_{1}^{+}}\mathbf{P}_{\perp}\big)}{\big(\boldsymbol{\ell}-\frac{k^{+}}{p_{1}^{+}}\mathbf{P}_{\perp}\big)^{2}}\frac{\boldsymbol{\ell}^{\bar{\lambda}}-\frac{p_{0}^{+}-k^{+}}{q^{+}}\mathbf{P}_{\perp}^{\bar{\lambda}}}{\boldsymbol{\ell}^{2}+\Delta_{\mathrm{P}}}\;,\\
 & =\frac{k^{+}}{2p_{1}^{+}}\mathbf{P}_{\perp}^{\bar{\lambda}}\mathcal{B}_{1}\big(0,\Delta_{\mathrm{P}},\frac{k^{+}}{p_{1}^{+}}\mathbf{P}_{\perp}\big)\\
 & -\frac{1}{2}\Big(\frac{1}{2}+\frac{p_{0}^{+}(p_{1}^{+}-k^{+})}{k^{+}q^{+}}\Big)\frac{k^{+}}{p_{1}^{+}}\mathbf{P}_{\perp}^{\bar{\lambda}}\\
 & \times\Bigg[\mathcal{B}_{0}(\Delta_{\mathrm{P}})+\mathcal{B}_{0}\big(\Delta_{\mathrm{P}},\frac{k^{+}}{p_{1}^{+}}\mathbf{P}_{\perp}\big)-\Big(\frac{k^{+}}{p_{1}^{+}}\Big)^{2}\mathbf{P}_{\perp}^{2}\mathcal{C}_{0}\big(\Delta_{\mathrm{P}},\frac{k^{+}}{p_{1}^{+}}\mathbf{P}_{\perp}\big)\Bigg]\;,
\end{aligned}
\label{eq:IV2_even}
\end{equation}
and
\begin{equation}
\begin{aligned}\epsilon^{\bar{\eta}\eta^{\prime}}\mathcal{I}_{\mathrm{V2}}^{\eta^{\prime}\bar{\eta}\lambda} & =\epsilon^{\bar{\eta}\eta^{\prime}}\int_{\boldsymbol{\ell}}\frac{\boldsymbol{\ell}^{\eta^{\prime}}}{\boldsymbol{\ell}^{2}}\frac{\boldsymbol{\ell}^{\bar{\eta}}-\frac{k^{+}}{p_{1}^{+}}\mathbf{P}_{\perp}^{\bar{\eta}}}{\big(\boldsymbol{\ell}-\frac{k^{+}}{p_{1}^{+}}\mathbf{P}_{\perp}\big)^{2}}\frac{\boldsymbol{\ell}^{\bar{\lambda}}-\frac{p_{0}^{+}-k^{+}}{q^{+}}\mathbf{P}_{\perp}^{\bar{\lambda}}}{\boldsymbol{\ell}^{2}+\Delta_{\mathrm{P}}}\\
 & =\frac{k^{+}}{p_{1}^{+}}\mathbf{P}_{\perp}^{\bar{\eta}}\epsilon^{\bar{\eta}\bar{\lambda}}\frac{1}{2}\frac{1}{D-3}\Bigg[\mathcal{B}_{1}\big(0,\Delta_{\mathrm{P}},\frac{k^{+}}{p_{1}^{+}}\mathbf{P}_{\perp}\big)\\
 & -\frac{1}{2}\mathcal{B}_{0}(\Delta_{\mathrm{P}})-\frac{1}{2}\mathcal{B}_{0}\big(\Delta_{\mathrm{P}},\frac{k^{+}}{p_{1}^{+}}\mathbf{P}_{\perp}\big)+\frac{1}{2}\Big(\frac{k^{+}}{p_{1}^{+}}\Big)^{2}\mathbf{P}_{\perp}^{2}\mathcal{C}_{0}\big(\Delta_{\mathrm{P}},\frac{k^{+}}{p_{1}^{+}}\mathbf{P}_{\perp}\big)\Bigg]\;.
\end{aligned}
\label{eq:IV2_odd}
\end{equation}
Neither of the above results contain an UV divergence. Moreover, the
infrared poles contained in the structures $\mathcal{B}_{0}(\Delta_{\mathrm{P}})$,
$\mathcal{B}_{0}\big(\Delta_{\mathrm{P}},\frac{k^{+}}{p_{1}^{+}}\mathbf{P}_{\perp}\big)$,
and $\mathcal{C}_{0}\big(\Delta_{\mathrm{P}},\frac{k^{+}}{p_{1}^{+}}\mathbf{P}_{\perp}\big)$
all cancel (see eq.(\ref{eq:DivInts})). Expressions (\ref{eq:IV2_even})
and (\ref{eq:IV2_odd}) are, therefore, finite.

\section{\label{subsec:app_Z}Quark field-strength renormalization}

We consider a dressed quark Fock state, and expand it perturbatively
to first order:
\begin{equation}
\begin{aligned}|\mathbf{q}(\vec{p}_{0})\rangle & \!=\!\sqrt{\mathcal{Z}}|\mathbf{q}(\vec{p}_{0})\rangle_{0}+\int\mathrm{PS}(\vec{p}_{1},\vec{k})\frac{_{0}\langle q(\vec{p}_{1}),g(\vec{k})|\hat{V}|q(\vec{p}_{0})\rangle_{0}}{p_{0}^{-}\!-\!k^{-}\!-\!p_{1}^{-}}|\mathbf{q}(\vec{p}_{1}),\mathbf{g}(\vec{k})\rangle_{0}\!+\!\mathcal{O}(g_{s}^{2})\;.\end{aligned}
\label{eq:dressedgluon-1}
\end{equation}
The interaction Hamiltonian sandwiched between the bare Fock states
yields:
\begin{equation}
\begin{aligned}_{0}\langle \mathbf{q}(\vec{p}_{1}),\mathbf{g}(\vec{k})|\hat{V}|\mathbf{q}(\vec{p}_{0})\rangle_{0} & =(2\pi)^{D-1}\delta^{(D-1)}\big(\vec{p}_{0}-\vec{k}-\vec{p}_{1}\big)g_{s}t^{c}\bar{u}(\vec{p}_{1})\cancel{\epsilon}_{\eta}^{*}(\vec{k})u(\vec{p}_{0})\;,\\
 & =(2\pi)^{D-1}\delta^{(D-1)}\big(\vec{p}_{0}-\vec{k}-\vec{p}_{1}\big)g_{s}t^{c}\\
 & \times\frac{\big(\mathbf{k}-\frac{k^{+}}{p_{0}^{+}}\mathbf{p}_{0}\big)^{\bar{\eta}}}{2(p_{0}^{+}-k^{+})}\mathrm{Dirac}^{\eta\bar{\eta}}\Big(2\frac{p_{0}^{+}}{k^{+}}-1\Big)\;,
\end{aligned}
\end{equation}
while the energy denominator gives: 
\begin{equation}
\begin{aligned}\frac{1}{p_{0}^{-}-k^{-}-p_{1}^{-}} & =-\frac{2k^{+}(p_{0}^{+}-k^{+})}{p_{0}^{+}\big(\mathbf{k}-\frac{k^{+}}{p_{0}^{+}}\mathbf{p}_{0}\big)^{2}}\;.\end{aligned}
\end{equation}
Combining the above results, we obtain the following result for the
first-order expansion of the dressed quark state, temporarily indicating all the spin- and color indices
\begin{equation}
\begin{aligned}|\mathbf{q}(\vec{p}_{0},s,i)\rangle & =\sqrt{\mathcal{Z}}|\mathbf{q}(\vec{p}_{0},s,i)\rangle_{0}\\
 & -\sum_{j,\tilde{s},h,c}\int\mathrm{PS}(\vec{p}_{1},\vec{k})(2\pi)^{D-1}\delta^{(D-1)}\big(\vec{p}_{0}-\vec{k}-\vec{p}_{1}\big)\\
 & \times g_{s}t_{ij}^{c}\frac{k^{+}}{p_{0}^{+}}\frac{\big(\mathbf{k}-\frac{k^{+}}{p_{0}^{+}}\mathbf{p}_{0}\big)^{\bar{\eta}}}{\big(\mathbf{k}-\frac{k^{+}}{p_{0}^{+}}\mathbf{p}_{0}\big)^{2}}\mathrm{Dirac}^{\eta\bar{\eta}}\Big(2\frac{p_{0}^{+}}{k^{+}}-1\Big)|\mathbf{q}(\vec{p}_{1},\tilde{s},j),\mathbf{g}(\vec{k},h,c)\rangle_{0}.
\end{aligned}
\label{eq:dressedgluon1-1}
\end{equation}
Per definition, we require the Fock states to be normalized in the
sense of a distribution:
\begin{equation}
\begin{aligned} \langle \mathbf{q}(\vec{p}_{0}^{\prime},s^{\prime},i^{\prime})|\mathbf{q}(\vec{p}_{0},s,i)\rangle & =2p_{0}^{+}\delta_{s^{\prime}s}\delta_{i^{\prime}i}(2\pi)^{D-1}\delta^{(D-1)}\big(\vec{p}_{0}-\vec{p}_{0}^{\prime}\big)\;,\end{aligned}
\end{equation}
or:
\begin{equation}
\frac{1}{2N_{c}}\sum_{s^{\prime}i^{\prime}}\begin{aligned}\int\mathrm{PS}(\vec{p}_{0}^{\prime}){}_{0}\langle \mathbf{q}(\vec{p}_{0}^{\prime},s^{\prime},i^{\prime})|\mathbf{q}(\vec{p}_{0},s,i)\rangle & =1\;.\end{aligned}
\label{eq:ketnorm-1}
\end{equation}
Since the above is supposed to hold to any order of perturbation theory,
we can write:
\begin{equation}
\begin{aligned}1 & =\mathcal{Z}+\frac{1}{2N_{c}}\int\mathrm{PS}(\vec{p}_{0}^{\prime},\vec{p}_{1},\vec{k},\vec{p}_{1}^{\prime},\vec{k}^{\prime})(2\pi)^{D-1}\delta^{(D-1)}\big(\vec{p}_{0}-\vec{k}-\vec{p}_{1}\big)\\
 & \times(2\pi)^{D-1}\delta^{(D-1)}\big(\vec{p}_{0}^{\prime}-\vec{k}^{\prime}-\vec{p}_{1}^{\prime}\big)g_{s}^{2}\mathrm{Tr}(t^{c}t^{c})\frac{k^{+}}{p_{0}^{+}}\frac{k^{\prime+}}{p_{0}^{\prime+}}\\
 & \times\frac{\big(\mathbf{k}-\frac{k^{+}}{p_{0}^{+}}\mathbf{p}_{0}\big)^{\bar{\eta}}}{\big(\mathbf{k}-\frac{k^{+}}{p_{0}^{+}}\mathbf{p}_{0}\big)^{2}}\frac{\big(\mathbf{k}^{\prime}-\frac{k^{\prime+}}{p_{0}^{\prime+}}\mathbf{p}_{0}^{\prime}\big)^{\eta^{\prime}}}{\big(\mathbf{k}^{\prime}-\frac{k^{\prime+}}{p_{0}^{\prime+}}\mathbf{p}_{0}^{\prime}\big)^{2}}\mathrm{Dirac}^{\eta\eta^{\prime}\dagger}\Big(2\frac{p_{0}^{\prime+}}{k^{\prime+}}-1\Big)\mathrm{Dirac}^{\eta\bar{\eta}}\Big(2\frac{p_{0}^{+}}{k^{+}}-1\Big)\\
 & \times{}_{0}\langle \mathbf{q}(\vec{p}_{1}^{\prime}),\mathbf{g}(\vec{k}^{\prime})|\mathbf{q}(\vec{p}_{1}),\mathbf{g}(\vec{k})\rangle_{0}\;,\\
 & =\mathcal{Z}+\alpha_{s}C_{F}\int_{\mathbf{k}}\frac{1}{\mathbf{k}^{2}}\times\int_{k_{\mathrm{min}}^{+}}^{p_{0}^{+}}\frac{\mathrm{d}k^{+}}{k^{+}}\frac{(k^{+})^{2}}{(p_{0}^{+})^{2}}\Big[\big(1-\frac{2p_{0}^{+}}{k^{+}}\big)^{2}+(D-3)\Big]\\
 & =\mathcal{Z}+\alpha_{s}C_{F}\int_{\mathbf{k}}\frac{1}{\mathbf{k}^{2}}\times\Big[-3+4\mathrm{ln}\frac{p_{0}^{+}}{k_{\mathrm{min}}^{+}}\Big]\\
 & =\mathcal{Z}+\frac{\alpha_{s}C_{F}}{2\pi}\Big(\frac{1}{\epsilon_{\mathrm{UV}}}-\frac{1}{\epsilon_{\mathrm{IR}}}\Big)\Big(-\frac{3}{2}+2\mathrm{ln}\frac{p_{0}^{+}}{k_{\mathrm{min}}^{+}}\Big)+\mathcal{O}(\epsilon_{\mathrm{UV}})\;,
\end{aligned}
\end{equation}
and finally:
\begin{equation}
\begin{aligned}\mathcal{Z} & =\Bigg[1-\frac{\alpha_{s}C_{F}}{2\pi}\Big(\frac{1}{\epsilon_{\mathrm{UV}}}-\frac{1}{\epsilon_{\mathrm{IR}}}\Big)\Big(-3+2\mathrm{ln}\frac{p_{0}^{+}}{k_{\mathrm{min}}^{+}}+2\mathrm{ln}\frac{p_{1}^{+}}{k_{\mathrm{min}}^{+}}\Big)\Bigg]\;.\end{aligned}
\label{eq:fieldrenormalization-4}
\end{equation}

\bibliographystyle{JHEP.bst}

\end{document}